\RequirePackage{lineno}
\documentclass[twocolumn,floatfix,showpacs,aps,prd,nofootinbib]{revtex4}

\usepackage{amsmath}
\usepackage{amssymb}
\usepackage{graphicx}
\usepackage{dcolumn}
\usepackage{bm}
\usepackage{enumitem}

\usepackage{verbatim}
\usepackage{subfigure}

\newcommand{\BABARPubYear}    {14}
\newcommand{\BABARPubNumber}  {007}

\newcommand{\SLACPubNumber} {16361}
\newcommand{\LANLNumber} {wwwwwwww}

\newcommand{\beq}{\begin{linenomath}
\begin{equation}}
\newcommand{\eeq}{\end{equation}
\end{linenomath}}
\newcommand{\beqn}{\begin{linenomath}
\begin{eqnarray}}
\newcommand{\eeqn}{\end{eqnarray}
\end{linenomath}}
\newcommand{\beqns}{\begin{linenomath}
\begin{eqnarray*}}
\newcommand{\eeqns}{\end{eqnarray*}
\end{linenomath}}

\newcommand{\vsp}{\vspace{0.25cm}}

\newcommand{\mumu}{\ensuremath{\mu^+\mu^-}\xspace}
\newcommand{\mmg}{\ensuremath{\mu^+\mu^-\gamma}\xspace}
\newcommand{\pipi}{\ensuremath{\pi^+\pi^-}\xspace}
\newcommand{\pipig}{\ensuremath{\pi^+\pi^-\gamma}\xspace}

\newcommand{\xxg}{\ensuremath{x^+x^-\gamma}\xspace}

\def\mcdot{\kern-0.2em\cdot\kern-0.4em} 
\def\cedilla#1{\setbox0=\hbox{#1}\ifdim\ht0=1ex \accent'30 #1%
  \else{\ooalign{\hidewidth\char'30\hidewidth\crcr\unbox0}}\fi}
\input babarsym
\def\sbabar{\mbox{\slshape{\small B\kern-0.1em{\smaller A}\kern-0.1em
    B\kern-0.1em{\smaller A\kern-0.2em R}}}}
\def\vsbabar{\mbox{\slshape{\footnotesize B\kern-0.1em{\smaller A}\kern-0.1em
    B\kern-0.1em{\smaller A\kern-0.2em R}}}}

\graphicspath{{figures/}}

\begin{document}

\begin{flushleft}
\babar-PUB-\BABARPubYear/\BABARPubNumber\\
SLAC-PUB-\SLACPubNumber\\
arXiv:\LANLNumber\ [hep-ex]\\[10mm]
~ \\
\end{flushleft}

\title{{\boldmath Measurement of ISR-FSR interference in the processes $\epem\to\mmg$ and $\epem\to\pipig$ }}

%
\author{J.~P.~Lees}
\author{V.~Poireau}
\author{V.~Tisserand}
\affiliation{Laboratoire d'Annecy-le-Vieux de Physique des Particules (LAPP), Universit\'e de Savoie, CNRS/IN2P3,  F-74941 Annecy-Le-Vieux, France}
\author{E.~Grauges}
\affiliation{Universitat de Barcelona, Facultat de Fisica, Departament ECM, E-08028 Barcelona, Spain }
\author{A.~Palano$^{ab}$ }
\affiliation{INFN Sezione di Bari$^{a}$; Dipartimento di Fisica, Universit\`a di Bari$^{b}$, I-70126 Bari, Italy }
\author{G.~Eigen}
\author{B.~Stugu}
\affiliation{University of Bergen, Institute of Physics, N-5007 Bergen, Norway }
\author{D.~N.~Brown}
\author{L.~T.~Kerth}
\author{Yu.~G.~Kolomensky}
\author{M.~J.~Lee}
\author{G.~Lynch}
\affiliation{Lawrence Berkeley National Laboratory and University of California, Berkeley, California 94720, USA }
% ...
\collaboration{The \babar\ Collaboration}
\noaffiliation


\begin{abstract}
Charge asymmetry in processes $\epem\to\mmg$ and $\epem\to\pipig$ is
measured 
using  $232\invfb$ of data collected with the \babar\ detector at
$\epem$ center-of-mass energies near $10.58\gev$. An observable is introduced 
and shown to be very robust against detector asymmetries while keeping a 
large sensitivity to the physical charge asymmetry that results from the interference 
between initial and final state radiation. The asymmetry is determined as a
function of the invariant mass of the final-state tracks from production threshold to 
a few \gevcc. It is
compared to the expectation from QED for $\epem\to\mmg$, and from theoretical
models for $\epem\to\pipig$.
A clear interference pattern is observed  in $\epem\to\pipig$, particularly 
in the vicinity of the $f_2(1270)$ resonance. The inferred
rate of lowest order FSR production is consistent with the QED expectation for
$\epem\to\mmg$, and is negligibly small for $\epem\to\pipig$.
\end{abstract}


\maketitle



\section{Introduction}

The radiative processes
\beqn
\label{radiative} \epem\to X \gamma 
\eeqn 
have been extensively studied by several
$\epem$ experiments and the cross sections for $\epem\to X$ have been
measured using the initial state radiation (ISR)
method~\cite{isr1,isr2,isr3,isr4}.  At {\babar}~\cite{babarISR}, the
cross sections have thus been determined in large energy ranges below the
total $\epem$ center-of-mass (c.m.) energy $\sqrt{s}\sim 10.58\gev$
available at the SLAC PEP-II collider. The state $X$ can be either
fully described by Quantum Electrodynamics (QED) such as $\mumu$, or any hadronic state 
with $J^{PC}=1^{--}$. 

In reaction~(\ref{radiative}) at lowest order (LO) the photon can be emitted
from either the incoming electron or positron, or from the final state
(final state radiation, or FSR). At {\babar}, the kinematic conditions are such that the
process is dominated by ISR photons, which justifies the ISR
method.  The LO FSR contribution to the hadronic radiative process is
neglected, as its theoretical estimates are well below the systematic
uncertainties of the cross section measurement. This is due to the fact that 
the available $\epem$ c.m. energy is
far beyond the domain of the hadronic resonances that dominate the
cross-section, so that hadronic form factors considerably reduce the probability
that the photon is emitted from the final state. However, the theoretical
estimations are model-dependent, and it is thus important to have a direct
experimental proof of the smallness of the FSR contribution to the hadronic
cross sections when high precision is at stake, as for the determination of
the hadronic contribution to the $g-2$ value of the muon~\cite{prd-pipi}. Because of the
point-like nature of the muon, the FSR reduction does not occur for the
$\epem\to\mmg$ process. The LO FSR contribution to the cross section is expected
to vanish at threshold and to increase with the invariant mass of the muon pair
($m_{\mu\mu}$). Still, the FSR fraction remains small for low di-muon mass
(less than 1\% for $m_{\mu\mu}<1\gevcc$). For the $\epem\to\mmg$ cross section measurement, a
correction is applied for the LO FSR contribution as a function
of $m_{\mu\mu}$, which is so far determined by turning off FSR in the Monte Carlo (MC)
generation.

While it is not possible to distinguish ISR from FSR photons on an
event-by-event basis, as the corresponding amplitudes are both present and
interfere, a measurement of the interference provides a sensitive and
quantitative determination of their relative strength.  
Measurement of the forward-backward asymmetry of the pions was first proposed  
in Ref.~\cite{isr3}, as a test of the underlying model for final state radiation.
In this paper, the
ISR-FSR interference for $\epem\to\mmg$ and $\epem\to\pipig$ is studied through
the charge asymmetry of the production of these events at various decay plane
angles. The comparison between the QED prediction and the measurement is done
for the charge asymmetry in $\epem\to\mmg$. 
Various FSR models are discussed
for $\epem\to\pipig$, and the most realistic quark-FSR model is compared to the
measurement of the charge asymmetry in that channel. 

This paper reports the first measurement of charge asymmetry in the $\epem\to\mmg$ 
process. For $\epem\to\pipig$, a preliminary measurement~\cite{KLOE}
of the forward-backward asymmetry has been reported
at low energies ($\sqrt{s}\sim 1\gev$). No previous result exists at high energies.

\section{ISR-FSR interference and charge asymmetry}

\subsection{Charge asymmetry}

\begin{figure*}[h!]
\centering 
\includegraphics[width=0.32\textwidth]{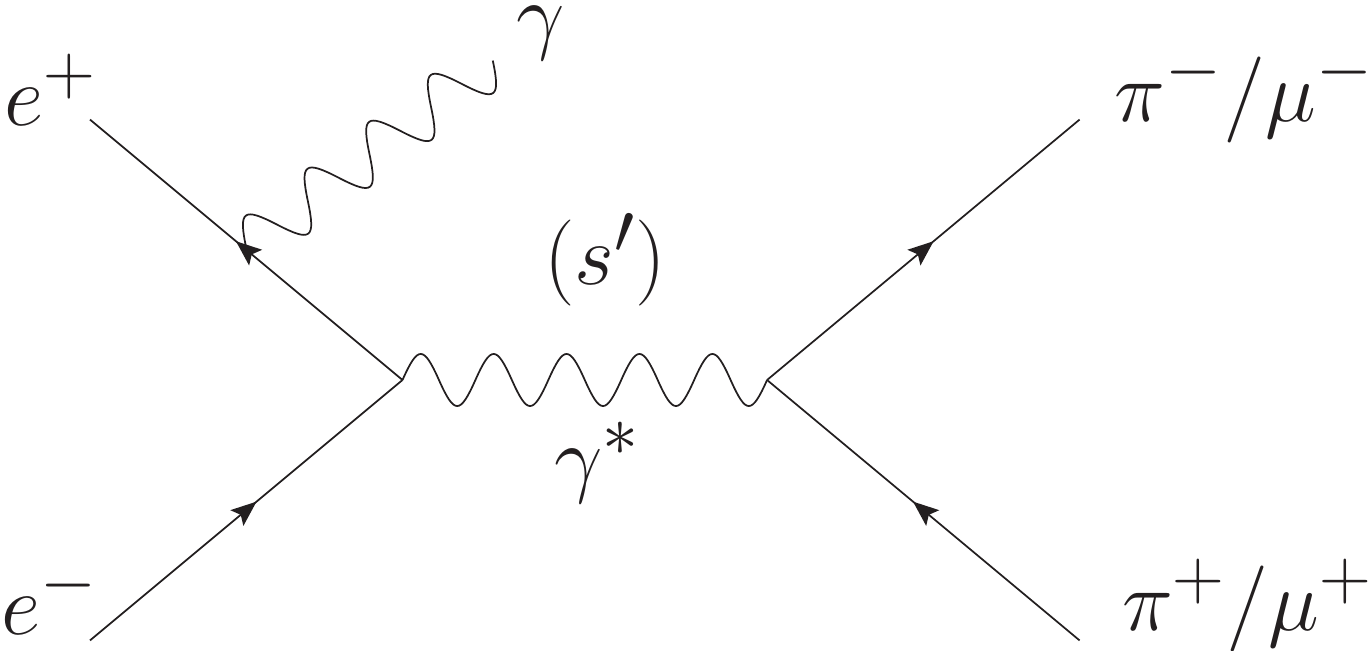} \hspace{10mm}
\includegraphics[width=0.32\textwidth]{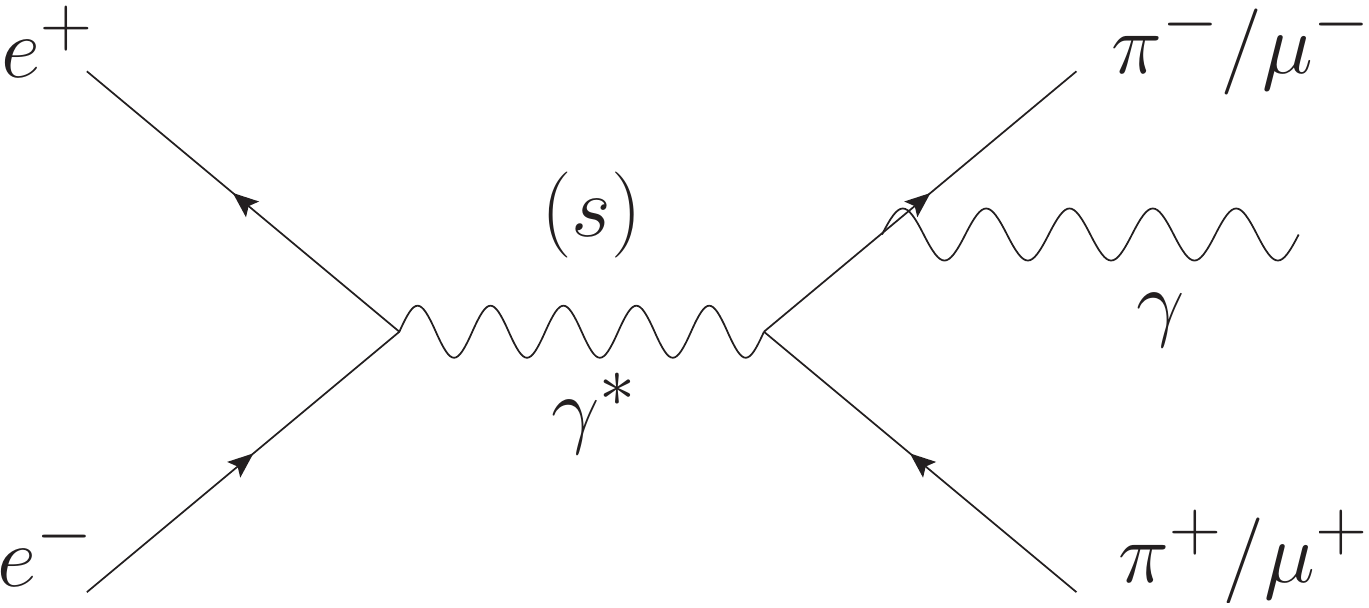}
\caption{Feynman diagrams for $\epem\to x^+x^-\gamma$~($x=\mu$, $\pi$), where the photon is
from lowest order initial state radiation (LO ISR, left) or lowest order final
state radiation (LO FSR, right).}
\label{fig:Feynman-gmm}
\end{figure*}

The Feynman diagrams for the LO ISR and LO FSR emission in the process
$\epem\to x^+x^-\gamma$~(where $x=\mu$ or $\pi$), are illustrated in
Fig.~\ref{fig:Feynman-gmm}.  The total LO amplitude $\mathcal{M}$ is
the sum of the corresponding amplitudes $\mathcal{M}_{\rm ISR}$ and
$\mathcal{M}_{\rm FSR}$, and the cross section for $\epem\to
x^+x^-\gamma$ is 
\beqn 
\label{eq:2}
\sigma\propto|\mathcal{M}|^2=|\mathcal{M}_{\rm ISR}|^2+|\mathcal{M}_{\rm FSR}|^2 
+2{\mathcal Re}(\mathcal{M}_{\rm ISR}\mathcal{M}_{\rm FSR}^*). \nonumber\\ 
\eeqn

If the photon is emitted from the initial~(final) state, the $x^+x^-$ pair is
produced with charge parity $C=-1~(+1)$, which implies that the interference term
changes sign if one interchanges $x^+$ and $x^-$. While the contribution of
the interference term to the total cross section vanishes when one integrates
over the kinematic variables of the final state, that term induces a
significant observable charge asymmetry in the differential cross section.

Charge asymmetry is defined as
\beqn
\nonumber  \mathcal{A} &=& \frac{|\mathcal{M}|^2-|\mathcal{M}_{x^+\leftrightarrow
             x^-}|^2}{|\mathcal{M}|^2+|\mathcal{M}_{x^+\leftrightarrow x^-}|^2} \\
             &=& \frac{2{\mathcal Re}(\mathcal{M}_{\rm ISR}\mathcal{M}_{\rm FSR}^*)}
                      {|\mathcal{M}_{\rm ISR}|^2+|\mathcal{M}_{\rm FSR}|^2},
\eeqn
where $x^+\leftrightarrow x^-$ means that $x^+$ and
$x^-$ are interchanged.

Although it is not possible to reconstruct $\mathcal{M}_{\rm ISR}$ or
$\mathcal{M}_{\rm FSR}$ from the charge asymmetry and the cross section, as the
relative phase between them remains unknown, information on the 
ratio $|\mathcal{M}_{\rm FSR} / \mathcal{M}_{\rm ISR}|$ can be
derived within the framework of specific models.

\subsection{Choice of kinematic variables}
\label{sec:4var}

Aside from an overall azimuthal rotation about the beam axis, the kinematic
topology of the $\xxg$ final state (where $x=\mu$ or $\pi$) is described by 
four variables, which are the muon-pair (pion-pair) invariant mass $m_{xx}$ (or
equivalently $E^*_{\gamma}$, the energy of the radiated photon in the $\epem$
c.m.) and three angular variables. At a given $m_{xx}$ mass, the
distribution of the three angular variables contains all the available information
on the ISR/FSR amplitudes.

\begin{figure*}
\centering
\subfigure{\includegraphics[width=0.42\textwidth]{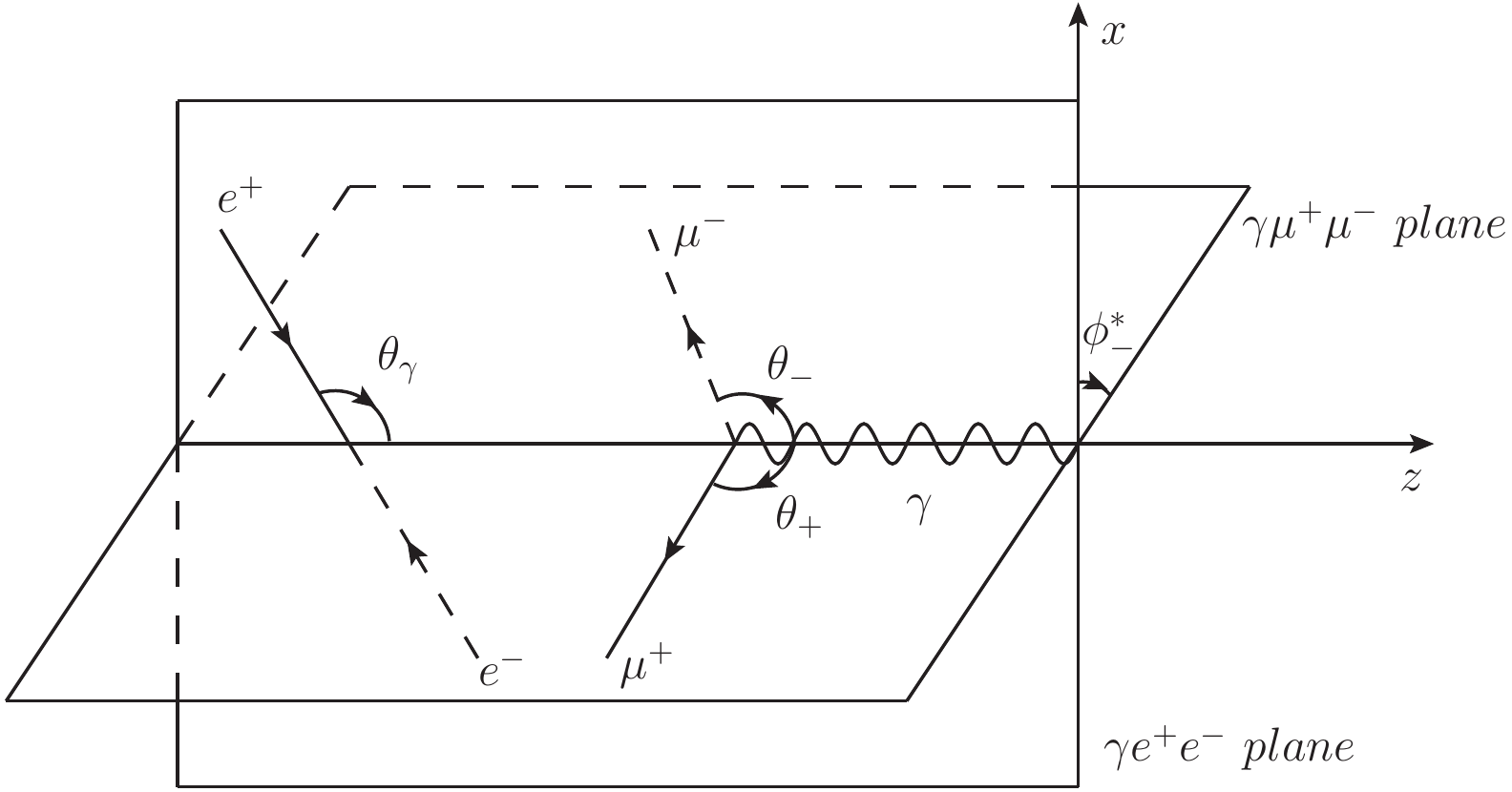}} \hspace{5mm}
\subfigure{\includegraphics[width=0.42\textwidth]{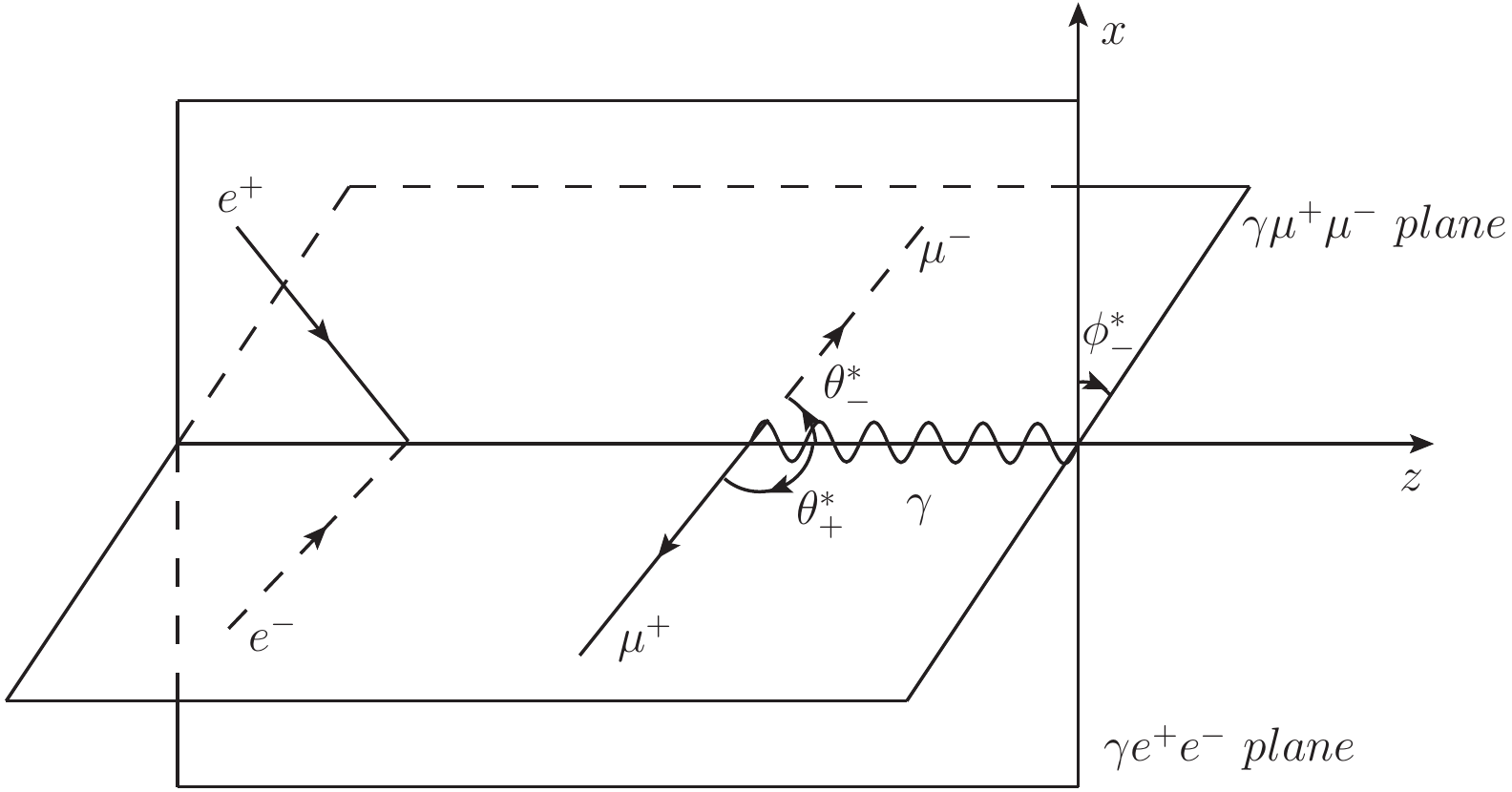}}
\caption{Definition of the angular variables describing the
kinematic topology of the final states of the process $\epem\to\xxg$ $(x=\mu, \pi)$
at a given $x^+x^-$ invariant mass: (left) in the $\epem$
c.m.s., (right) in the $x^+x^-$ c.m.s.}
\label{fig:phisDefinition}
\end{figure*}

At variance with the definition of forward-backward asymmetry used in Ref.~\cite{czyz}, 
which refers to the polar angle of $x^-$ with respect to the incoming electron
in the $\epem$ c.m. system (c.m.s.),
this analysis introduces the set of angular variables illustrated in
Fig.~\ref{fig:phisDefinition}. These are found to be more sensitive
observables to measure the ISR-FSR interference:
\begin{itemize}
\item $\theta^*_{\gamma}$ --- polar angle of the radiated photon in the $\epem$
c.m.s. (with respect to the $\epem$ axis),
\item $\theta^*$        --- polar angle of $x^-$ with respect to the photon axis in the
$x^+x^-$ c.m.s.,
\item $\phi^*$          --- azimuthal angle of $x^-$ with respect to the $\gamma e^+e^-$ plane 
in the $x^+x^-$ c.m.s. (or the $\epem$ c.m.s.)
\end{itemize}

Since $x^+\leftrightarrow x^-$ interchange means reversal of the $x^-$ direction to 
its opposite in the $x^+x^-$ c.m.s. system,
the charge asymmetry, for fixed $m_{xx}$ and $\theta^*_{\gamma}$, is equal to
\beqn\label{eq_A}
\mathcal{A}(\theta^*,\phi^*)=\frac
{\sigma(\theta^*,\phi^*)-\sigma(\pi-\theta^*,\pi+\phi^*)}
{\sigma(\theta^*,\phi^*)+\sigma(\pi-\theta^*,\pi+\phi^*)}.
\eeqn

\begin{figure*}
\centering 
\includegraphics[width=0.3\textwidth]{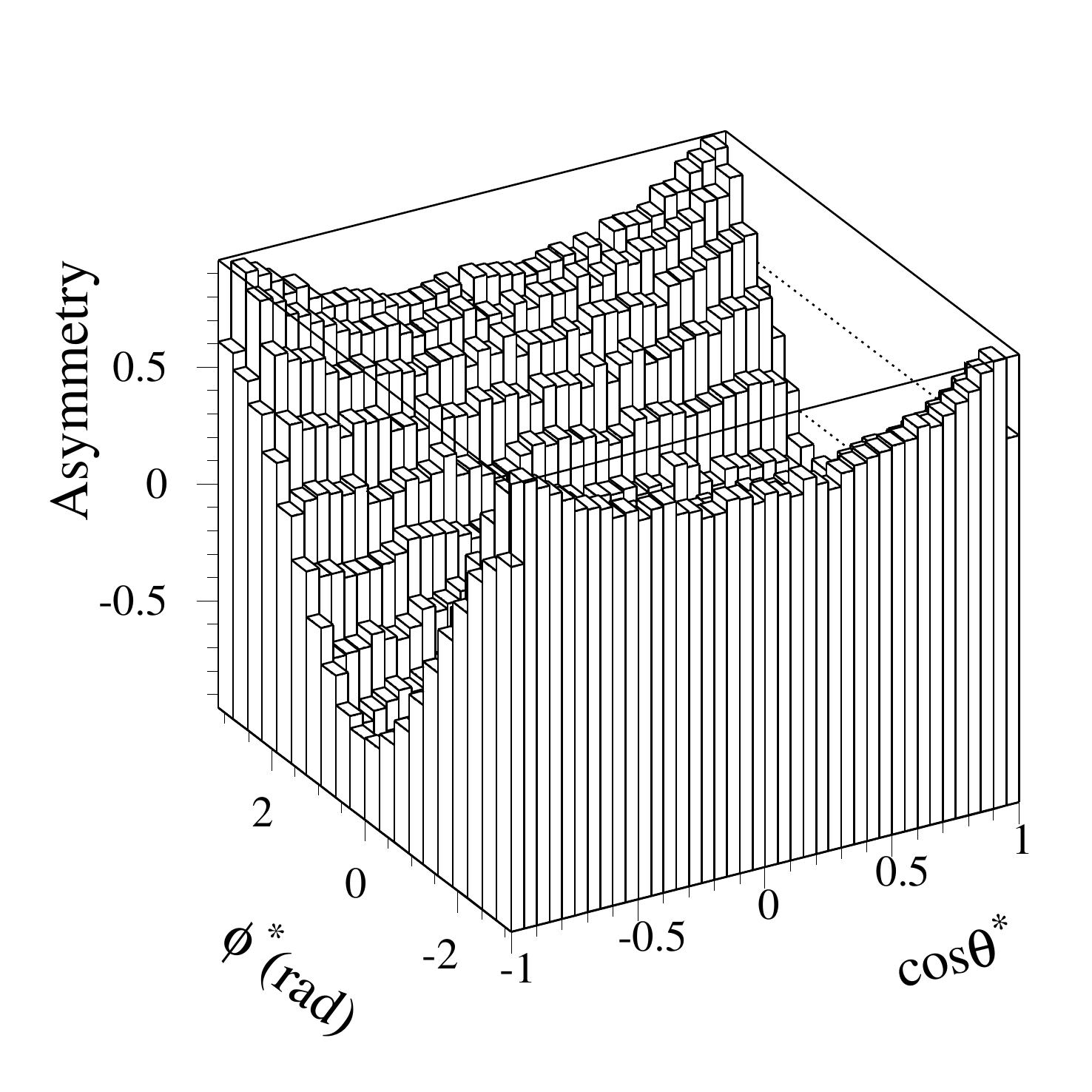}
\includegraphics[width=0.3\textwidth]{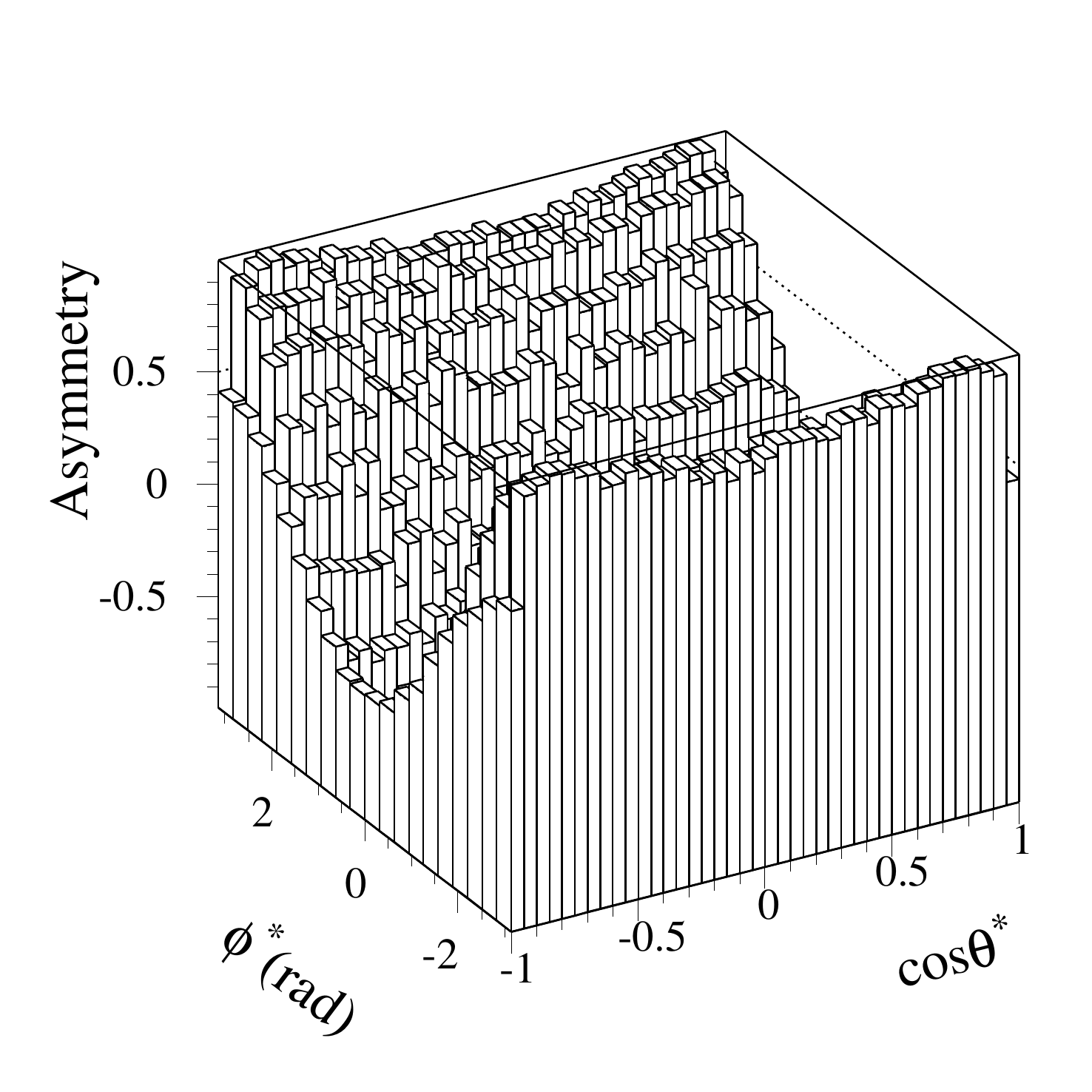}
\includegraphics[width=0.3\textwidth]{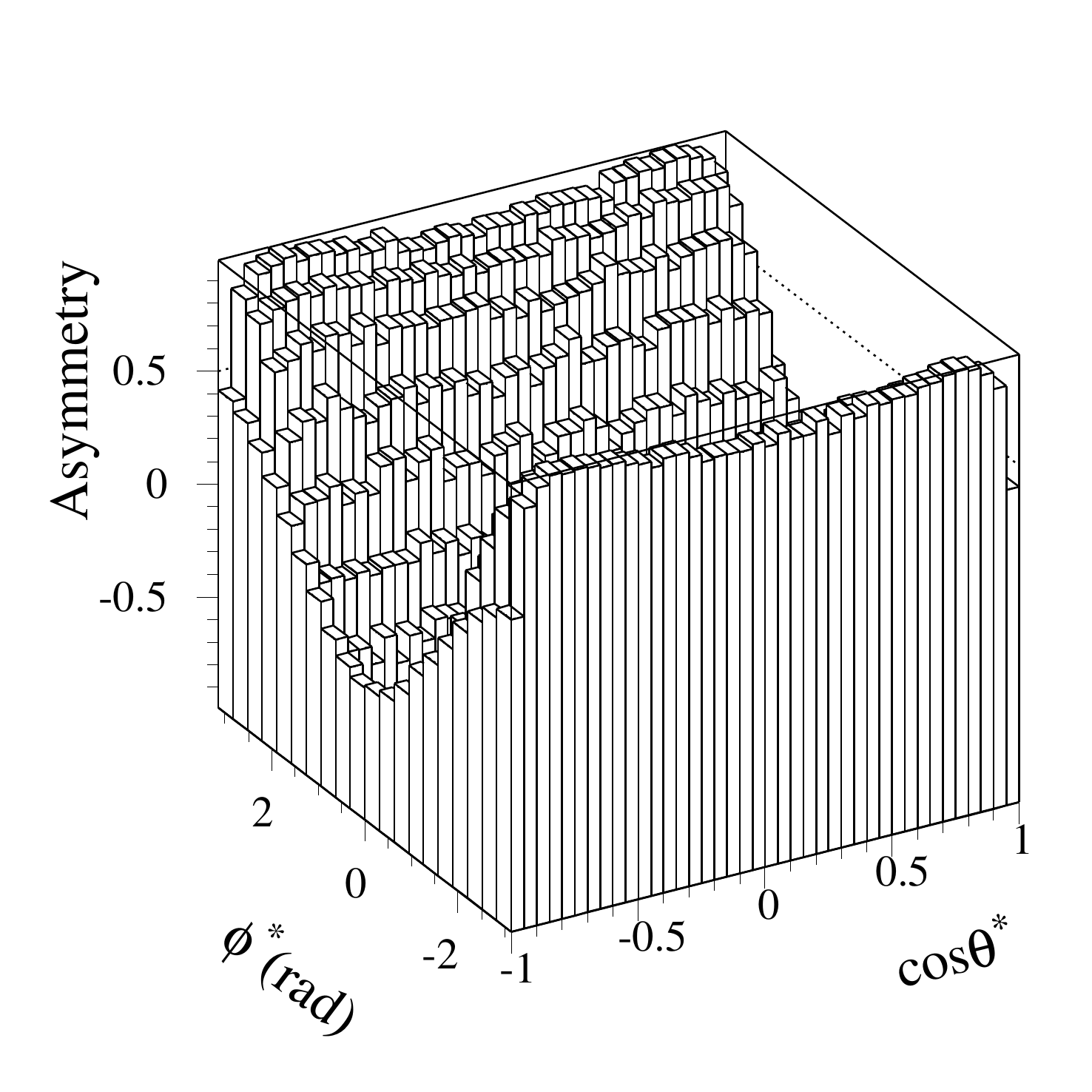}
\caption{Charge asymmetry at generator level in $\epem\to\mmg$ simulation, 
as a function of $\cos \theta^*$ and $\phi^*$ for the
same $m_{\mu\mu}$ interval ($6.5<m_{\mu\mu}<7.0\gevcc$) and various
$\cos\theta^*_{\gamma}$ ranges: 
(left) $-1< \cos\theta^*_{\gamma}<-0.6$,
(middle) $-0.6 < \cos\theta^*_{\gamma} < -0.4$,
(right) $-0.4 < \cos\theta^*_{\gamma} < 0.$}
\label{fig:A-cosths-phis-gmm-1}
\end{figure*}
For the $\epem\to\mmg$ process, the charge asymmetry as a function of
$\cos\theta^*$ and $\phi^*$, studied with the AfkQed generator (see Sec.~\ref{sec:MC}), is shown in
Fig.~\ref{fig:A-cosths-phis-gmm-1}.
The FSR amplitude is dominant at $|\cos\theta^*|\sim1$, when one of
the charged-particle tracks is very close to the radiated
photon. However, Fig.~\ref{fig:A-cosths-phis-gmm-1}
shows that $\phi^*$ is a more sensitive variable to measure the
ISR/FSR content over the full phase space, with sign reversal of the
charge asymmetry. After
integration over $\cos\theta^*_{\gamma}$ and integration over
symmetrical $\cos\theta^*$ intervals, the distribution of the
integrated charge asymmetry $A(\cos\phi^*)$ suggests a simple linear
dependence 
\beqn
\label{eq:A_linear} A(\cos\phi^*)=A_0\cos\phi^*.
\eeqn 
From the expressions of the differential cross section detailed in the next section, 
it results that the slope $A_0$ is an estimator of the ISR-FSR
interference, sensitive to the ratio $|\mathcal{M}_{\rm
FSR} / \mathcal{M}_{\rm ISR}|$ in each $m_{\mu\mu}$
interval. Moreover, it will be shown in Sec.~\ref{sec:effectAcc} 
that the measurement of $A_0$ is barely affected by detector charge asymmetries.

\section{Theoretical predictions for the charge asymmetry}

\subsection{\boldmath QED prediction for the $\epem\to\mmg$ process}

In the massless limit~\cite{GW}, the differential cross section of the QED $\epem\to\mmg$
process, written as a function of the four
kinematic variables defined above (Sec.~\ref{sec:4var}), implies that the differential
charge asymmetry is proportional to $\cos\phi^*$:
\begin{widetext}
\beqn
\label{eq:A_qed_massless}
    \mathcal{A}_{\epem\to\mmg}(m_{\mu\mu},\theta^*_\gamma,\theta^*,\phi^*) =
    -\frac{2\sqrt{s}\,m_{\mu\mu}\sin\theta^*_\gamma\sin\theta^*\cos\phi^*}
    {s\,\sin^2\theta^*+m^2_{\mu\mu}\sin^2\theta^*_\gamma}.
\eeqn
\end{widetext}

\begin{figure*}[t]
  \includegraphics[width=0.45\textwidth]{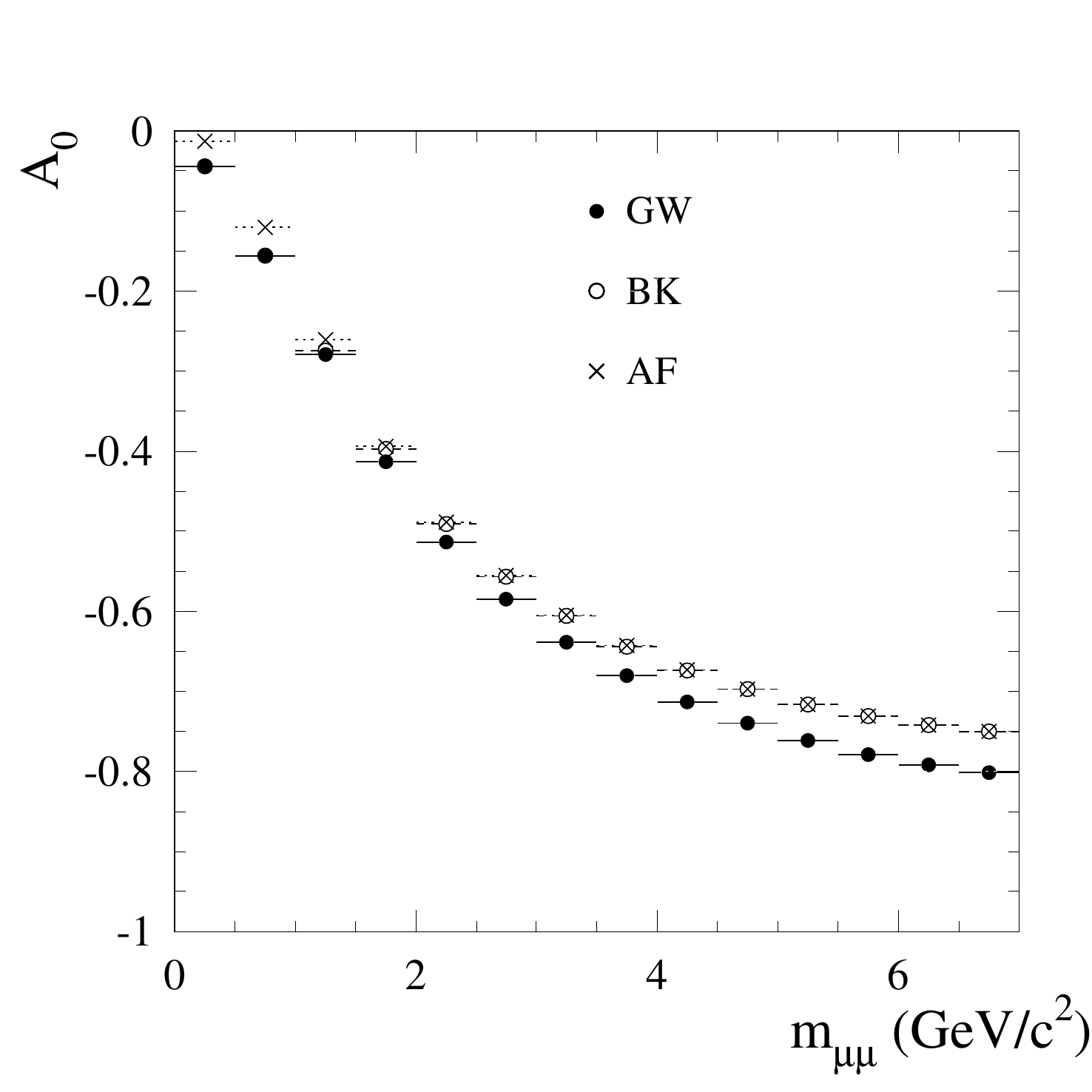}
  \includegraphics[width=0.45\textwidth]{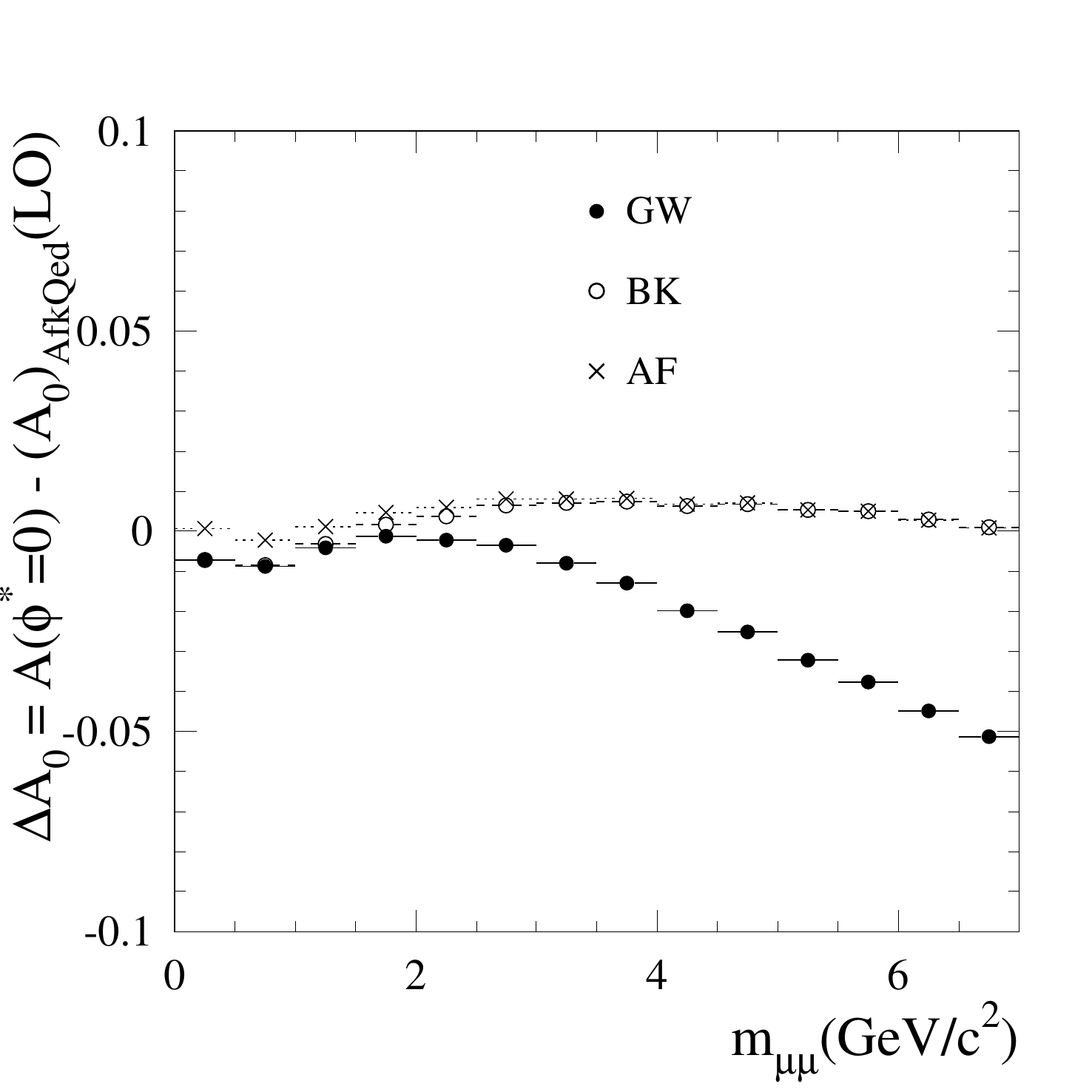}
  \caption{(left) Charge asymmetry at $\phi^*=0$, $A_0$, as a function of
  $m_{\mu\mu}$, obtained by numerical integration according to three
  different theoretical predictions (see text), with the condition
  $20^\circ<\theta^*_\gamma<160^\circ$ applied. (right) The difference between the 
  prediction and the AfkQed LO value. Results labeled GW, BK, AF are obtained from
  references~\protect\cite{GW},\protect\cite{BK},\protect\cite{AF}, respectively.}
  \label{fig:AphiZNum}
\end{figure*}

When the masses are taken into account, the effect from the
electron/positron mass is found to be negligible for radiated photons away from the
beams. The effect from the muon mass is sizeable, especially at large
$m_{\mu\mu}$ when the radiated photon is close to one of the muons.
Predictions for the charge asymmetry in the massive case are obtained by
numerical integration of several variants of the QED differential cross section
~\cite{GW,BK,AF}. The phase space considered in those calculations is limited to
the experimental acceptance $20^\circ<\theta^*_\gamma<160^\circ$, and the results
are shown in Fig.~\ref{fig:AphiZNum} as a function of  $m_{\mu\mu}$.  Predictions
differ at the physical threshold ($m_{\mu\mu}=2m_\mu$), where only the charge asymmetry
based on Ref.~\cite{AF} extrapolates to
zero as expected, suggesting that the validity of formulae in Ref.~\cite{GW,BK} 
does not extend to small $m_{\mu\mu}$. At large mass ($m_{\mu\mu}>3\gevcc$), the 
prediction from Ref.~\cite{GW} differs from the others by up to a few percent.  The
formula of the differential LO cross section implemented in the AfkQed
generator, which is used in this analysis for simulation (see Sec.~\ref{sec:MC}), is the one
by Arbuzov {\it et al.}~\cite{AF}, which has the most reliable behavior over
the full $m_{\mu\mu}$ range.

\subsection{\boldmath FSR models for the $\epem\to\pipig$ process}

\begin{figure*}
\centering 
\subfigure[Initial state radiation]
  {\includegraphics[width=0.28\textwidth]{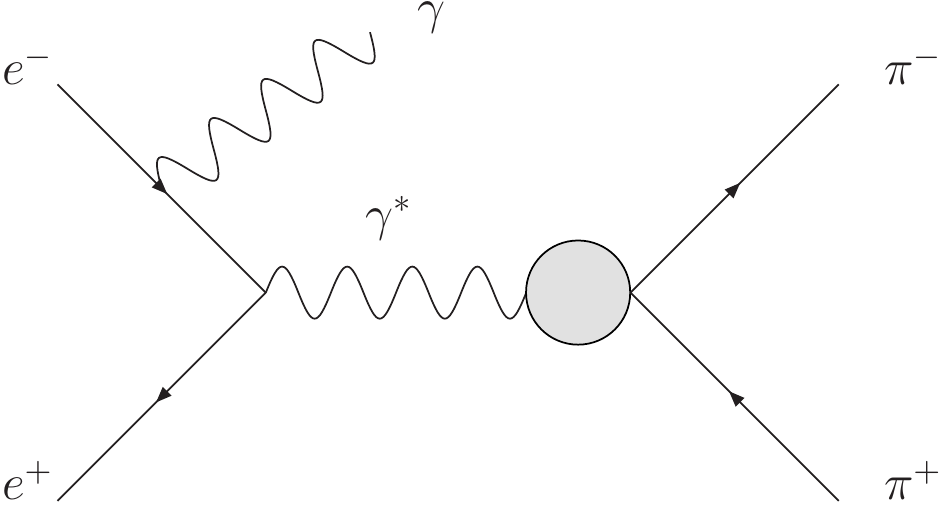}} \hspace{2mm}
\subfigure[Final state radiation with pions treated as point-like particles
(FSR model 1)]
  {\includegraphics[width=0.28\textwidth]{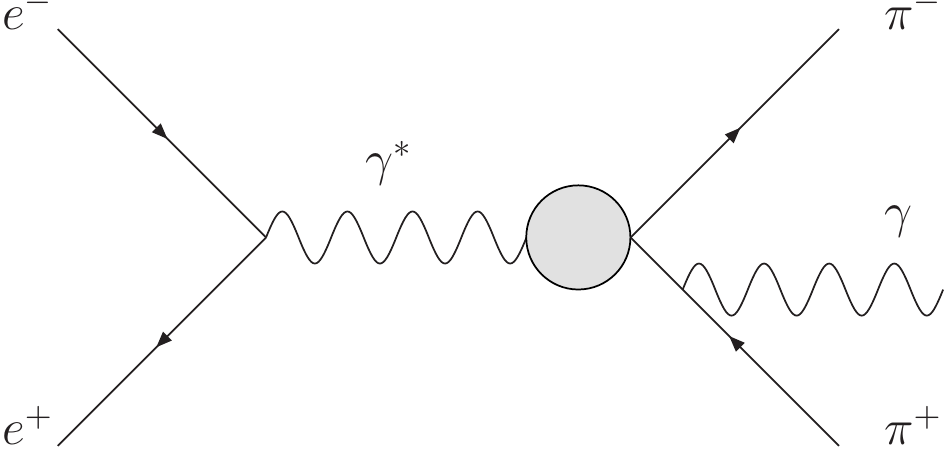}} \hspace{2mm}
\subfigure[Final state radiation at quark level 
(FSR model 2)]
  {\includegraphics[width=0.29\textwidth]{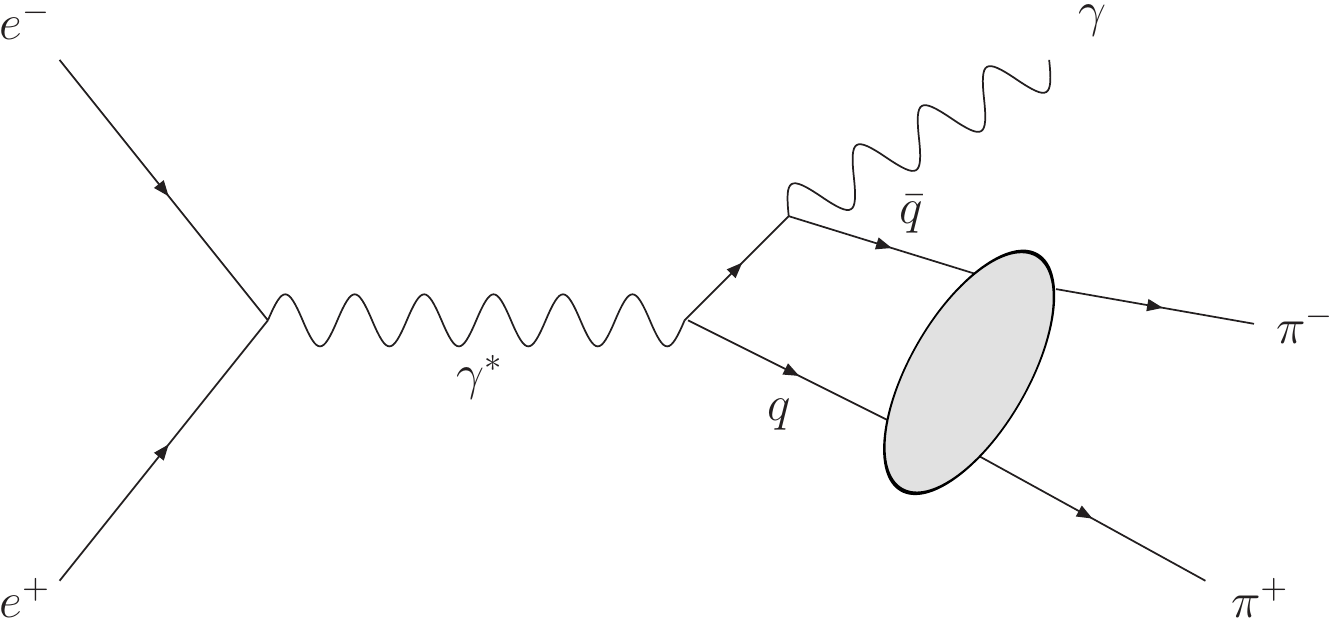}} 
\caption{Feynman diagrams for $\epem\to\pipig$.}
\label{fig:Feynman-gpipi}
\end{figure*}

As in the $\epem\to\mmg$ process,  ISR and FSR contribute to
$\epem\to\pipig$ (Fig.~\ref{fig:Feynman-gpipi}).  However, the charge
asymmetry is expected to be much smaller in the latter process because the
FSR contribution is strongly reduced by the pion form factor at large
$\sqrt{s}$. In addition, its estimate is model-dependent.

\subsubsection{FSR from point-like pions (model 1)}
\label{sec:FSRmd1}

In the FSR model shown in Fig.~\ref{fig:Feynman-gpipi}(b), the photon is emitted
from one of the final state pions, where the pion is treated as a point-like
particle.  In this hypothesis, the FSR amplitude $\mathcal{M}_{\rm FSR}$ is
proportional to the pion form factor at the collision energy squared $s$, namely
$F_\pi(s)$.  The ISR amplitude $\mathcal{M}_{\rm ISR}$ shown in
Fig.~\ref{fig:Feynman-gpipi}(a) is proportional to the pion form factor
$F_{\pi}(s')$ at a reduced energy squared $s'=s(1-2E^*_\gamma/\sqrt{s})$.
According to this FSR model, the charge asymmetry to be measured at {\babar}
reflects the relative magnitude of the pion form factor at
$\sqrt{s}=10.58\gev$ and at low energy. It is consequently negligibly small, since
$F_{\pi}(s')$, dominated by the $\rho$ resonance in the $s'=m^2_{\pi\pi}$ domain
accessible to the experiment, is three orders of magnitude larger than
$|F_{\pi}(10.58^2\gev^2)|\sim 0.01$, as estimated from an extrapolation of
existing  data~\cite{cleo-ff,prd-pipi} using a $1/s$ dependence. 
This model is studied with the {\small PHOKHARA}~4.0~\cite{phok} generator, in which
the FSR current has a point-like Lorentz structure, including a contact term, 
globally multiplied by the pion form factor.
In this model, the $A_0(m_{\pi\pi})$ distribution is
expected to increase quadratically with mass on the $\rho$ resonance, 
with a change of sign at the $\rho$ mass
\beqn\label{eq:A0-M-gpipi}
A_0\sim2\times 10^{-3}(m_{\pi\pi}^2-m_{\rho}^2),
\eeqn
with values well below the sensitivity of this analysis because of the large
pion form factor suppression at $10.58\gev$.
 
\subsubsection{FSR from quarks (model 2)}
\label{sec:FSRmd2}

In the {\it a priori} more realistic FSR model for $\epem\to\pipig$ depicted in 
Fig.~\ref{fig:Feynman-gpipi}(c), 
the FSR photon is emitted from the quarks,
which subsequently hadronize into a pion pair~\cite{piInt}. The
dominant ISR and FSR contributions, and their interference, are written in terms 
of the variables defined in Sec.~\ref{sec:4var}:
\begin{widetext}
\beqn 
\label{eq:sigma_pi_ISR}
 \frac{d\sigma^{\rm ISR}_{\epem\to\pipig}}{dm^2_{\pi\pi} \,d\!\cos\theta^*_\gamma \,d\!\cos\theta^* \,d\phi^*}
 &=& \frac{\alpha^3\beta^3}{16\pi
 s^2m^2_{\pi\pi}(s-m^2_{\pi\pi})}|F_\pi(m^2_{\pi\pi})|^2 \nonumber\\
 &\times&\bigg\{(s^2+m^4_{\pi\pi})\frac{1+\cos^2\theta^*_\gamma}{\sin^2\theta^*_\gamma}\sin^2\theta^*
 + 4sm^2_{\pi\pi}\cos^2\theta^* \nonumber\\
 & & +2\sqrt{s}m_{\pi\pi}(s+m^2_{\pi\pi})(\tan\theta^*_\gamma)^{-1}\sin2\theta^*\cos\phi^*
 -2sm^2_{\pi\pi}\sin^2\theta^*\cos2\phi^* \bigg\},
\eeqn 
where $\alpha$ and $\beta$ are the QED fine structure constant and the pion velocity 
$\beta=\sqrt{1-4m_\pi^2/m^2_{\pi\pi}}$, respectively.
The FSR contribution is
\beqn 
\label{eq:sigma_pi_FSR}
 \frac{d\sigma^{\rm FSR}_{\epem\to\pipig}}{dm^2_{\pi\pi} \,d\!\cos\theta^*_\gamma \,d\!\cos\theta^* \,d\phi^*}
 &=&
 \frac{\alpha^3\beta(s-m^2_{\pi\pi})}{64\pi s^3}(1+\cos^2\theta^*_\gamma)|V(m^2_{\pi\pi},\theta^*)|^2,
\eeqn 
and the interference term
\beqn 
\label{eq:sigma_pi_I}
 \frac{d\sigma^I_{\epem\to\pipig}}{dm^2_{\pi\pi} \,d\!\cos\theta^*_\gamma \,d\!\cos\theta^* \,d\phi^*}
      &=& \frac{\alpha^3\beta^2}{16\pi
      s^2\sqrt{s}m_{\pi\pi}}\textrm{Re}\{F_\pi^*(m^2_{\pi\pi})V(m^2_{\pi\pi},\theta^*)\} \nonumber\\
      &\times&\bigg\{-\sqrt{s}m_{\pi\pi}\cos\theta^*_\gamma\cos\theta^*+
      [(1+\cos^2\theta^*_\gamma)s+m^2_{\pi\pi}\sin^2\theta^*_\gamma]
      \frac{\sin\theta^*\cos\phi^*}{2\sin\theta^*_\gamma}
      \bigg\},
\eeqn 
where 
\beqn 
  V      =& \sum_q e^2_qV_q
        &=  \sum_q e^2_q
          \int_0^1dz\frac{2z-1}{z(1-z)}\Phi_q^+(z,m^2_{\pi\pi},\cos\theta^*) ~~~ (q=u,d),
\eeqn
\end{widetext}
and $\Phi_q^+(z,m^2_{\pi\pi},\cos\theta^*)$ is the C-even part of the 2-pion generalized 
distribution amplitudes (GDA). The pion time-like form factor
$F_\pi(m^2_{\pi\pi})$ is taken from a fit to {\babar } data~\cite{prd-pipi} with
a vector dominance model.

So far, there is no implementation of this model in an MC generator to describe the ISR-FSR
interference in the $\epem\to\pipig$ process.  In order to predict the charge
asymmetry numerically, we take the following GDA model, which is a modified
version of the model found in Ref.~\cite{piDGP}:
\begin{widetext}
\beqn 
\label{eq:PhiqToFit}
\Phi_u^+(z,m^2_{\pi\pi},\cos\theta^*)&=&\Phi_d^+(z,m^2_{\pi\pi},\cos\theta^*) \nonumber\\
                            &=&10z(1-z)(2z-1)\left[
                             c_{0}\frac{3-\beta^2}{2}e^{i\delta_0(m_{\pi\pi})}
                            +c_{2}\beta^2{\rm BW}(m_{\pi\pi})P_2(\cos\theta^*)\right],
\eeqn 
\end{widetext}
where $c_0$  and $c_2$ are the magnitudes of the S-wave and D-wave contributions, 
respectively. As the scalar 
sector is known to involve wide resonances, the S-wave contribution is approximated 
by a constant amplitude with a 
mass-dependent phase $\delta_0(m_{\pi\pi})$ taken from pion-pion phase-shift 
analyses~\cite{deltapipi} in the region below $1.6\gevcc$. This model 
incorporates the rapid phase variation across the $f_0(980)$ resonance. 
Using $c_0=-0.5$~\cite{piDGP} yields an $A_0$ value of about $-1\%$ near the $\rho$ 
resonance and nearly flat with mass.
For the D-wave tensor contribution, we use a Breit-Wigner 
form (BW) for the $f_2(1270)$ resonance in order to take properly into account the mass 
dependence of the amplitude, the phase variation being given by the 
BW form in agreement with the measured $\delta_2(m_{\pi\pi})$ 
values~\cite{deltapipi}. The angular dependence in the $\pi\pi$
center-of-mass is given by the Legendre polynomial $P_2(\cos\theta^*)$, which
assumes the dominance of helicity 0 for $f_2(1270)$ production.

\subsection{Other sources of charge asymmetry}
\label{sec:NLO}

Next-to-leading order (NLO) corrections including additional photons (soft and
hard) and loops are expected to affect the lowest-order (LO) predictions for 
the charge asymmetry. For the $\mu^+ \mu^- \gamma$ process these corrections 
have been computed recently~\cite{NLO-muons} and implemented in the 
{\small PHOKHARA} 9.0 generator~\cite{phok-2013}. As discussed in 
Sec.~\ref{sec:syst_mu}, the effects are found to be small, at the percent 
level for the experimental conditions of the present analysis, and to be
well accounted for by the simpler structure function approach implemented
in AfkQed. 
No exact NLO calculation is available for the $\pi^+ \pi^- \gamma$ process. 
In this case, since the LO charge asymmetry is expected to be small because 
the FSR amplitude is suppressed, NLO corrections could play a 
relatively more important role. The soft and virtual photon
contributions to the Born process $e^+ e^- \to \pi^+ \pi^-$ are
known~\cite{brown-mikaelian,arbuzov-pipi} to generate an asymmetry of the
pion production, with asymmetry values at the percent level at a $\pi\pi$ mass of $1\gevcc$. 
However, it is unclear if the above result can be used in the conditions of the 
present process $e^+ e^- \to \pi^+ \pi^- \gamma$, where one of the incoming 
electrons is highly off-shell after emission of a hard ISR photon. Furthermore, 
such an asymmetry would vanish because of the symmetrical integration in $\cos\theta^*$.
NLO corrections as implemented in AfkQed have indeed no effect on the charge asymmetry. 
No correction on the measured charge asymmetry is therefore applied for
the $\pi^+ \pi^- \gamma$ process.

Another potential source of charge asymmetry comes from $Z$
exchange. This contribution is strongly suppressed by the $Z$
propagator, especially for the ISR diagrams where $m_{xx}^2/M_Z^2
\sim 10^{-4}$. Therefore one expects this effect to be negligible for
$\pi^+ \pi^- \gamma$. The contribution is larger for the FSR
diagrams for $\mu^+ \mu^- \gamma$ since here the relevant
ratio is $s/M_Z^2 = 1.4\%$. The contribution of $Z$ exchange is studied with the 
{\small KKMC} generator~\cite{kk2f}. As reported in Sec.~\ref{sec:syst_mu}, the
effect is at the level of a few per mille.

\section{Experimental analysis}

\subsection{The {\babar} detector and data samples}

The analysis is based on $232\invfb$ of data~\cite{babar-lumi} collected with the {\babar}
detector at the SLAC National Accelerator Laboratory at the \pep2
asymmetric-energy $\epem$ collider operated at the $\FourS$ resonance. About 10\% 
of the data was collected 40\mev below the resonance. The
{\babar} detector is described in detail elsewhere~\cite{detector}.
Charged-particle tracks are measured with a five-layer double-sided silicon
vertex tracker (SVT) together with a 40-layer drift chamber (DCH), both inside a
1.5~T superconducting solenoid. Photons are assumed to originate from the
primary vertex defined by the charged-particle tracks of the event, and their energy and
position are measured in a CsI(Tl) electromagnetic calorimeter (EMC).
Charged-particle identification (PID) uses the ionization energy loss $\dedx$ in
the SVT and DCH, the Cherenkov radiation detected in a ring-imaging device
(DIRC), the shower energy deposit ($E_{\rm cal}$) in the EMC, and the shower shape
in the instrumented flux return (IFR) of the magnet. The IFR system is made of
modules of resistive plate chambers (RPC) interspaced with iron slabs, arranged
in a layout with a barrel and two endcaps. Collision events are recorded and
reconstructed if they  pass three levels of trigger (hardware, online
software, and offline  filter), each using complementary information
from the sub-detectors.

\subsection{Monte Carlo generators and simulation}
\label{sec:MC}

Signal and background processes $\epem \to X\gamma$ are simulated with  the
AfkQed event generator, which is based on QED for $\epem \to \mmg$ 
and Ref.~\cite{eva} for hadronic production.  LO ISR and FSR
emission is simulated for $\epem \to \mmg$, while LO FSR is neglected for
hadronic processes. The main photon (hereafter called `ISR' photon) is emitted within
the angular range $20^\circ < \theta^*_{\gamma} < 160^\circ$ in the $\epem$
c.m. system, bracketing the photon detection range with a margin for resolution .
Additional ISR photons are generated with the structure function
method~\cite{struct-fct}, and additional FSR photons with the {\small
PHOTOS}~\cite{photos} program. Additional ISR photons are emitted along the
$e^+$ or $e^-$ beam particle direction. A minimum mass $m_{X\gamma_{\rm
ISR}}>8\gevcc$ is imposed at generation, which puts an upper bound on the
additional ISR photon energy. Samples corresponding to 5 to 10 times the data
are generated for the signal $\epem \to \mmg$ and $\epem \to \pipig$ channels,
as well as large samples of  backgrounds from the other two-prong and
multi-hadron ISR processes.
Background processes $\epem\to\qqbar$ ($q=u,d,s,c$) are generated with the
{\small JETSET}~\cite{jetset} generator, and $\epem\to\tau^+\tau^-$ with the
{\small KORALB}~\cite{koralb} program. The response of the {\babar} detector is
simulated using the {\small GEANT}4~\cite{geant} package.

\subsection{Event selection}

Event selection follows the same procedure as the selection of
two-charged particle ISR events used for cross section
measurements~\cite{prd-pipi}. It requires a photon with energy
$E_\gamma^*>3\gev$  in the $\epem$ c.m.\ and laboratory polar angle with respect
to the $e^-$ beam in the range [0.35--2.4]\rad, and exactly two tracks of
opposite charge, each with momentum $p>1\gevc$ and within the angular range
[0.40--2.45]\rad. If more than one photon is detected, the candidate with the 
highest $E_\gamma^*$ is taken
to be the `ISR' photon. To ensure a rough momentum
balance at an early stage of the selection, the `ISR' photon is required to lie
within $0.3\rad$ of the missing momentum of the charged particles (or of the
tracks plus the other photons).  The tracks are required to have at least 15
hits in the DCH, to originate within $5\mm$ of the collision axis and within 
$6\cm$ from the beam spot along
the beam direction, and to extrapolate to the DIRC and IFR
active areas in order to exclude low-efficiency regions. Both tracks are
required to be identified either as muons or as pions. To suppress the
background to $\pipig$ at threshold due to the $\epem\to\gamma\gamma$ process
followed by a photon conversion and misidentification of both electrons as
pions, it is further required that the distance in transverse plane $V_{xy}$
between the vertex of the two tracks and the beam collision point be less than
0.5\cm for $m_{\pi\pi}<0.5\gevcc$. Electron background to $\mmg$ is negligible
over the full mass range.

In order to suppress multi-hadron ISR events and reduce higher order radiative
processes, the selected two-prong candidates are subjected to a one-constraint
kinematic fit to the $\epem\to x^+x^-\gamma$ hypothesis ($x=\mu, \pi$), in which only the
two good charged-particle tracks are taken as input and the corresponding missing mass is
constrained to the null photon mass. The $\chi^2$ value of the kinematic fit is
required to be less than 15.

\subsection{Charge asymmetry calculation}

For a complete topology of the final states, the azimuth $\phi^*$ defined in
Sec.~\ref{sec:4var} should cover the
$2\pi$ range. However, the event sample with $x^-$ azimuth $\phi^*_-\in[0,\pi]$ is
complementary to the sample with $x^+$ azimuth $\phi^*_+\in[0,\pi]$, since
$\phi^*_+=\pi+\phi^*_-~(\rm mod \, 2\pi)$ in every event. This allows to
restrict  $\phi^*$ to the range $[0,\pi]$ with no loss of phase space.
After integrating over $\theta^*_\gamma$
and $\theta^*$, the total event sample in a fixed $m_{xx}$
interval subdivides into two subsamples: one with
$\phi^*_-\in[0,\pi]$ ($N_-$), the other with $\phi^*_+\in[0,\pi]$ ($N_+$). 

We obtain separately the distributions in $\cos\phi^*$ of the $N^{\rm{obs}}_\pm$ samples 
in data, namely $N^{\rm{obs}}_-(\cos\phi^*)$ with $\phi^*_-=\phi^*$ and 
$N^{\rm{obs}}_+(\cos\phi^*)$ with $\phi^*_+=\phi^*$. 
Distributions of background events $N^{\rm{BG}}_\pm(\cos\phi^*)$ are determined 
separately for each subsample, as described below. Likewise, efficiencies are split 
into $\epsilon_\pm(\cos\phi^*)$ 
and computed using the full simulation of $\epem\to x^+x^-\gamma$ ($x=\mu, \pi$) 
events, with corrections for the differences between data and simulation (see
Sec.~\ref{sec:syst_mu} and~\ref{sec:syst_pi}).
In a given $m_{\mu\mu}$ ($m_{\pi\pi}$) interval, the asymmetry at a given $\cos\phi^*$
is derived from the difference between the $N_-(\cos\phi^*)$ and $N_+(\cos\phi^*)$ yields, 
corrected for efficiency, and is obtained from the following expression
\begin{widetext}
\beqn 
\label{eq:A_cosphis}
  A(\cos\phi^*)
 =\frac{N_-(\cos\phi^*)/\epsilon_-(\cos\phi^*)-N_+(\cos\phi^*)/\epsilon_+(\cos\phi^*)}
       {N_-(\cos\phi^*)/\epsilon_-(\cos\phi^*)+N_+(\cos\phi^*)/\epsilon_+(\cos\phi^*)},
\eeqn
\end{widetext}
where $N_\pm(\cos\phi^*)=N^{\rm{obs}}_\pm(\cos\phi^*)-N^{\rm{BG}}_\pm(\cos\phi^*)$.  

Note that distributions of $N_\pm(\cos\phi^*)$ 
can be obtained in each $(m_{xx}, \theta^*_\gamma,\cos\theta^*)$ cell of the phase space, 
and the asymmetry defined by Eq.~(\ref{eq:A_cosphis}) can be
calculated. However, this one-dimensional quantity $A(\cos\phi^*)$ 
is a valid definition of charge asymmetry only when the variable
$\cos\theta^*$ is integrated within a symmetric range. This is a consequence of
the fact that the $x^+\leftrightarrow x^-$ interchange means both
$\phi^*\rightarrow \pi+\phi^*$ and $\cos\theta^*\rightarrow -\cos\theta^*$,
and therefore a non-null value of $A(\cos\phi^*)$ in an arbitrary
$\cos\theta^*$ interval is not an intrinsic signature of ISR-FSR interference.  

\subsection{Event samples and backgrounds after selection}
\label{sec:kineFit_BG}

\begin{figure*}
  \centering
  \includegraphics[width=0.25\textwidth]{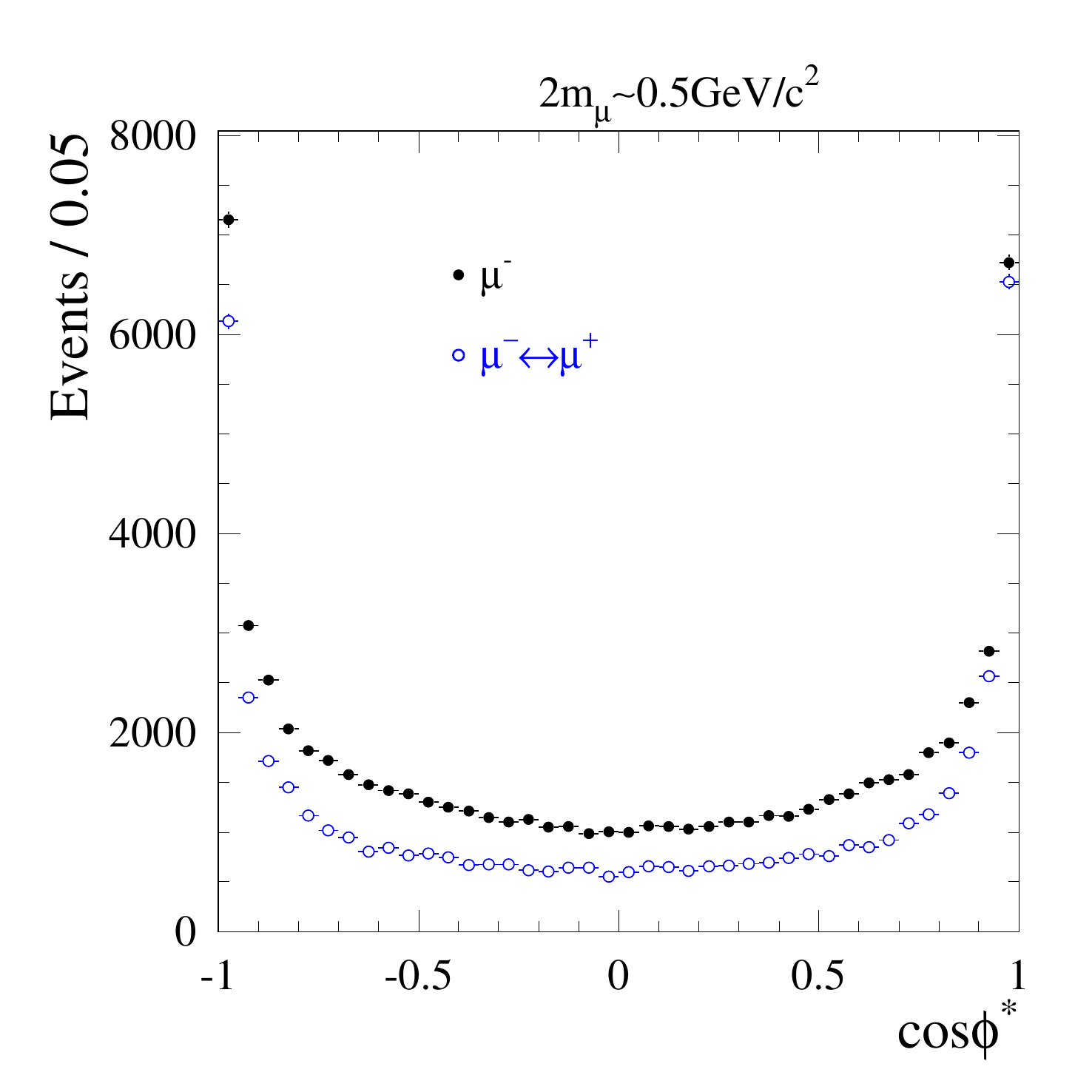}
  \includegraphics[width=0.25\textwidth]{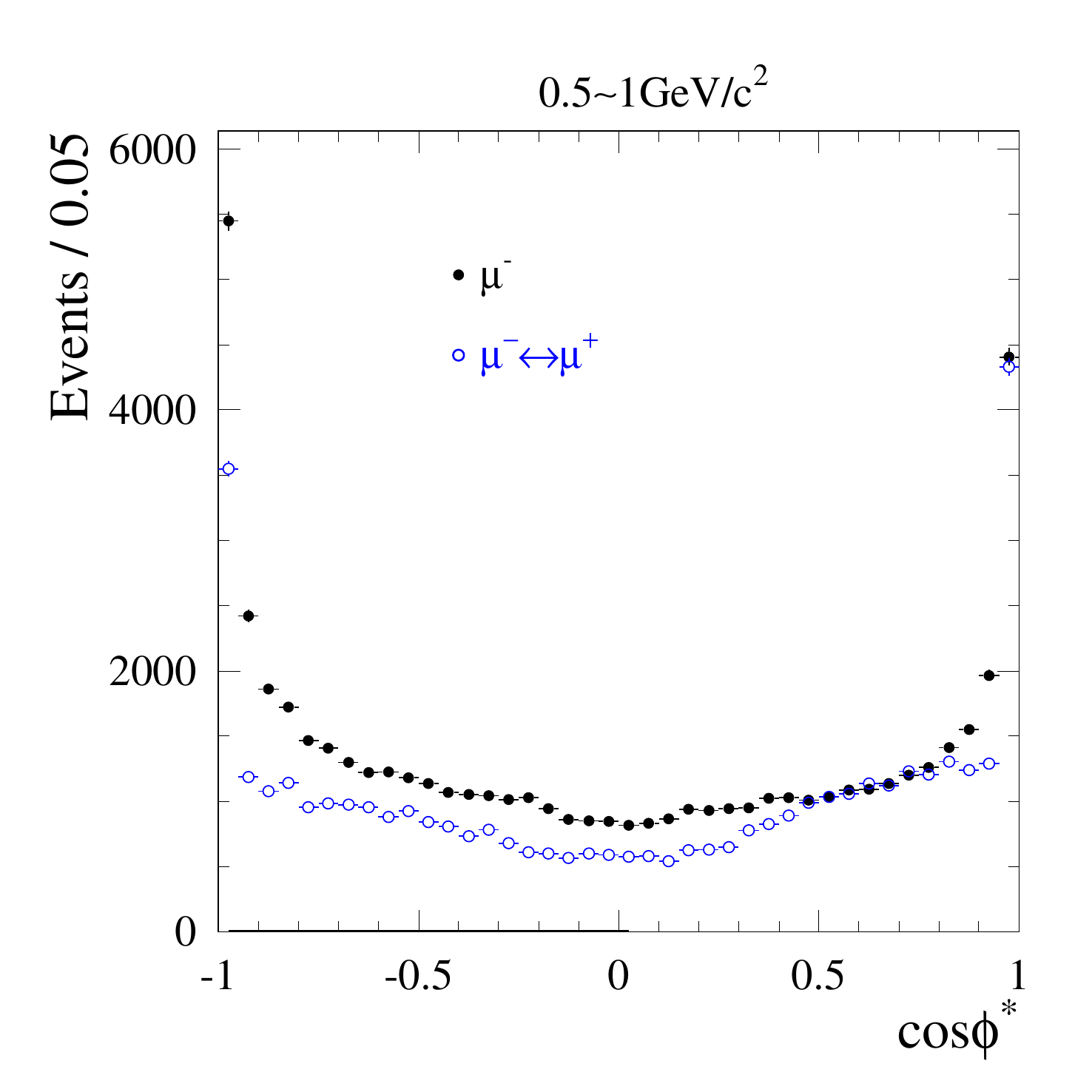}
  \includegraphics[width=0.25\textwidth]{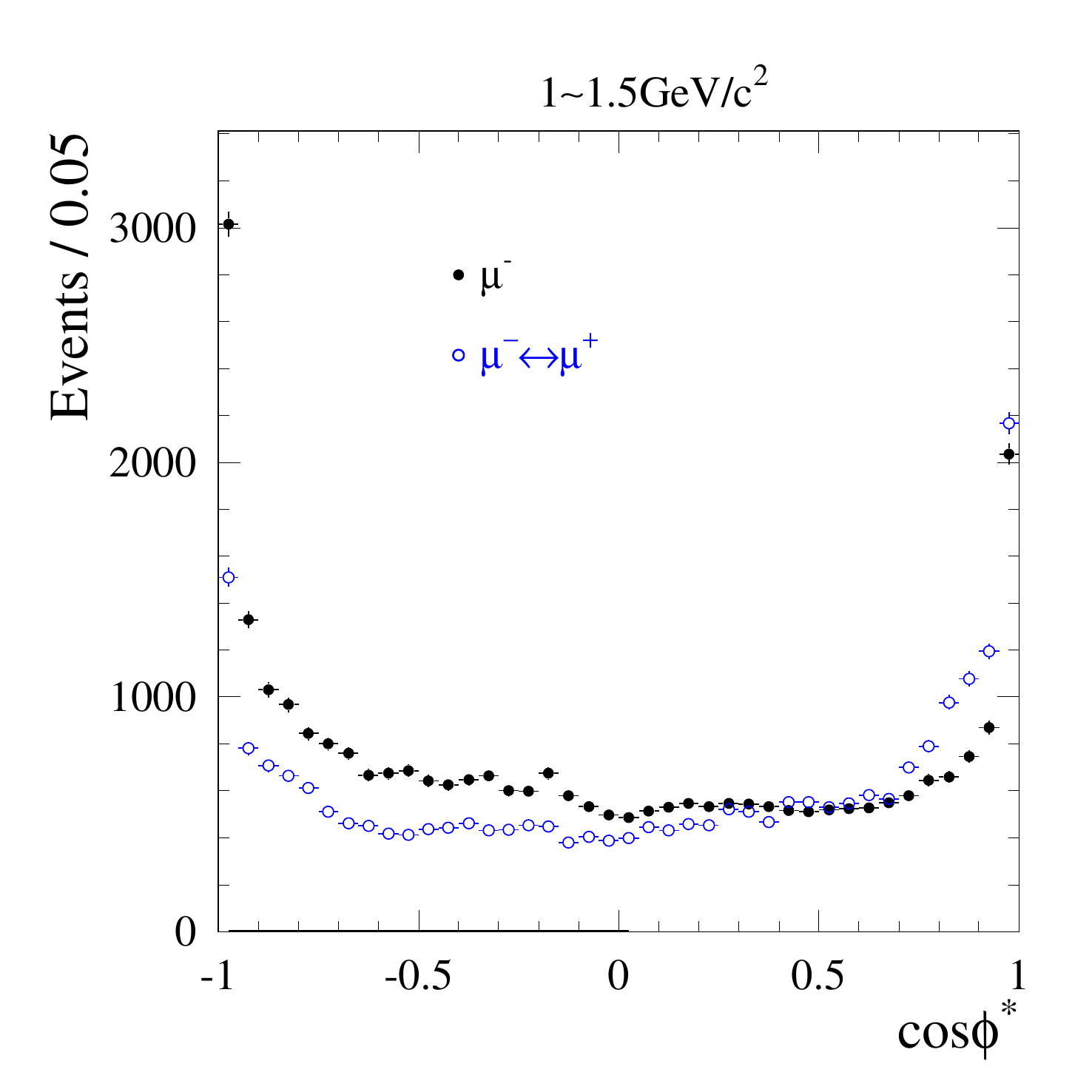}
  \includegraphics[width=0.25\textwidth]{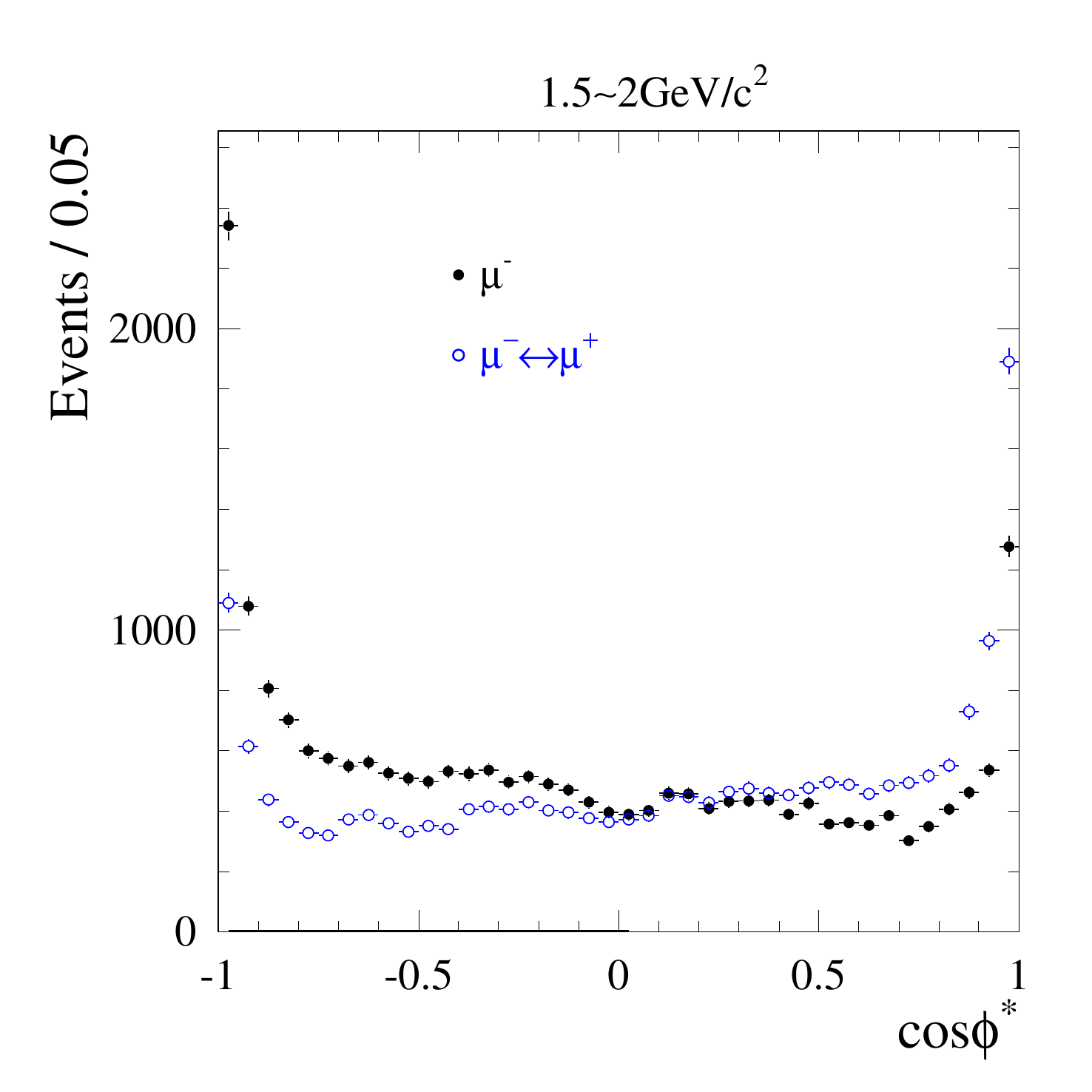}
  \includegraphics[width=0.25\textwidth]{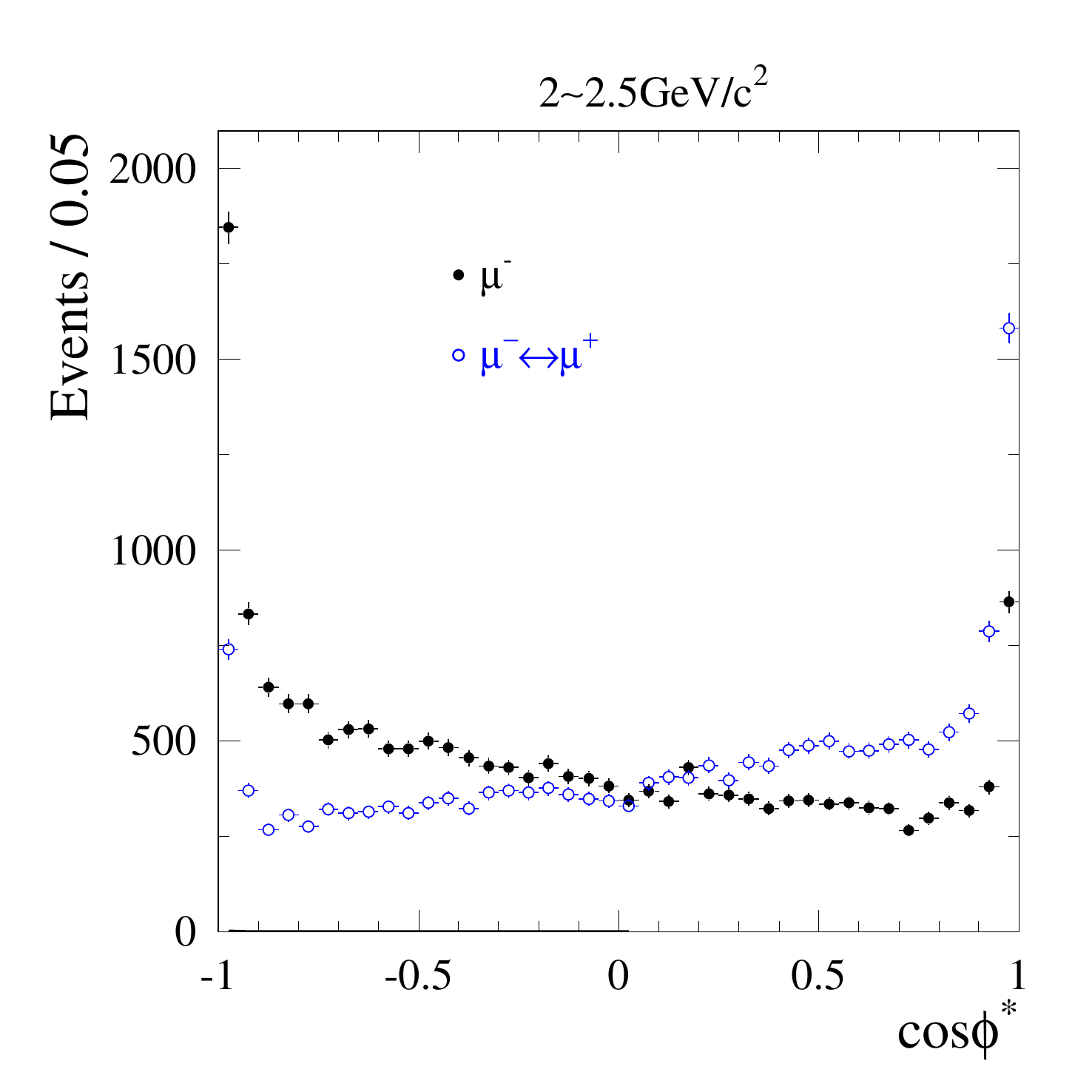}
  \includegraphics[width=0.25\textwidth]{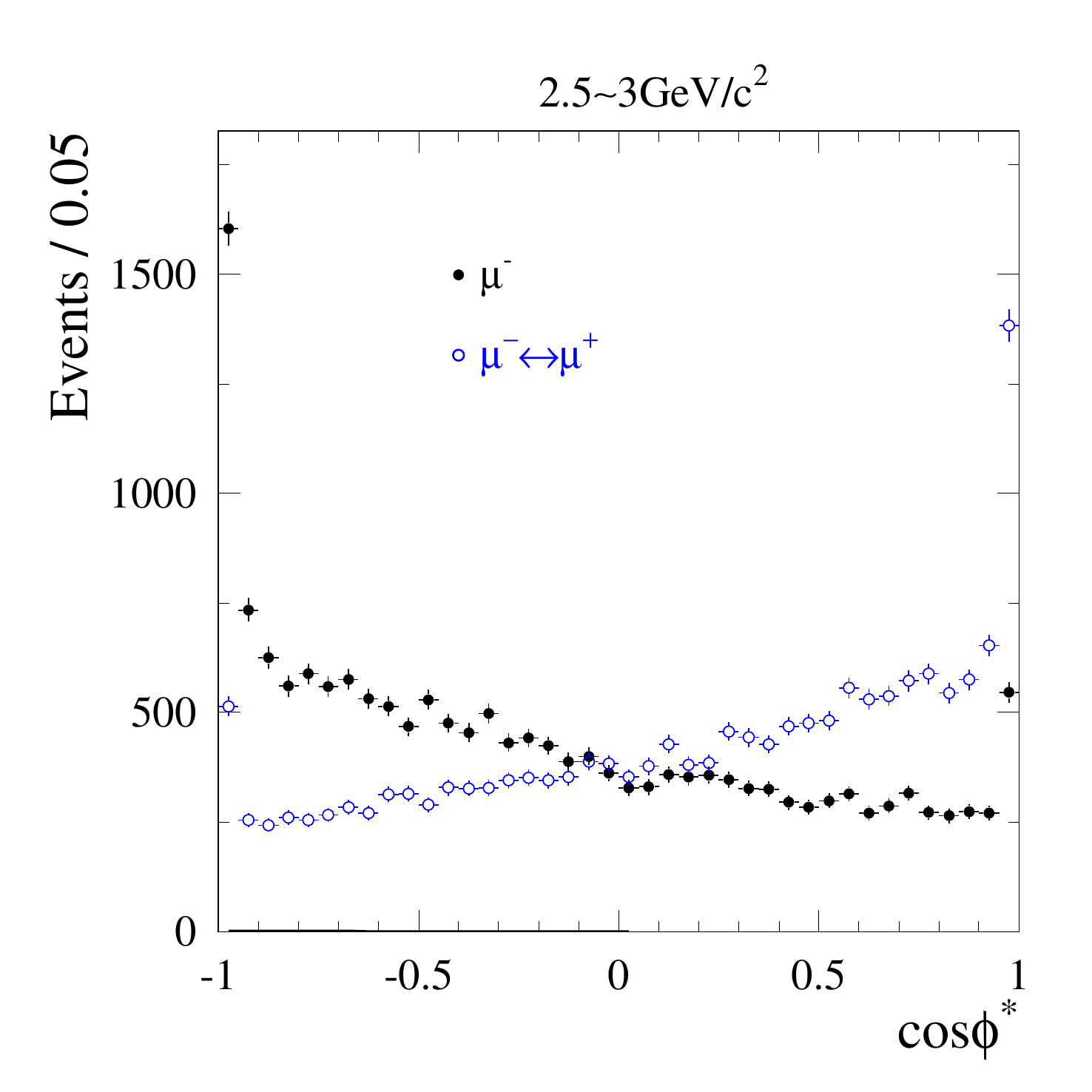}
  \includegraphics[width=0.25\textwidth]{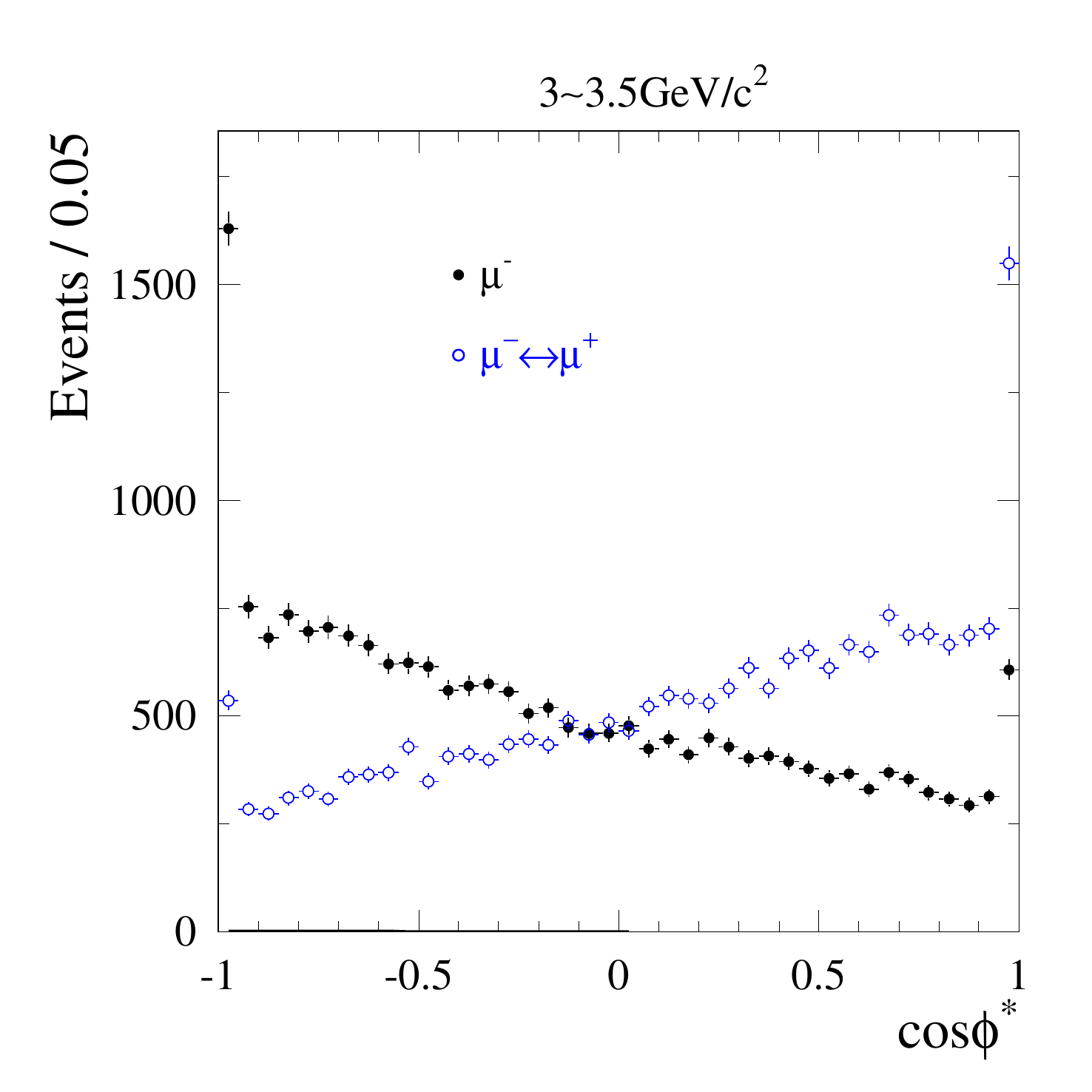}
  \includegraphics[width=0.25\textwidth]{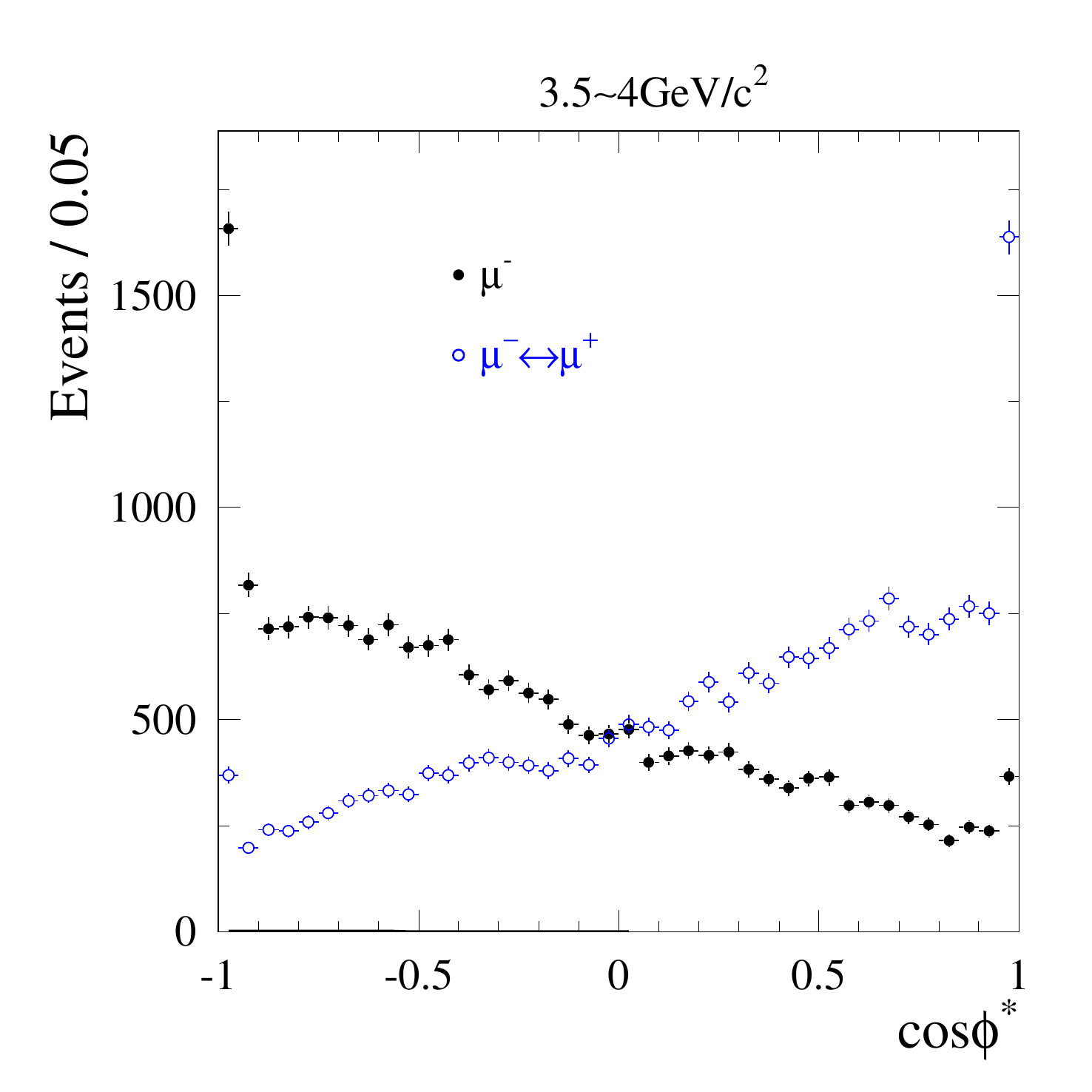}
  \includegraphics[width=0.25\textwidth]{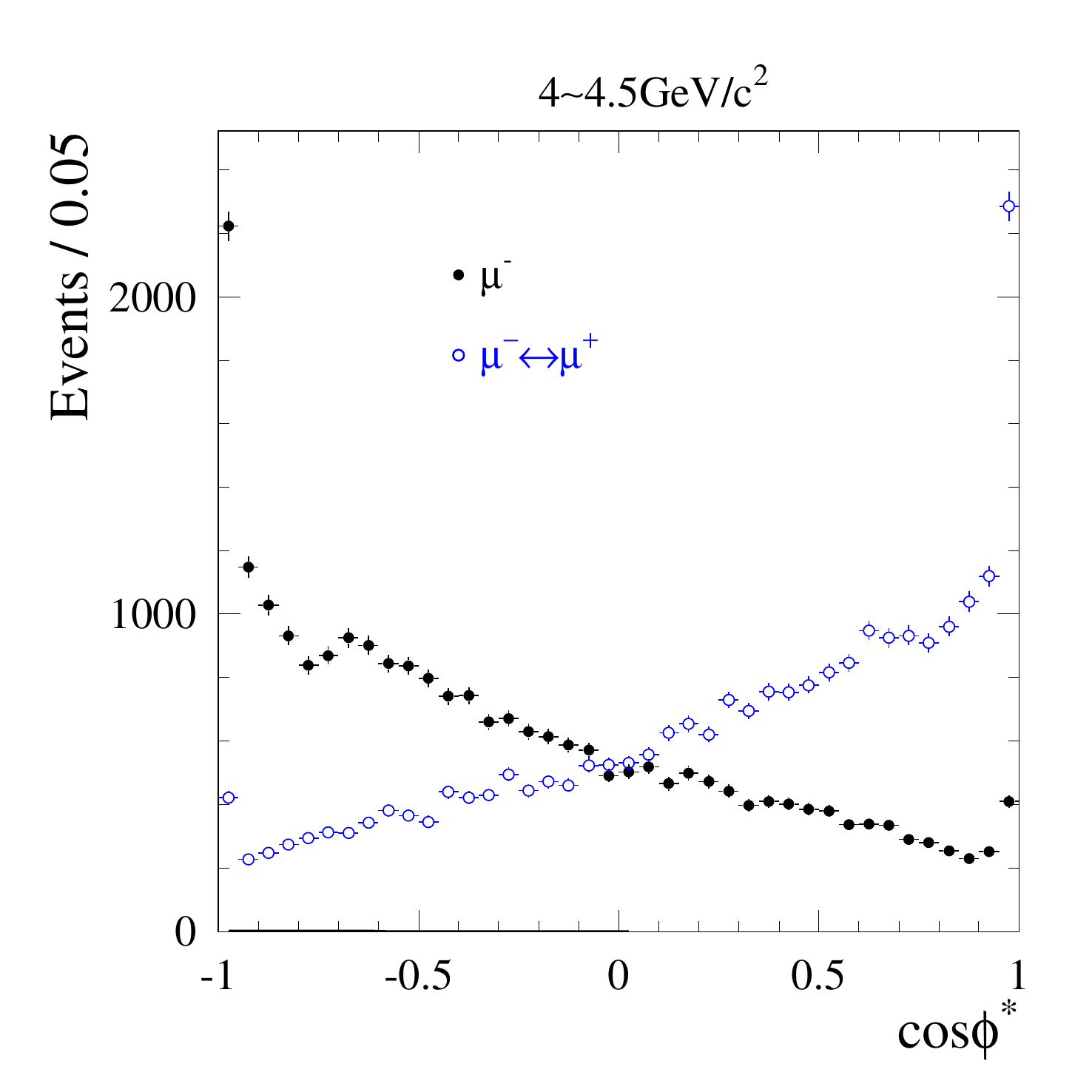}
  \includegraphics[width=0.25\textwidth]{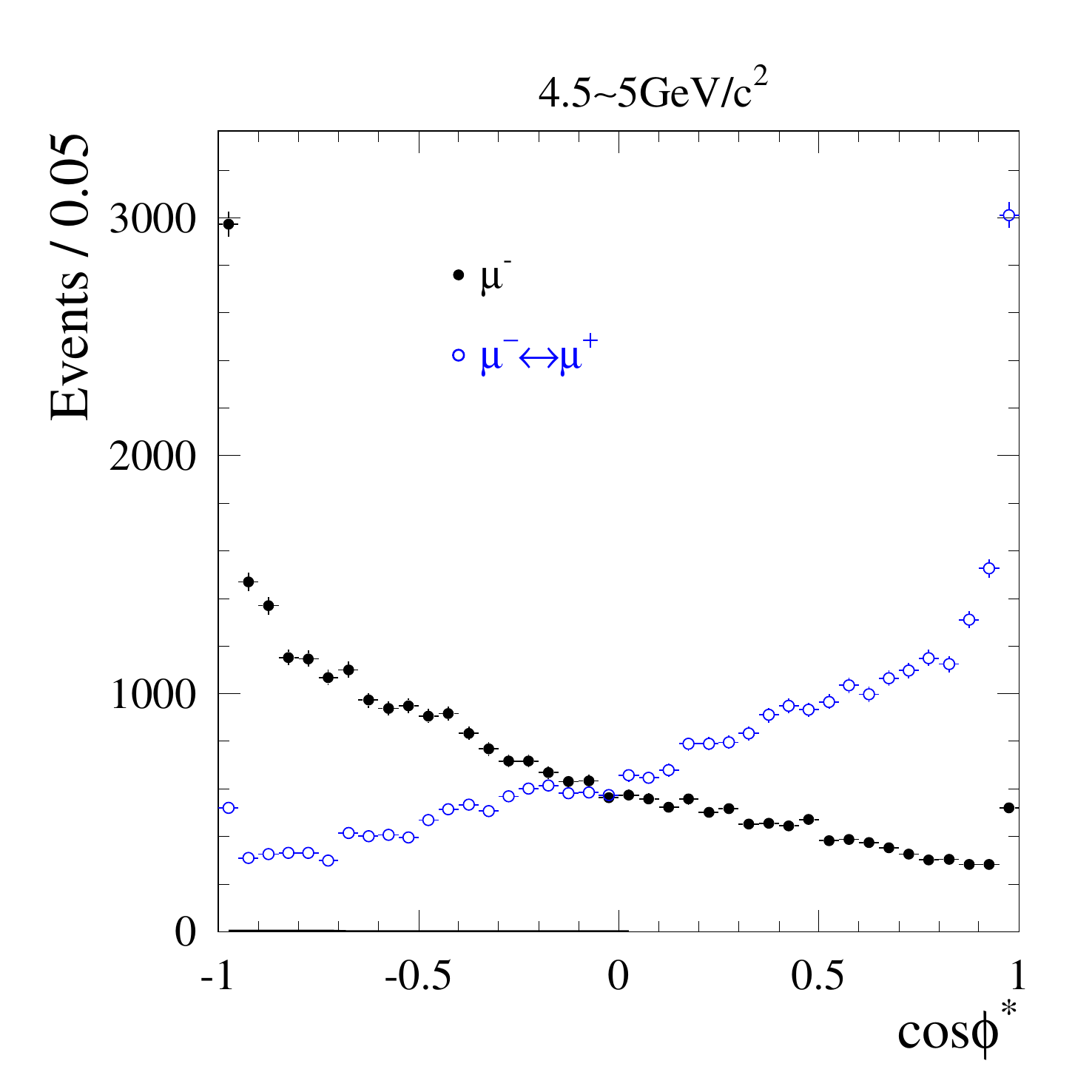}
  \includegraphics[width=0.25\textwidth]{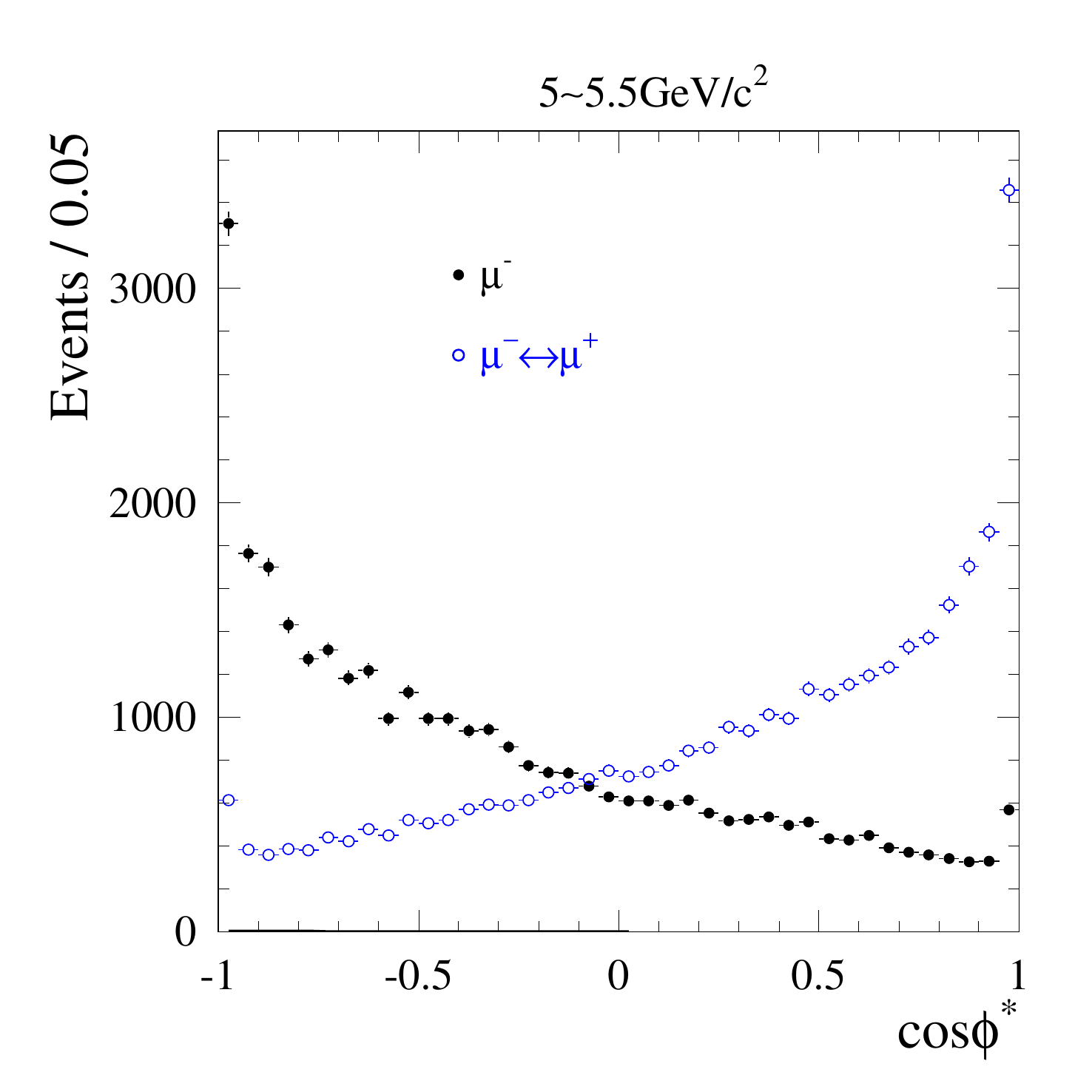}
  \includegraphics[width=0.25\textwidth]{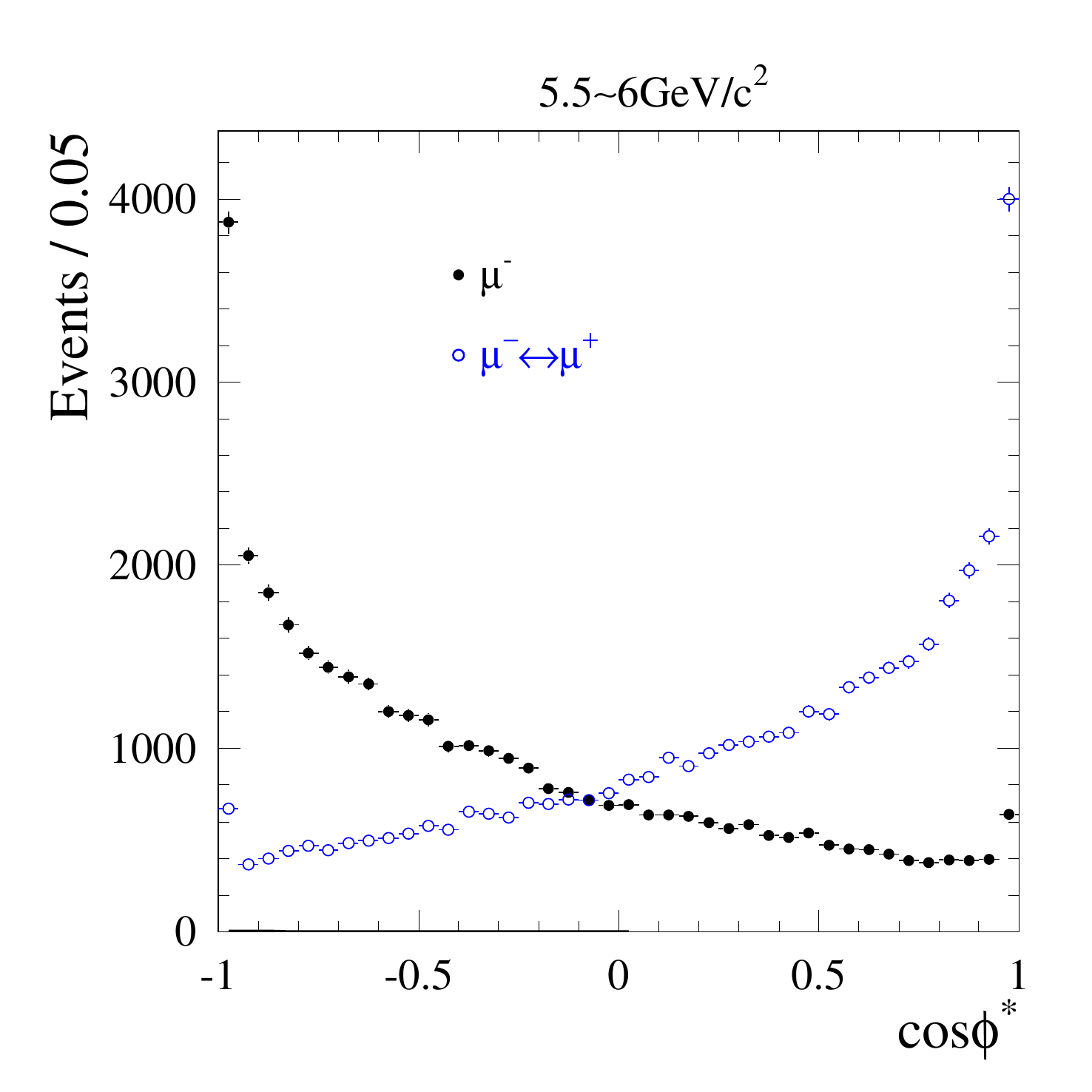}
  \includegraphics[width=0.25\textwidth]{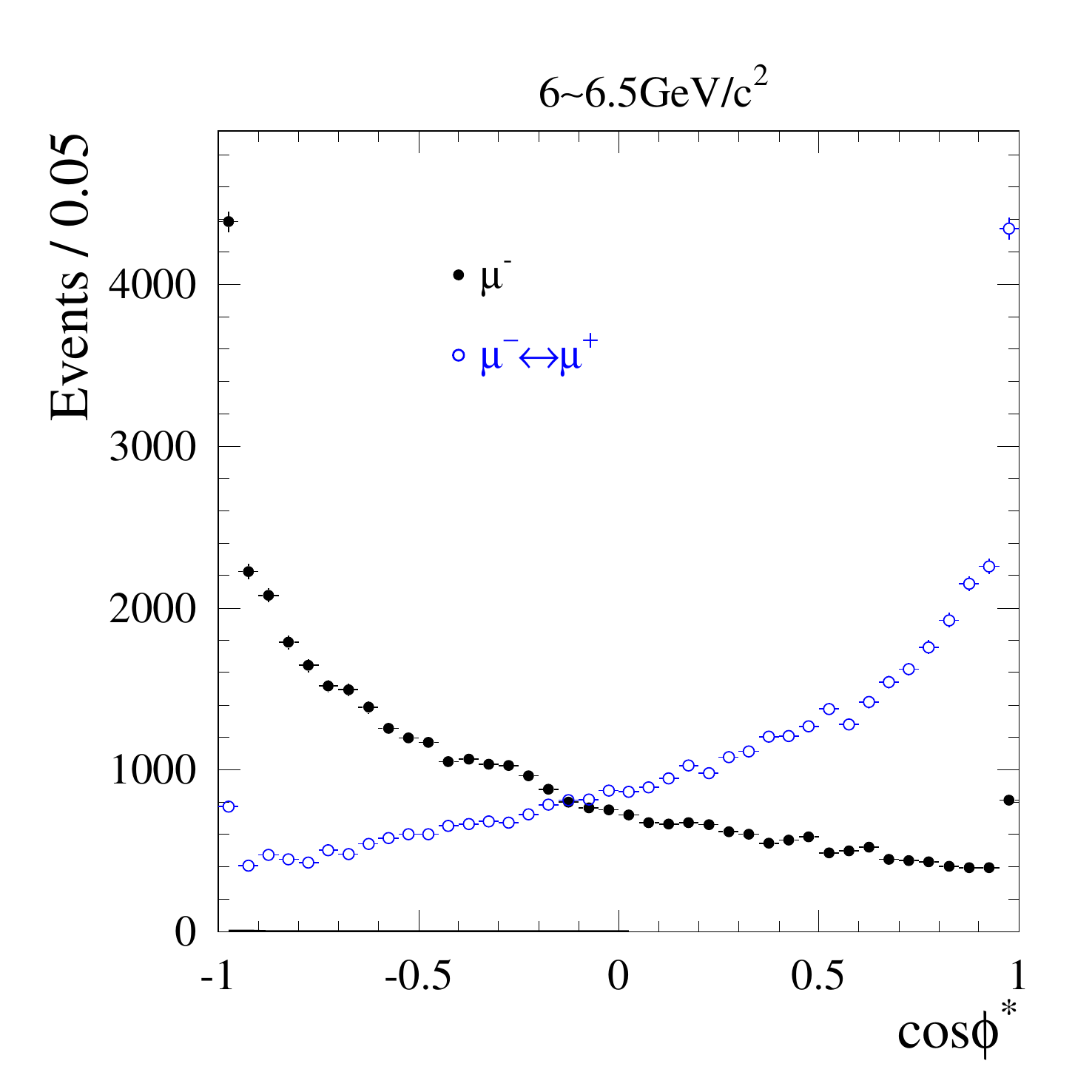}
  \includegraphics[width=0.25\textwidth]{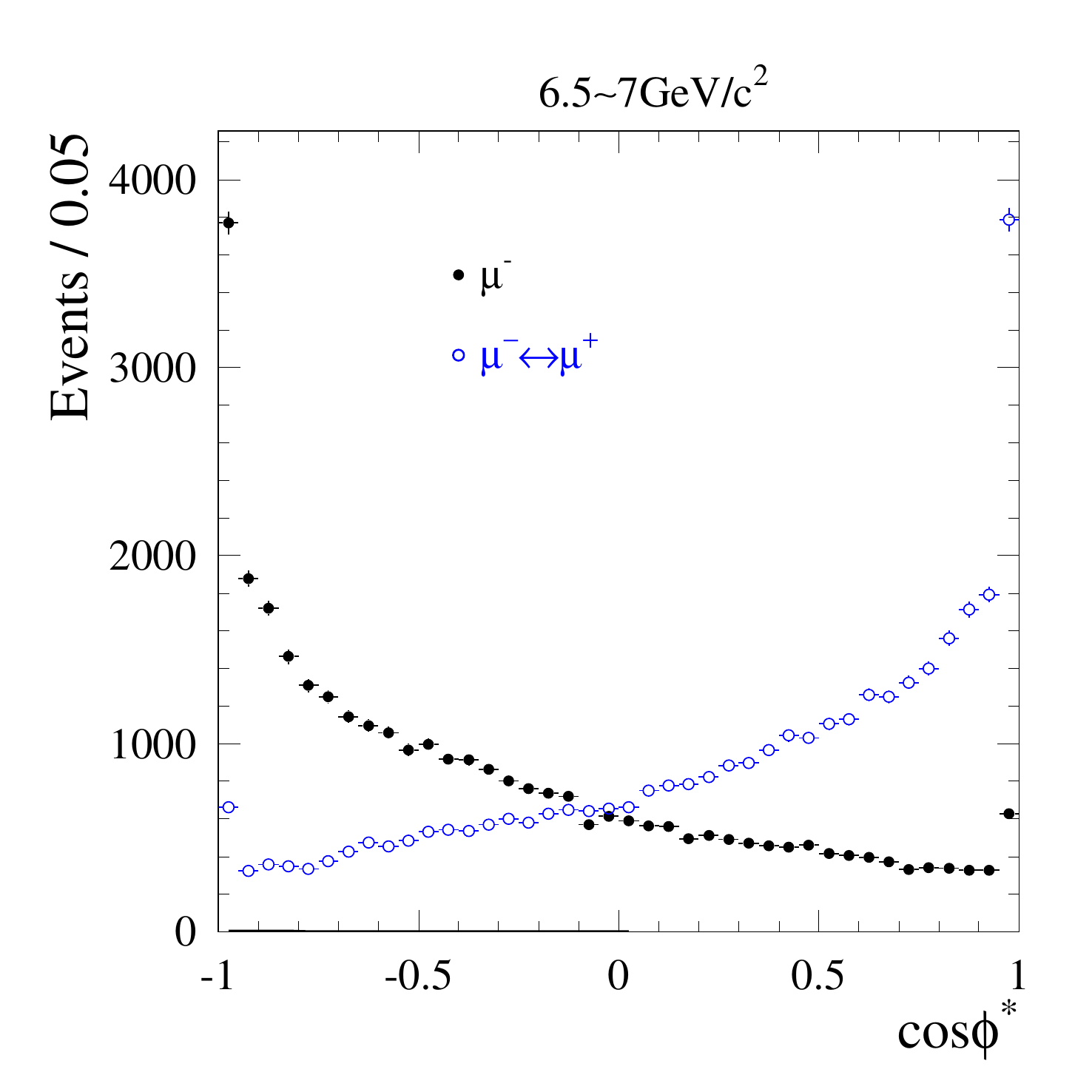}
  \caption{The $\cos\phi^*$ distributions for $\epem\to\mmg$ data in $0.5\gevcc$
  $m_{\mu\mu}$ intervals.  The points labeled `$\mu^-$'
  refer to the configurations with $\phi^*_-\in[0,\pi]$, while the points
  labeled `$\mu^-\leftrightarrow\mu^+$' correspond to $\phi^*_+\in[0,\pi]$.
}
  \label{fig:cosphis_mmg}
\end{figure*}

\begin{figure*}
  \centering
  \includegraphics[width=0.25\textwidth]{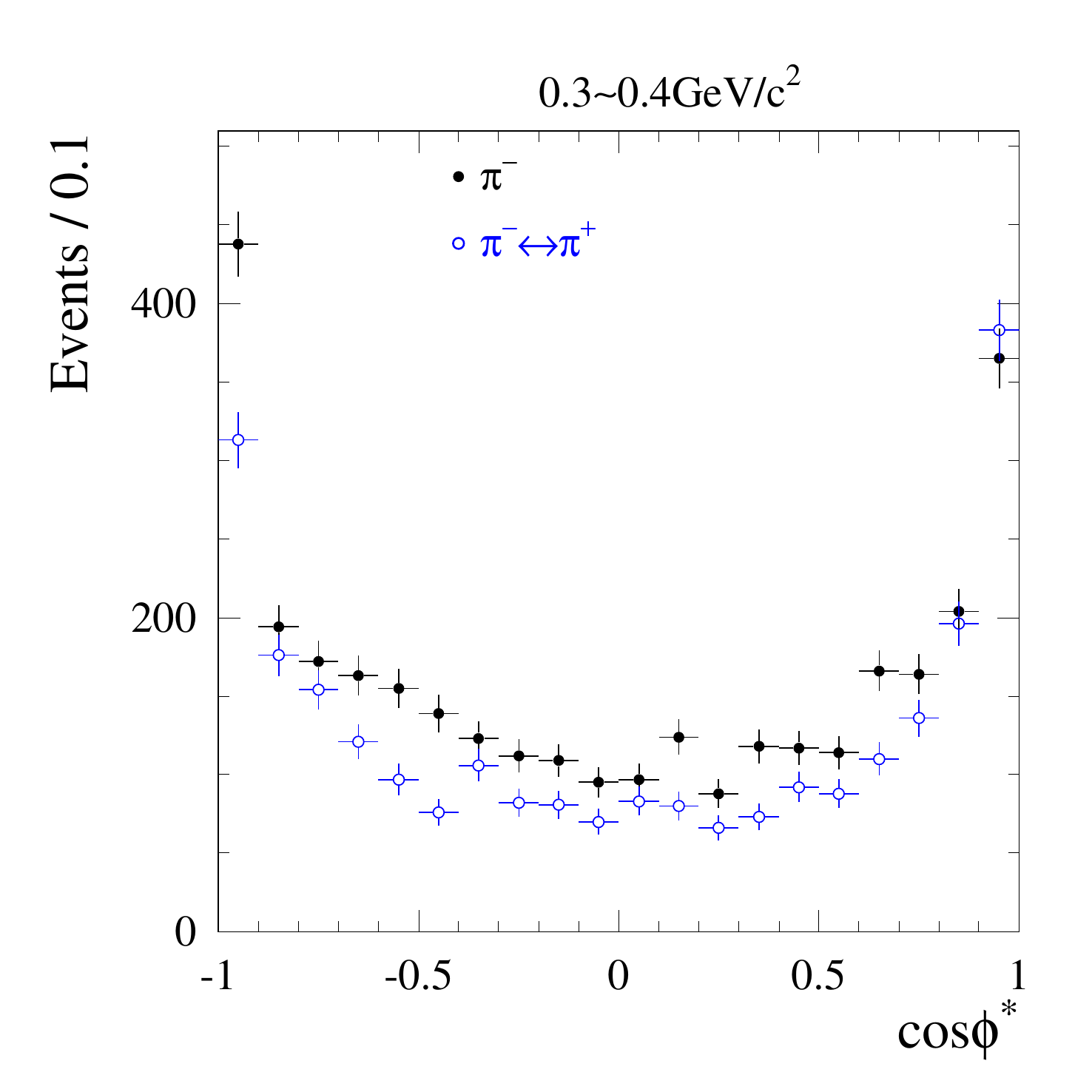}
  \includegraphics[width=0.25\textwidth]{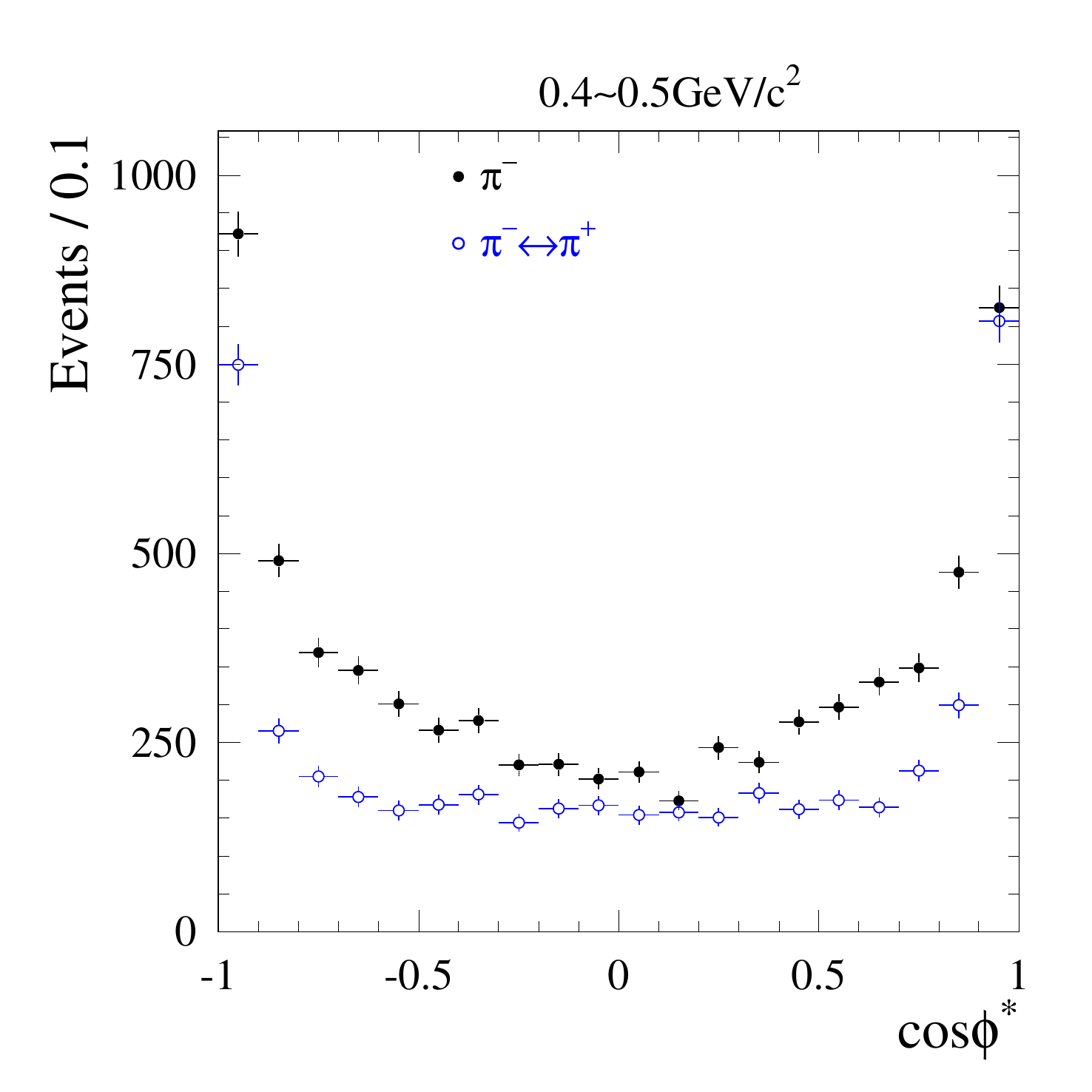}
  \includegraphics[width=0.25\textwidth]{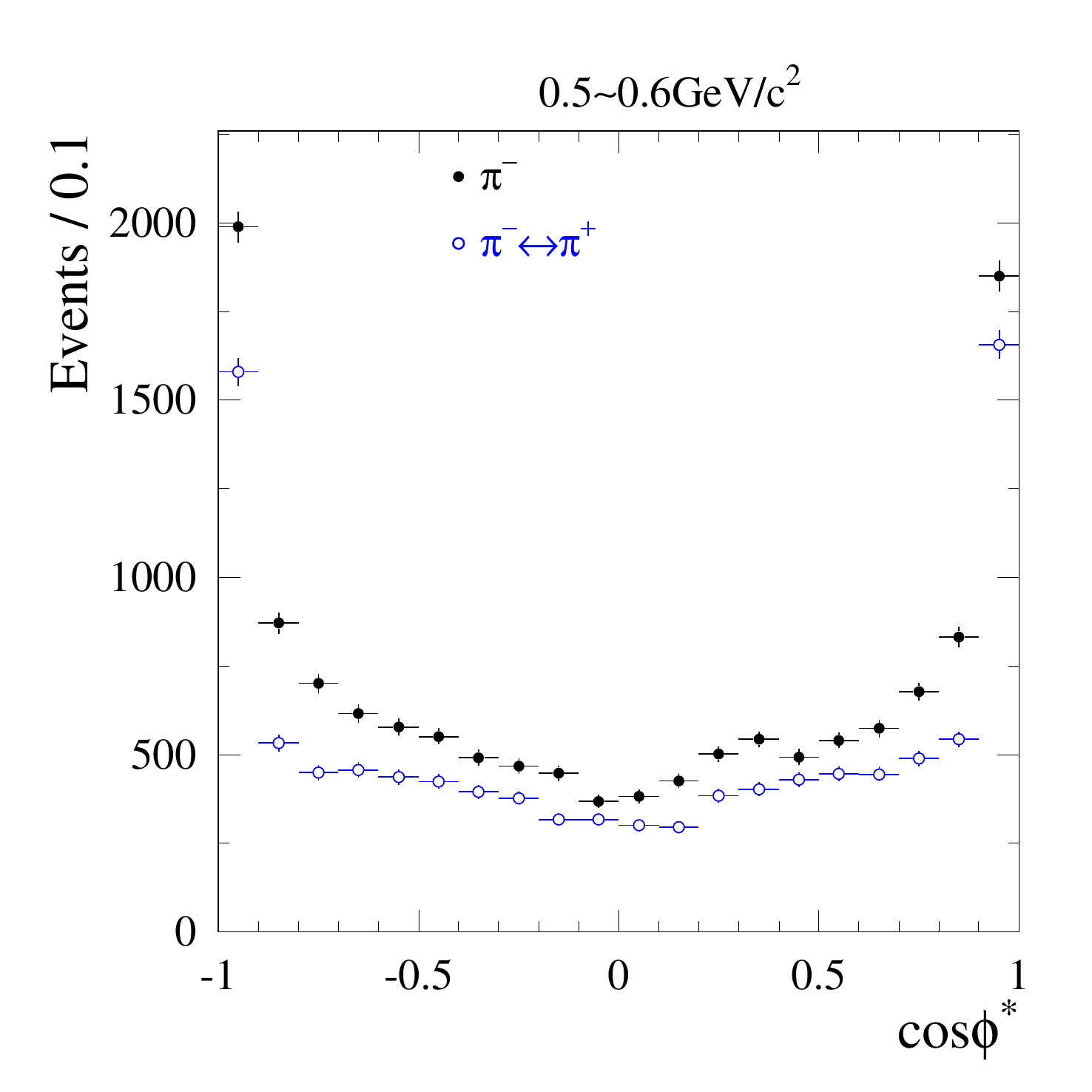}
  \includegraphics[width=0.25\textwidth]{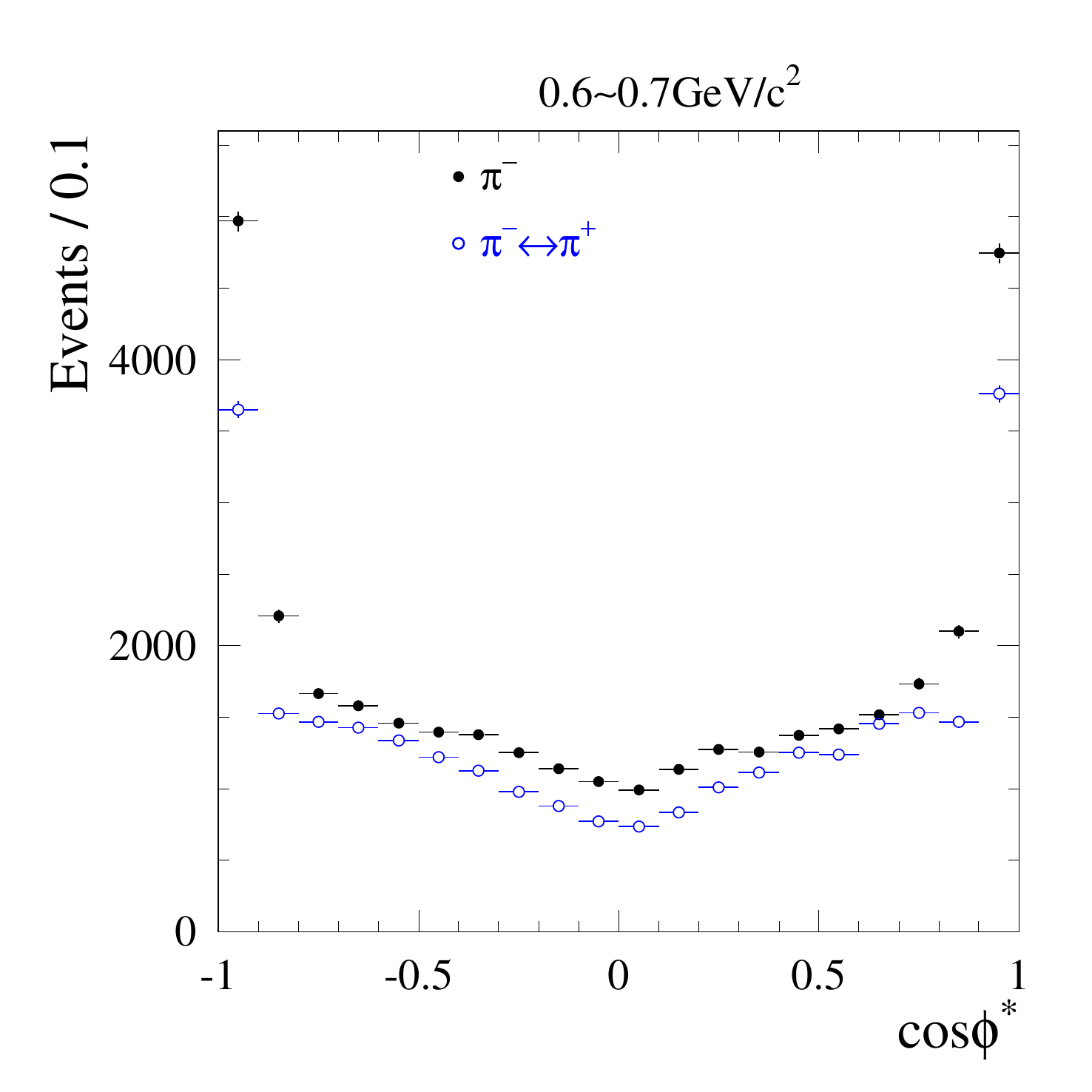}
  \includegraphics[width=0.25\textwidth]{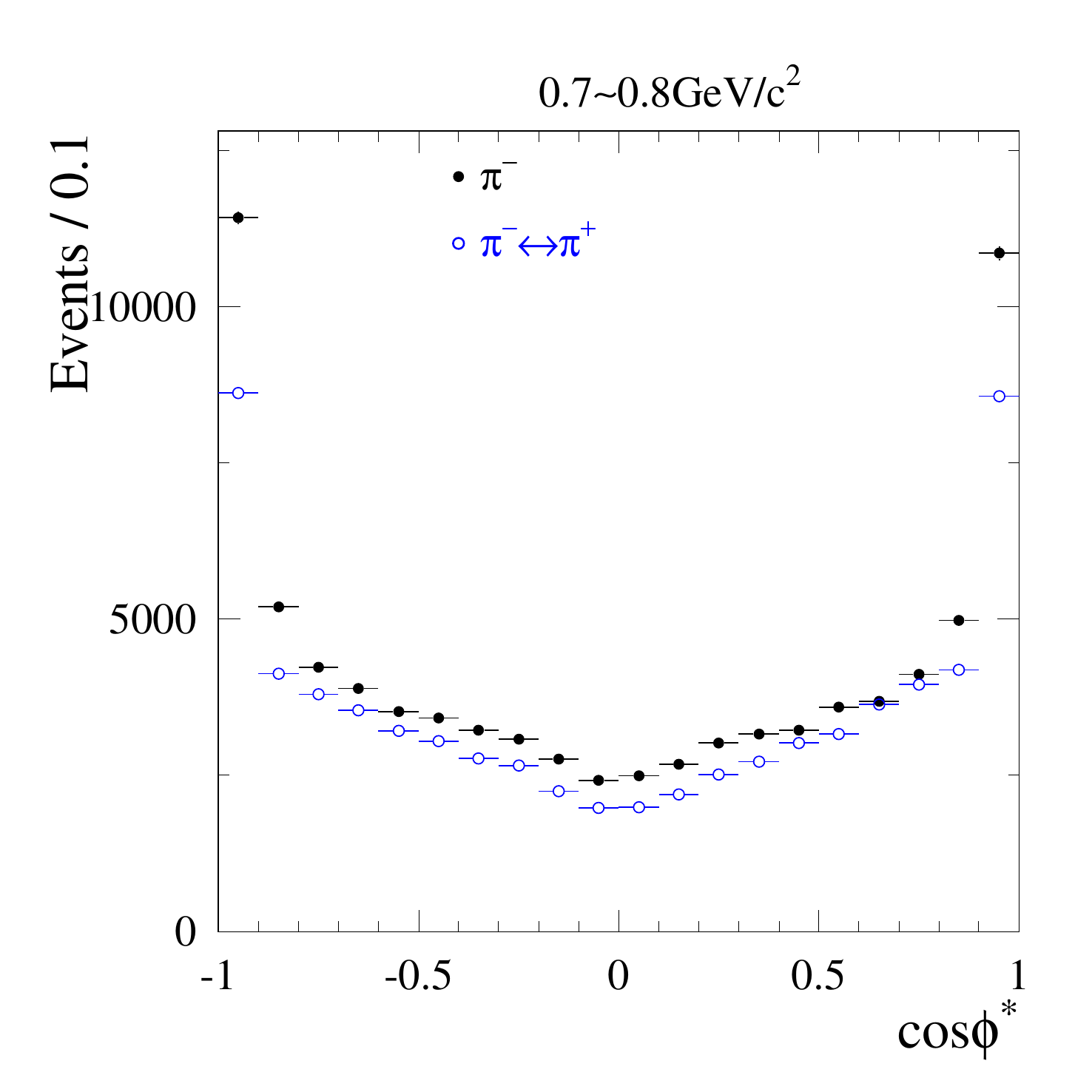}
  \includegraphics[width=0.25\textwidth]{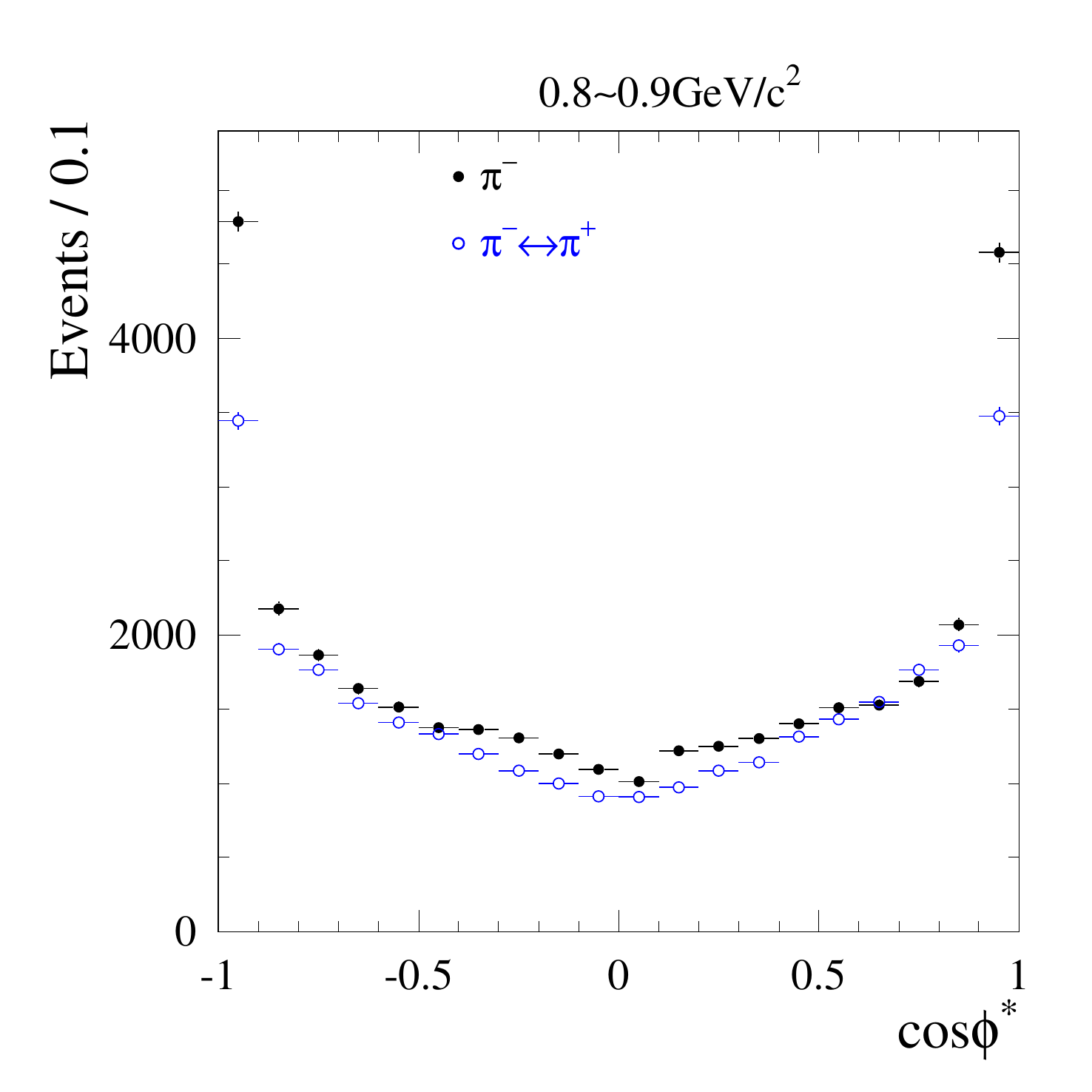}
  \includegraphics[width=0.25\textwidth]{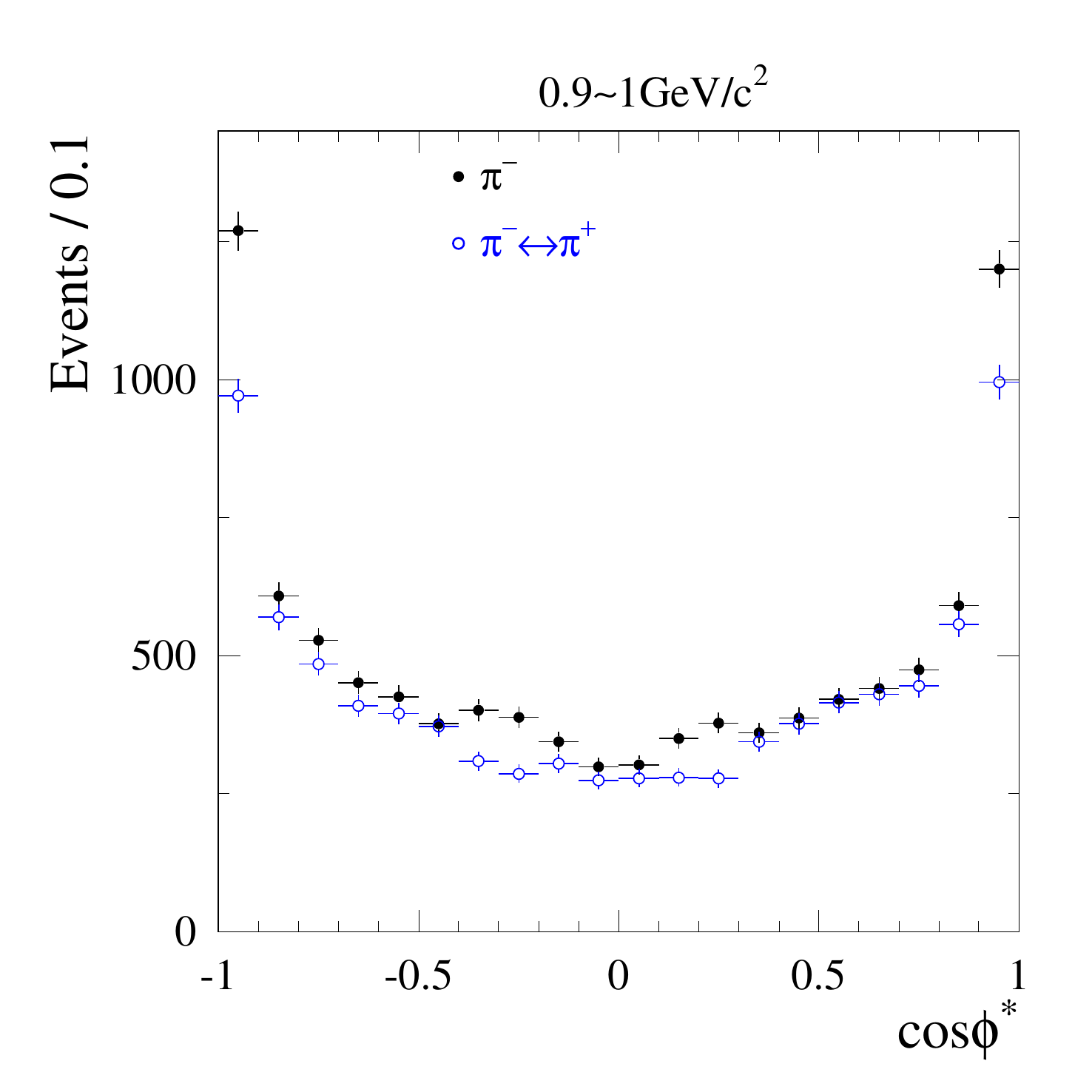}
  \includegraphics[width=0.25\textwidth]{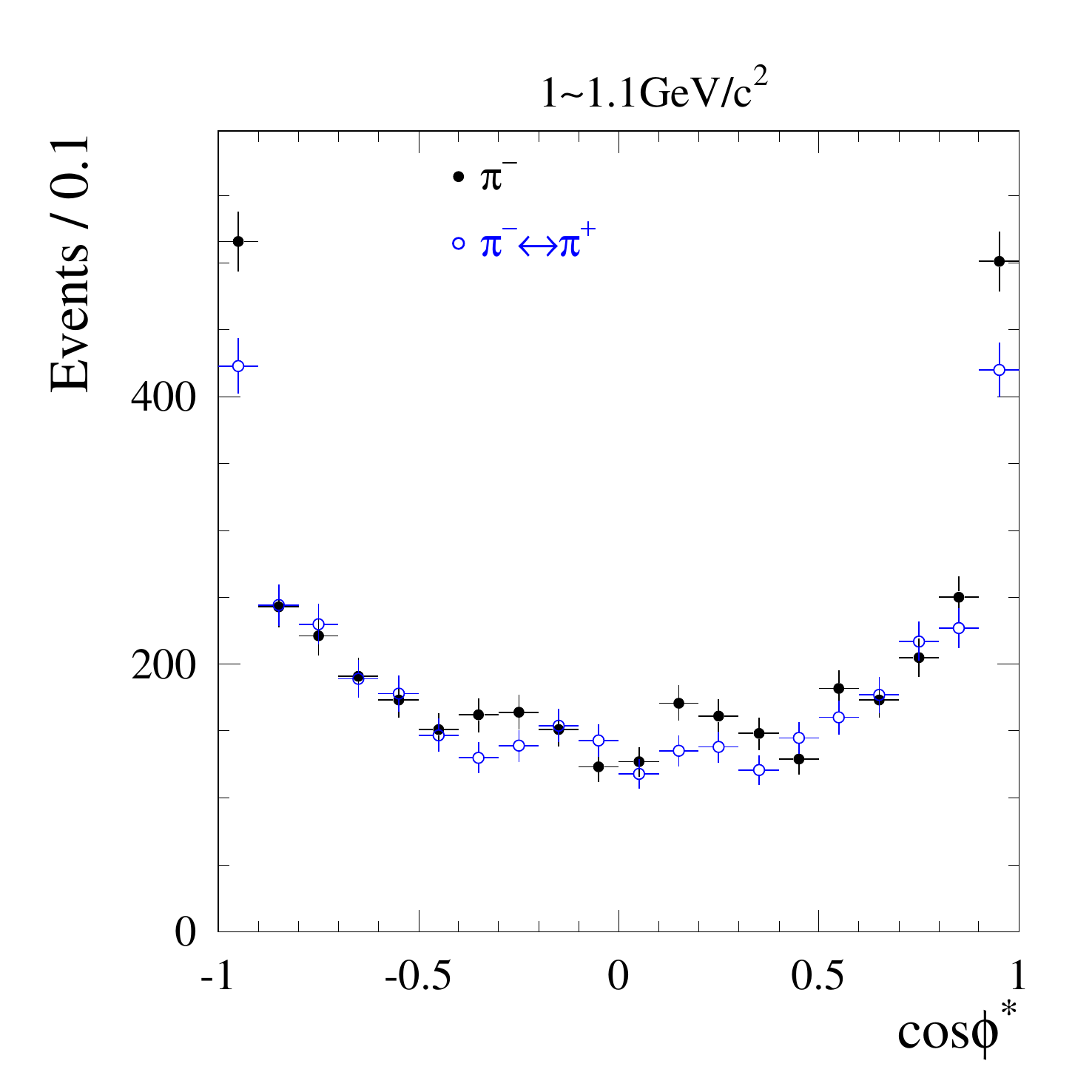}
  \includegraphics[width=0.25\textwidth]{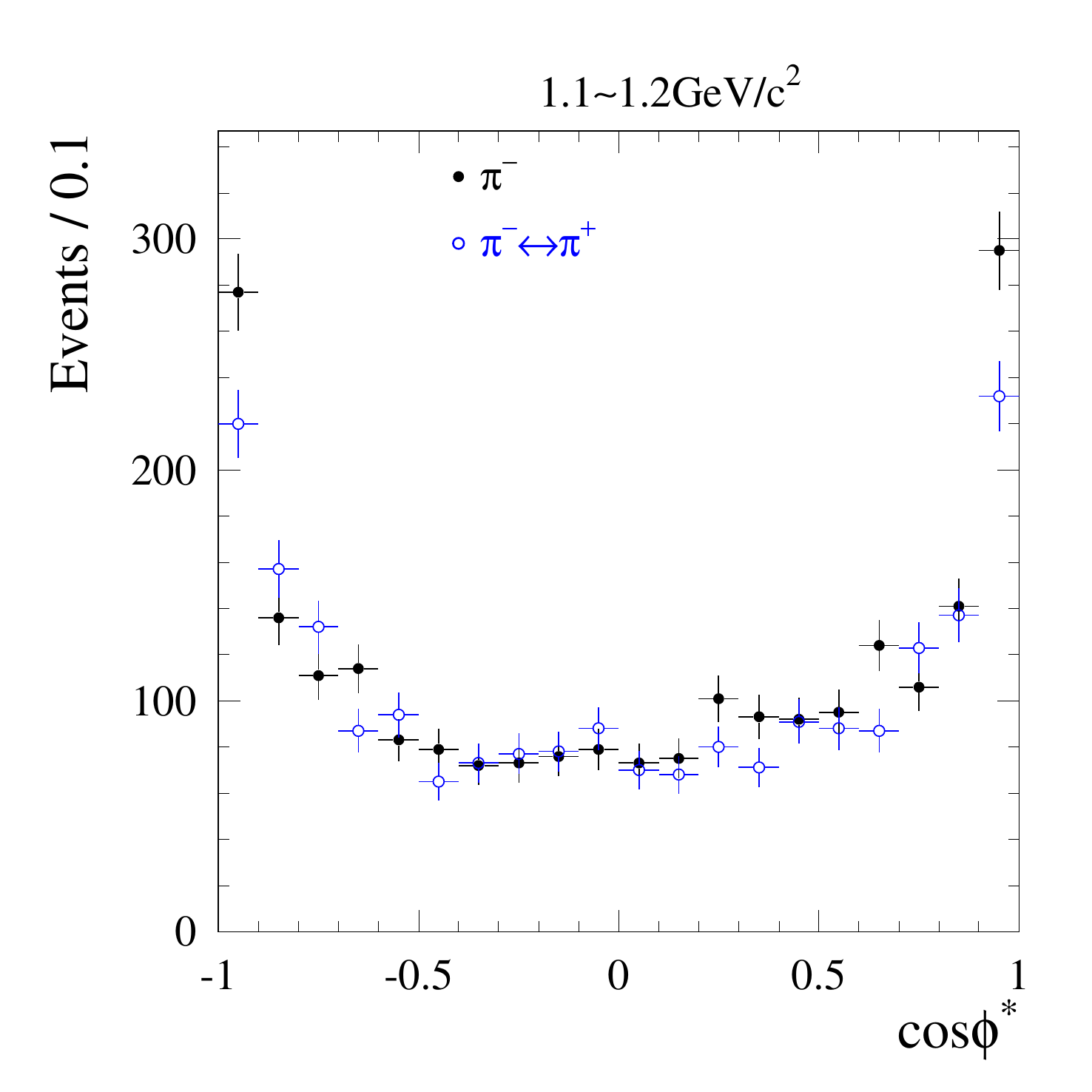}
  \includegraphics[width=0.25\textwidth]{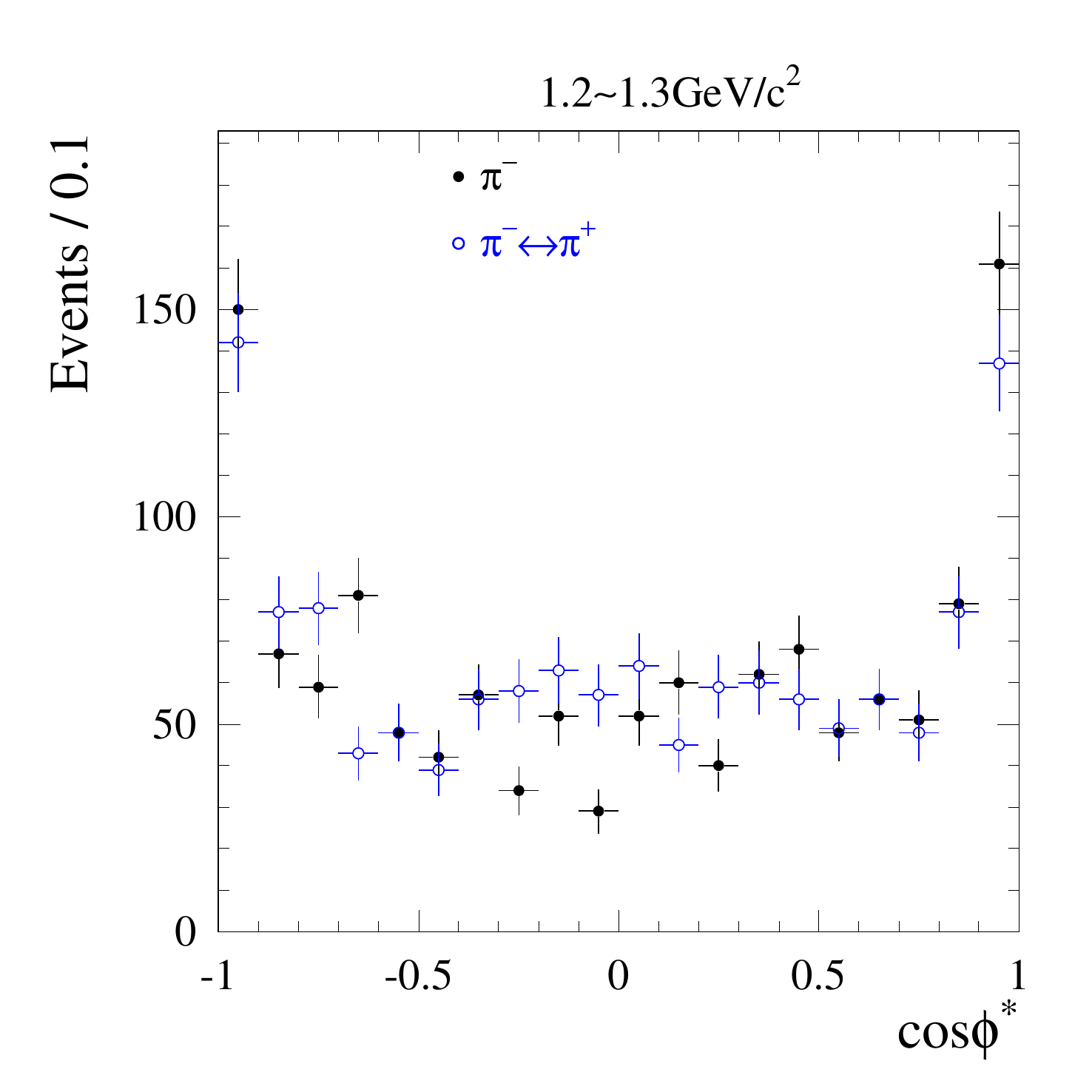}
  \includegraphics[width=0.25\textwidth]{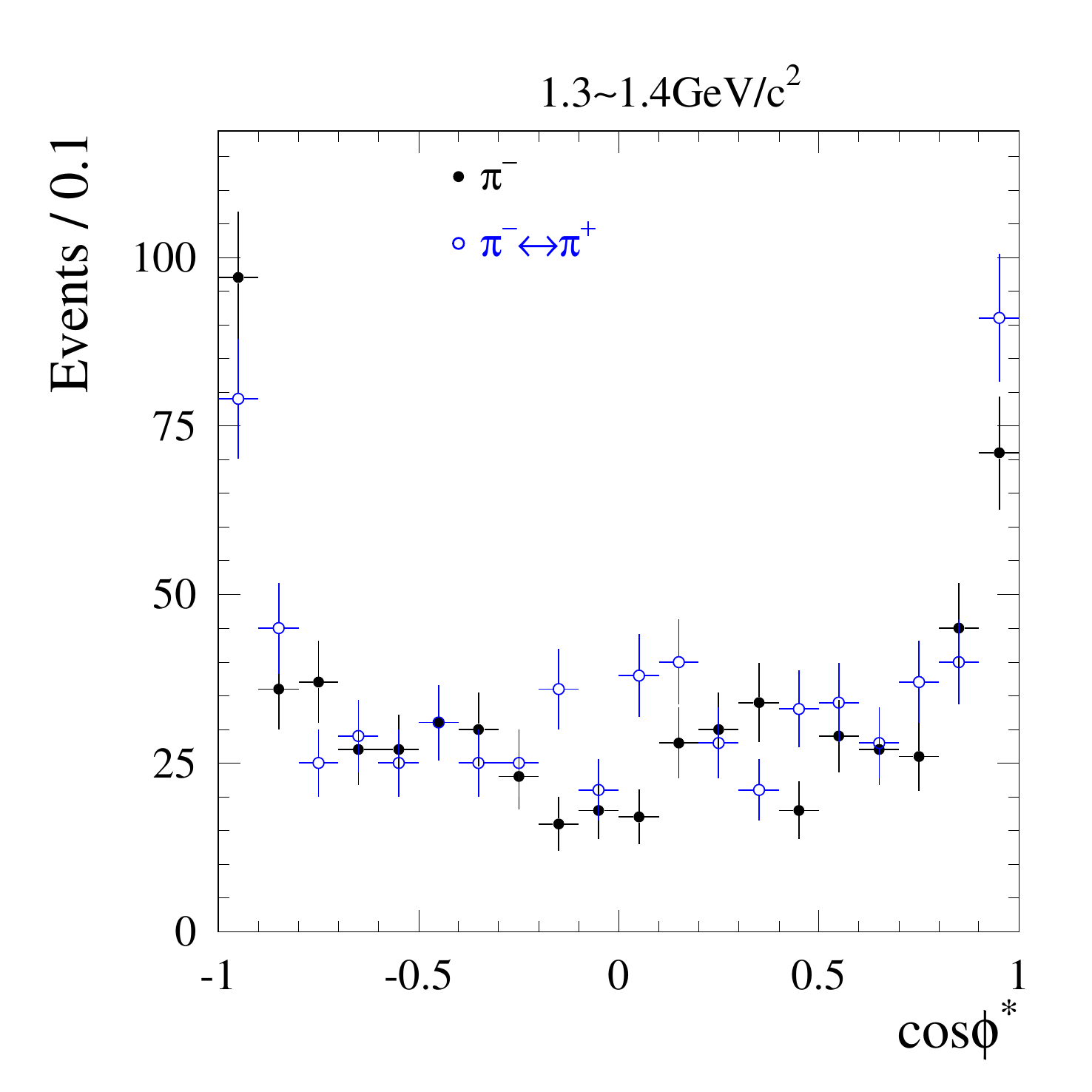}
  \includegraphics[width=0.25\textwidth]{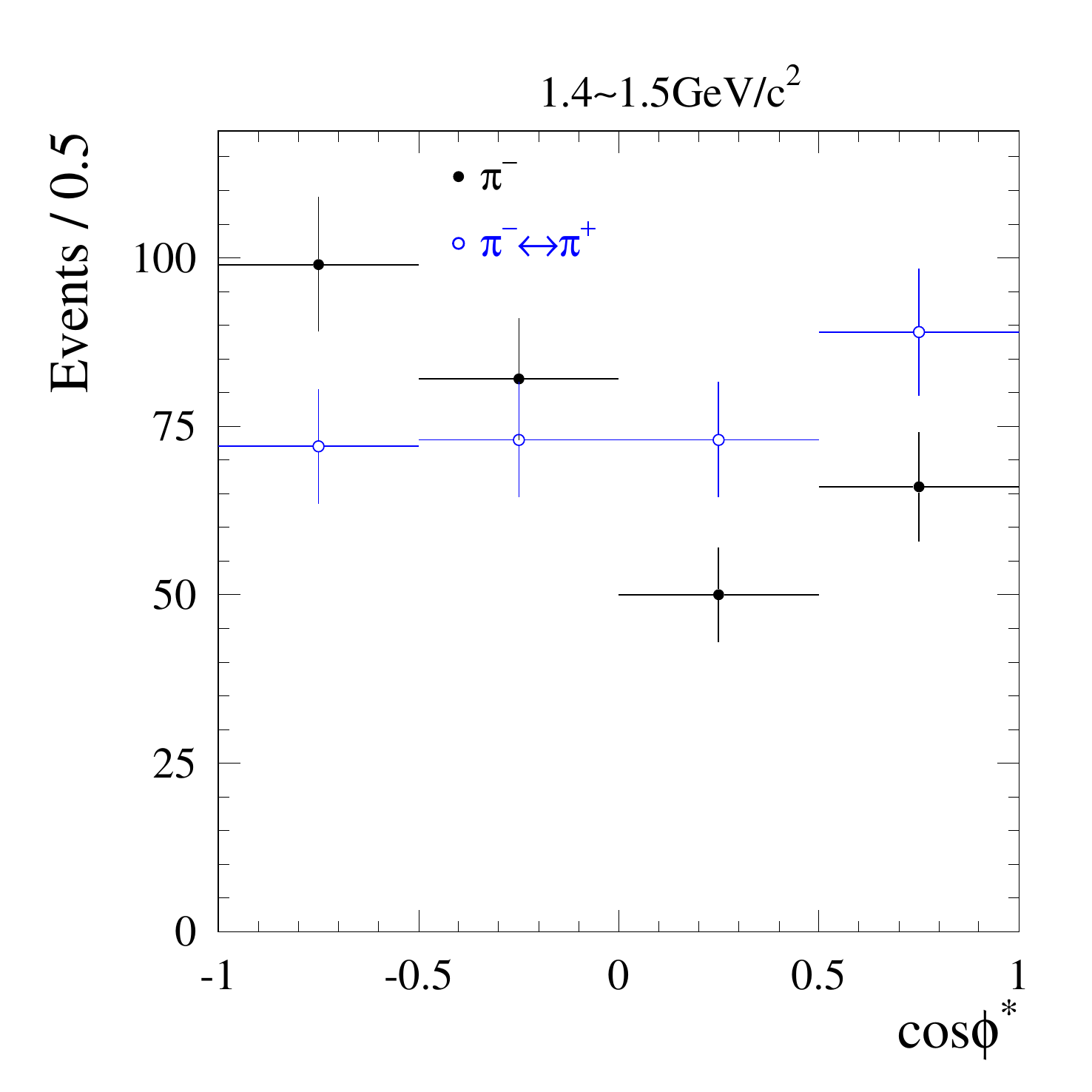}
  \includegraphics[width=0.25\textwidth]{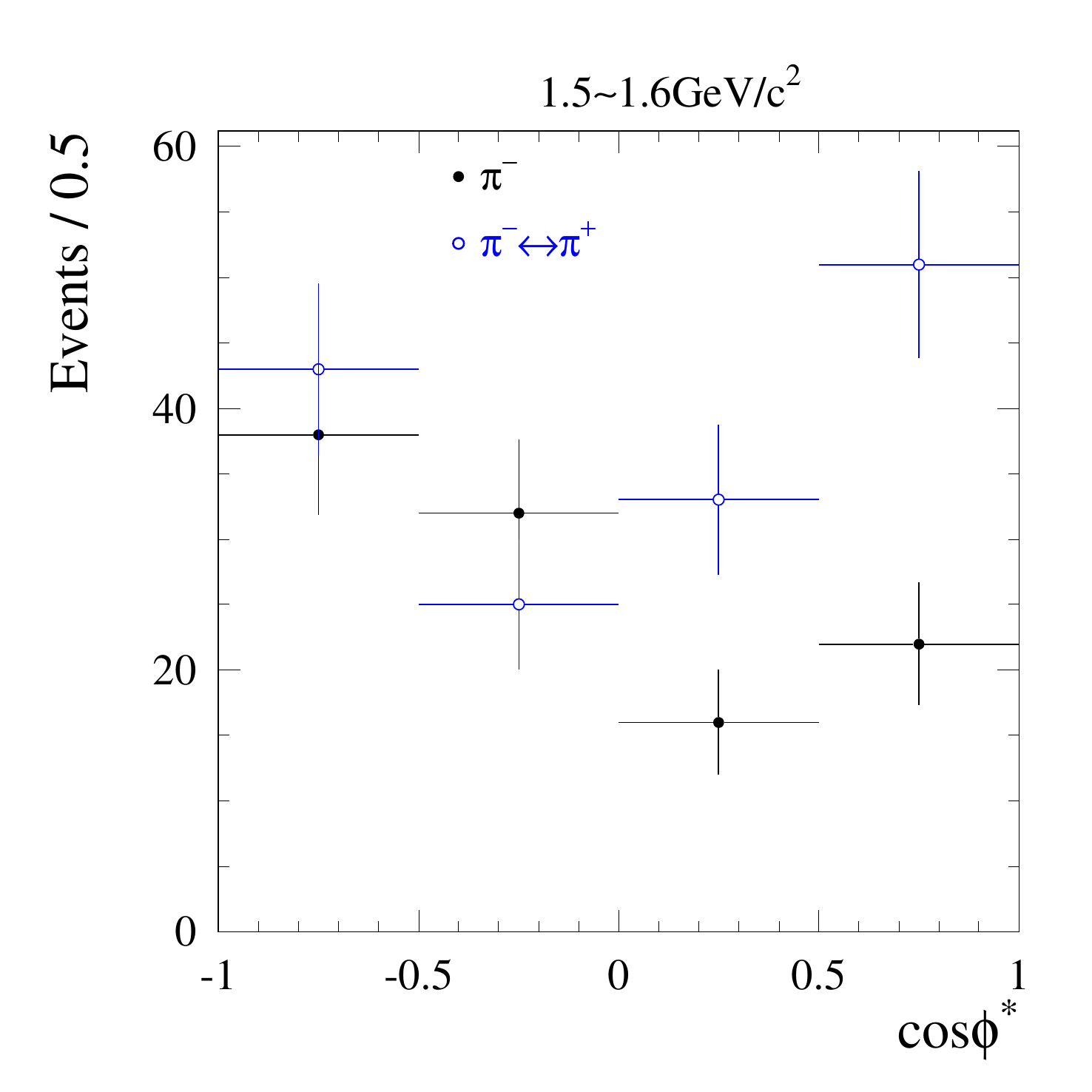}
  \includegraphics[width=0.25\textwidth]{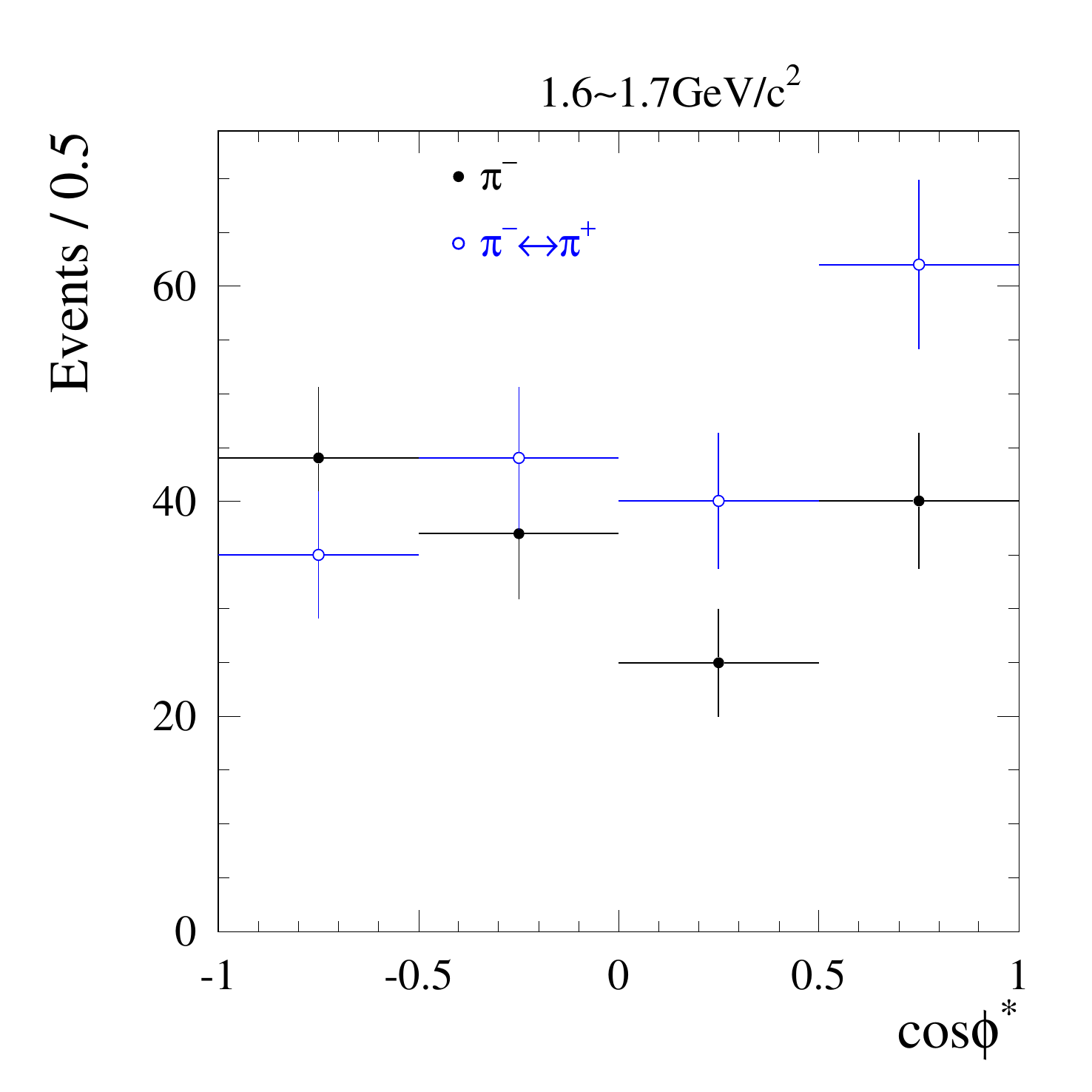}
  \includegraphics[width=0.25\textwidth]{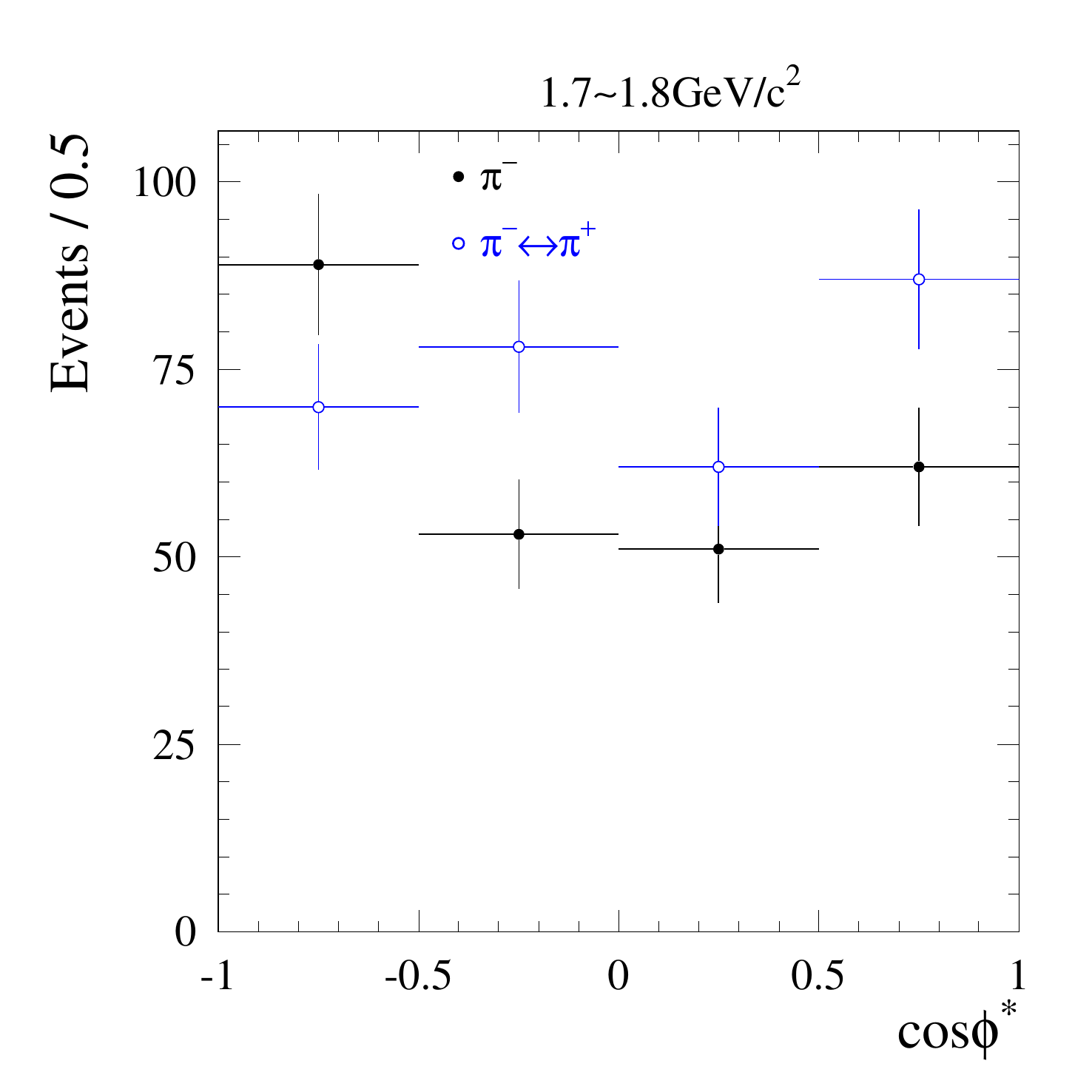}
  \caption{The $\cos\phi^*$ distributions for $\epem\to\pipig$ data in $0.1\gevcc$ 
   $m_{\pi\pi}$ intervals. The points labeled `$\pi^-$'
  refer to the configurations with $\phi^*_-\in[0,\pi]$, while the points
  labeled `$\pi^-\leftrightarrow\pi^+$' correspond to $\phi^*_+\in[0,\pi]$.}
  \label{fig:cosphis_pipig}
\end{figure*}

The $\cos\phi^*$ distributions $N^{\rm{obs}}_\pm(\cos\phi^*)$ for $\epem\to\mmg$ 
obtained in data after the overall event selection are shown in
Fig.~\ref{fig:cosphis_mmg}, for $\mu\mu$ mass intervals ranging from threshold 
to $7\gevcc$.  The event distributions $N^{\rm{obs}}_\pm(\cos\phi^*)$ obtained for 
$\epem\to\pipig$ in data are shown in Fig.~\ref{fig:cosphis_pipig}, in $0.1\gevcc$
mass intervals ranging from $0.3\gevcc$ to $1.8\gevcc$.

\begin{figure*}
  \centering 
  \includegraphics[width=0.32\textwidth]{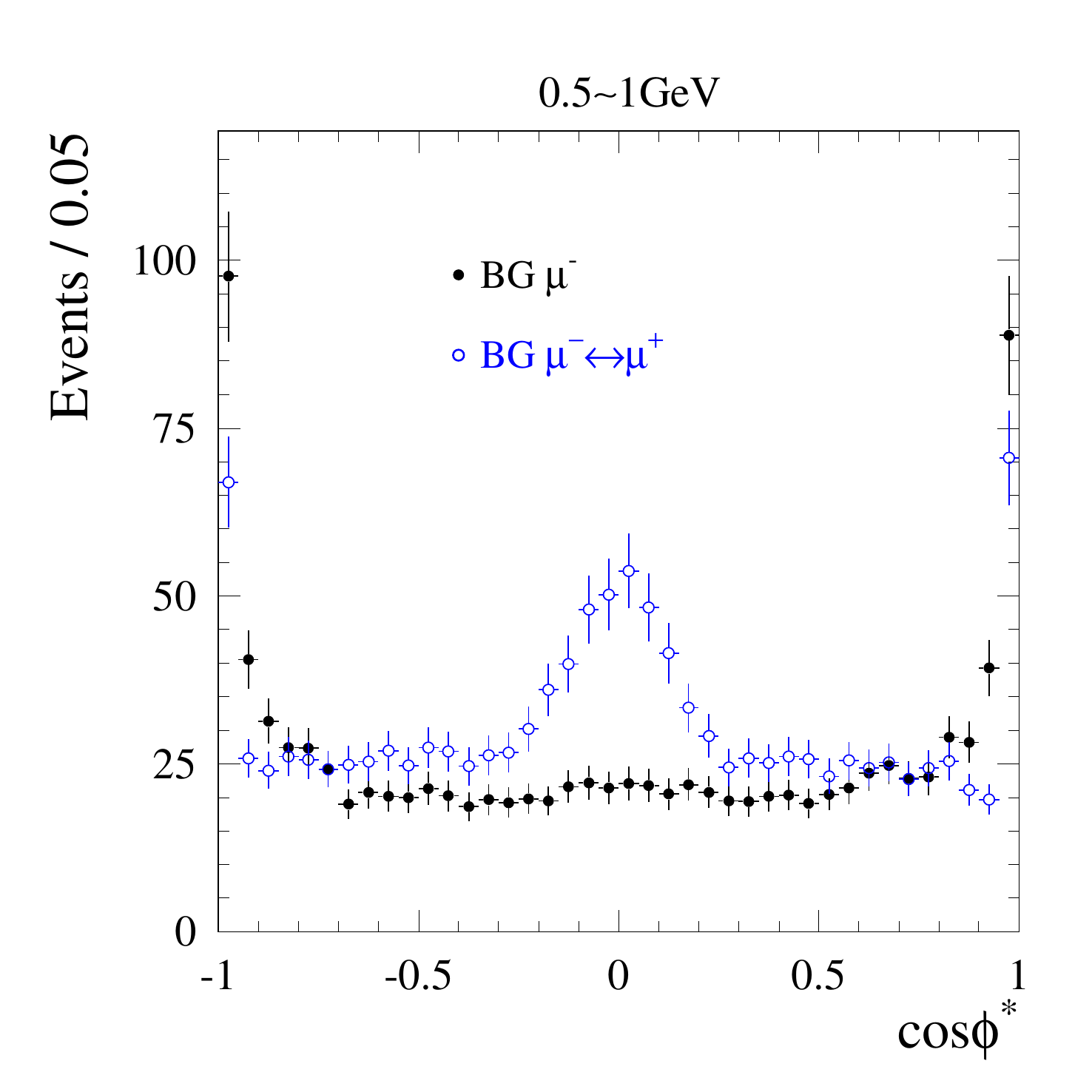}
  \includegraphics[width=0.32\textwidth]{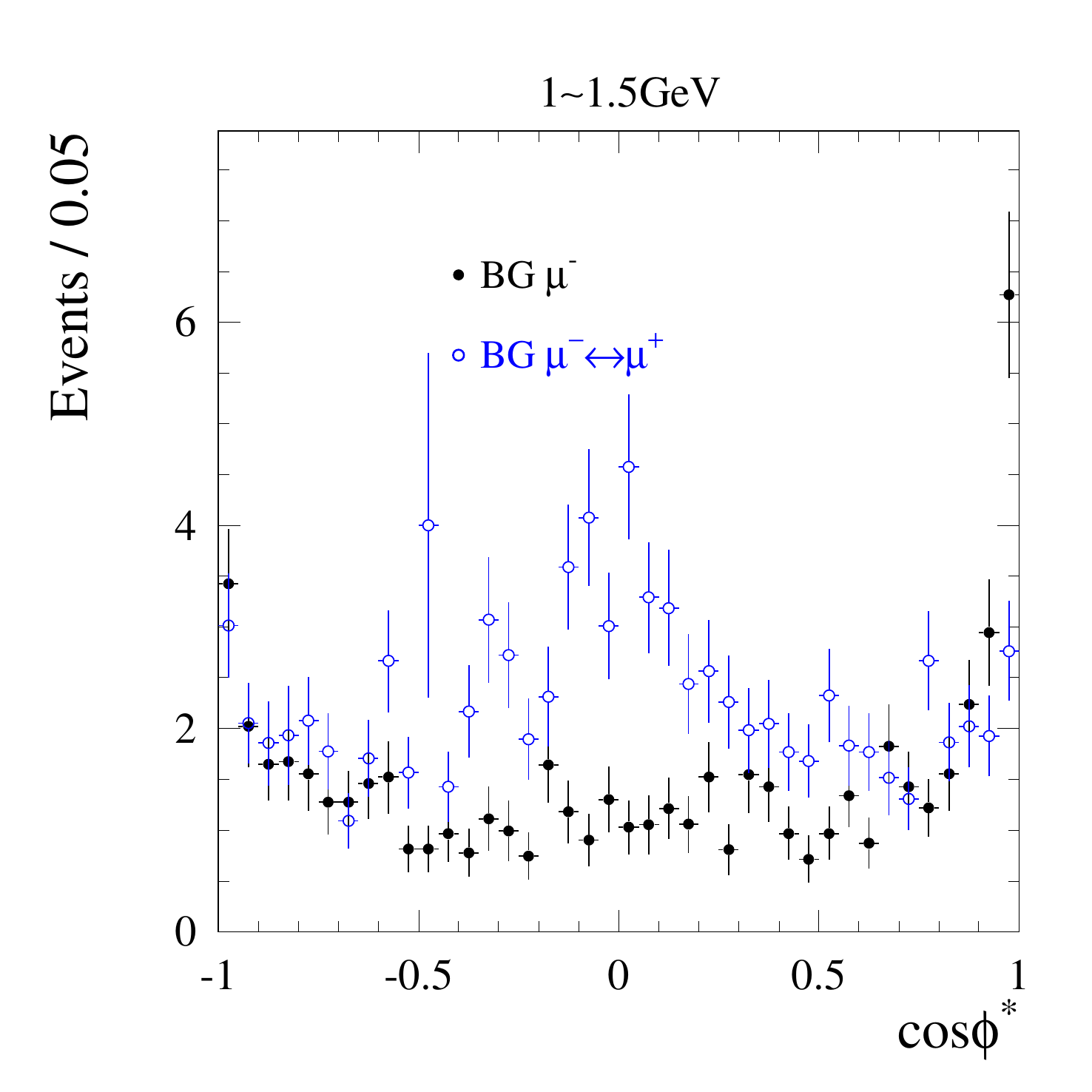}
  \includegraphics[width=0.32\textwidth]{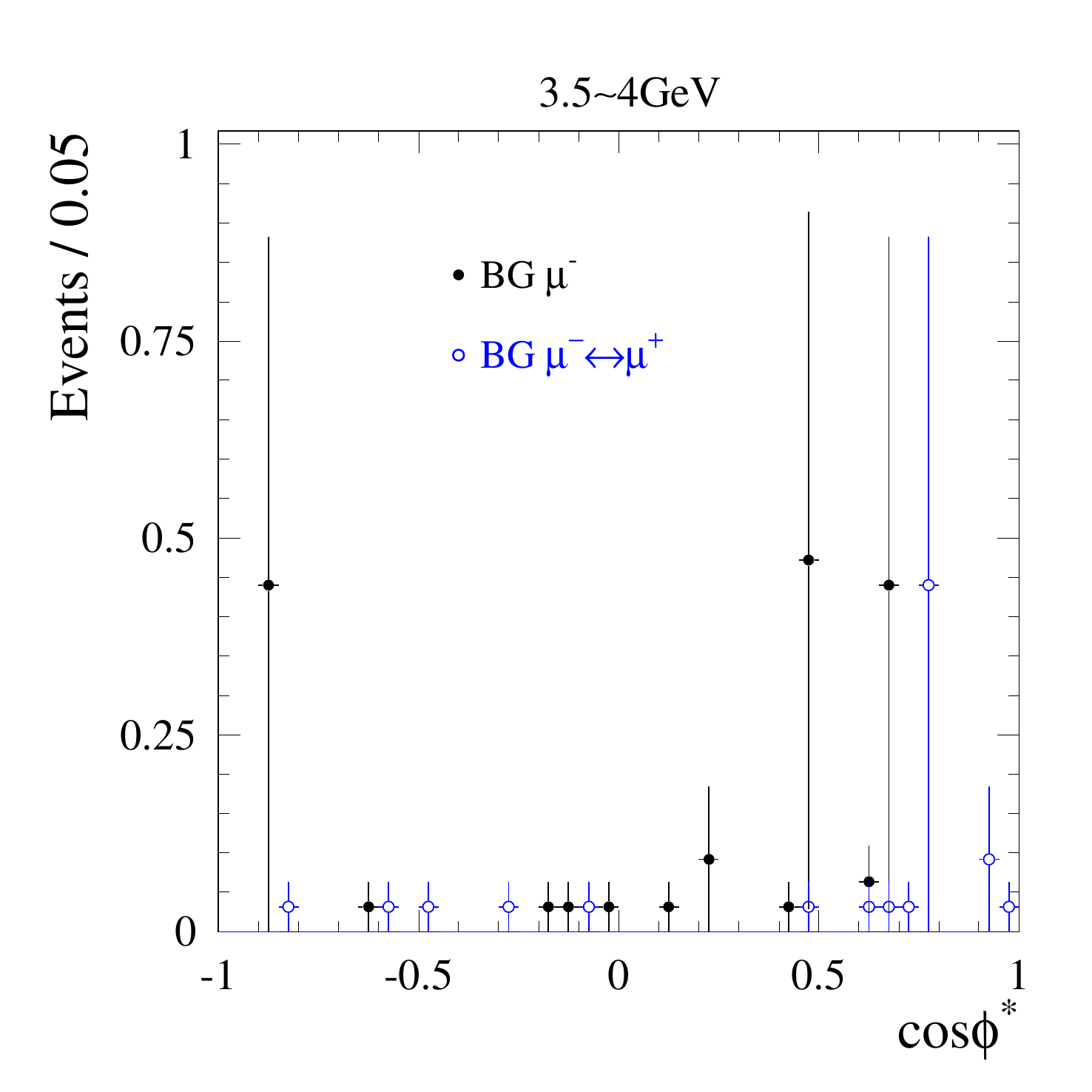}
  \caption{Backgrounds estimated with MC for $\epem\to\mmg$ as a function of
  $\cos\phi^*$ in selected $m_{\mu\mu}$ intervals. 
  The points labeled `$\mu^-$'
  refer to the configurations with $\phi^*_-\in[0,\pi]$, while the points
  labeled `$\mu^-\leftrightarrow\mu^+$' correspond to $\phi^*_+\in[0,\pi]$.
}
  \label{fig:bg_mu}
\end{figure*}

\begin{figure*}
  \centering 
  \includegraphics[width=0.32\textwidth]{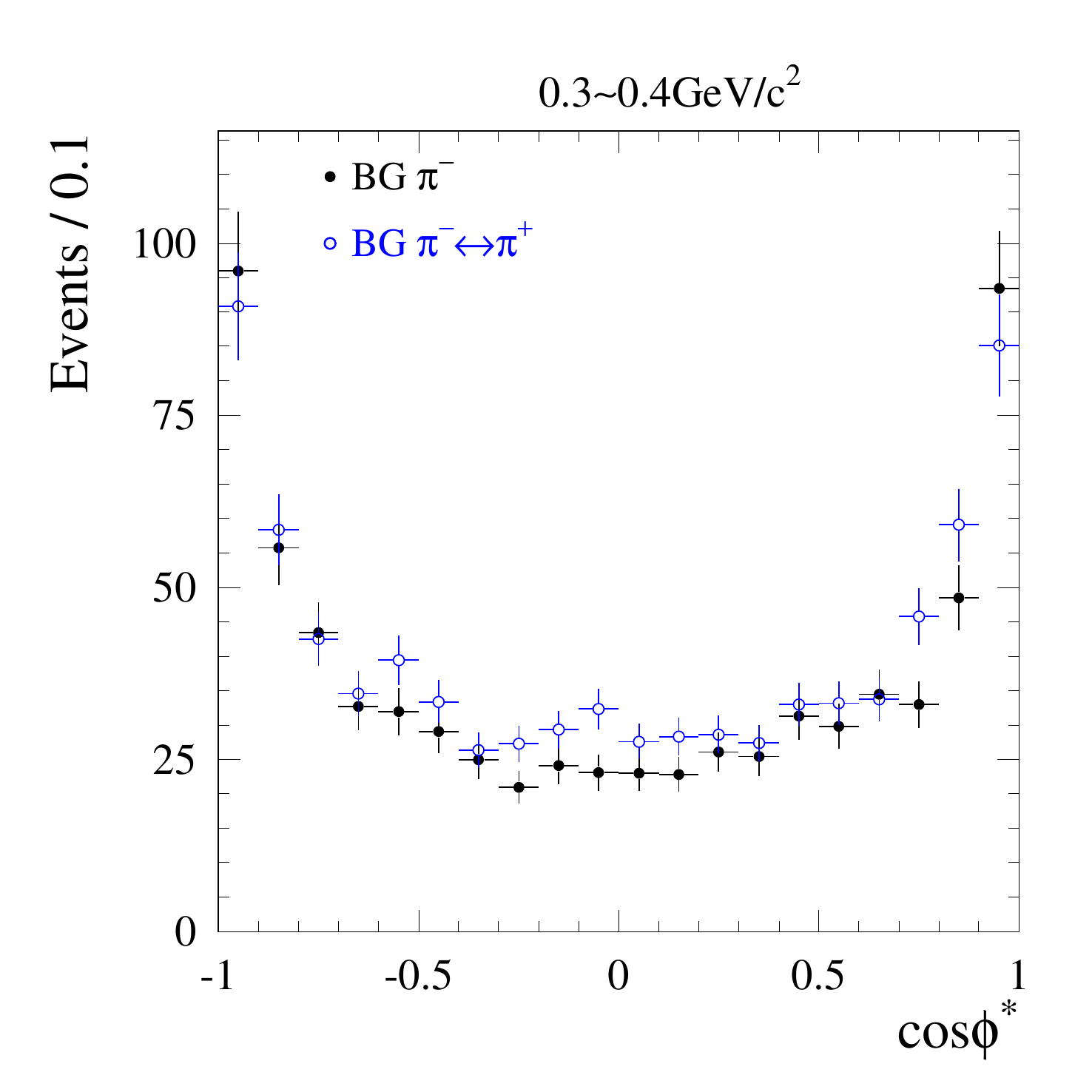}
  \includegraphics[width=0.32\textwidth]{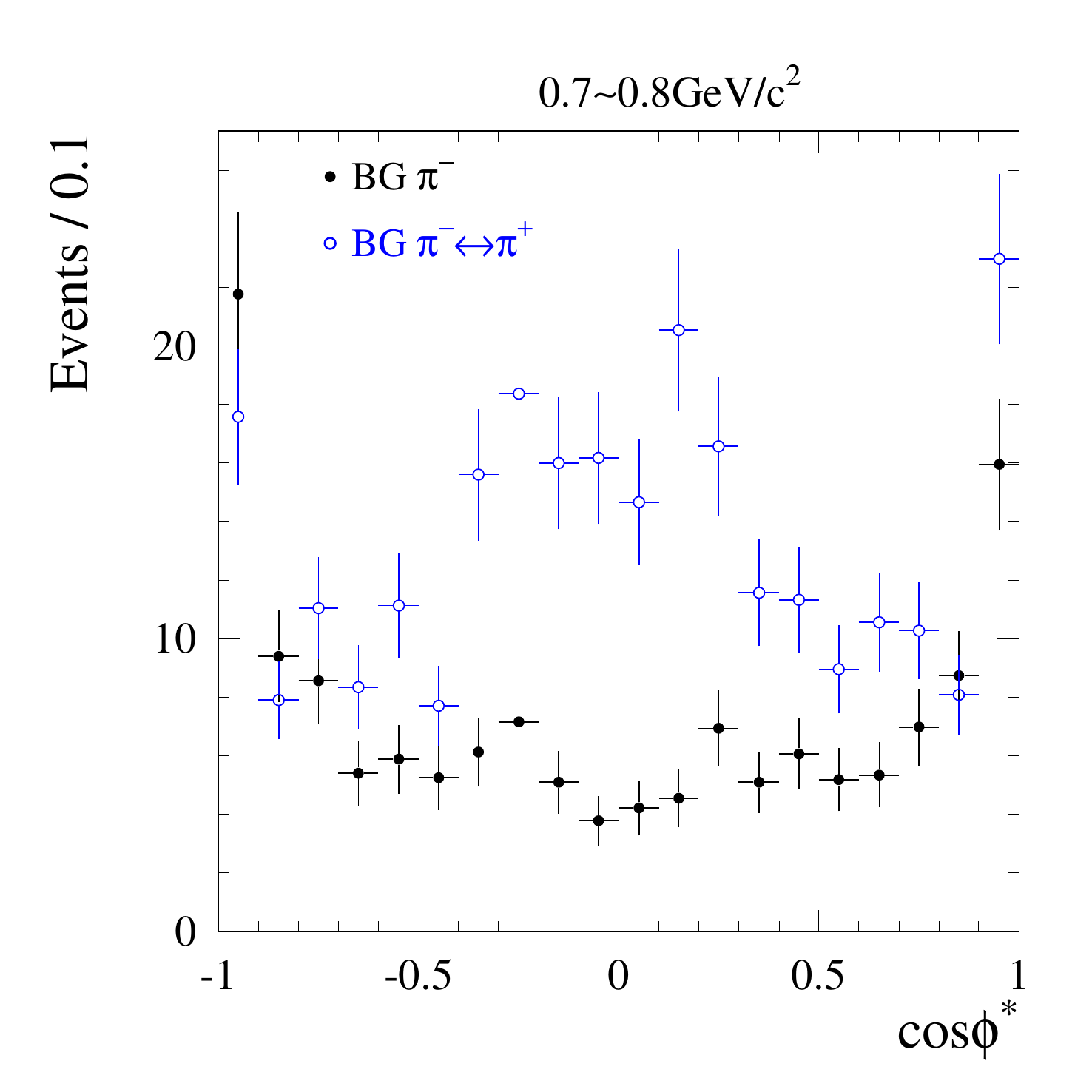}
  \includegraphics[width=0.32\textwidth]{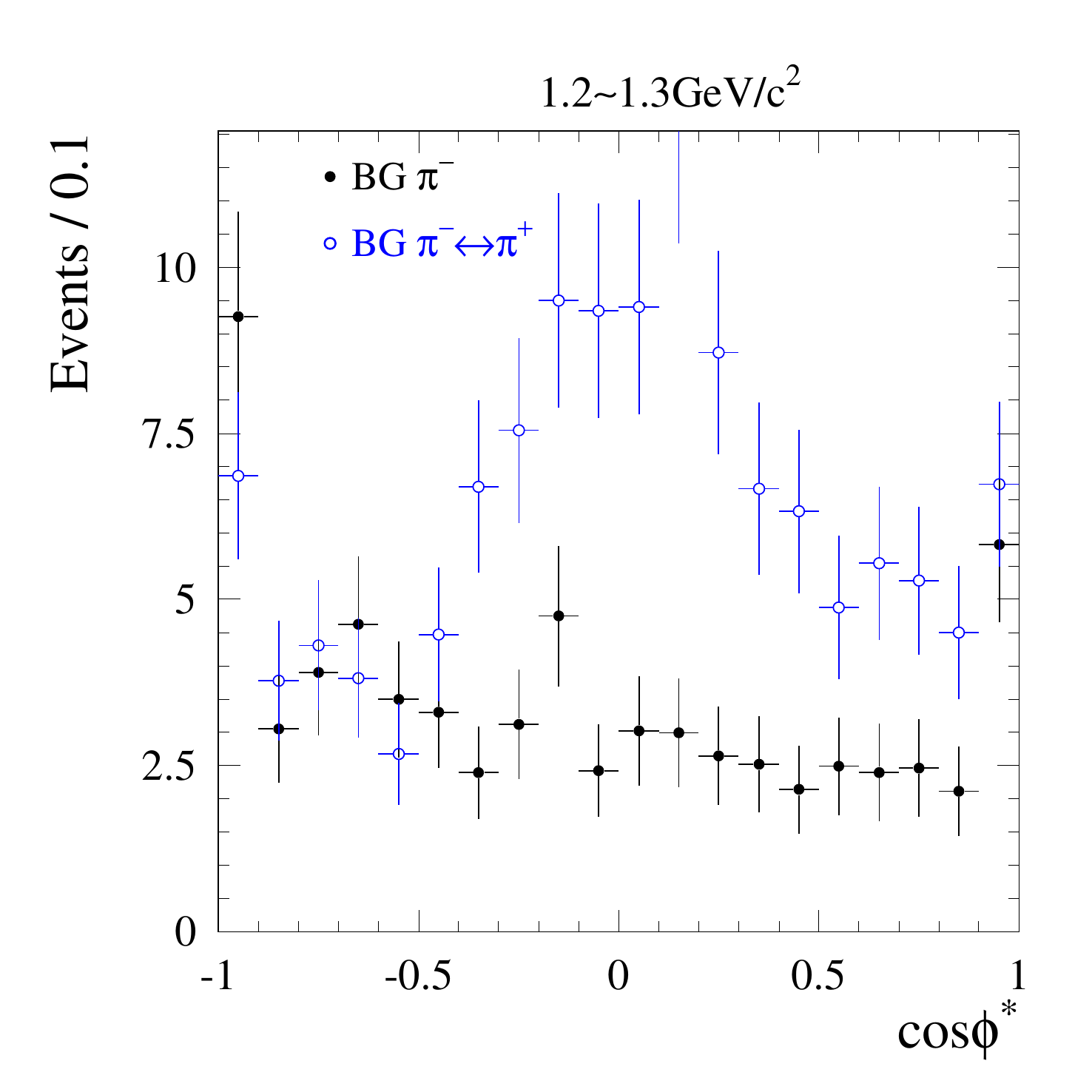}
  \caption{Backgrounds estimated with MC for $\epem\to\pipig$ as a function of
  $\cos\phi^*$ in selected $m_{\pi\pi}$ intervals.   
  The points labeled `$\pi^-$'
  refer to the configurations with $\phi^*_-\in[0,\pi]$, while the points
  labeled `$\pi^-\leftrightarrow\pi^+$' correspond to $\phi^*_+\in[0,\pi]$.}
  \label{fig:bg_pi}
\end{figure*}

The backgrounds remaining after selection are estimated using the full
simulation, normalized to the data luminosity, of the non-signal two-prong ISR
events, multi-hadron events produced through ISR, $\epem \to q\bar{q}$ events,
and $\tau^+\tau^-$ events. The expected contamination for $\epem\to\mmg$ as a
function of $\cos\phi^*$ in typical $m_{\mu\mu}$ intervals is shown in
Fig.~\ref{fig:bg_mu}, where the total error is the quadratic sum of the
statistical error and 10\% systematic uncertainty on normalization~\cite{prd-pipi}.
Likewise, the estimated backgrounds for $\epem\to\pipig$ in typical $m_{\pi\pi}$ intervals are shown
in Fig.~\ref{fig:bg_pi}. 

\section{Acceptance and detector efficiency effects on the charge asymmetry}
\label{sec:effectAcc}

The charge asymmetry measurement is affected by the event reconstruction and
selection. These experimental effects are investigated using the full simulation
of signal events through changes of the raw charge asymmetry, defined as
$A^{\rm raw}=(N_- - N_+)/(N_- + N_+)$, which are observed after each selection step as a 
function of $\cos\phi^*$. 

\subsection{Study of the effects with the muon simulation}
\label{sec:effectAcc_mu}

\subsubsection{Kinematic acceptance}
\label{sec:kinAccep}

The kinematic acceptance includes the angular acceptance for the primary photon
and the two charged-particle tracks, and the momentum restriction ($p>1\gevc$) applied to
charged-particle tracks. Each kinematic selection is found to modify the slope of the raw charge asymmetry
significantly, though the total effect on the slope from the kinematic requirements
altogether turns out to be small due to accidental cancelations.

It is worthwhile to note that the kinematic selection in itself is
charge-symmetric. Hence the observed bias on the measured raw charge asymmetry is a
cross effect of physical charge-asymmetric kinematics and charge-symmetric
detector acceptance. It does vanish for a null physical charge asymmetry.  As
checked with a $\mmg$ simulated sample produced by ISR only, no fake charge asymmetry
emerges from the kinematic selection.

\subsubsection{Software trigger and tracking}

\begin{figure*}
  \centering
\includegraphics[height=0.32\textwidth]{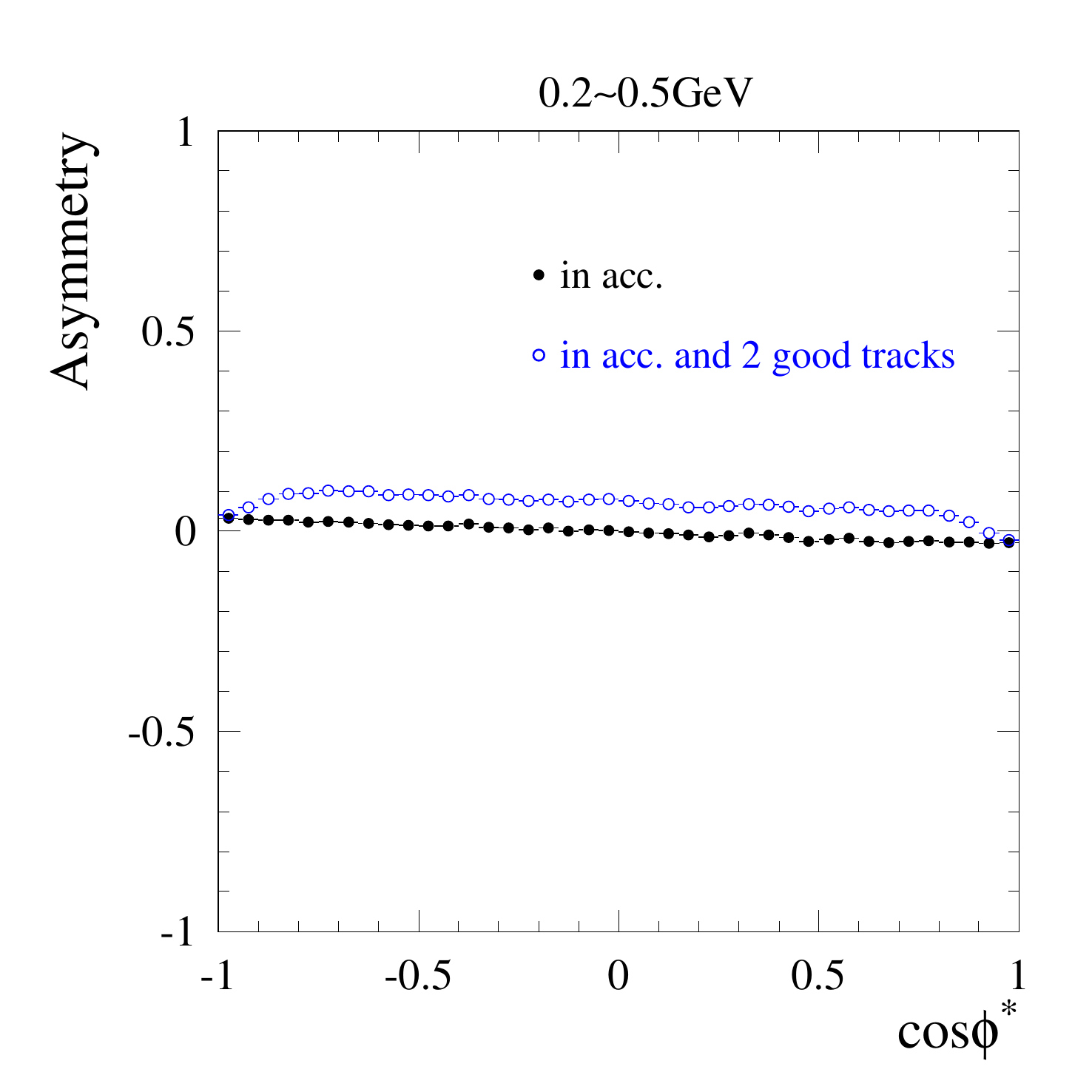}
\includegraphics[height=0.32\textwidth]{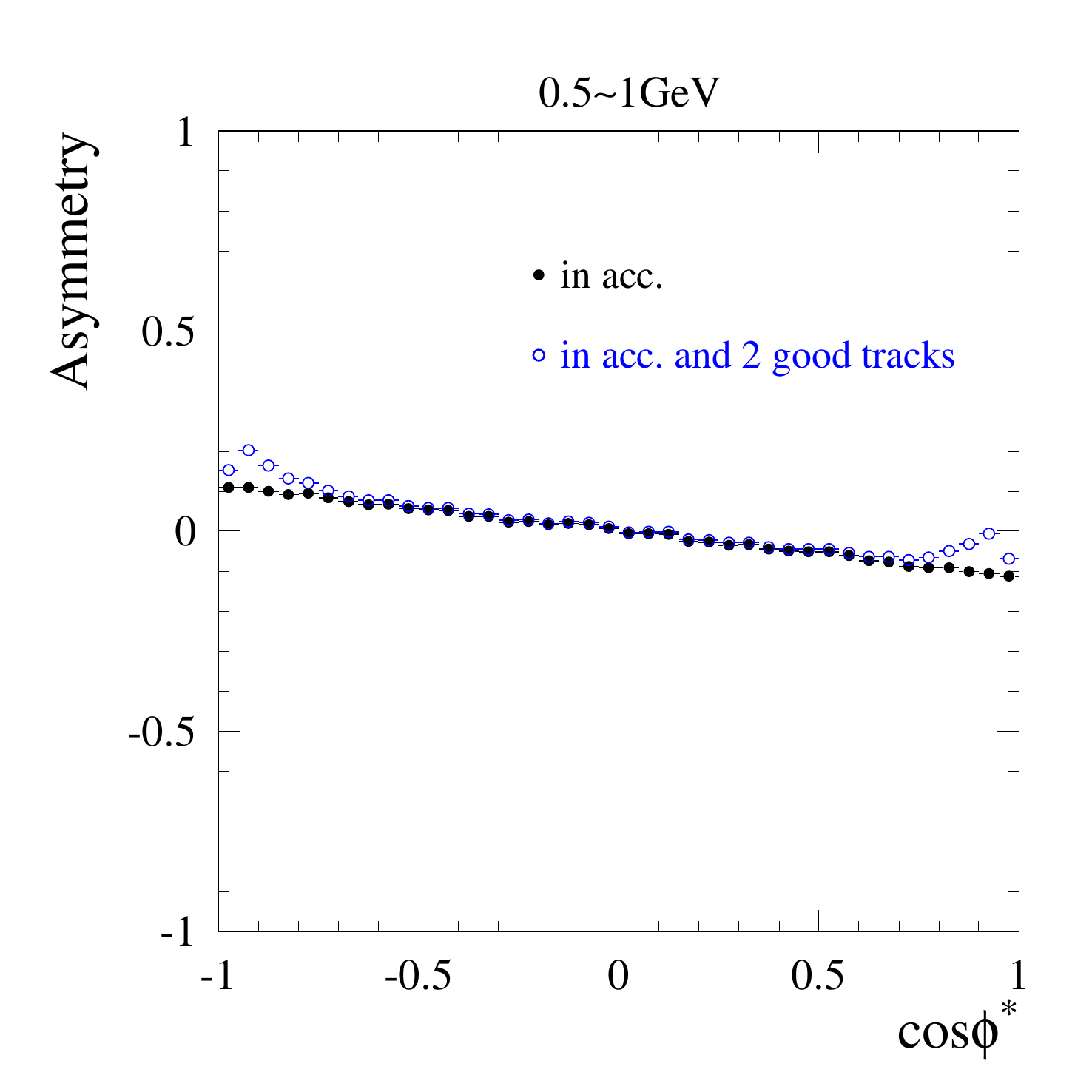}
\includegraphics[height=0.32\textwidth]{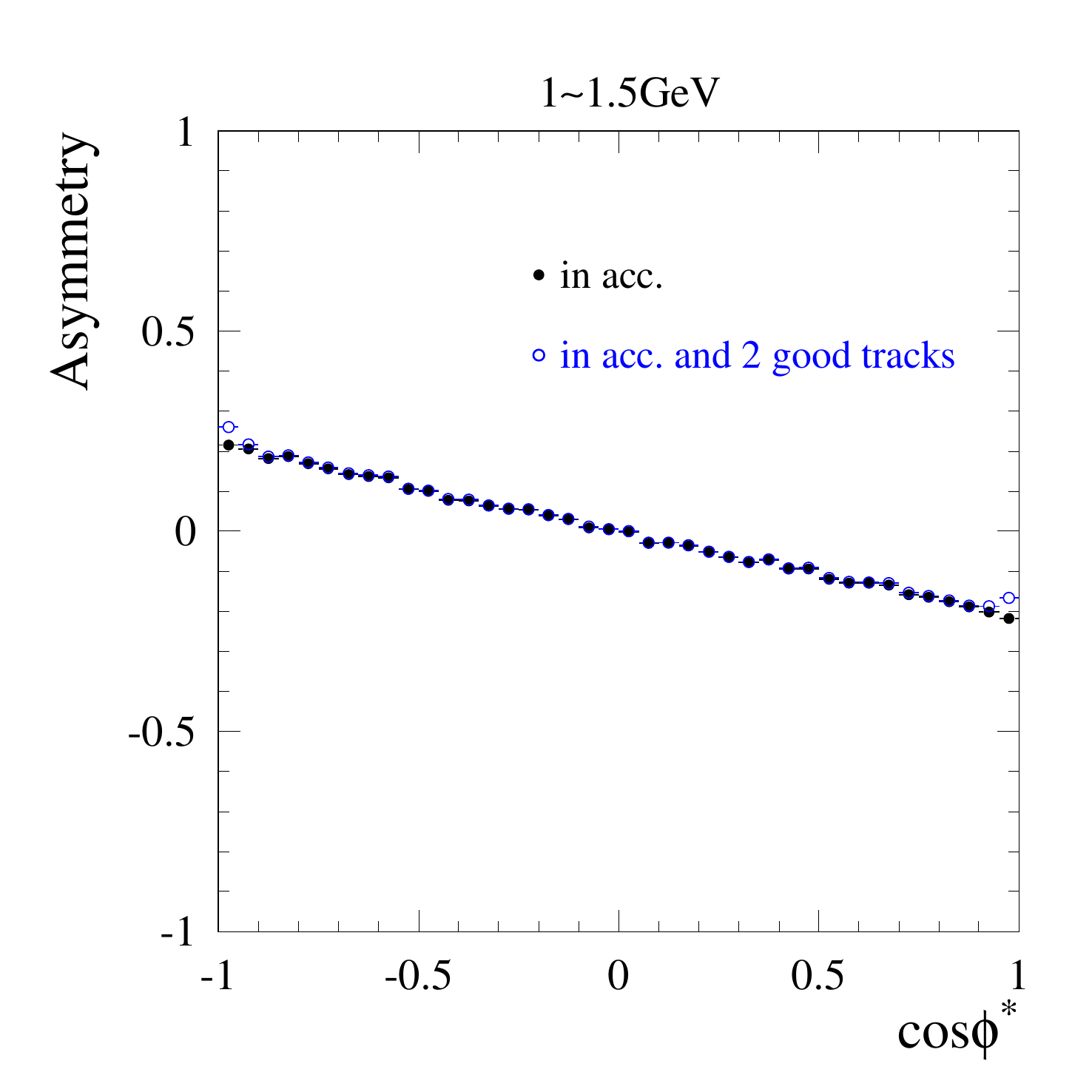}
  \caption{Raw charge asymmetry as a function of $\cos\phi^*$  in selected $m_{\mu\mu}$ intervals, 
    for $\epem\to\mmg$ MC events with ($\circ$) and without ($\bullet$) reconstruction
    of two good charged-particle tracks, where the events are already required to be within
    the kinematic acceptance.  }
  \label{fig:A-MCTruth-trking}
\end{figure*}

Biases on the raw charge asymmetry measurement originate from the software trigger and
the track reconstruction. They are observed in the low mass region, as illustrated in 
Fig.~\ref{fig:A-MCTruth-trking} 
and vanish
at high mass ($m_{\mu\mu}>1.5\gevcc$).

The common origin of the mass-dependent trigger and tracking inefficiencies
is geometrical and has been thoroughly studied for the $\pipi$ cross section
measurement~\cite{prd-pipi}.  Converging trajectories in the DCH of oppositely
deflected tracks emitted in close-by directions confuses the track
reconstruction and causes both the software trigger and the final
tracking inefficiencies. In the charge-conjugate configuration, in
which the positive and negative tracks are interchanged, tracks
diverge in the magnetic field and are well separated in the transverse
plane, although with the same absolute azimuthal opening angle.  The
efficiencies are consequently charge-asymmetric, sharply reduced for
overlapping tracks in low $m_{\mu\mu}$ regions, at $\Delta\phi$ values
close to zero but always positive, where $\Delta\phi$ is the signed
angular difference between the azimuths of the positive and negative
tracks 
\beqn \Delta\phi=(\phi_+-\phi_-)\in[-\pi,\pi].  \eeqn


\subsubsection{`ISR' photon reconstruction}
\label{sec:ISRphoton}

\begin{figure*}
\centering
\includegraphics[width=0.32\textwidth]{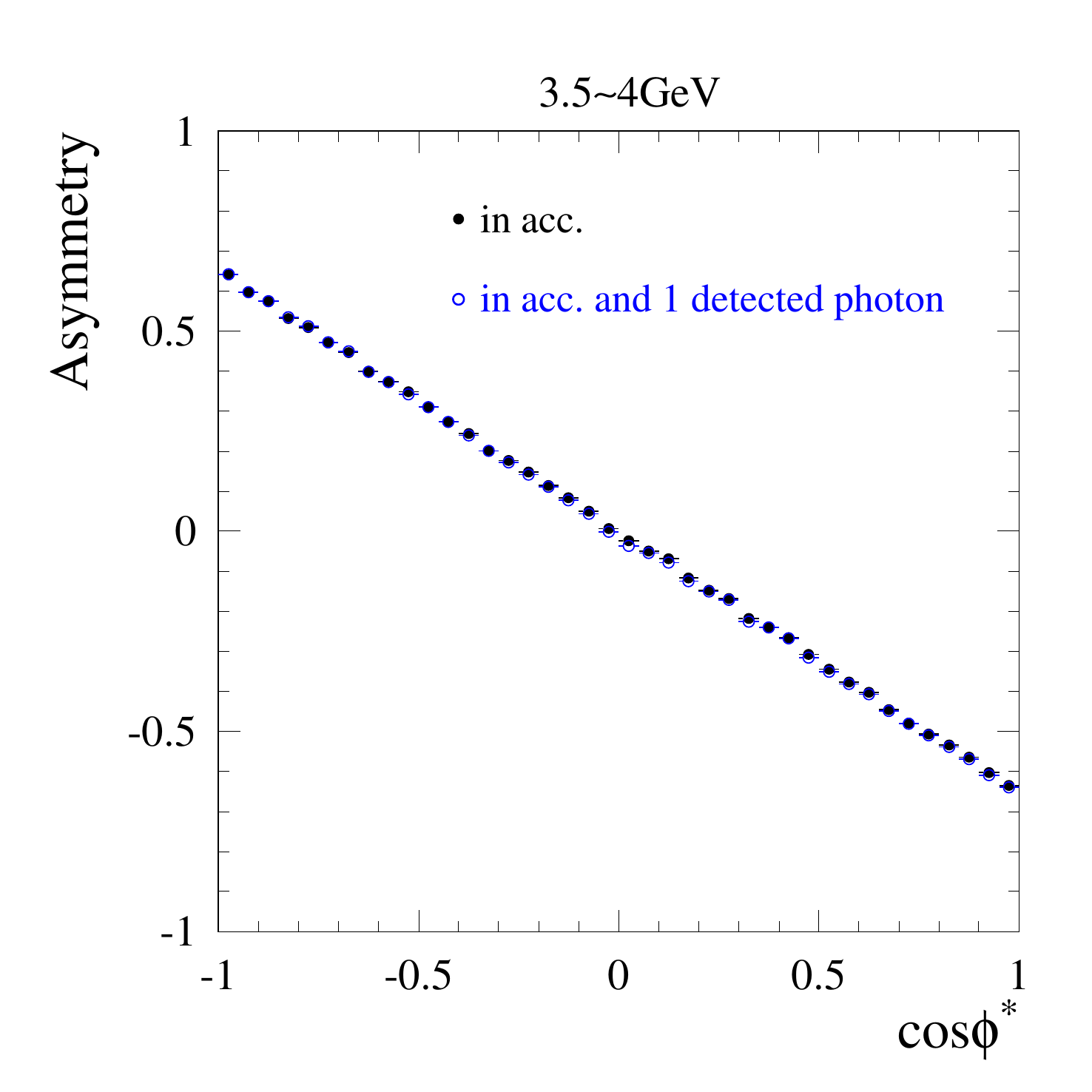}
\includegraphics[width=0.32\textwidth]{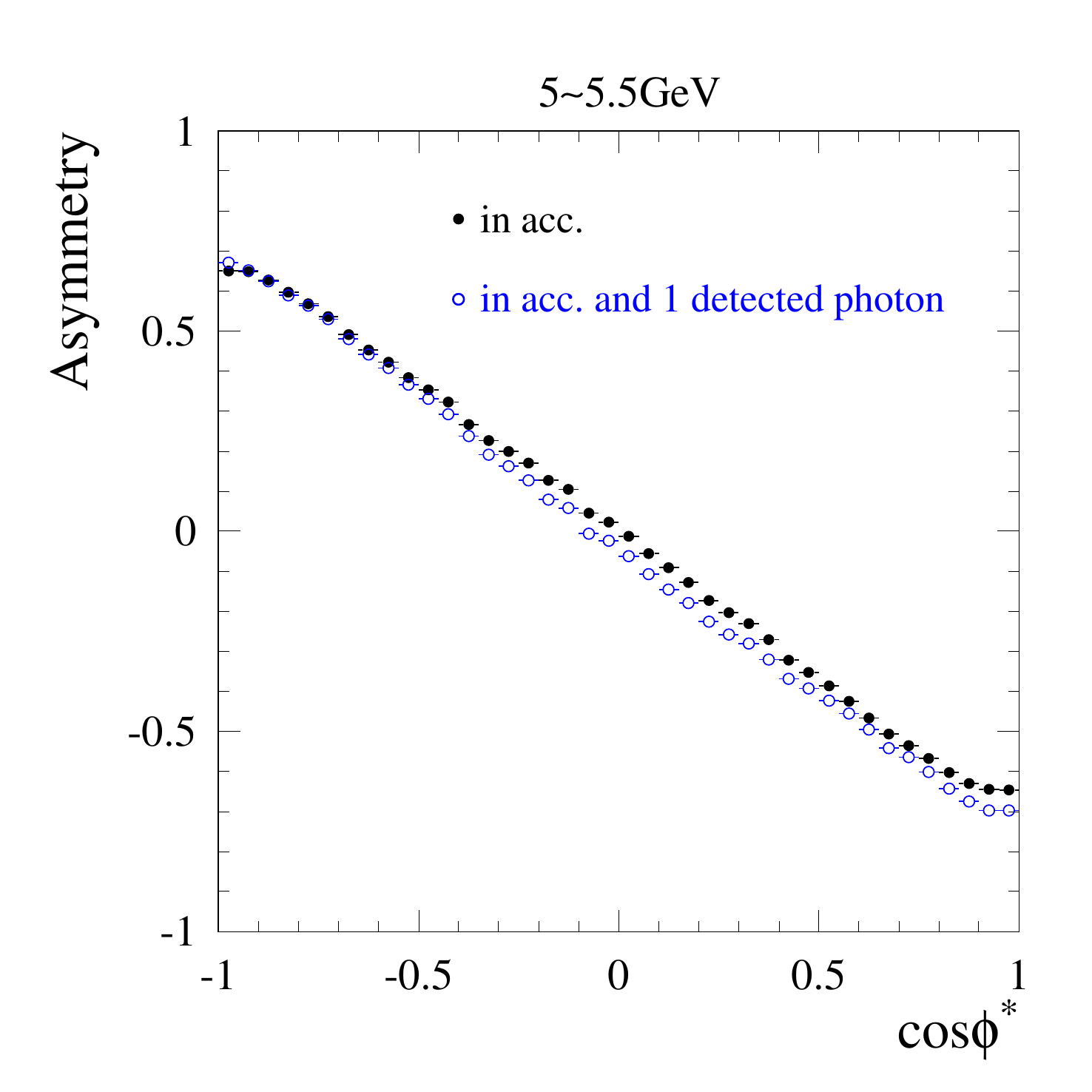}
\includegraphics[width=0.32\textwidth]{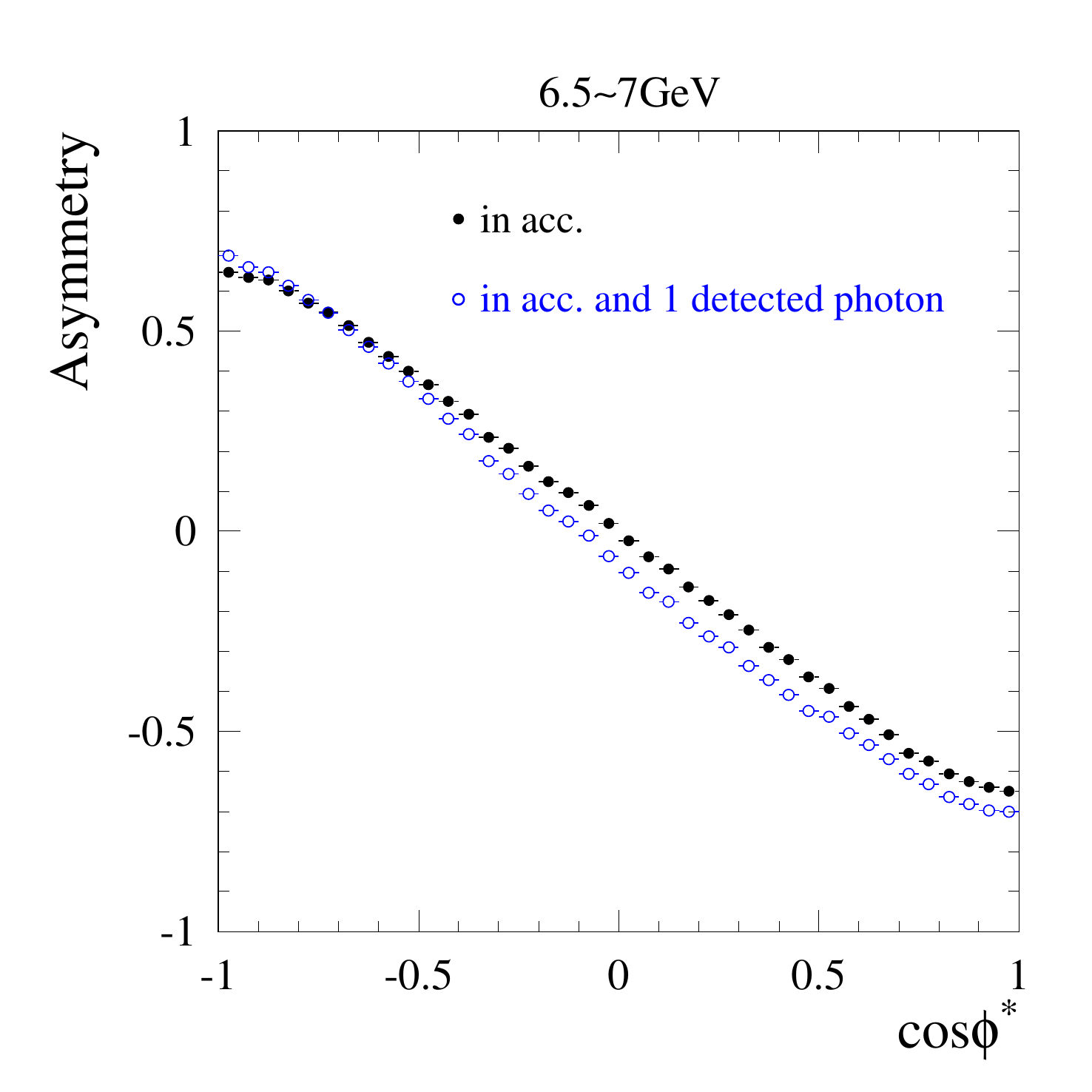}
\caption{Raw charge asymmetry as a function of $\cos\phi^*$ in selected $m_{\mu\mu}$ intervals,
for $\epem\to\mmg$ MC events with ($\circ$) and without ($\bullet$) the requirement
of the reconstruction of the `ISR' photon, where the events are already required
to be within the kinematic acceptance.}
\label{fig:A-MCTruth-isrG}
\end{figure*}

The event selection requires that an `ISR' photon with $E^*_\gamma>3\gev$ 
be measured in the EMC. The raw charge asymmetries for the fully simulated events
with and without the requirement of the `ISR' photon being reconstructed are shown in
Fig.~\ref{fig:A-MCTruth-isrG}, where the events are already required to be
within the kinematic acceptance. Effects are observed in high $m_{\mu\mu}$
regions.

The origin of a charge-asymmetric photon reconstruction inefficiency is again
geometrical. In case one of the charged-particle tracks and the `ISR' photon overlap in the
EMC, the shower produced by the `ISR' photon is mistakenly associated to the
charged-particle track, and the `ISR' photon is lost.  In the charge-conjugate
configuration, no overlap occurs because of the opposite deflection of the
charged-particle track in the magnetic field. The overlap happens at
$\Delta\phi=-\pi+\varepsilon$ where $\varepsilon$ is a small positive quantity,
and as a consequence, the `ISR' photon reconstruction efficiency is
charge-asymmetric, strongly reduced around  $\Delta\phi\sim-\pi$. As the overlap
of one charged-particle track and the `ISR' photon occurs preferentially at high mass, due
to phase space, the corresponding effects are only observed at
$m_{\mu\mu}>3.5\gevcc$.

\subsubsection{Muon identification}

\begin{figure*}
\centering
\includegraphics[height=0.32\textwidth]{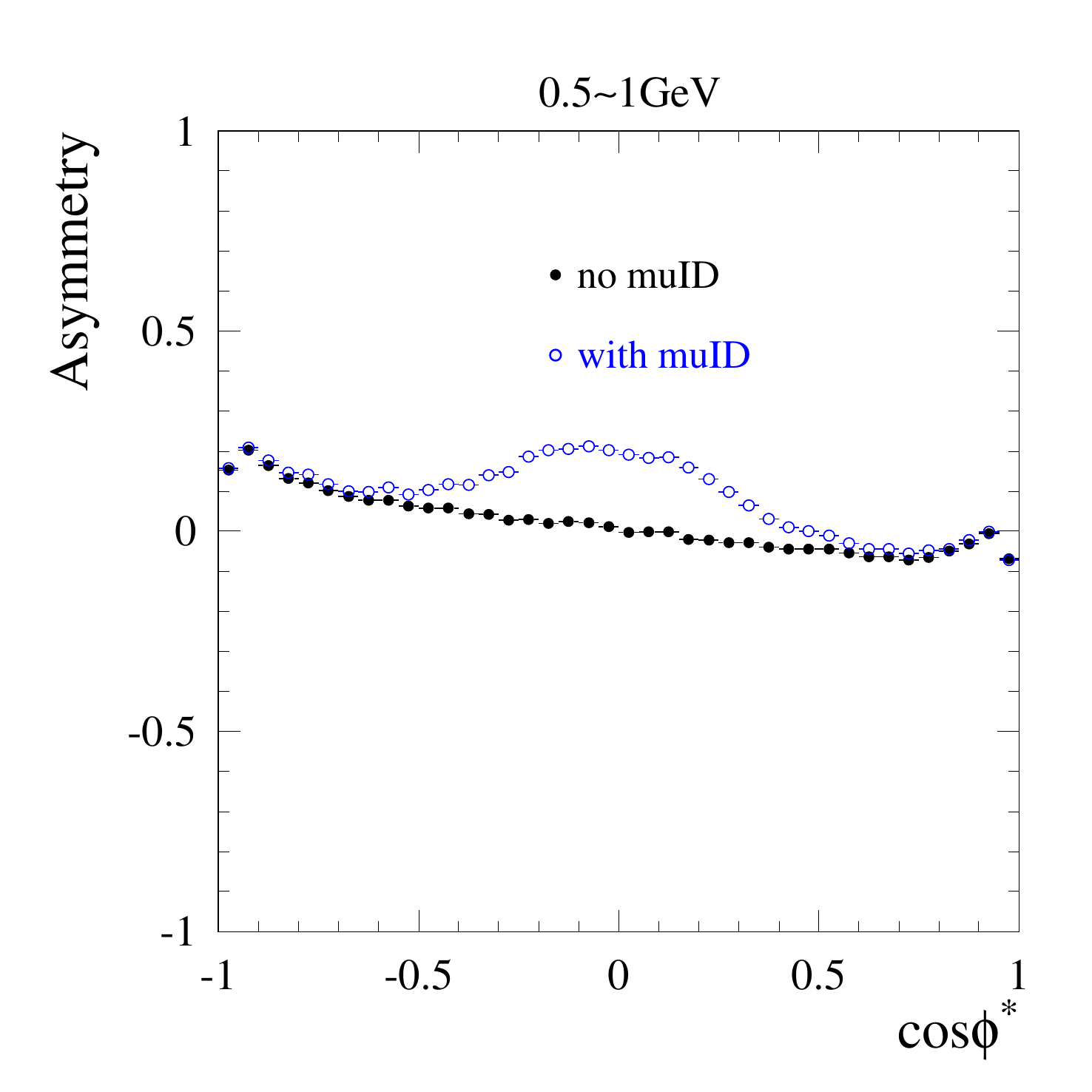}
\includegraphics[height=0.32\textwidth]{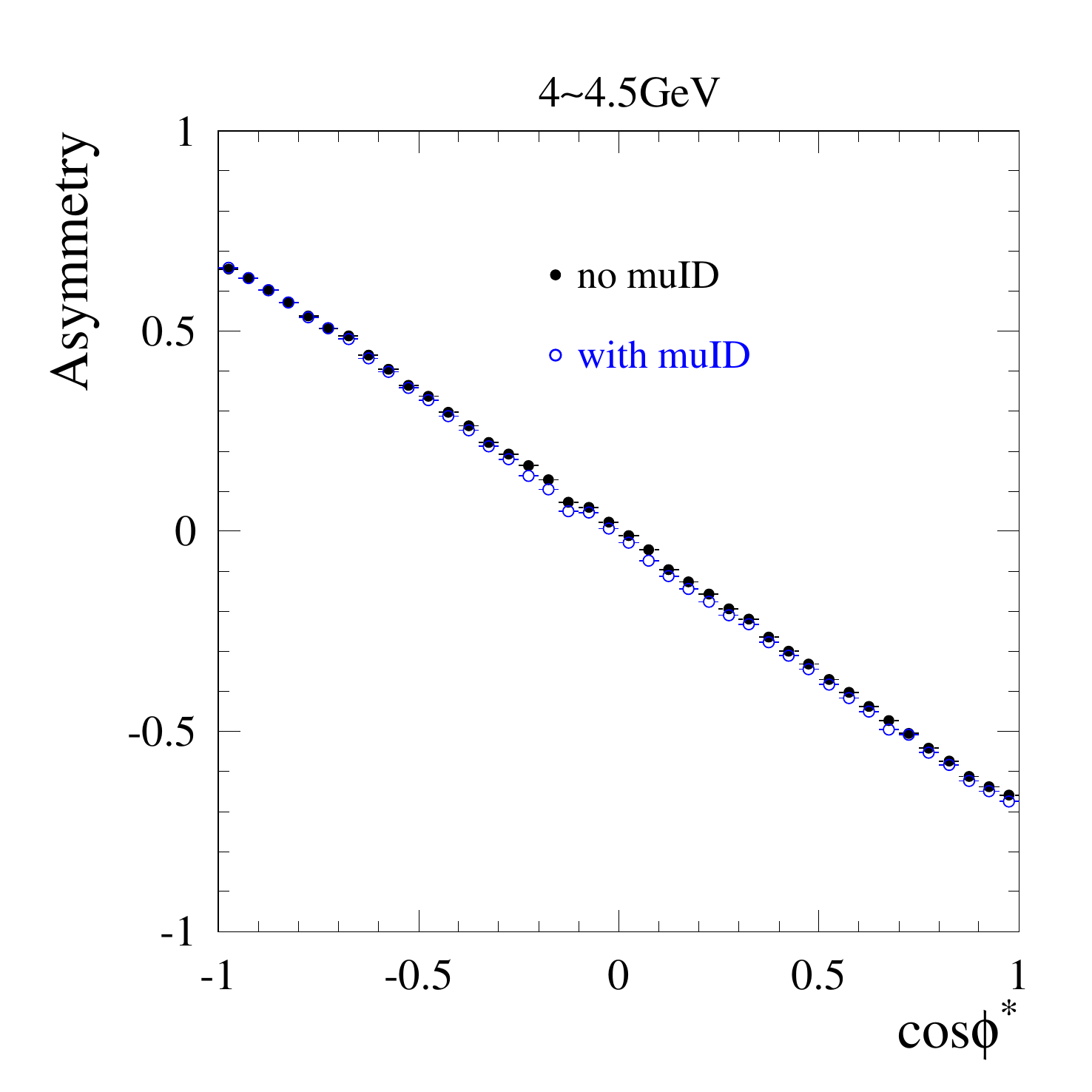}
\includegraphics[height=0.32\textwidth]{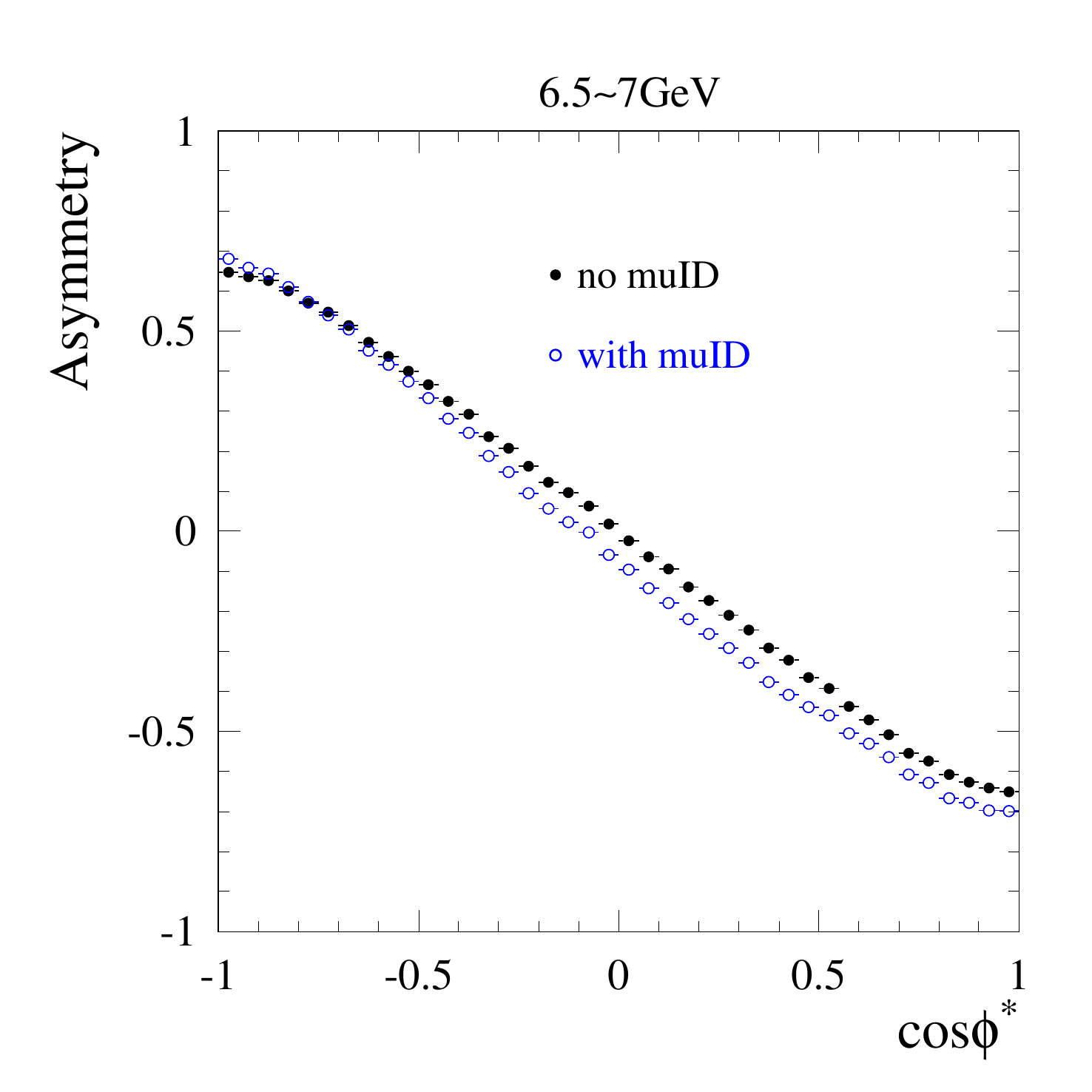}
\caption{Raw charge asymmetry as a function of $\cos\phi^*$ in selected $m_{\mu\mu}$ intervals,
for $\epem\to\mmg$ MC events with ($\circ$) and without ($\bullet$) the two-muon
identification,  where the events are already required to be within the
kinematic acceptance and have two good charged-particle tracks reconstructed.}
\label{fig:A-MCTruth-pid}
\end{figure*}

The charge asymmetry measurement is also affected by muon identification, in the
low and high $m_{\mu\mu}$ regions, as shown in Fig.~\ref{fig:A-MCTruth-pid}.
Charge-asymmetric inefficiency of muon identification results again from event
topologies. The first cause, which affects low $m_{\mu\mu}$ regions, is the
overlap of the two charged-particle tracks at the IFR, which confuses the muon
identification algorithm. The second cause is the partial overlap of one charged-particle
track and the `ISR' photon at the EMC, which makes the track look unlike a
muon. The latter effect is more pronounced at high mass, when a muon and the
`ISR' photon are emitted in close-by directions. The efficiency of muon
identification as a function of $\Delta\phi$
exhibits a sharp dip at positive $\Delta\phi$ at low mass, and at
$\Delta\phi\simeq-\pi$ at high mass.

\subsubsection{Summary of the acceptance and detector efficiency effects in the $\mmg$ process}

The overall efficiencies $\epsilon_{\pm}$ needed to correct the $N_{\pm}$ event yields
entering the charge asymmetry measurement (Eq.~(\ref{eq:A_cosphis}))
are the overall result of the acceptance-induced and detector asymmetries discussed above. They are
determined using the full simulation, separately for the $N_{\pm}$ samples.

\begin{figure*}
  \centering
  \includegraphics[width=0.45\textwidth]{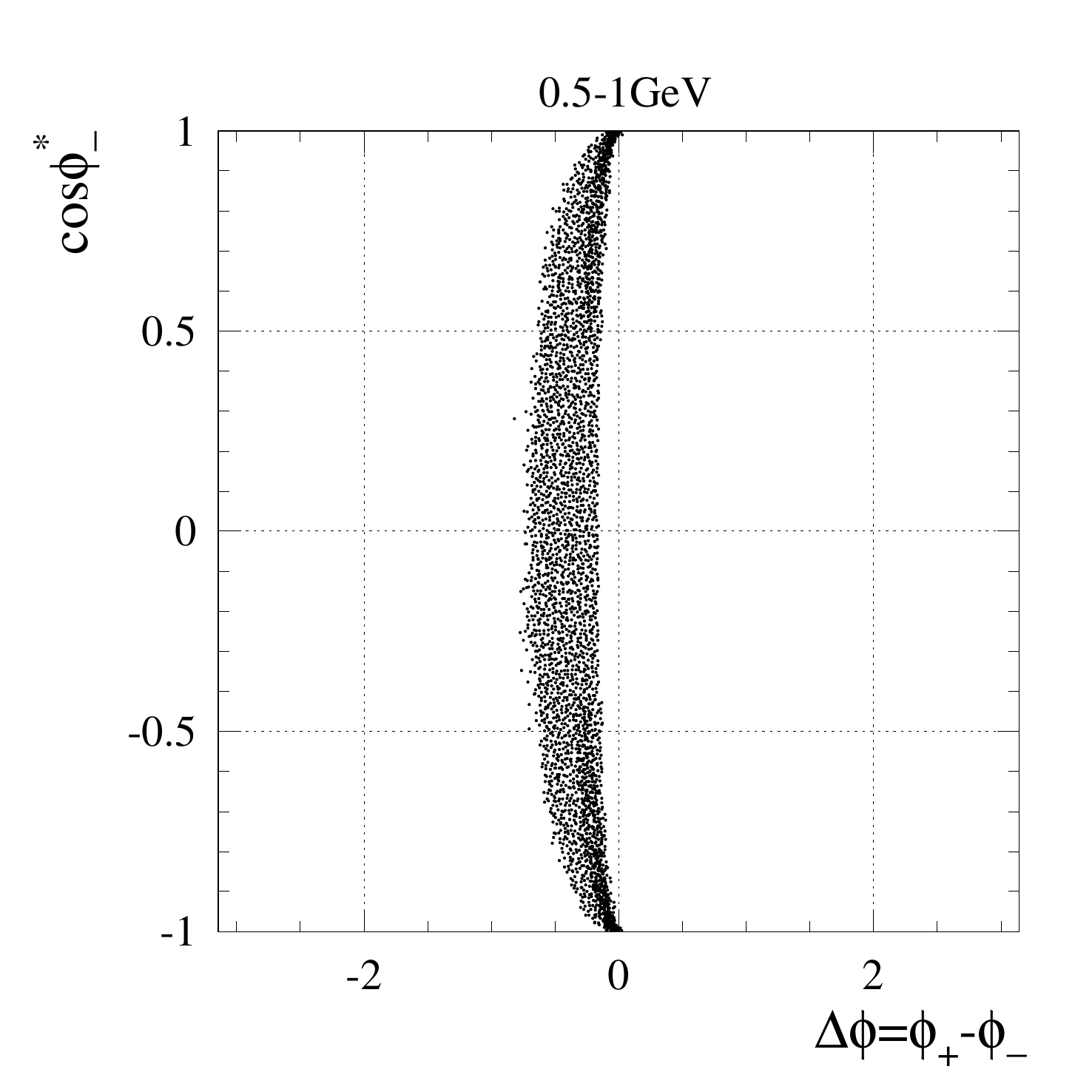}
  \includegraphics[width=0.45\textwidth]{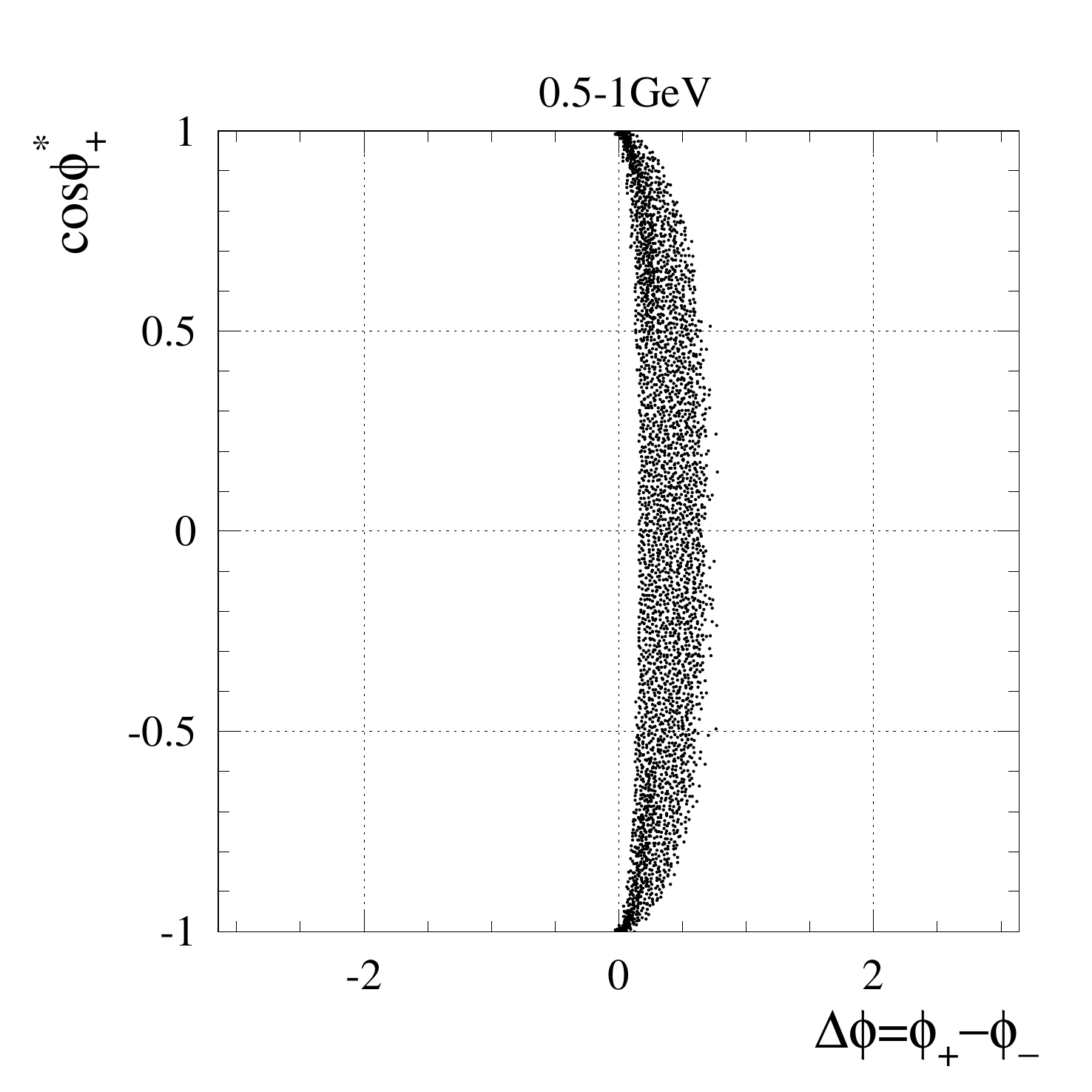}
  \caption{(left) Distribution of $\cos \phi^*_-$ vs. $\Delta\phi$ for the $N_-$ 
    sample ($\phi^*_-\in[0,\pi]$) in the (0.5--1.0)\gevcc mass interval for the $\mmg$ MC;
    (right) the same for the $N_+$ sample ($\phi^*_+\in[0,\pi]$)}
  \label{fig:phis-dphi}
\end{figure*}

As previously discussed, the detector inefficiencies are mostly caused by the spatial
overlap of trajectories occuring in the detector: 2-track overlaps in the DCH and 
the IFR, respectively for $\Delta\phi=0.1\pm0.1$ and $0.5\pm0.2$ and
affecting masses below 2\gevcc, and the photon-muon overlap in the EMC,
for $\Delta\phi \gtrsim -\pi$ and affecting masses above 4\gevcc.
These various overlap effects contribute very asymmetrically to the two
$N_{\pm}$ samples, due to a complete correlation between the 
$\cos\phi^*_{\pm}$ and $\Delta\phi$ variables. This is demonstrated in 
Fig.~\ref{fig:phis-dphi}, which shows 
that the $N_+$ ($N_-$) sample corresponds to $\Delta\phi >0$ ($\Delta\phi <0$). 
Since the 2-track overlaps occur for $\Delta\phi >0$, one expects  
$\epsilon_- >\epsilon_+$. For the photon-muon overlap with nearly opposite 
tracks the situation is reversed.

As a consequence, as summarized in
Fig.~\ref{fig:A_data_vs_MC}  (Sec.~\ref{sec:mmgResults}),
the acceptance and detector inefficiencies induce a change $\Delta A$
in the observed charge asymmetry magnitude and also distort the linear
dependence on $\cos\phi^*$. The dominant effects are from geometric
acceptance, `ISR' photon reconstruction and the track momentum
requirement $p>1\gevc$.

However, although the detector is not completely charge symmetric, Fig.~\ref{fig:phis-dphi} shows
that the effects producing an asymmetry are nearly symmetric in $\cos\phi^*$.
The $A_0$ observable introduced in Eq.~(\ref{eq:A_linear}) is thus expected to be robust 
against such effects. Overall, the slope of the asymmetry
is barely affected by detector inefficiencies and event selection. 
In the $\mmg$ process, the maximum effects, of a few $10^{-2}$,
take place around $m_{\mu\mu}\sim 2-4\gevcc$. In the low mass
region [0.5--1.0]\gevcc, the effect from the overall selection is at the level of a few $10^{-3}$. 

\subsection{Study of the effects with the pion simulation}

The acceptance and detector effects are also studied with the simulated $\epem\to\pipig$
events. 
The overall effect around the $\rho$ resonance
($m_{\pi\pi}\in[0.4,~1.2]\gevcc$) is $(0.30\pm0.07)\times 10^{-2}$ in average.

As the charge asymmetry is null for
$\epem\to\pipig$ MC events, generated with no LO FSR, the acceptance effects on the
slope of the charge asymmetry are quite small for any selection requirement, including
the kinematic ones. This is in contrast with the $\mmg$ case, where the
individual kinematic requirements induce large effects. However this conclusion holds
only if the charge asymmetry in the data is actually null. If a sizeable asymmetry
is measured, the bias introduced by the cross effect of acceptance and asymmetry
has to be evaluated and corrected (Sec.~\ref{sec:results-reweightedMC}).


\section{\boldmath Results on the charge asymmetry in the $\epem\to\mmg$ process}

\label{sec:mmgResults}

\begin{figure*}
    \centering 
    \includegraphics[width=0.25\textwidth]{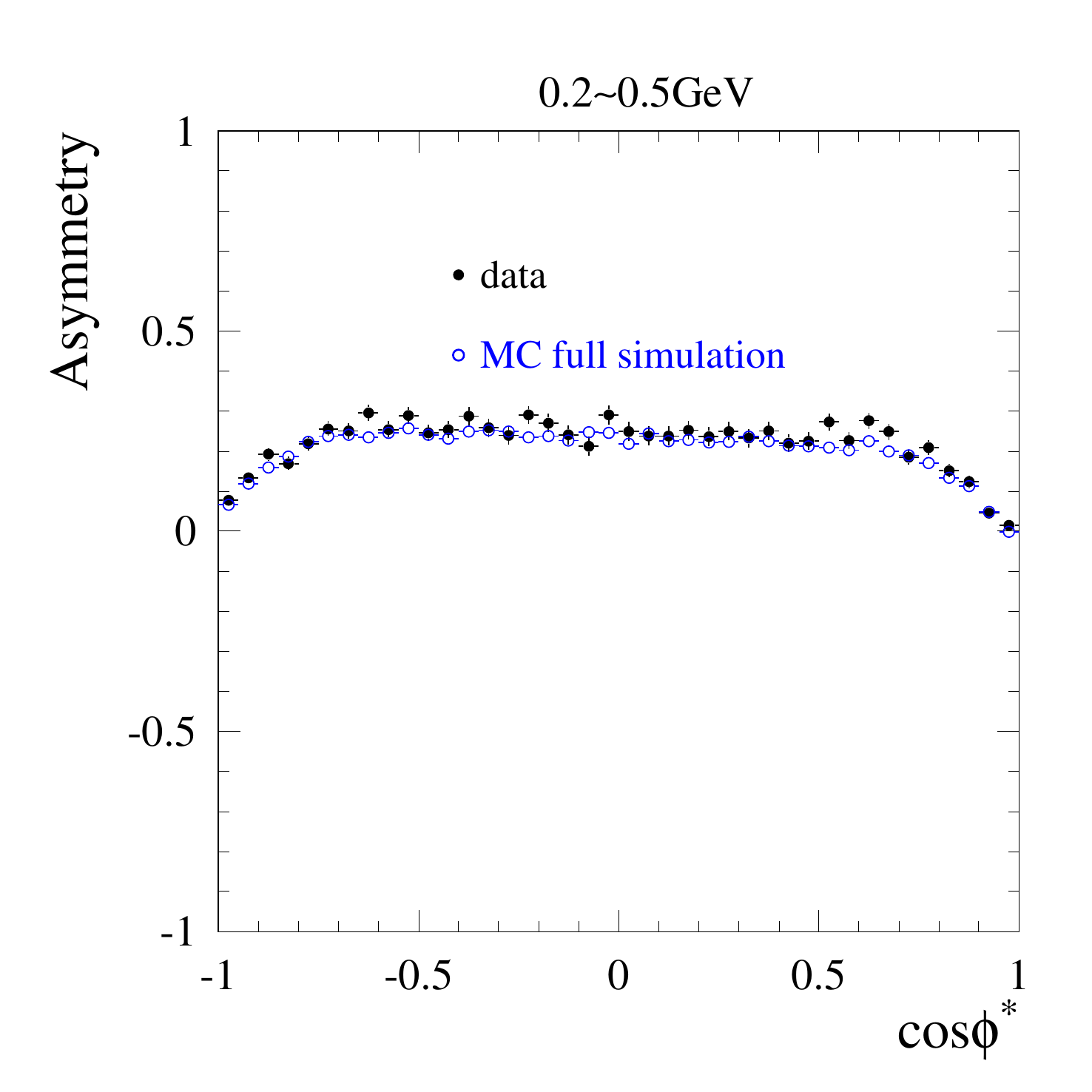}
    \includegraphics[width=0.25\textwidth]{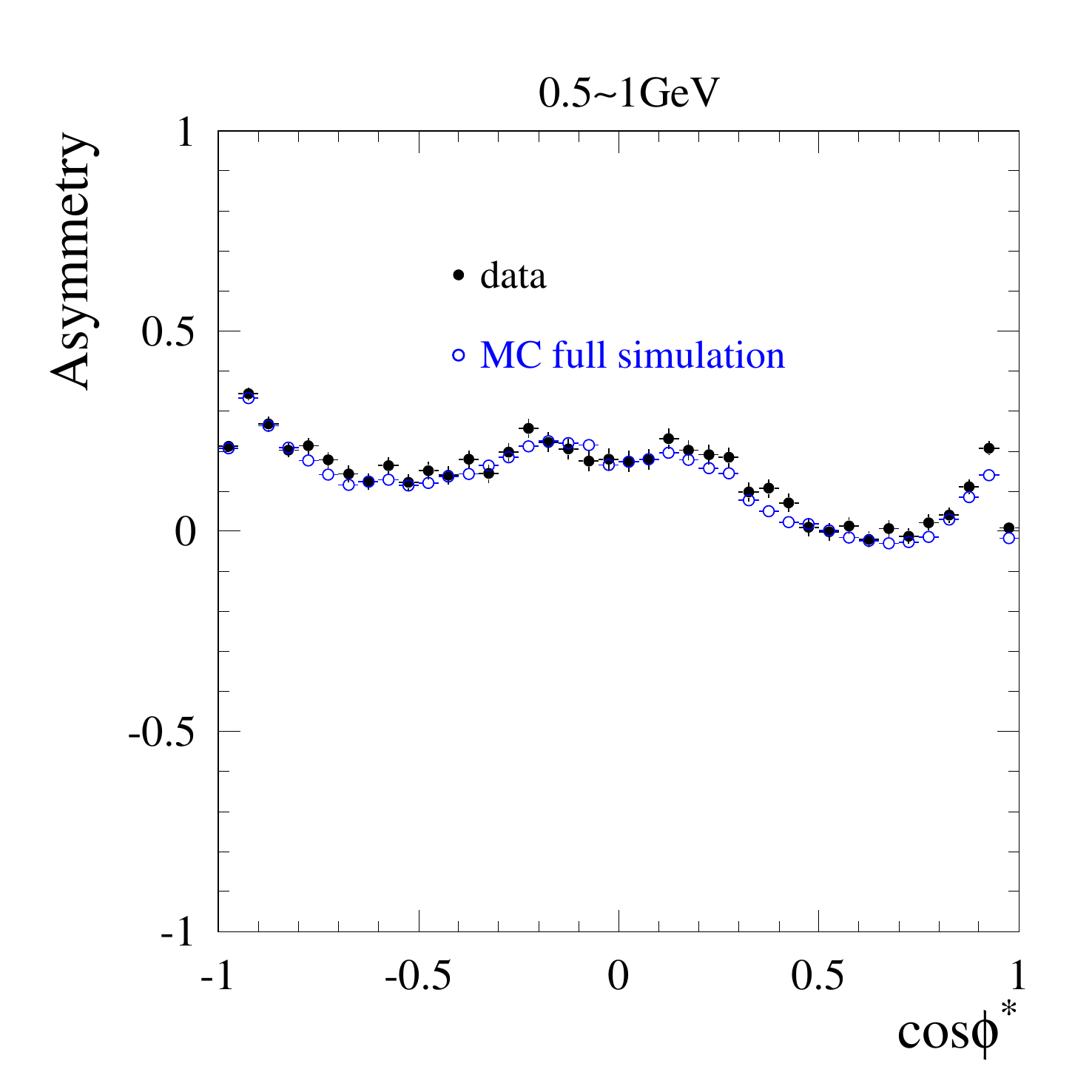}
    \includegraphics[width=0.25\textwidth]{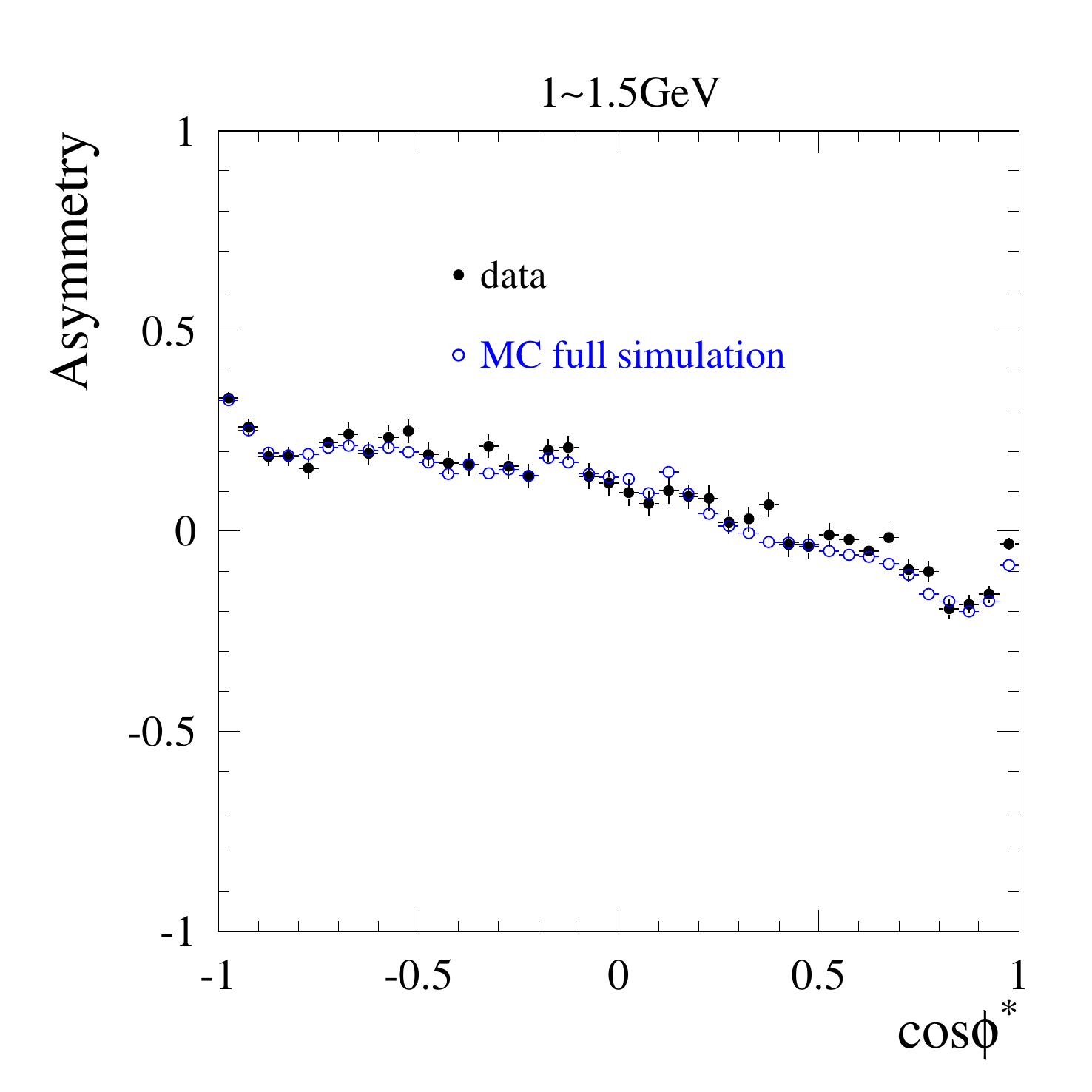}
    \includegraphics[width=0.25\textwidth]{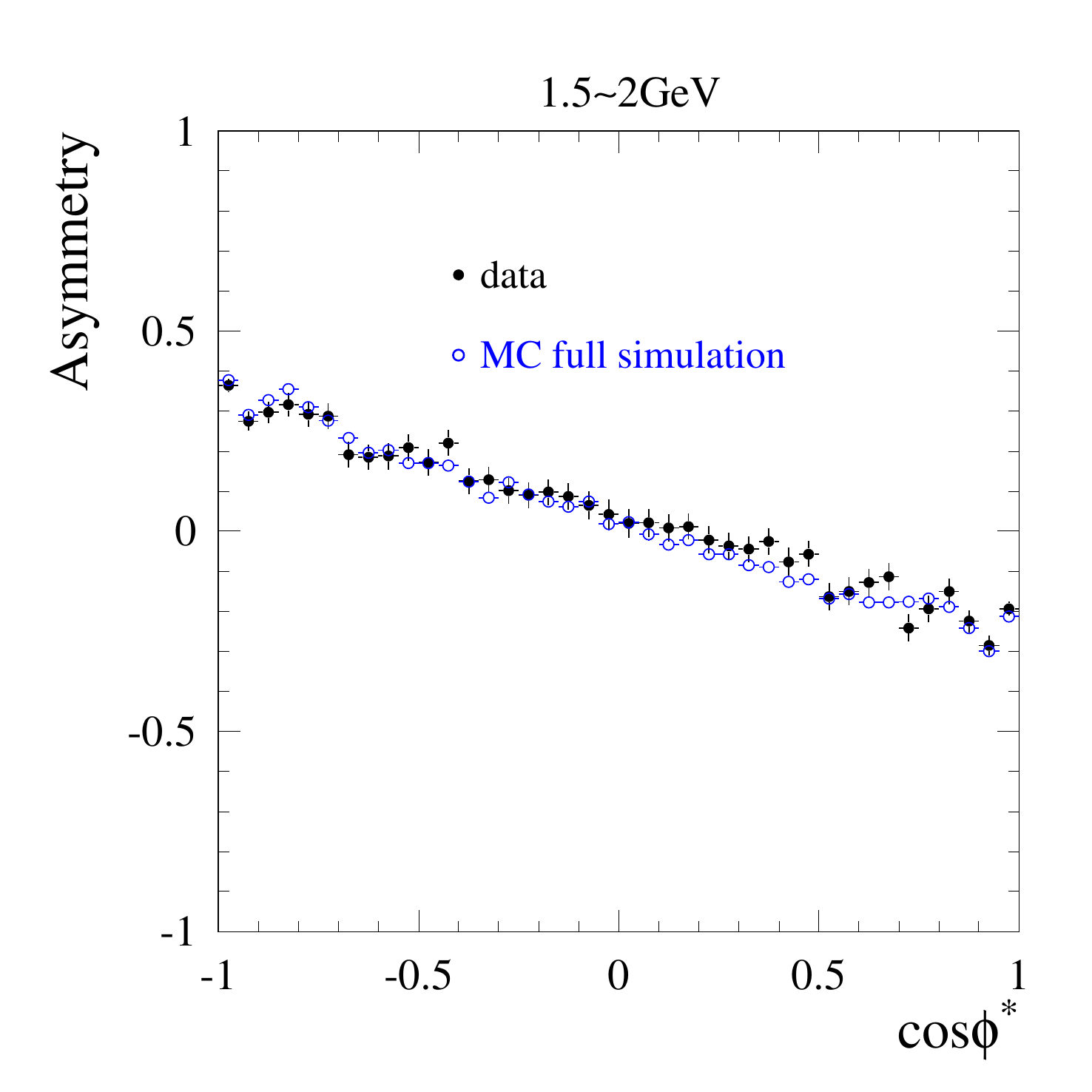}
    \includegraphics[width=0.25\textwidth]{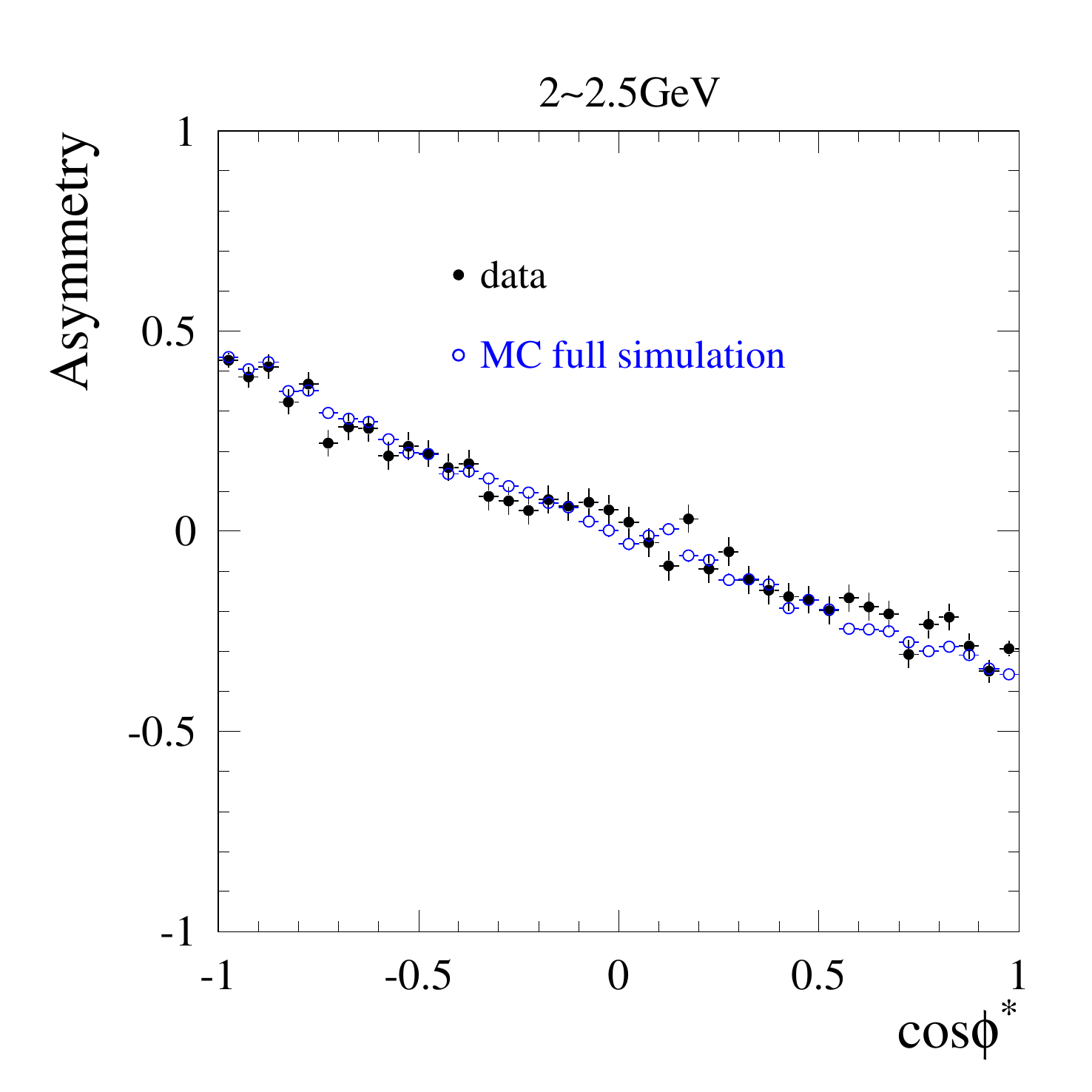}
    \includegraphics[width=0.25\textwidth]{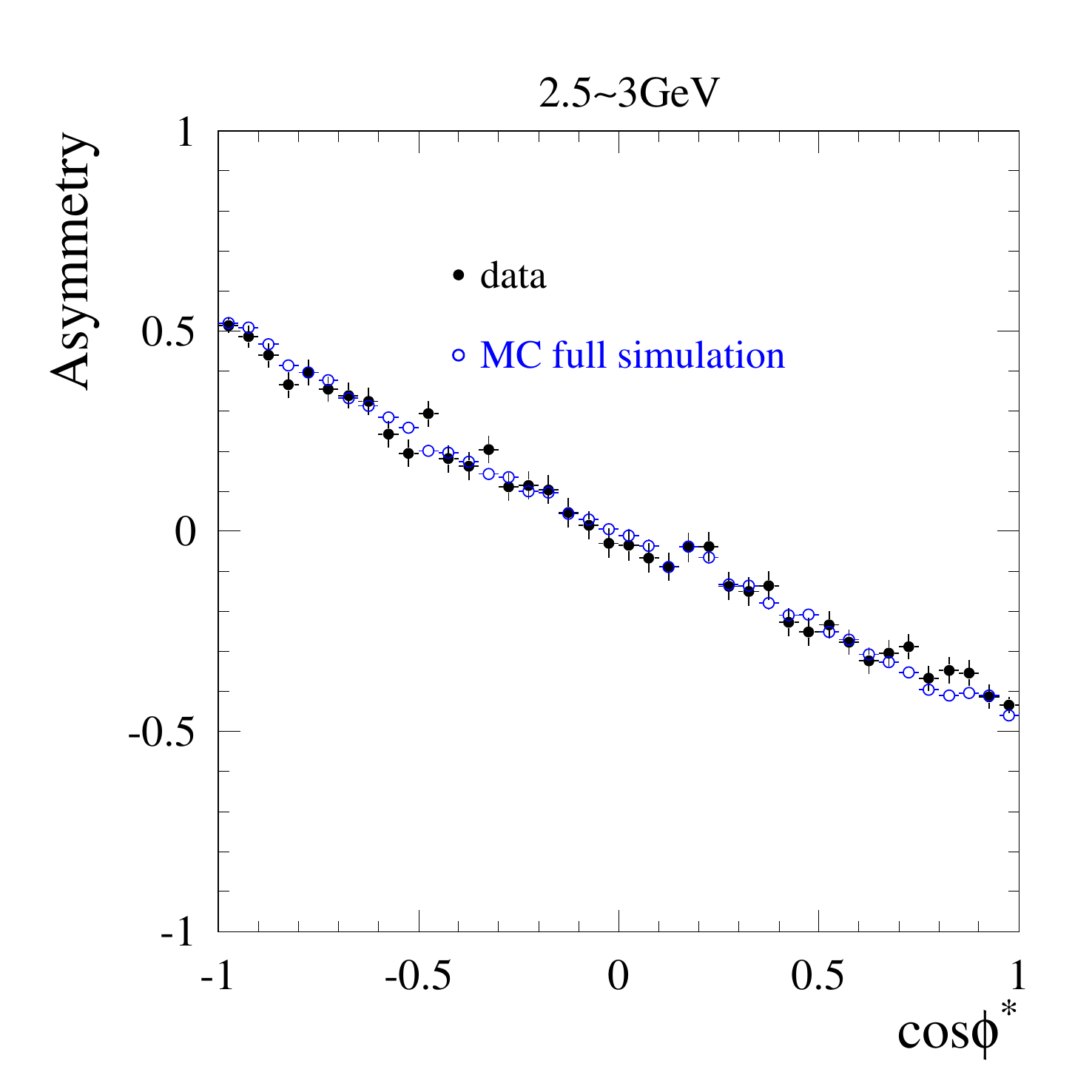}
    \includegraphics[width=0.25\textwidth]{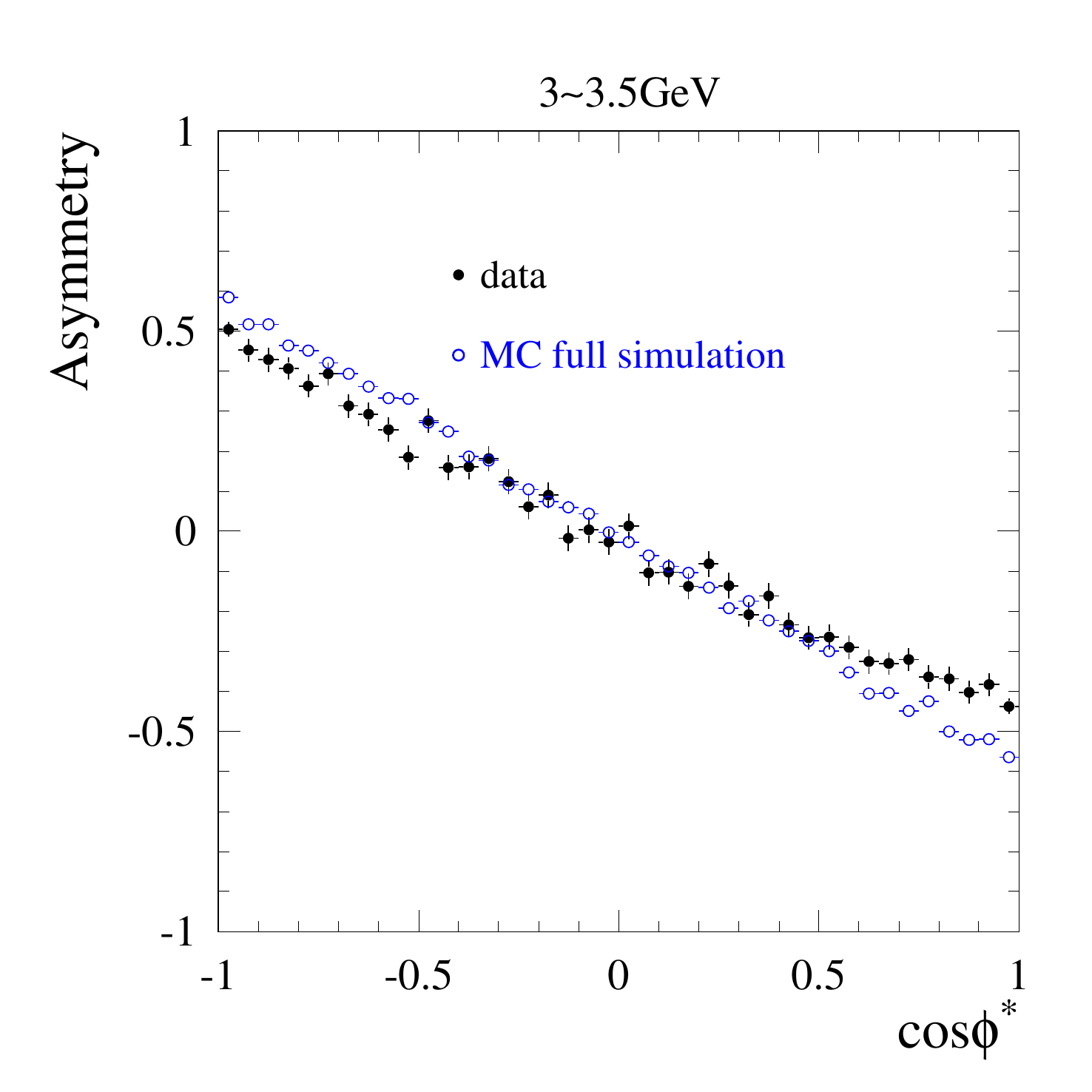}
    \includegraphics[width=0.25\textwidth]{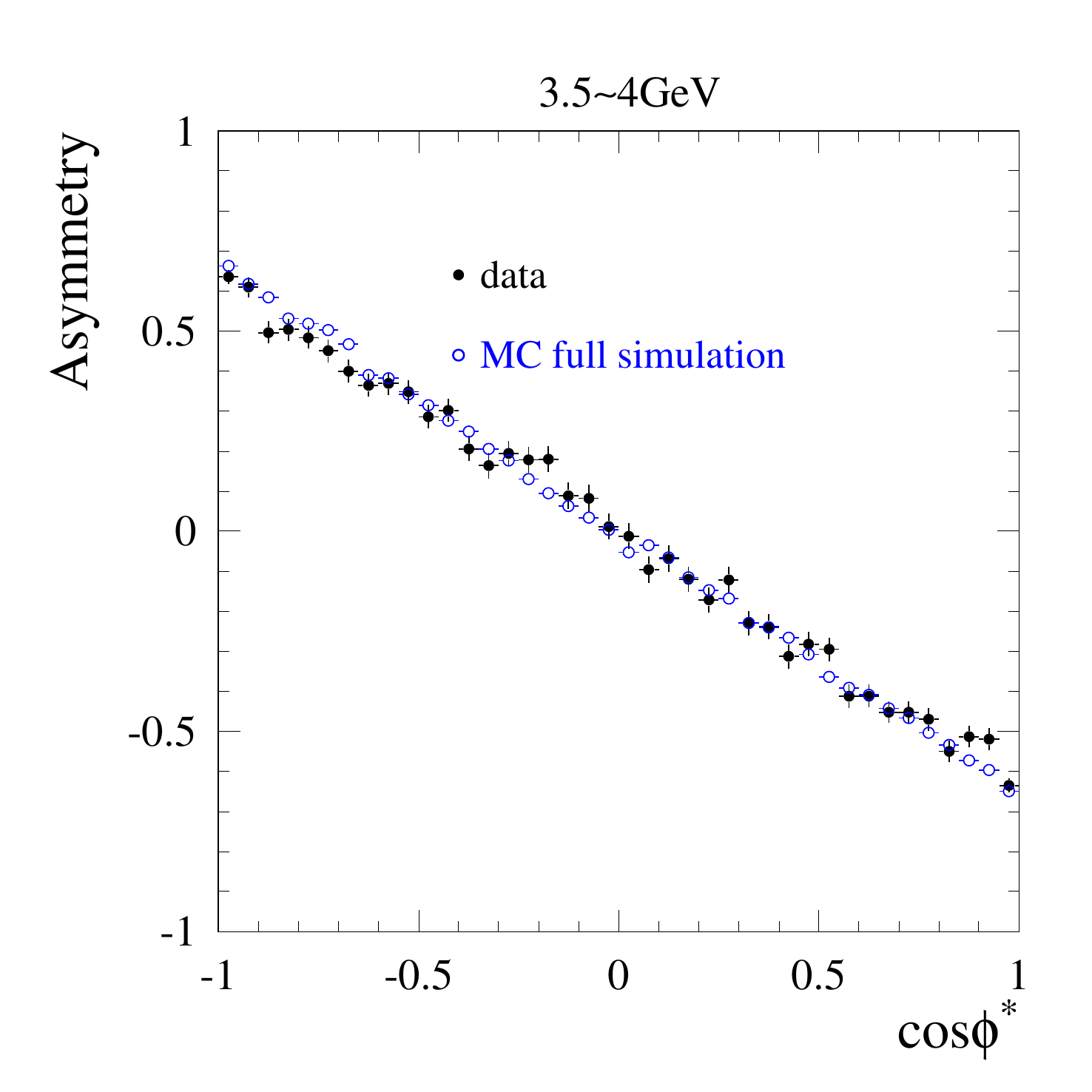}
    \includegraphics[width=0.25\textwidth]{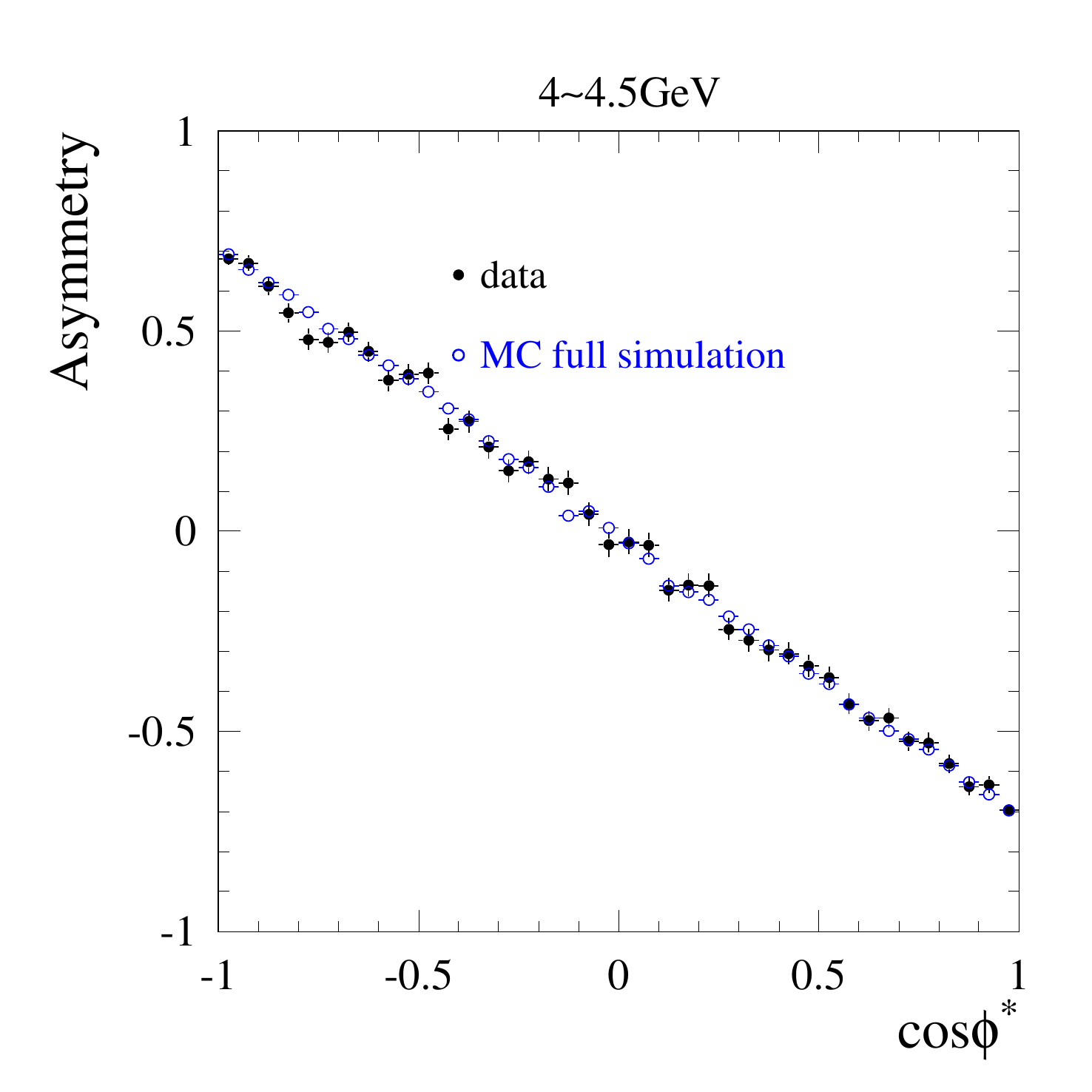}
    \includegraphics[width=0.25\textwidth]{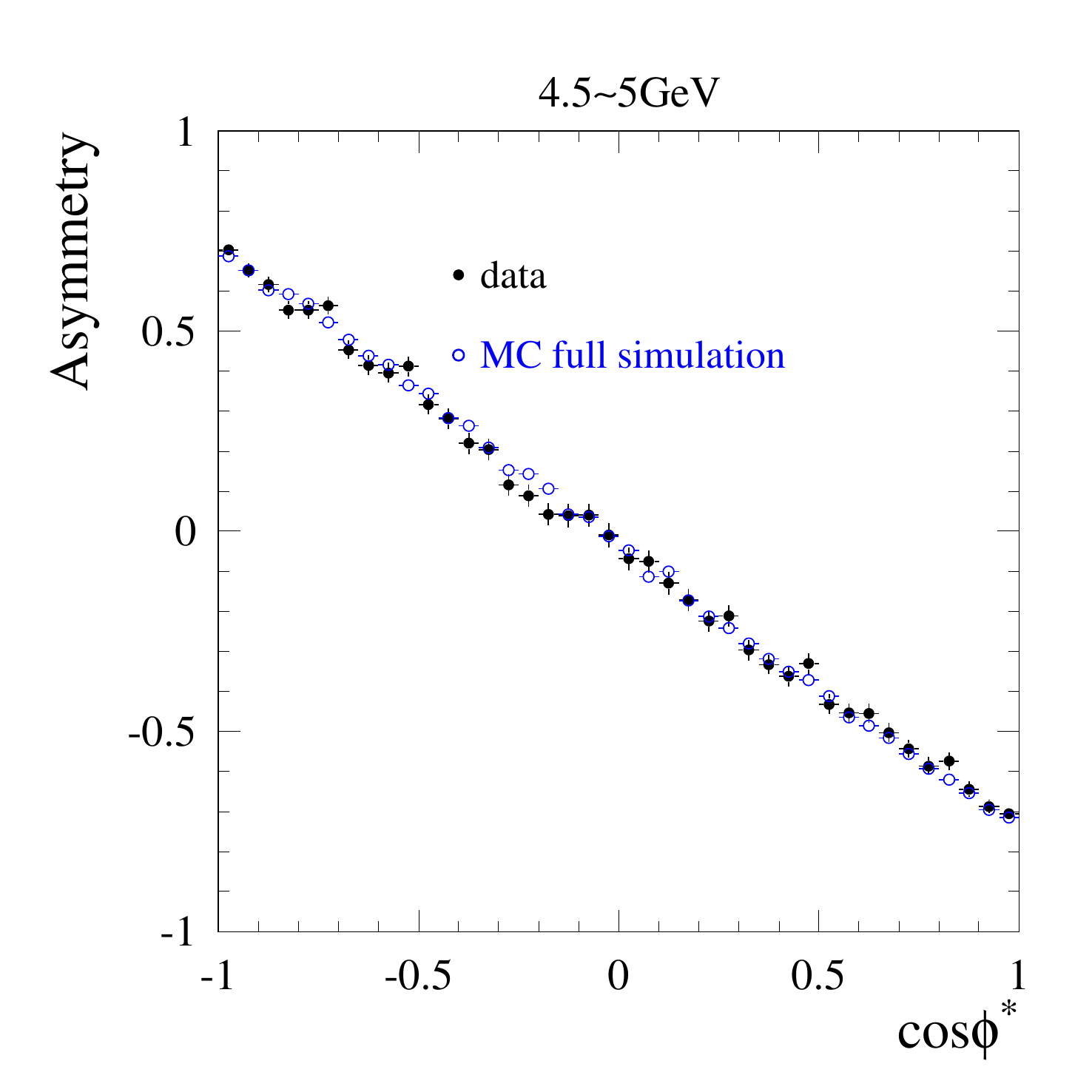}
    \includegraphics[width=0.25\textwidth]{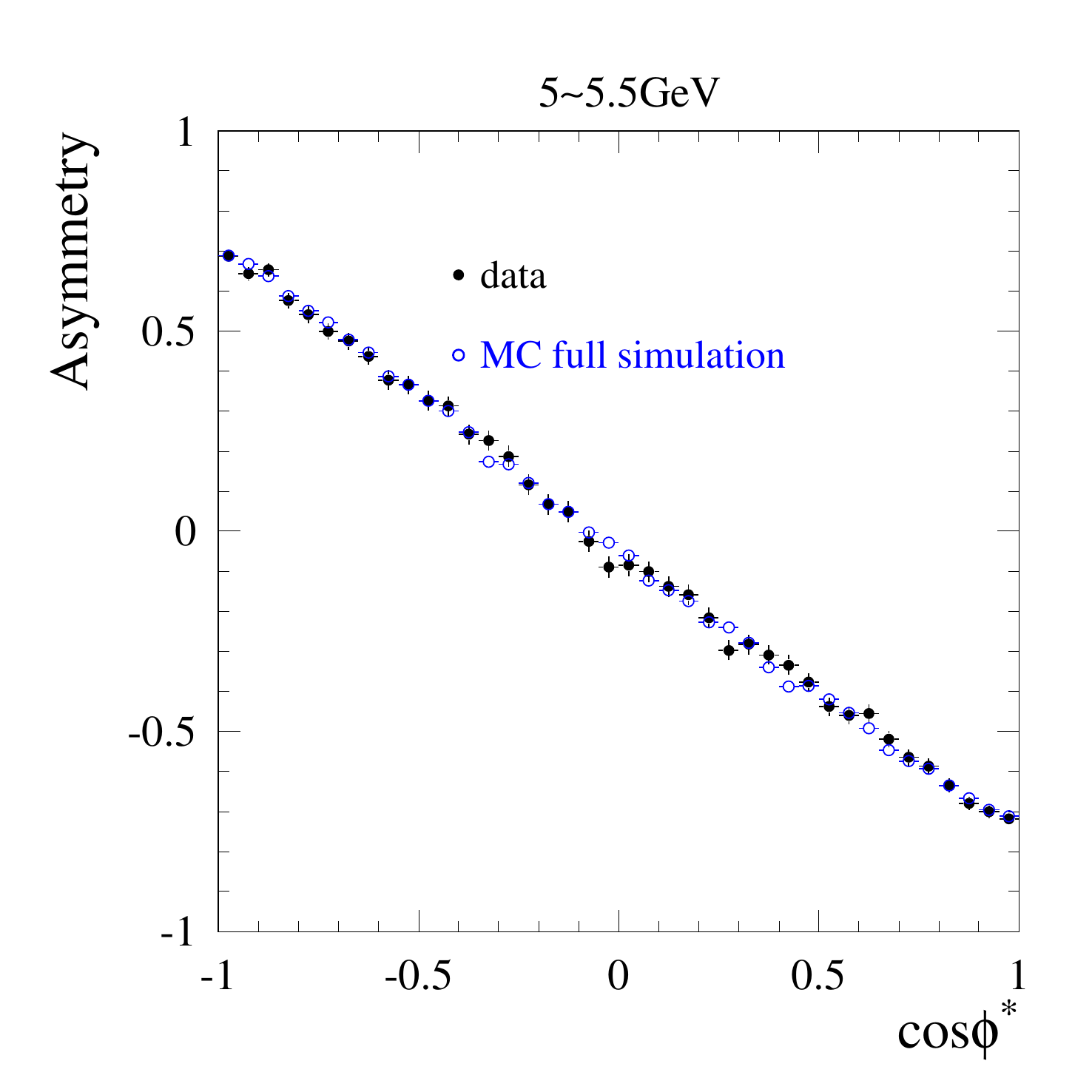}
    \includegraphics[width=0.25\textwidth]{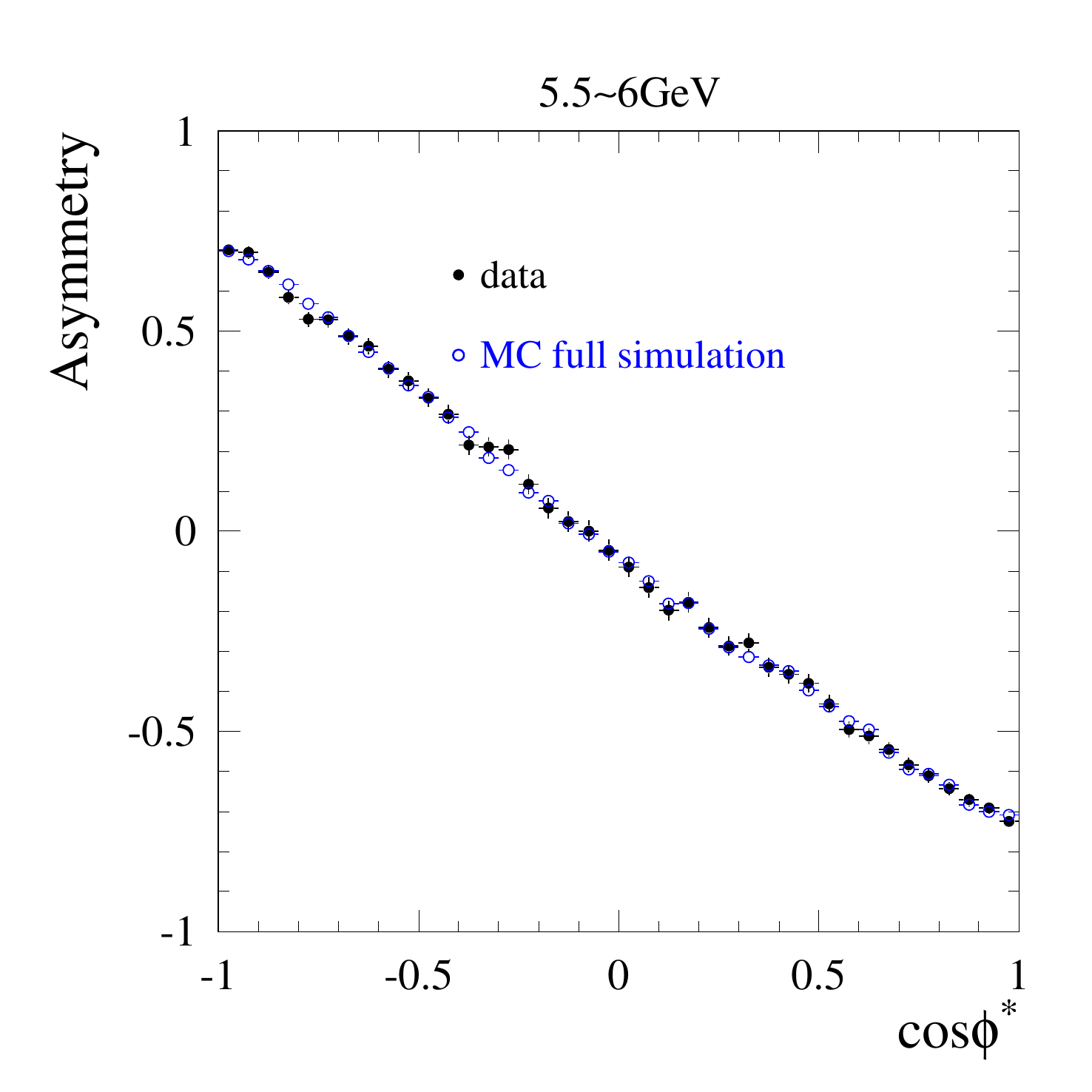}
    \includegraphics[width=0.25\textwidth]{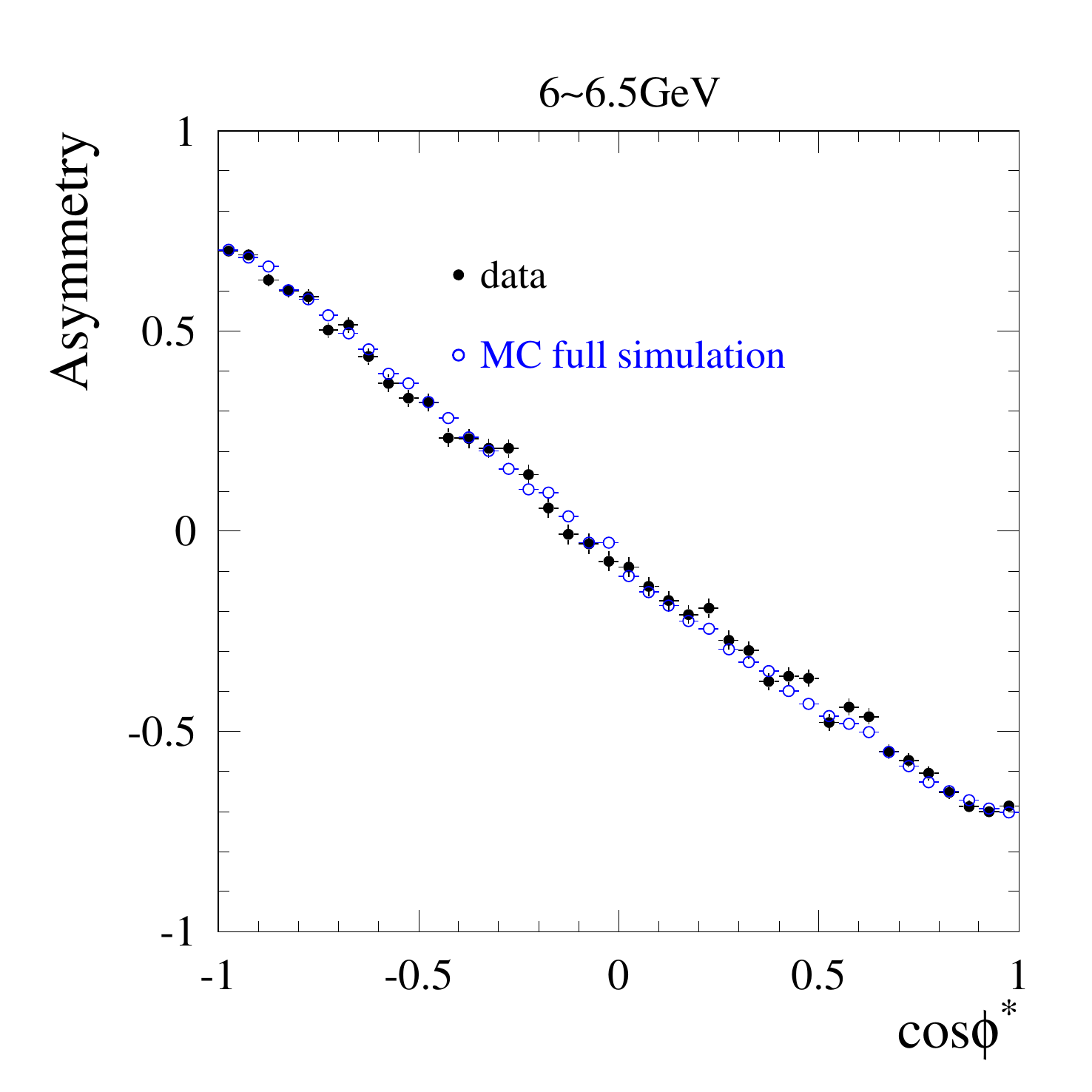}
    \includegraphics[width=0.25\textwidth]{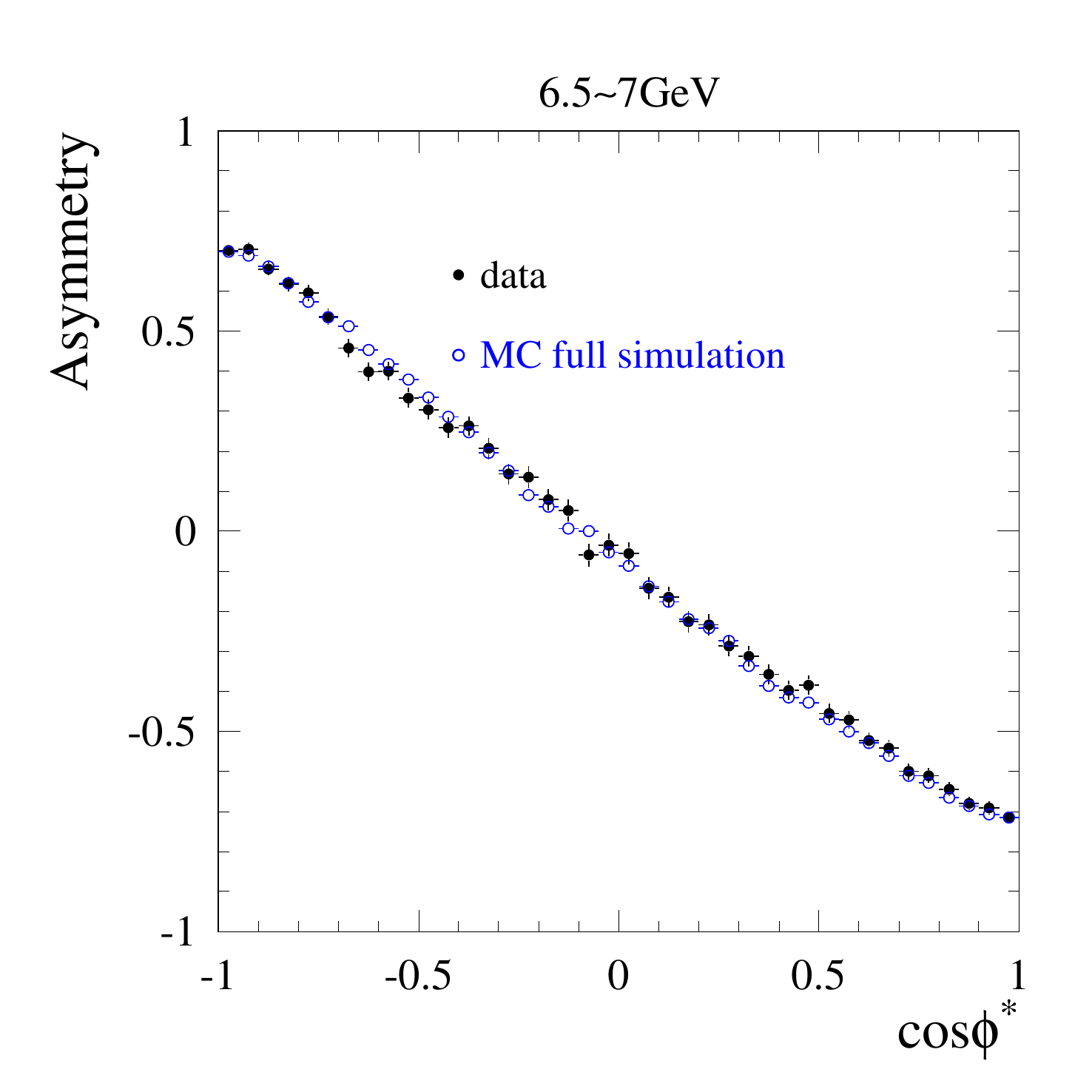}
    \caption{Raw charge asymmetry as a function of $\cos\phi^*$ for $\mmg$ 
      events in data ($\bullet$) and MC ($\circ$), in various $m_{\mu\mu}$ intervals, after 
      the complete event selection.}
    \label{fig:A_data_vs_MC}
  \end{figure*}

The measured raw charge asymmetry after the complete event selection for the data is
obtained as a function of $\cos\phi^*$ in various $m_{\mu\mu}$ intervals, and
shown in Fig.~\ref{fig:A_data_vs_MC}. 
It is consistent to the first order with
the full simulation of $\epem\to\mmg$ events, except in the mass interval
populated by the $J/\psi$ resonance (3.0--3.5\gevcc). This is expected as
$J/\psi$ production is not considered in AfkQed.

The physical charge asymmetry for $\epem\to\mmg$ is obtained from the measured
$\cos\phi^*$ distributions after background subtraction and efficiency
correction. The background dependence on $\cos\phi^*$ is estimated with the
simulation for each of the samples $\phi^*_\pm=\phi^* \in [0,\pi]$, as explained
above. Similarly, the overall efficiency  $\epsilon_\pm$ is obtained with fully
simulated $\epem\to\mmg$ events, and corrected for data/MC differences in
detector response.  The efficiency differences between data and simulation have
been studied extensively for the cross section measurements of $\epem\to\mmg$
and $\epem\to\pipig$~\cite{prd-pipi}. They are parameterized as a function of
the azimuthal opening angle $\Delta\phi$ between the two muons, and projected
onto  the $\cos\phi^*$ variable by sampling with MC.

\begin{figure*}
\centering 
\includegraphics[width=0.25\textwidth]{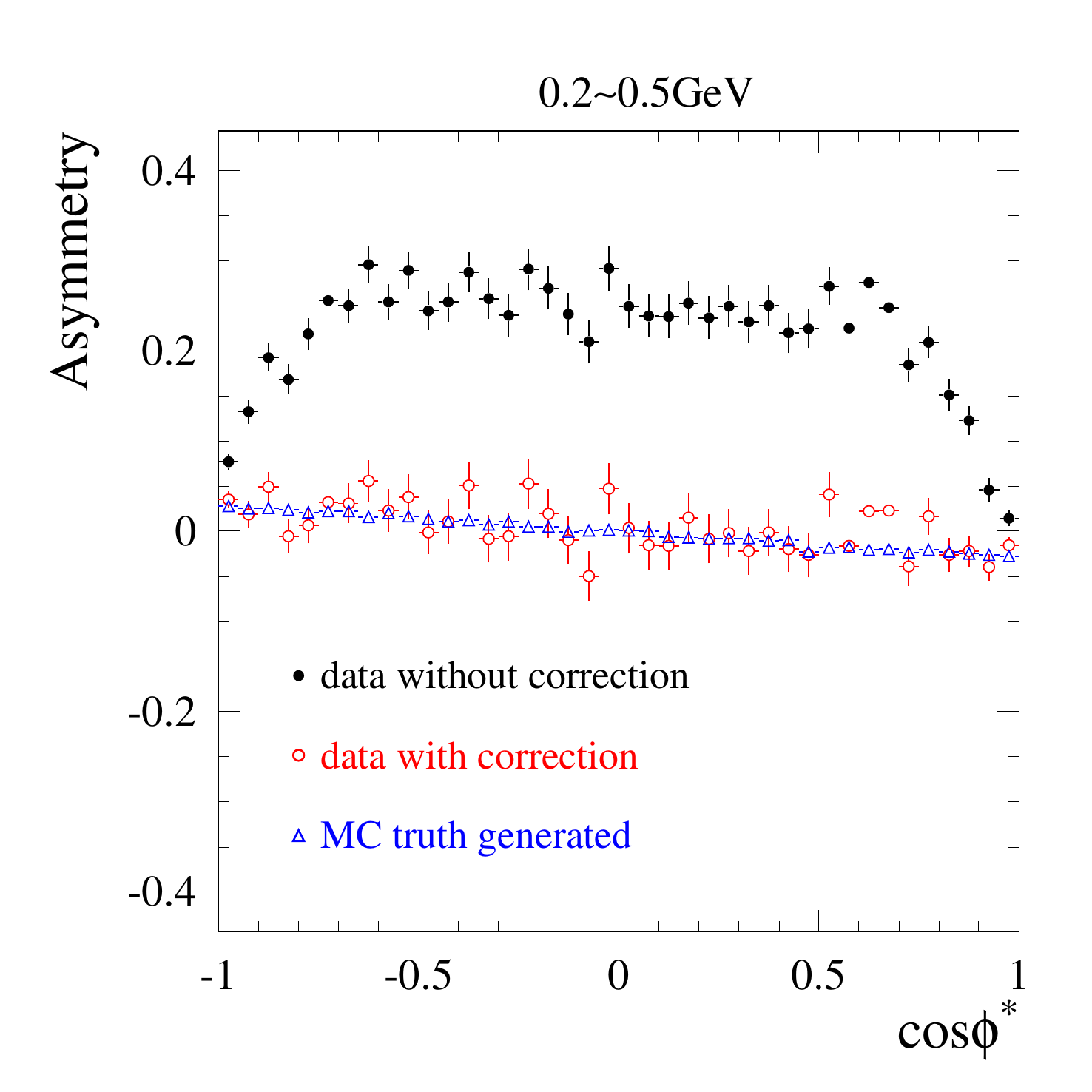}
\includegraphics[width=0.25\textwidth]{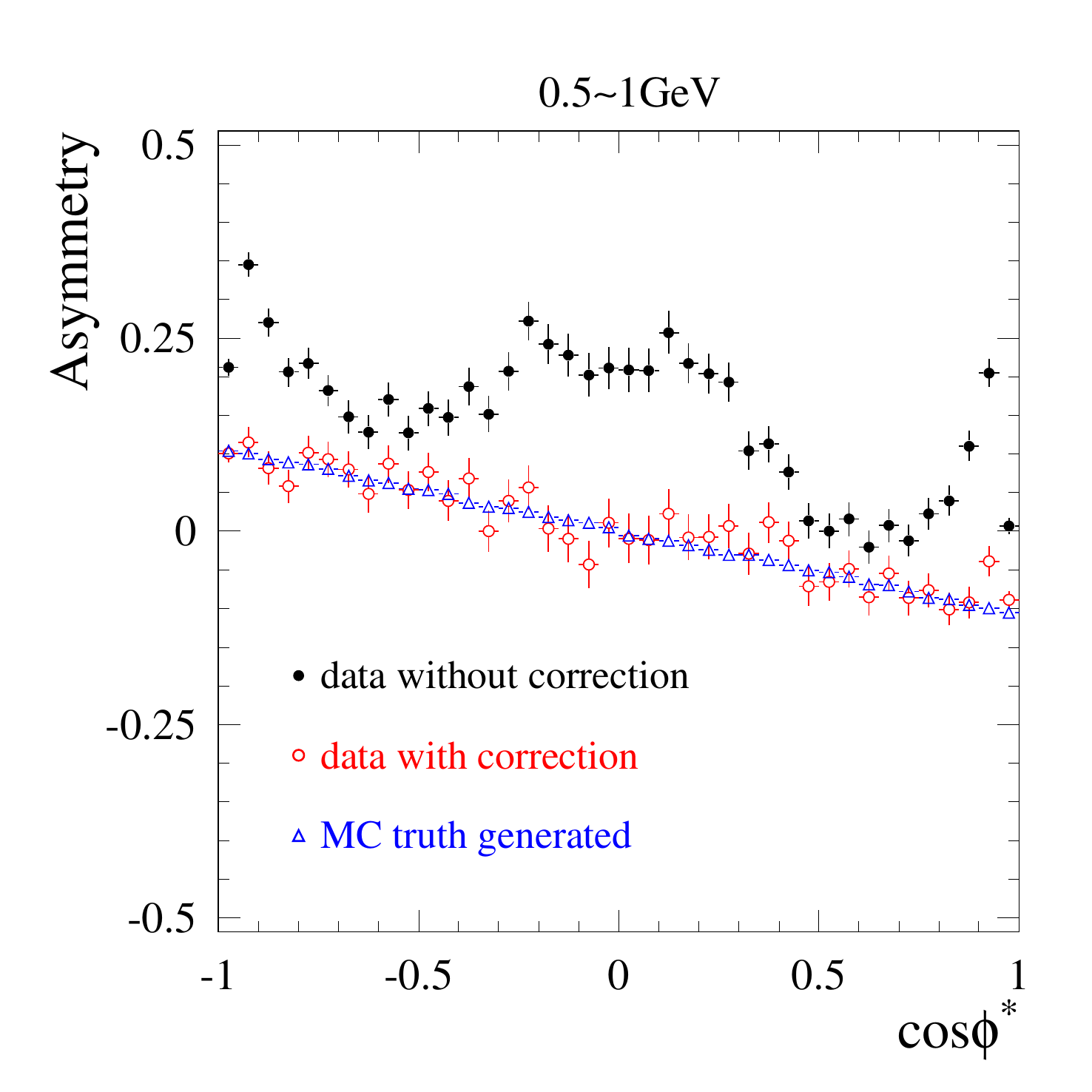}
\includegraphics[width=0.25\textwidth]{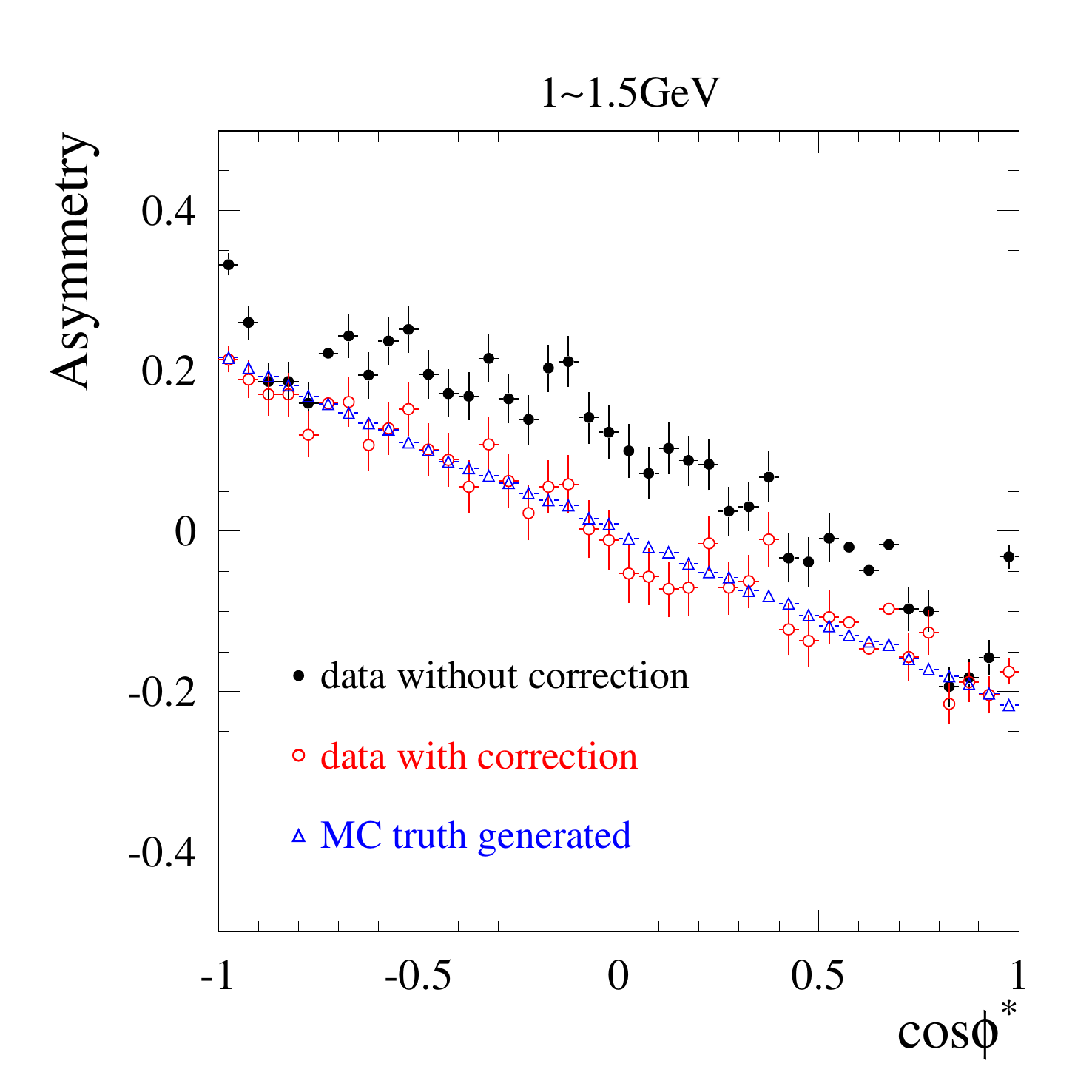}
\includegraphics[width=0.25\textwidth]{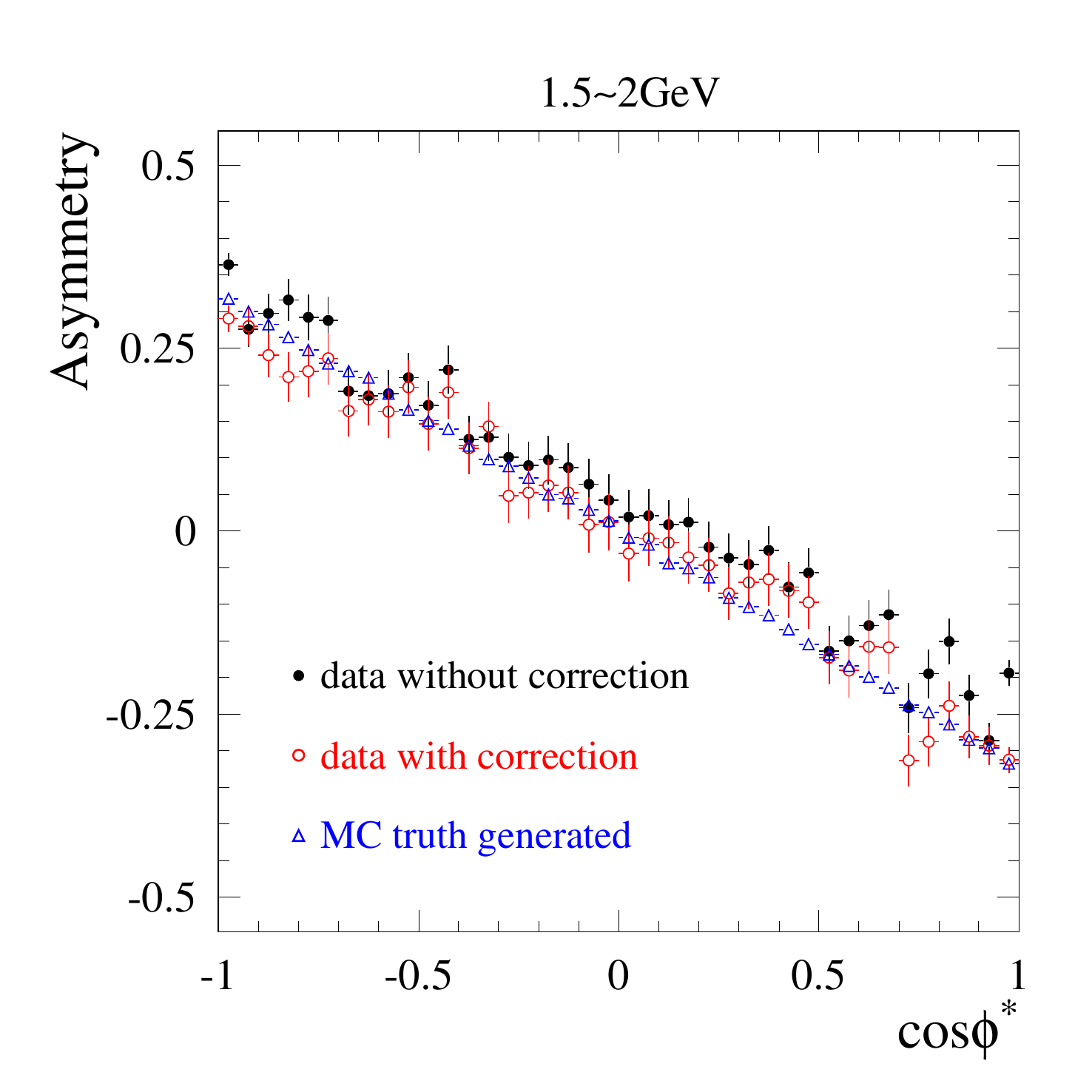}
\includegraphics[width=0.25\textwidth]{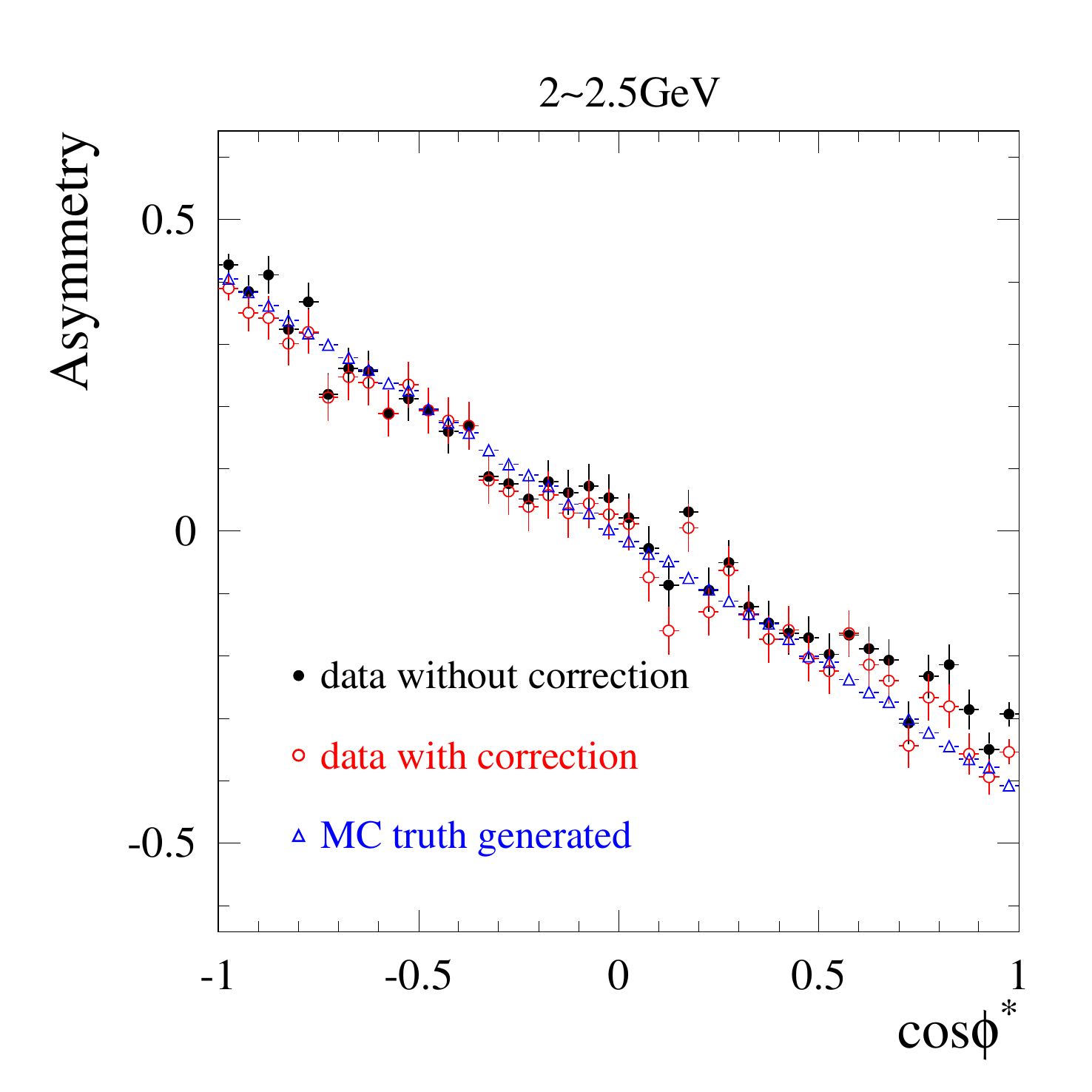}
\includegraphics[width=0.25\textwidth]{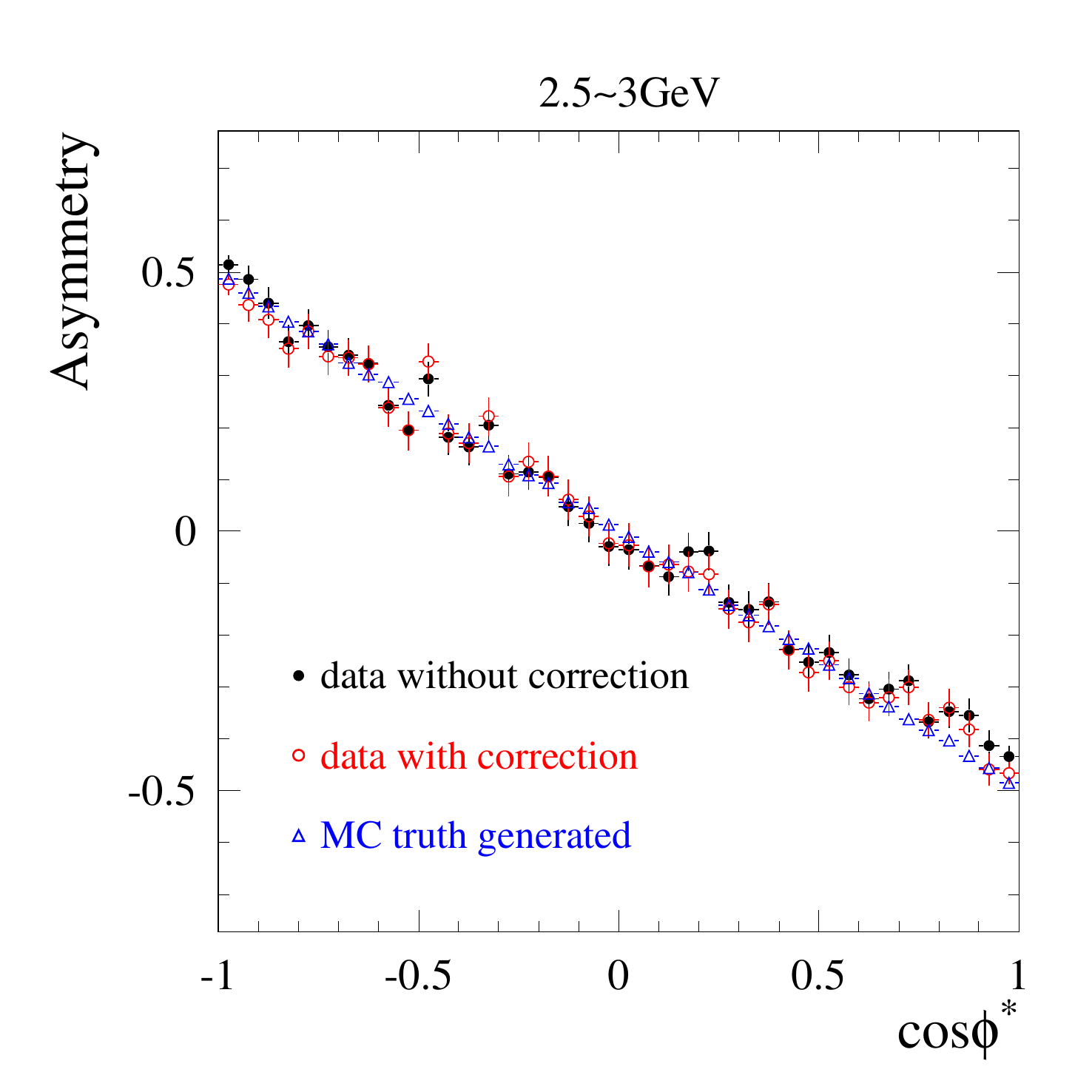}
\includegraphics[width=0.25\textwidth]{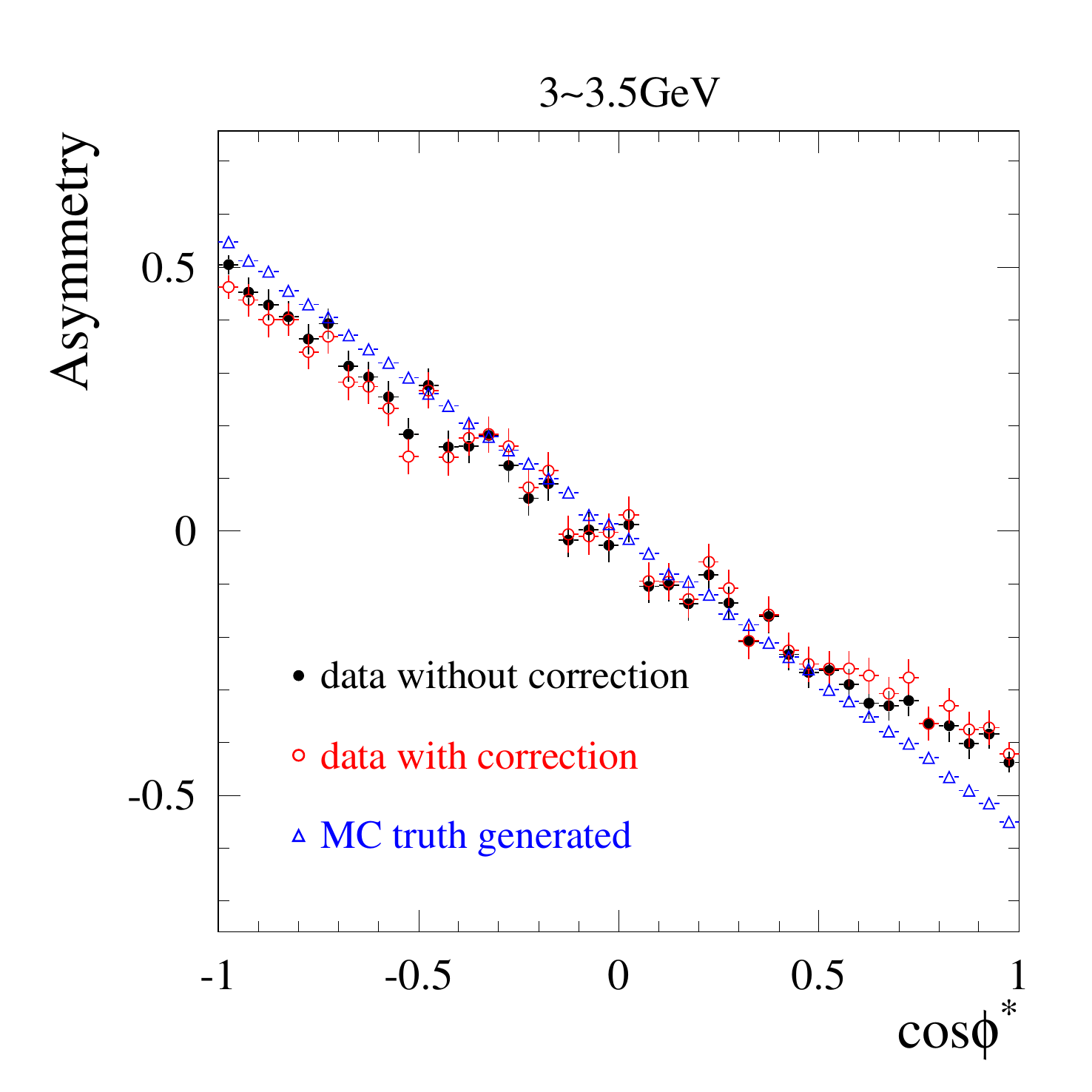}
\includegraphics[width=0.25\textwidth]{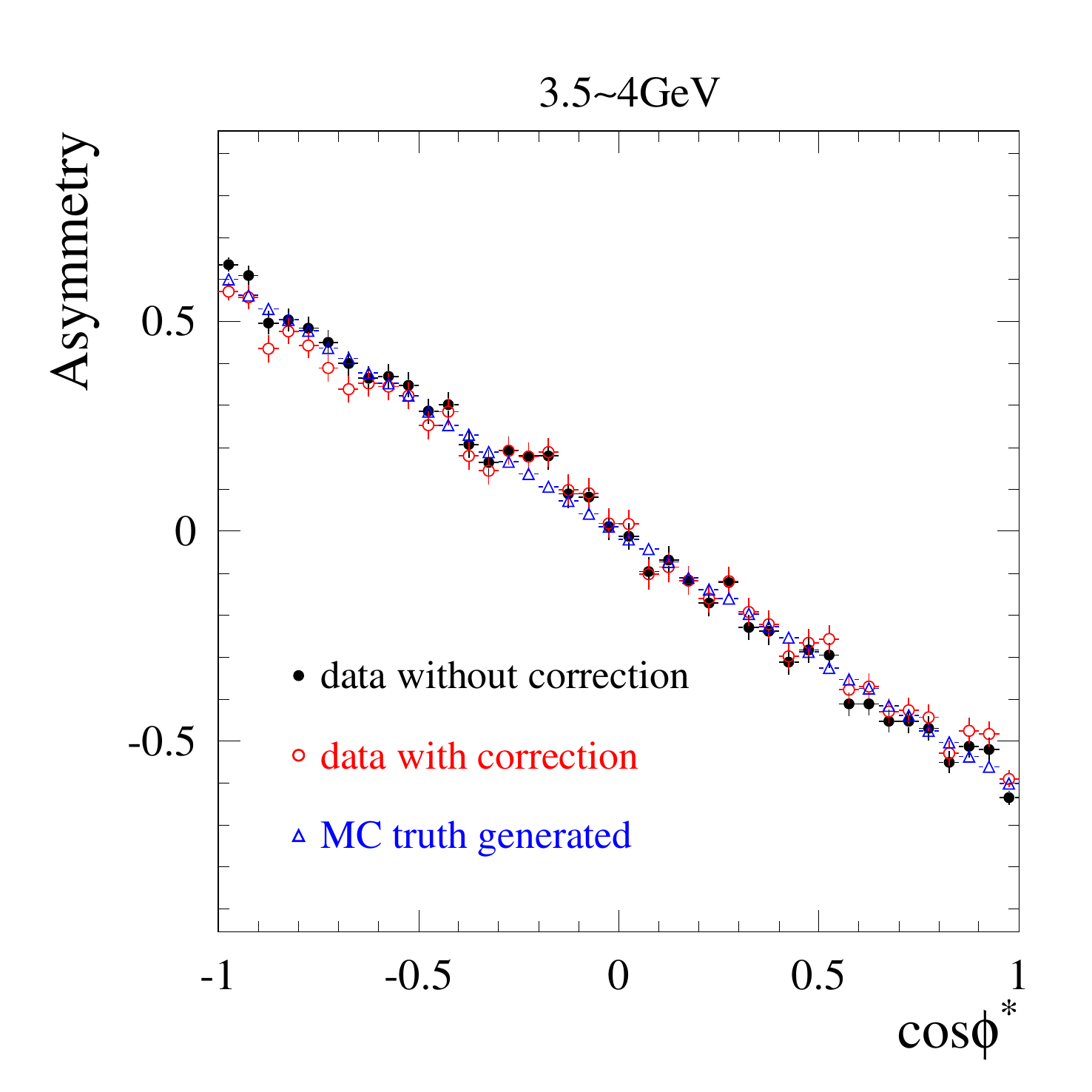}
\includegraphics[width=0.25\textwidth]{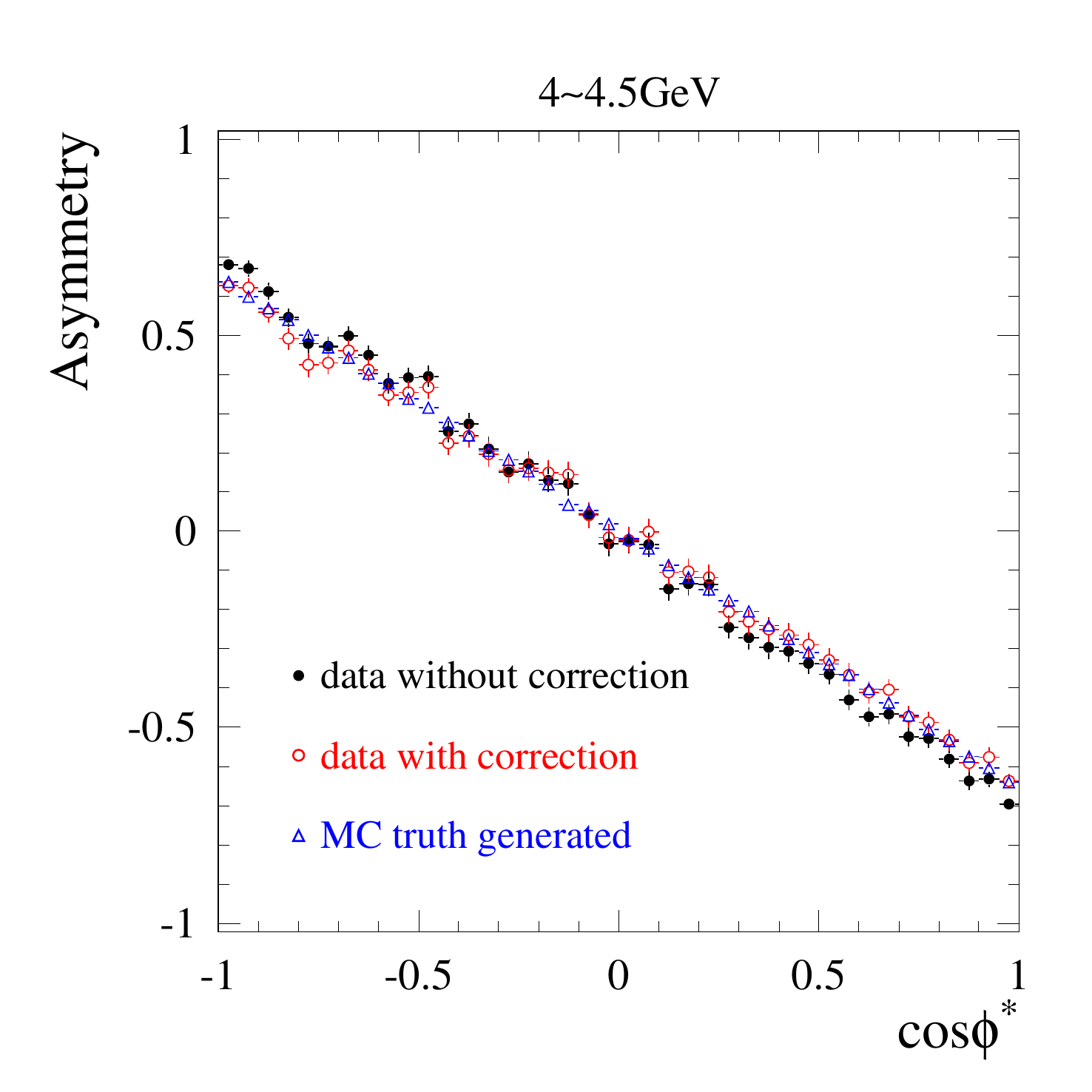}
\includegraphics[width=0.25\textwidth]{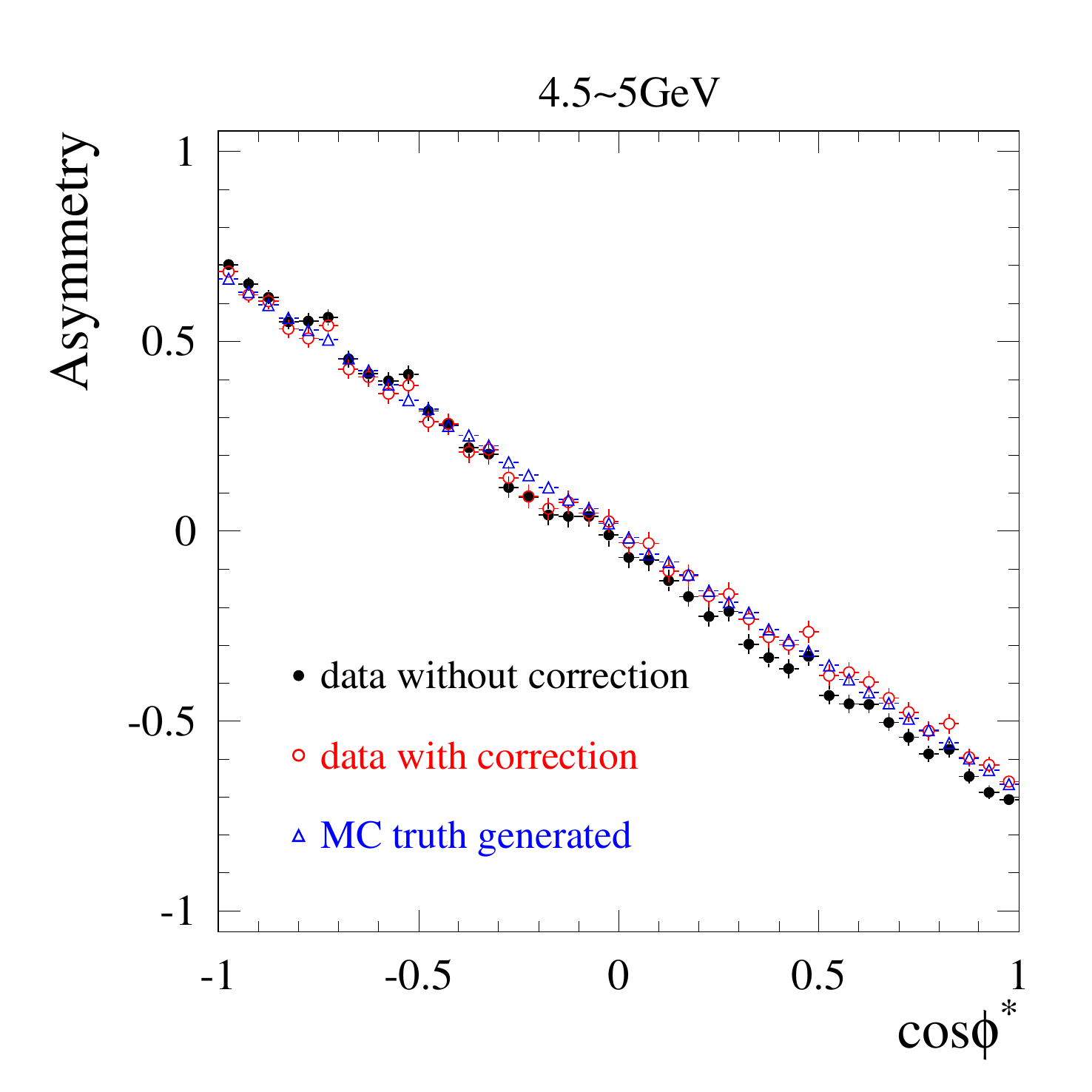}
\includegraphics[width=0.25\textwidth]{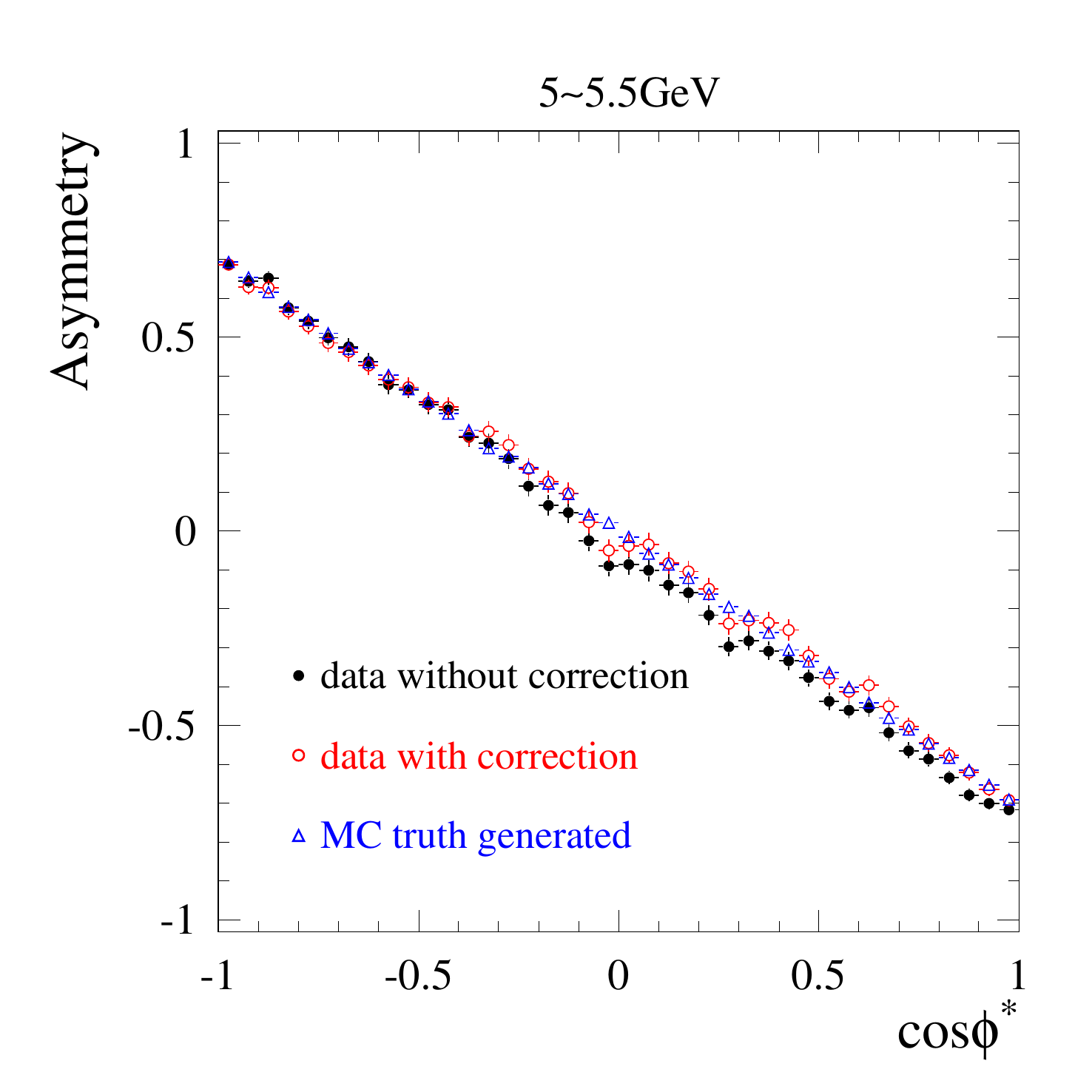}
\includegraphics[width=0.25\textwidth]{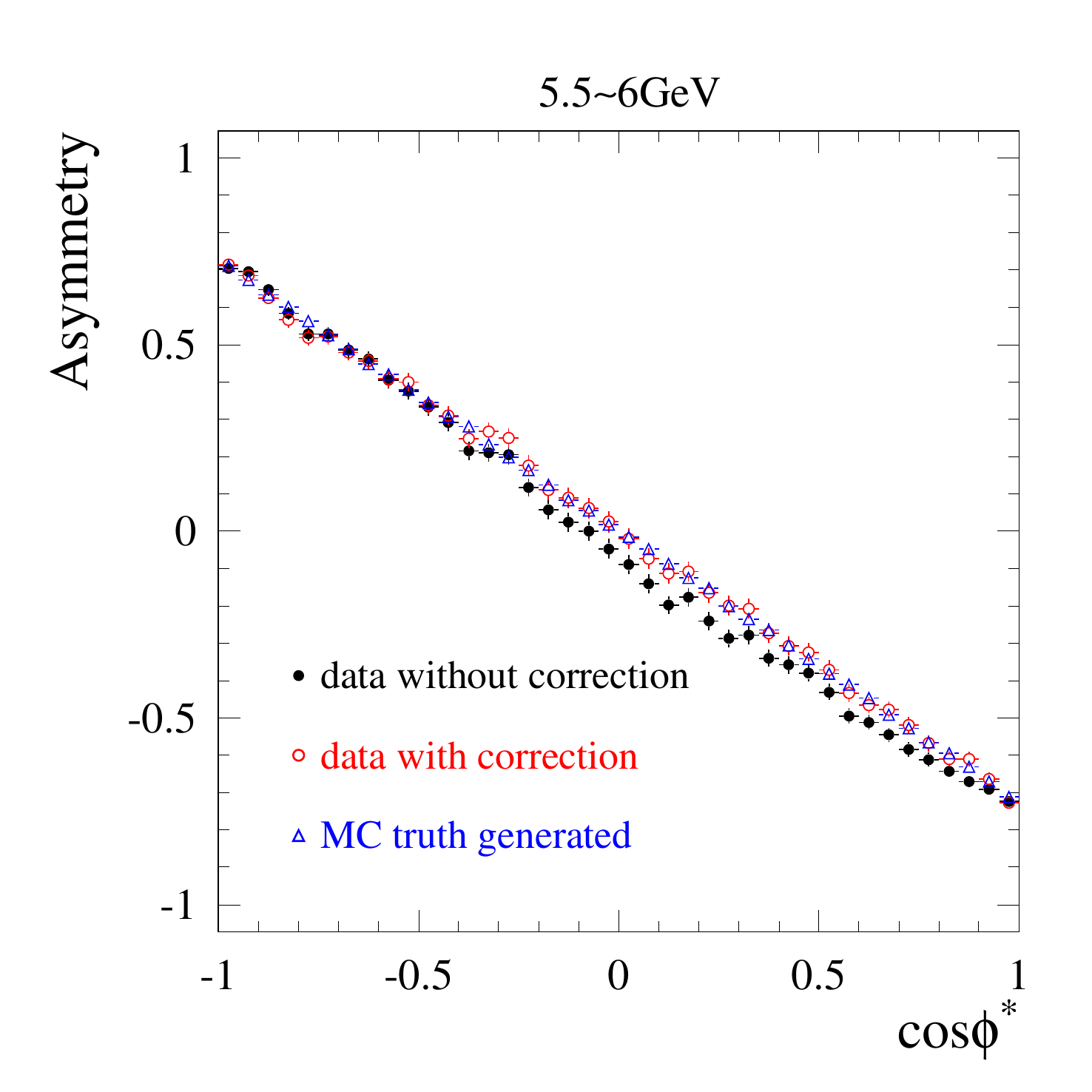}
\includegraphics[width=0.25\textwidth]{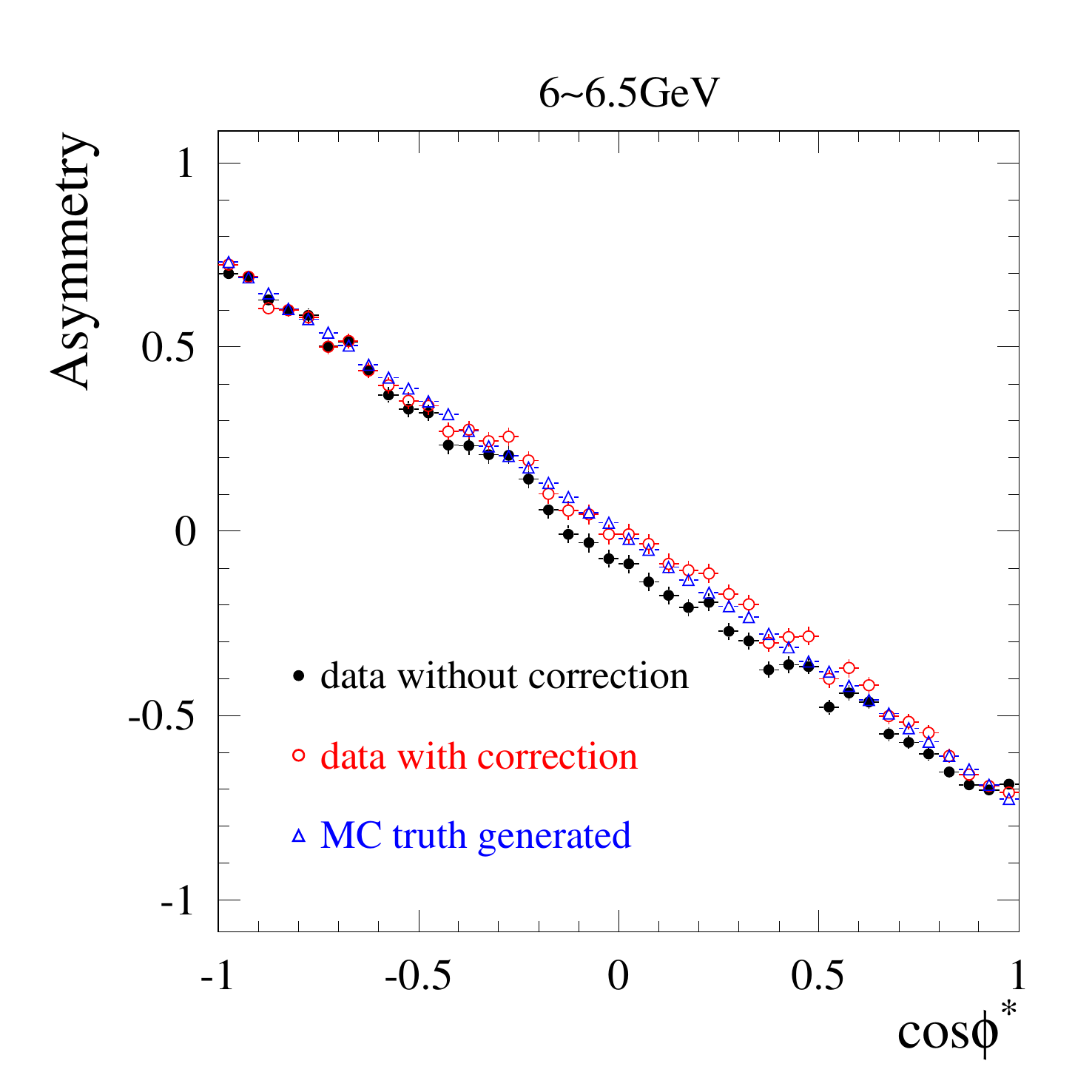}
\includegraphics[width=0.25\textwidth]{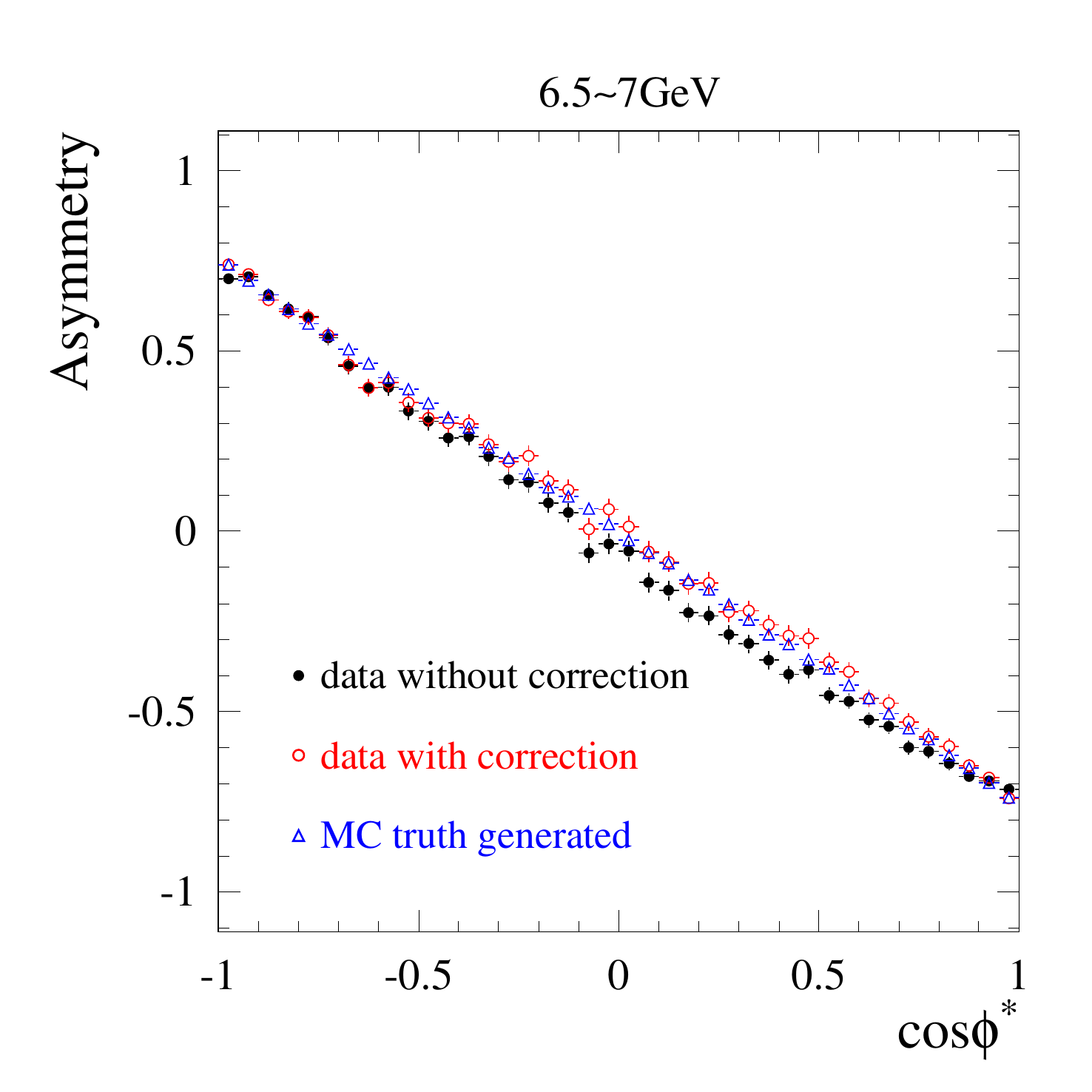}
\caption{Charge asymmetry in $\epem\to\mmg$ data before ($\bullet$) and after ($\circ$) background subtraction and 
efficiency corrections, and for MC ($\triangle$) at generation level. }  
\label{fig:A-dataCorr-mctruth-gmm}
\end{figure*}

The charge asymmetry distributions for $\epem\to\mmg$ data after
background subtraction and efficiency correction, as well as the charge
asymmetry for $\epem\to\mmg$ MC at generation level, are shown in
Fig.~\ref{fig:A-dataCorr-mctruth-gmm}. While the $\cos\phi^*$ dependence  of the
measured raw charge asymmetry is not linear, the corrected data distributions
are quite consistent with the MC distributions at generation level.
The slopes of charge asymmetry in various mass intervals are obtained by
fitting the background-subtracted efficiency-corrected charge asymmetry
distributions to $A_0\cos\phi^*$. 

\subsection{\boldmath Test of the charge asymmetry with $J/\psi\to\mumu$ events in data}
\label{sec:Jpsi}

Since $\epem\to\gamma J/\psi$ with on-shell $J/\psi\to
\mumu$ is a pure ISR process at $\Upsilon(4S)$ energies, the $J/\psi\to\mumu$ 
sample in the data provides a test of fake asymmetries that could arise
in the analysis.

\begin{figure*}\centering
  \includegraphics[width=0.45\textwidth]{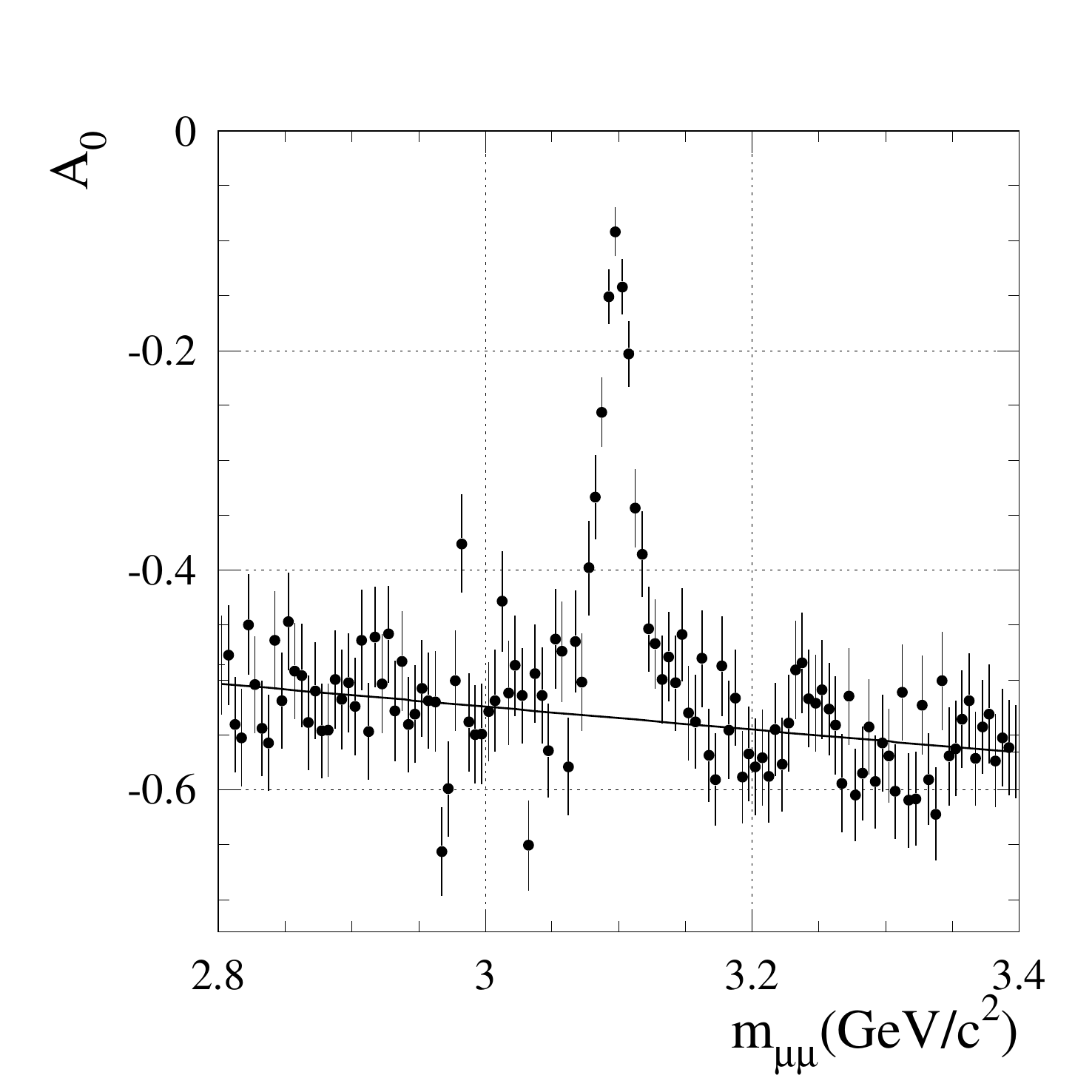}
  \includegraphics[width=0.45\textwidth]{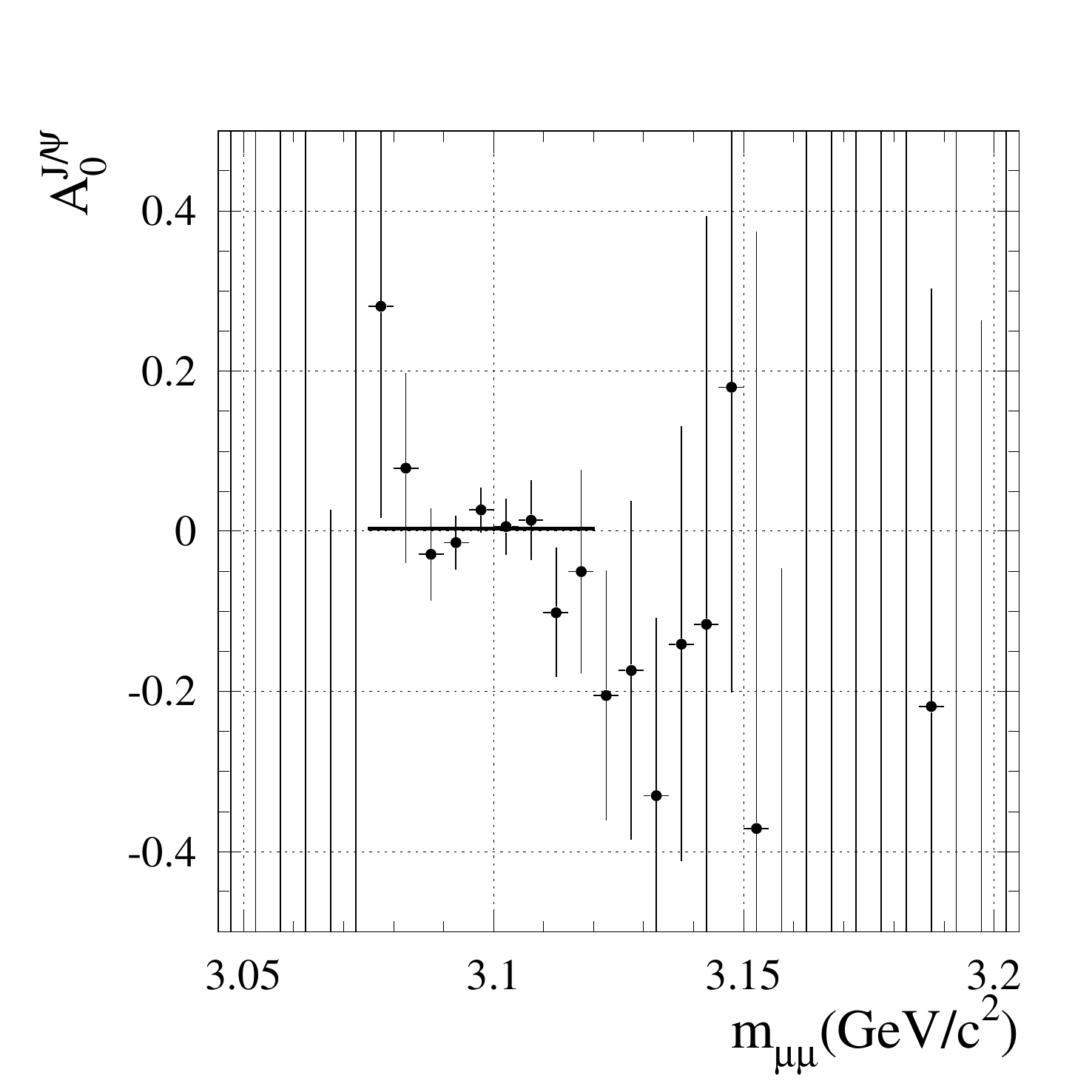}
  \caption{(left) The measured charge asymmetry as a function of $\mu\mu$ mass near the
  $J/\psi$ resonance. (right) The derived charge asymmetry for $\mmg$ from 
  $\epem\to\gamma J/\psi$ }
  \label{fig:A0_aroundJpsi}
\end{figure*}

To overcome the limited statistics of the  $J/\psi\to\mumu$ sample, a loosened
event selection is applied, with the muon identification requirement removed, which
provides a gain in statistics by a factor of about 4, with no significant
increase  of the hadronic background.  The $m_{\mu\mu}$ spectrum shows a clear
$J/\psi$ peak, over a linear QED background.  
Defining $A_0^{J/\psi}$ and $A_0^{\rm QED}$, the respective slopes of the
charge asymmetry for $J/\psi$ and underlying QED events, the slope $A_0$ measured in the vicinity of
the $J/\psi$ resonance is the average
\beqn\label{eq:A0_aroundJpsi}
    A_0=\frac{A_0^{J/\psi}N_{J/\psi}+A_0^{\rm QED}N_{\rm QED}}{N_{J/\psi}+N_{\rm QED}},
\eeqn
where $N_{J/\psi}$ and $N_{\rm QED}$ are the yields from $J/\psi$ and QED,
respectively. 
The quantities $N_{\rm
QED}$ and $N_{J/\psi}$ are obtained by fitting the mass spectrum with a sum of
a linear QED component and a Gaussian $J/\psi$ signal, with fixed width equal 
to the mass resolution at the $J/\psi$ and centered at the nominal $J/\psi$ 
mass. 
The slope $A_0^{\rm QED}$ is obtained by fitting the charge asymmetry
in the $J/\psi$ sidebands. The measured slope of the charge asymmetry as a
function of $m_{\mu\mu}$ is shown in Fig.~\ref{fig:A0_aroundJpsi}: the expected
behaviour is clearly observed, with a smooth variation from the QED continuum
with a large negative value and a sharp peak approaching a null slope on the
$J/\psi$ resonance.  The specific $A_0^{J/\psi}$ slope is obtained as a function
of $m_{\mu\mu}$ in Fig.~\ref{fig:A0_aroundJpsi} according to Eq.~(\ref{eq:A0_aroundJpsi}): its
value is stable across the $J/\psi$ peak and a fit to a constant between 3.07
and 3.12\gevcc yields
\beqn\label{res:A0_Jpsi}
    A_0^{J/\psi}=(0.3\pm1.6)\times 10^{-2},
\eeqn
which is consistent with zero, as expected from the ISR-only $J/\psi$ production.

\subsection{Comparison to QED}
\label{QED-comp-muon}

\begin{figure*}
\centering 
\includegraphics[height=0.45\textwidth]{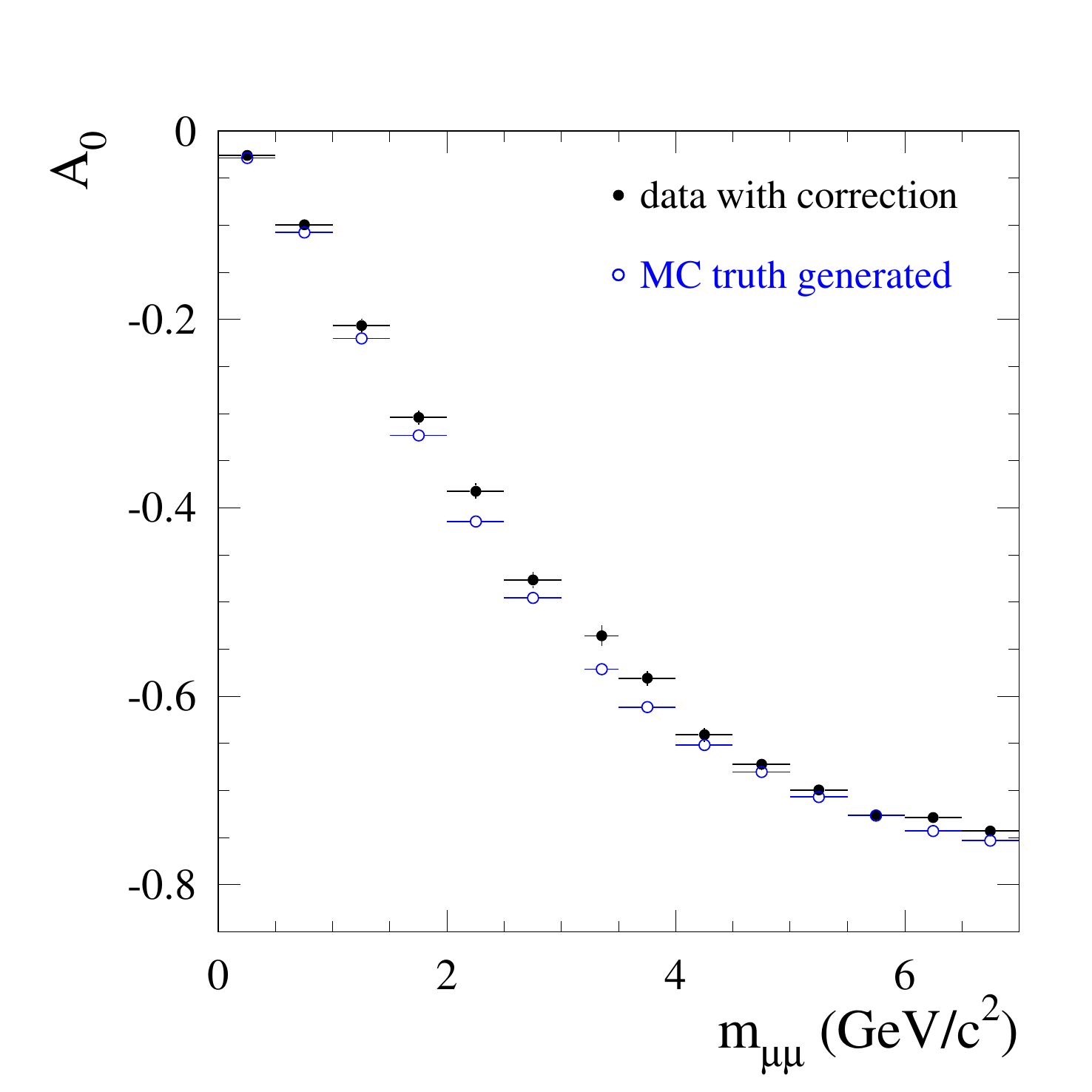}
\includegraphics[height=0.45\textwidth]{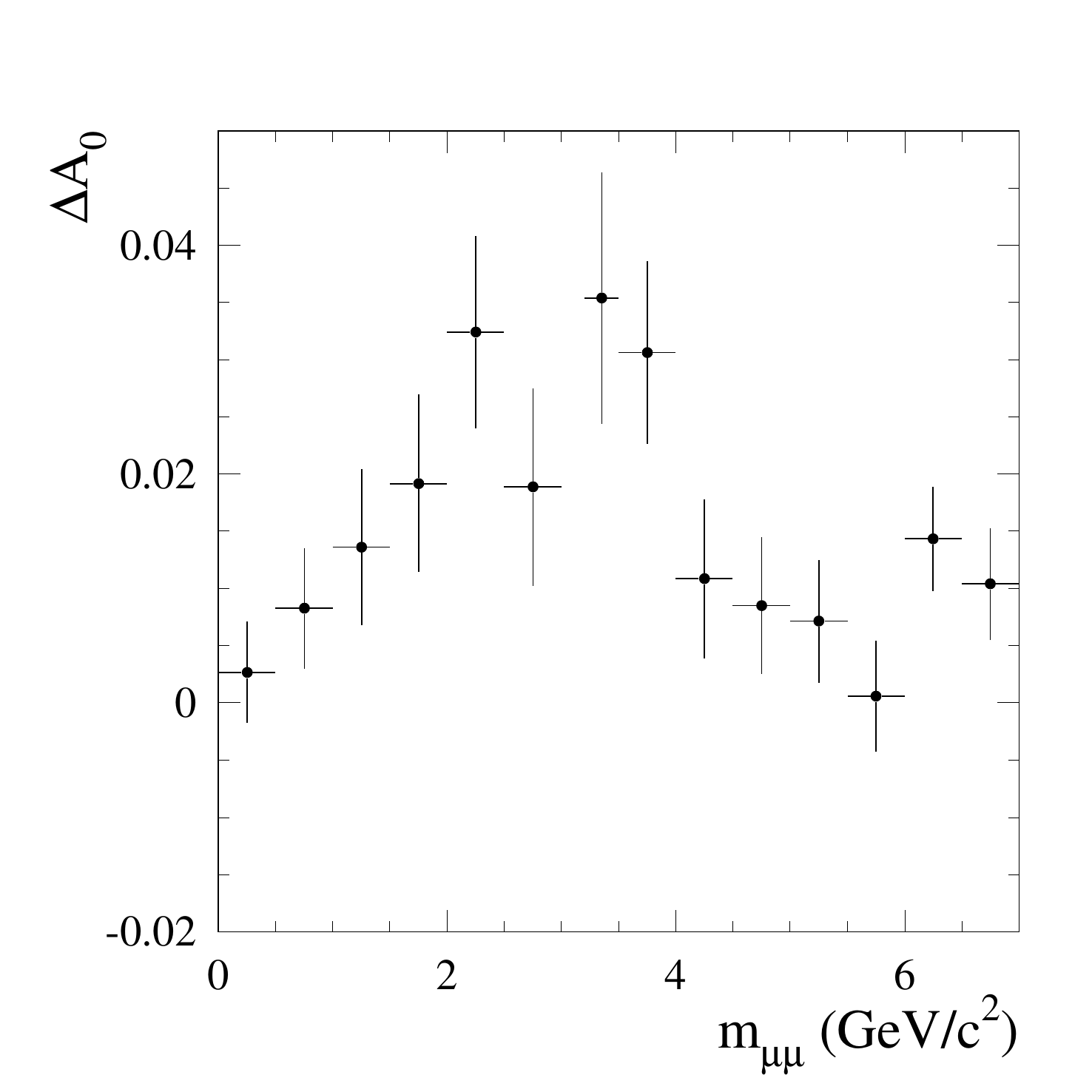}
\caption{The slope $A_0$ of charge asymmetry as a function of $m_{\mu\mu}$ in 
$\epem\to\mmg$ data and in MC at generation level (left), and the absolute difference
between them (right). The $J/\psi$ mass region (3.0-3.2\gevcc) is excluded.
Only statistical uncertainties are shown.}
\label{fig:A0_gmm_DT_MC}
\end{figure*}

The final slope $A_0$ as a function of $m_{\mu\mu}$ measured on the data is
shown in Fig.~\ref{fig:A0_gmm_DT_MC}, together with the asymmetry at the
MC generation level, and the difference between them. 
The mass interval containing the
pure-ISR contribution from the $J/\psi$ ($\epem\to\gamma_{\rm
ISR}J/\psi,~J/\psi\to\mumu$), discussed in detail in Sec.~\ref{sec:Jpsi},
is excluded.
The absolute difference between data and MC $\Delta A_0=A_0^{\rm data}-A_0^{\rm MC}$ 
is at a few percent level (0--3\%).

The measured slope of charge asymmetry is negative throughout the mass
range under study, and its magnitude increases with mass, reaching
values as large as $-0.7$ at $5\gevcc$, in agreement with the trend predicted by QED. 
However, while data and LO QED agree
within $10^{-2}$ at mass less than $1\gevcc$ and above $5\gevcc$, a small but
significant
discrepancy shows up for intermediate mass, reaching $\sim3 \times 10^{-2}$ between
1.5 and $4\gevcc$. Investigations of systematic uncertainties, both at the experimental 
and theoretical levels, are reported in the next 
section.  

\begin{figure} \centering 
  \includegraphics[width=0.49\textwidth]{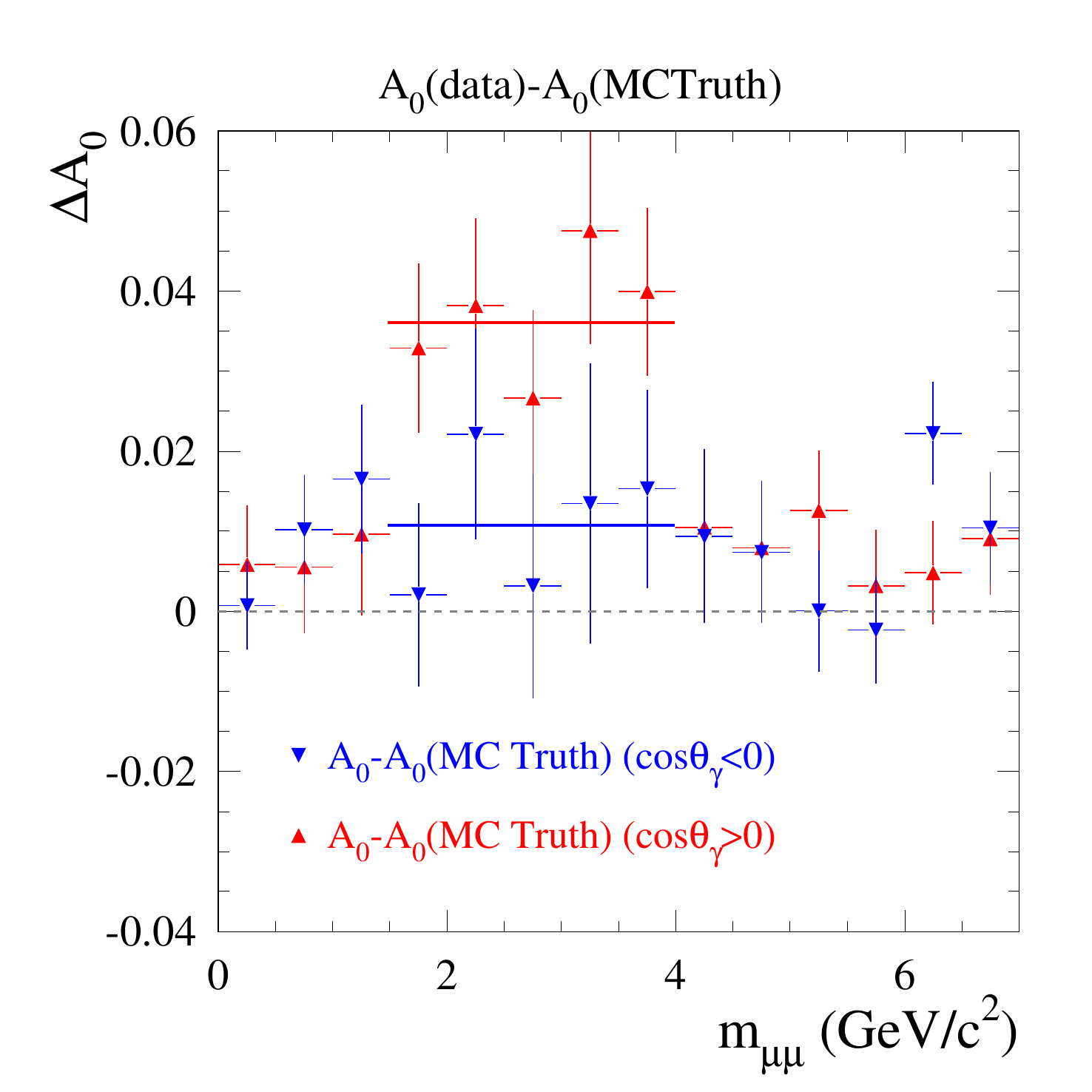} 
   \caption{The difference between the measured asymmetry slope in the
  $\mu^+\mu^-\gamma$ process and the AfkQed prediction, as a function of
  $m_{\mu\mu}$, excluding the $J/\psi$ 3.0-3.2\gevcc region.  Forward
  ($\cos\theta^*_\gamma>0$) and backward ($\cos\theta^*_\gamma<0$)
  hemispheres are analysed separately. 
Statistical uncertainties only.}
  \label{fig:FB-test}
\end{figure}

Due to the asymmetry of the beam energies at \pep2, independent
charge asymmetry measurements in two different kinematic regimes are provided
by splitting the data into a forward ($\cos\theta^*_\gamma>0$) sample
and a backward ($\cos\theta^*_\gamma<0$) sample. The full analysis,
including background subtraction and efficiency correction, is
redone on each sample separately. The results are shown in
Fig.~\ref{fig:FB-test}.
A significant discrepancy between data and AfkQed is observed in the forward
region, in the 1.5--4\gevcc mass region, while in the backward hemisphere
data and AfkQed are consistent. The differences are quantified in 
Table~\ref{tab:forwback}. No significant forward-backward difference 
is expected from the generator.

\begin{table*}
  \centering  
  \setlength{\extrarowheight}{1.5pt}
  \setlength{\tabcolsep}{5pt}
  \caption{The difference (in $10^{-2}$ units) between the measured $A_0$ and the AfkQed 
prediction for the two mass intervals 1.5--4 and 4--7 \gevcc in different
$\cos{\theta^*_\gamma}$ regions. The last line gives the difference between the two
regions $\cos{\theta^*_\gamma}>0$ and $\cos{\theta^*_\gamma}<0$.
Statistical uncertainties only.}   
  \vspace{0.1cm}
  \begin{tabular}{|c|c|c|} \hline\hline\noalign{\vskip2pt} 
   ($10^{-2}$) & $1.5<m_{\mu\mu}<4$\gevcc &  $4<m_{\mu\mu}<7$\gevcc \\
  \hline  
  all $\cos{\theta^*_\gamma}$   &  $2.65\pm0.38$  &  $0.86\pm0.22$  \\   
  $\cos{\theta^*_\gamma}>0$     &  $3.61\pm0.50$  &  $0.76\pm0.30$  \\
  $\cos{\theta^*_\gamma}<0$     &  $1.08\pm0.60$  &  $0.82\pm0.31$  \\
  difference                   &  $2.50\pm0.78$  &  $-0.05\pm0.42$  \\
  \hline\hline   
  \end{tabular}   
\label{tab:forwback}
\end{table*}

\subsection{Systematic uncertainties}
\label{sec:syst_mu}

\subsubsection{Experimental systematic effects}
The primary sources of systematic uncertainty are the background estimation,
data/MC differences in detector response, and differences between the physical
charge asymmetry in the data and in the generated MC events.


The difference
$\Delta A_0$ between the results with and without background subtraction is
found to be well below
$10^{-3}$ except in the 0.5--1.0\gevcc range, where it reaches 2$\times
10^{-3}$ because of the larger $\rho$ background with two pions misidentified
as muons. Since the background level is known with better than 10\%
accuracy~\cite{prd-pipi}, the corresponding systematic uncertainty on the
asymmetry slope is at most 2$\times 10^{-4}$ throughout the studied mass range.


The trigger, tracking and $\mu$-ID induce charge-asymmetric data/MC corrections, 
as overlap effects are not perfectly reproduced by simulation. However, the 
data/MC corrections have small
effects on the charge asymmetry slope, at most 3$\times 10^{-3}$. Since the corrections have
been measured with a precision of 10\% or better~\cite{prd-pipi}, the
corresponding systematic uncertainty on the asymmetry slope is less than
3$\times 10^{-4}$.


As explained in Sec.~\ref{sec:kinAccep}, the effects from the kinematic acceptance on the measured
slope of the charge asymmetry depend on the physical charge asymmetry
itself. The possible bias on the acceptance correction, induced by the
physical charge asymmetry in the generator inaccurately reproducing the data, is studied using
a sample of re-weighted $\epem\to\mmg$ MC events where weights are adjusted to
yield the same asymmetry as measured in the data in 
each ($m_{\mu\mu}$, $\theta^*_\gamma$, $\theta^*$, $\cos\phi^*$) phase-space 
cell. The expected bias in the measurement from a difference of charge asymmetry 
between data and MC is found to be less than 5$\times 10^{-3}$, which is taken as a
systematic uncertainty on the $A_0$ measurement.

\subsubsection{Effects from imperfect simulation}


Since a simple linear fit $A(\cos\phi^*)=A_0\cos\phi^*$ might be
questionable, we perform an alternate two-parameter fit on the charge asymmetry
after efficiency corrections $A(\cos\phi^*)=A_0\cos\phi^*+ B_0$. 
The $B_0$ values obtained in data are a few $10^{-3}$ at most, 
while the asymmetry slopes $A_0$ deviate from the final values, which use the one-parameter fit,
by less than $10^{-4}$.

\begin{figure*} \centering 
  \includegraphics[width=0.49\textwidth]{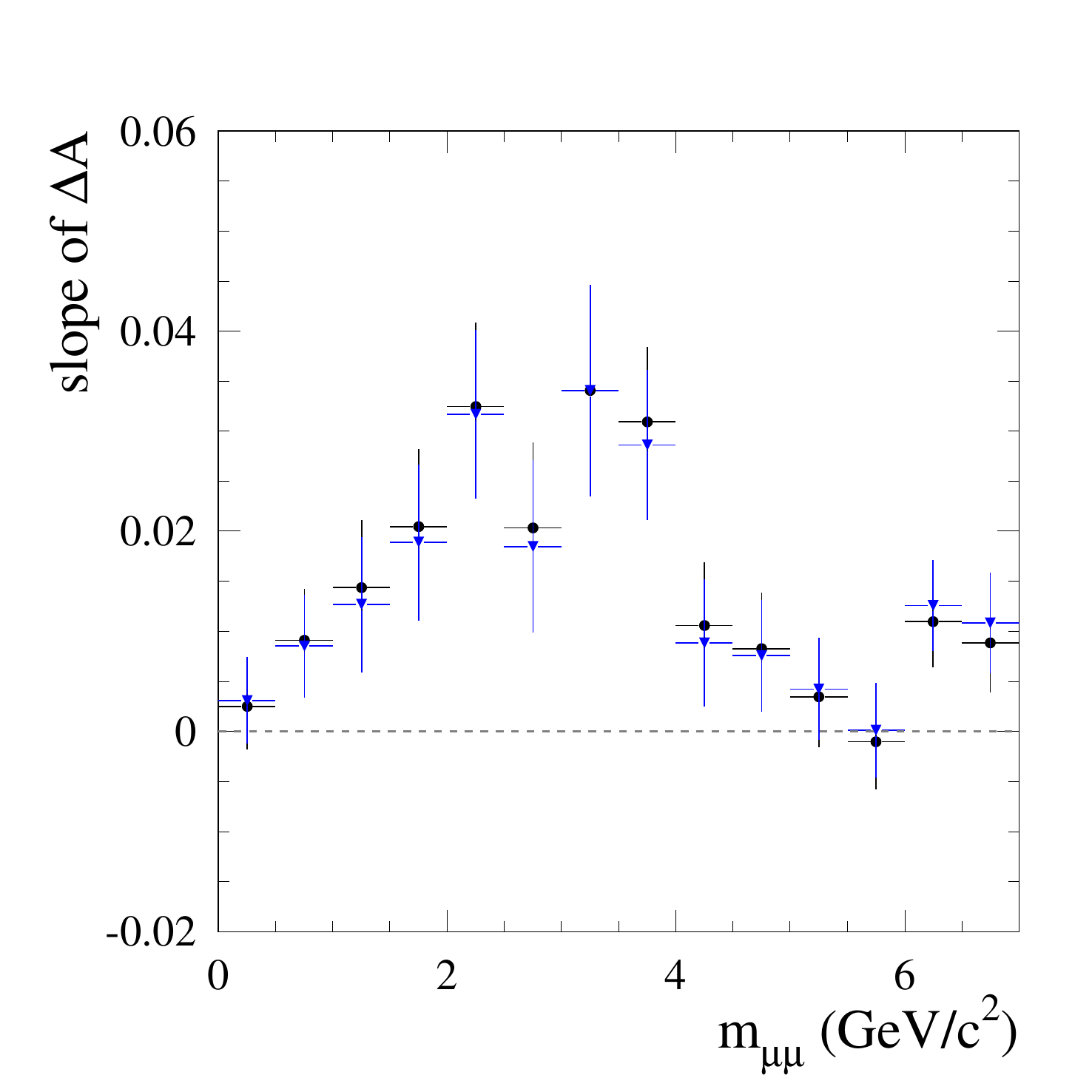} 
  \includegraphics[width=0.49\textwidth]{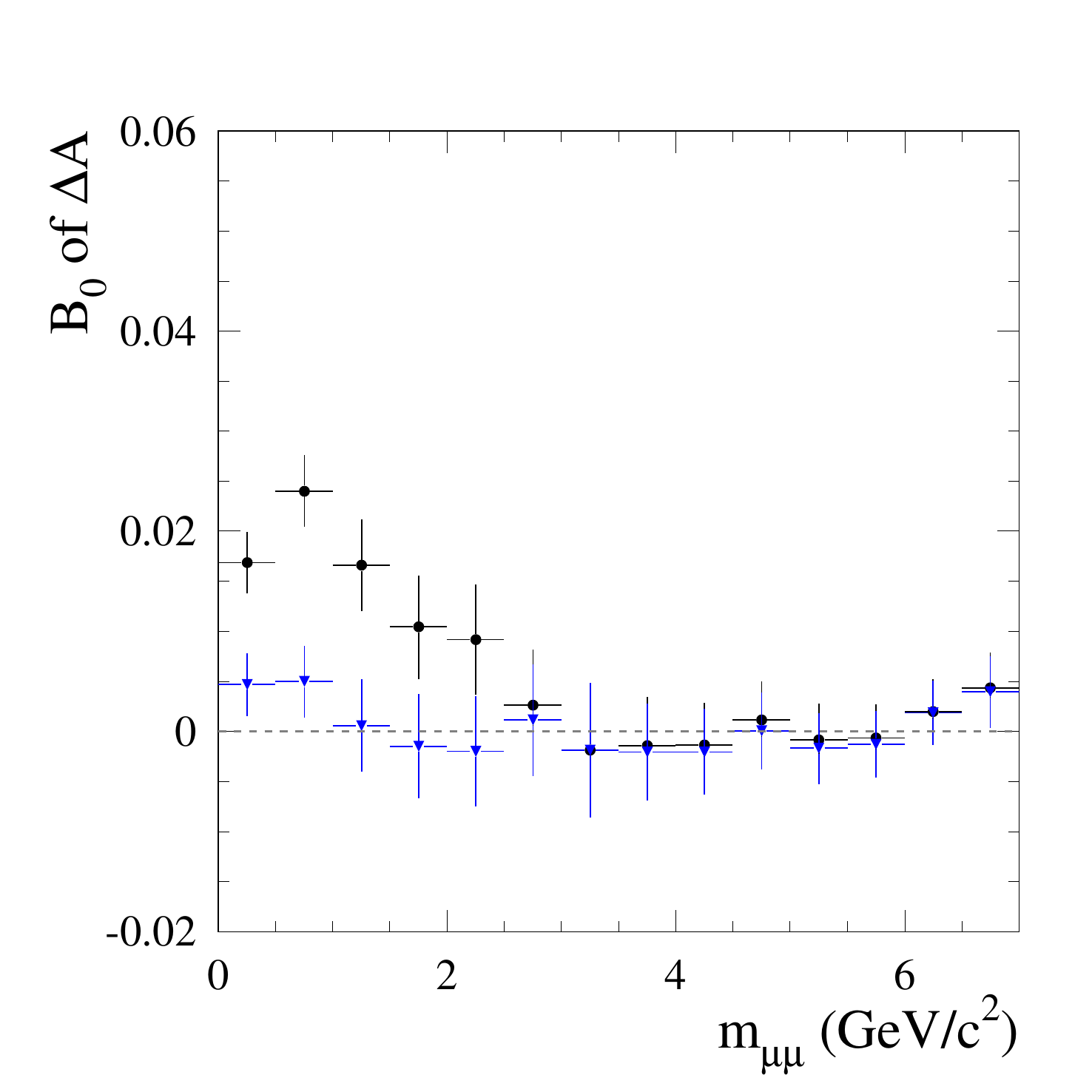} 
  \caption{(left) Slope of the difference $\Delta A^{\rm raw}(\cos\phi^*)$  for $\mu^+\mu^-\gamma$ 
  between the raw
  asymmetries observed in data and in MC before (black points) and after (blue triangles)
  the data/MC corrections,  
  as a function of
  $m_{\mu\mu}$. (right) Constant term of the two-parameter linear fits of
  $\Delta A^{\rm raw}(\cos\phi^*)$.  
Statistical uncertainties only.}
  \label{fig:rawdA-test}
\end{figure*}

To investigate whether the observed discrepancy results from the efficiency
corrections, we study the difference $\Delta A^{\rm raw}(\cos\phi^*)$  between the raw 
asymmetries observed in data and MC after full event selection.
Although the raw asymmetry itself is not linear, especially at low
mass (Fig.~\ref{fig:A_data_vs_MC}), the difference $\Delta A^{\rm raw}(\cos\phi^*)$ 
in each mass interval is observed to be
linear with $\cos\phi^*$. In particular, there are no edge effects in
the  vicinity of $|\cos\phi^*|\simeq 1$, which could have resulted
from different resolutions in data and MC. 
The results of the fits are shown on Fig.~\ref{fig:rawdA-test} (black points).  
The values of the slope of $\Delta A^{\rm raw}$ are
insensitive whether the linear fit is a one-parameter or a
two-parameter fit.
The data-MC discrepancy in the 1.5--4\gevcc mass region is already 
observed at the raw level, which excludes efficiency or resolution bias.  
The $B_0$ values returned by the two-parameter fit over $\Delta A^{\rm raw}$
depart from zero by up to 2.5$\times 10^{-2}$ at low mass, as expected
from imperfect detector simulation. When data/MC corrections of detector
efficiencies are applied to the simulated raw data, the results are shown on 
Fig.~\ref{fig:rawdA-test} (blue triangles). 
The $B_0$ values are reduced to a few $10^{-3}$, while the slope of $\Delta A^{\rm raw}$
is not changed by more than 3$\times 10^{-3}$.

When performed on the forward and backward samples independently, the
study at the raw data level confirms that the data-MC discrepancy in
the 1.5--4\gevcc mass interval is confined to the forward region, where
the slope of $\Delta A^{\rm raw}$ is significantly non-null by $\sim 5\sigma_{\rm stat}$ 
while it is consistent with zero within $2\sigma_{\rm stat}$ over the full mass
range for the $\cos\theta^*_\gamma<0$ sample. In contrast, the fitted
$B_0$  values  are consistent with each other in the two samples,
except at very low mass ($m_{\mu\mu}<1\gevcc$).


Comparisons of the data and MC distributions of event variables
entering the asymmetry analysis are performed, in particular near the
acceptance boundaries. A sizeable departure is observed in the high
$E_\gamma$, low  $\theta_\gamma$ region, in the forward
$\cos\theta^*_\gamma>0$ hemisphere. However, the asymmetry measurement is found
to be insensitive to this discrepancy at very forward photon angles.
To investigate whether different resolutions in data and MC might bias
the efficiency corrections and the event assignment to the $N_+$ or
$N_-$ samples, the analysis is fully redone with tighter acceptance
requirements.  The change of asymmetry slope is small, $(4 \pm 3)\times 10^{-3}$.
Conservatively, a systematic uncertainty of 7$\times 10^{-3}$ is assigned to account for
imperfect simulation near the edges of the selected phase space. 

The studies above first show that, although acceptance and detector inefficiency 
effects are important, they are well accounted for in the simulation. Data/MC corrections
are found to significantly reduce the symmetric component of the asymmetry ($B_0$), but
most importantly, the studies demonstrate that the  measurement of the asymmetry slope is 
robust against uncertainties in the efficiency corrections.

\begin{figure}
  \centering 
  \includegraphics[width=0.49\textwidth]{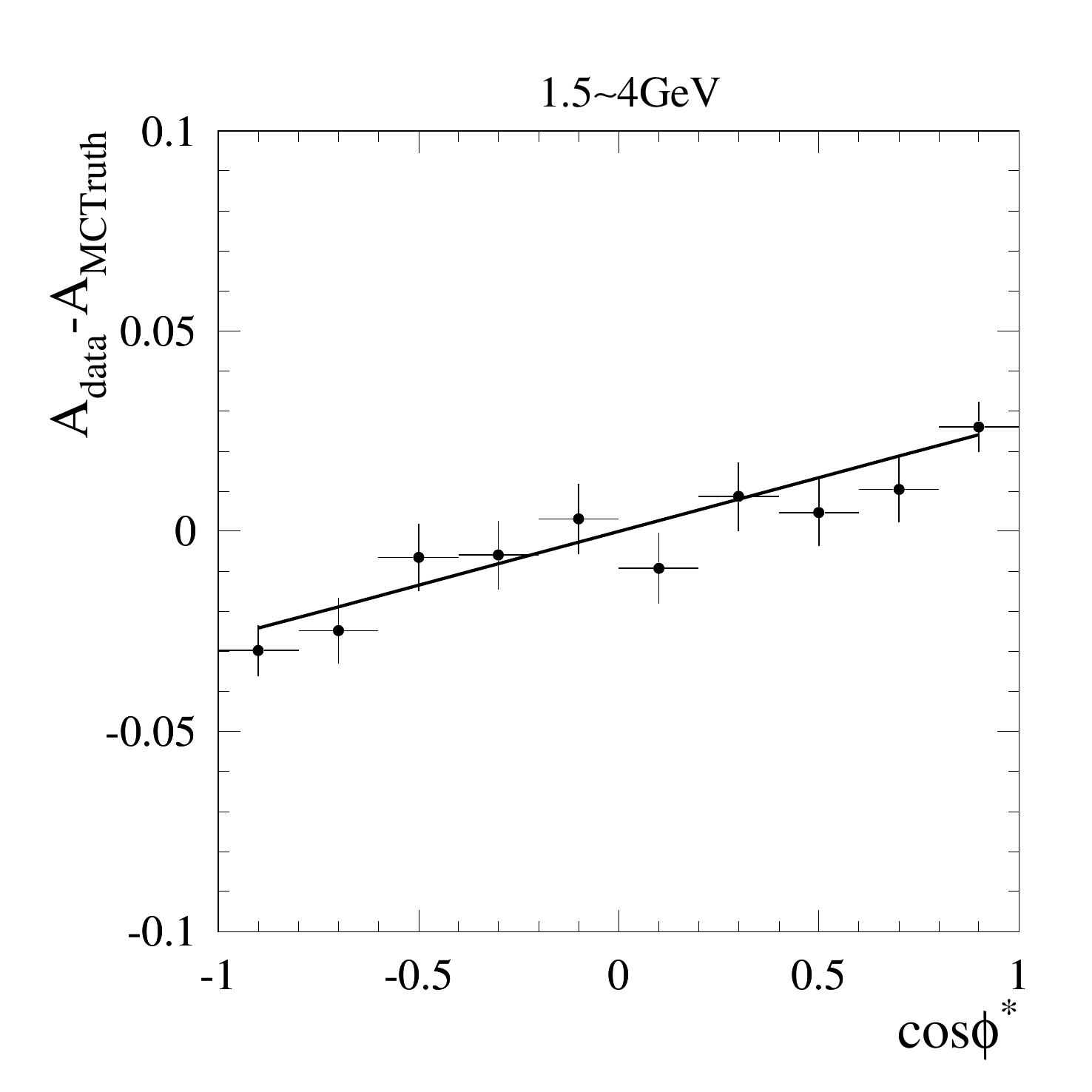} 
  \caption{The difference between the measured asymmetry in the 
$\mu^+\mu^-\gamma$ process and the QED prediction, as implemented in AfkQed,
as a function of $\cos\phi^*$ for the events in the mass interval 1.5--4\gevcc, 
excluding the $J/\psi$ 3.0--3.2\gevcc region. The result of a linear fit is 
shown by the solid line.}
  \label{fig:linearity-test}
\end{figure}

As a global test to differentiate between an uncorrected experimental
bias and a true deviation from the QED prediction, the difference
between the measured asymmetry and the theoretical one, as
implemented in AfkQed, is studied as a function of $\cos\phi^*$ for
the events in the mass interval 1.5--4\gevcc where the deviation is
the largest (excluding the $J/\psi$ 3.0--3.2\gevcc region). As shown in 
Fig.~\ref{fig:linearity-test},
a linear dependence is indeed observed. This supports the assertion that 
the deviation we observe does not originate from a detector effect unaccounted for 
in the simulation. 

\subsubsection{Theoretical systematic effects}


The AfkQed event generator only includes the LO QED interference
between  ISR and FSR amplitudes. Additional photons generated
independently for  ISR and FSR, induce a change in asymmetry through
kinematics.  The NLO contributions to the QED interference are studied
with the latest  version of the {\small PHOKHARA}
generator~\cite{phok-2013},  which includes a full matrix element
computed at NLO.  NLO contributions are found to affect the charge
asymmetry by 1-2$\times 10^{-2}$ in the mass range covered by the present
analysis, where events are generated with the highest energy photon
in the  $20^o <\theta^*_\gamma< 160^o$ reference
range~\cite{czyz-nlo}.  The {\small PHOKHARA} results with fully implemented
NLO  corrections are consistent with the AfkQed results with
independent extra photons, with some small discrepancy up to $10^{-2}$ for masses 
larger than 4\gevcc.
This shows that the small difference
between LO and NLO asymmetry originates essentially from kinematic
effects due to the extra photon.


The contribution from $Z^0$ exchange is investigated with the {\small KKMC} 
generator~\cite{kk2f}, either processed in the QED-only 
configuration, or including
the full $\gamma+Z^0$ exchange diagrams. As in AfkQed, extra photons are
generated independently in the initial state 
and final state (with {\small PHOTOS}).
Electroweak (EW) effects are found to be at a few $10^{-3}$ level, averaging 
over the full mass range.

A significant difference of  $(0.81\pm0.16)\times 10^{-2}$ is observed between the asymmetry slopes in 
{\small KKMC} and AfkQed, with an asymmetry slope $A_0$ larger (in absolute
value) in AfkQed than in {\small KKMC}. The conclusion holds if one
considers the forward and backward hemispheres separately. As already
observed for AfkQed, the  asymmetries expected
from {\small KKMC} in the two hemispheres are consistent with each other.

Comparison of the asymmetry slope measured in data, after acceptance
correction, to the full QED+$Z$  expectation, as implemented in {\small
KKMC}, confirms that a significant difference of 2$\times 10^{-2}$
remains in the 1.5--4\gevcc mass interval,  mostly in the forward
hemisphere ($\cos\theta^*_\gamma>0$).

\subsubsection{Conclusion on systematic uncertainties}

In the large number of tests, both experimental and theoretical, that have been performed,
the antisymmetric part ($A_0$) of the charge asymmetry is found to be remarkably stable. 
It is immune
to all detector effects taken into account in the simulation, unlike the symmetric part ($B_0$).
The simulation properly corrects known effects after data/MC
adjustment of separate sources. The experimental absolute systematic uncertainties 
on $A_0$ are estimated to be 0.5$\times 10^{-2}$ from MC reweighting, 0.3$\times 10^{-2}$ from data/MC efficiency 
corrections, 0.7$\times 10^{-2}$ from acceptance edge effects, which sum up to 0.9$\times 10^{-2}$. 
In view of the observed differences on $A_0$ using AfkQed, {\small PHOKHARA} 9.0 (LO and NLO) and  
{\small KKMC} (with and without EW corrections), 
we conservatively set a 1.0$\times 10^{-2}$ systematic uncertainty on the theoretical prediction.
Adding experimental and theoretical uncertainties quadratically a total absolute
systematic uncertainty of 1.4$\times 10^{-2}$ is obtained.

Although we have been unable to find a bias producing the observed shape as a function
of mass of the difference between the measured $A_0$ and the QED predictions, all data
points are within the estimated systematic uncertainty, except for 5 out of 14 points 
near 3\gevcc that exceed the systematic uncertainty by about 1-2 statistical standard deviations.

\section{\boldmath Results on the charge asymmetry in the $\epem\to\pipig$ process}

\begin{figure*}
  \centering 
  \includegraphics[height=0.25\textwidth]{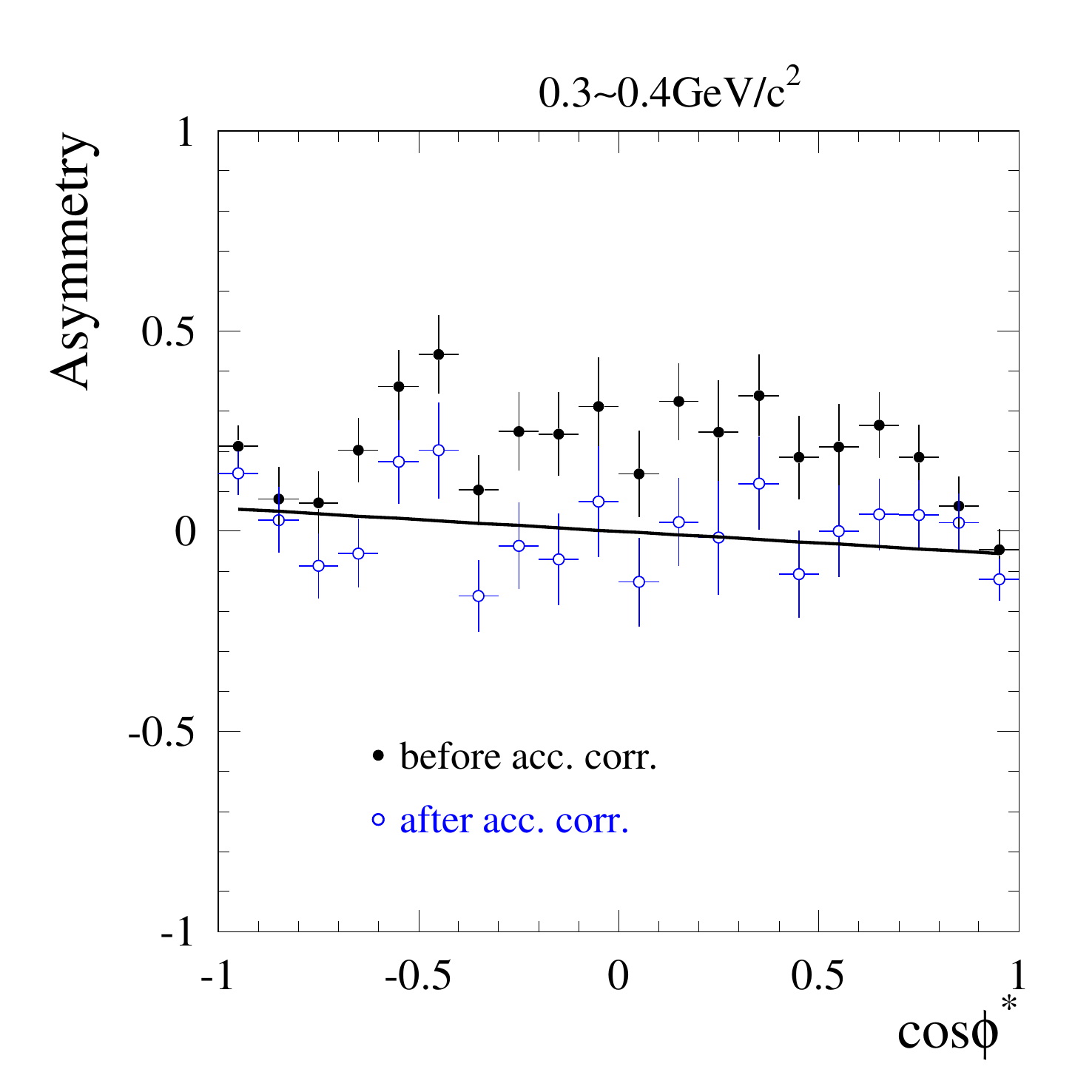}
  \includegraphics[height=0.25\textwidth]{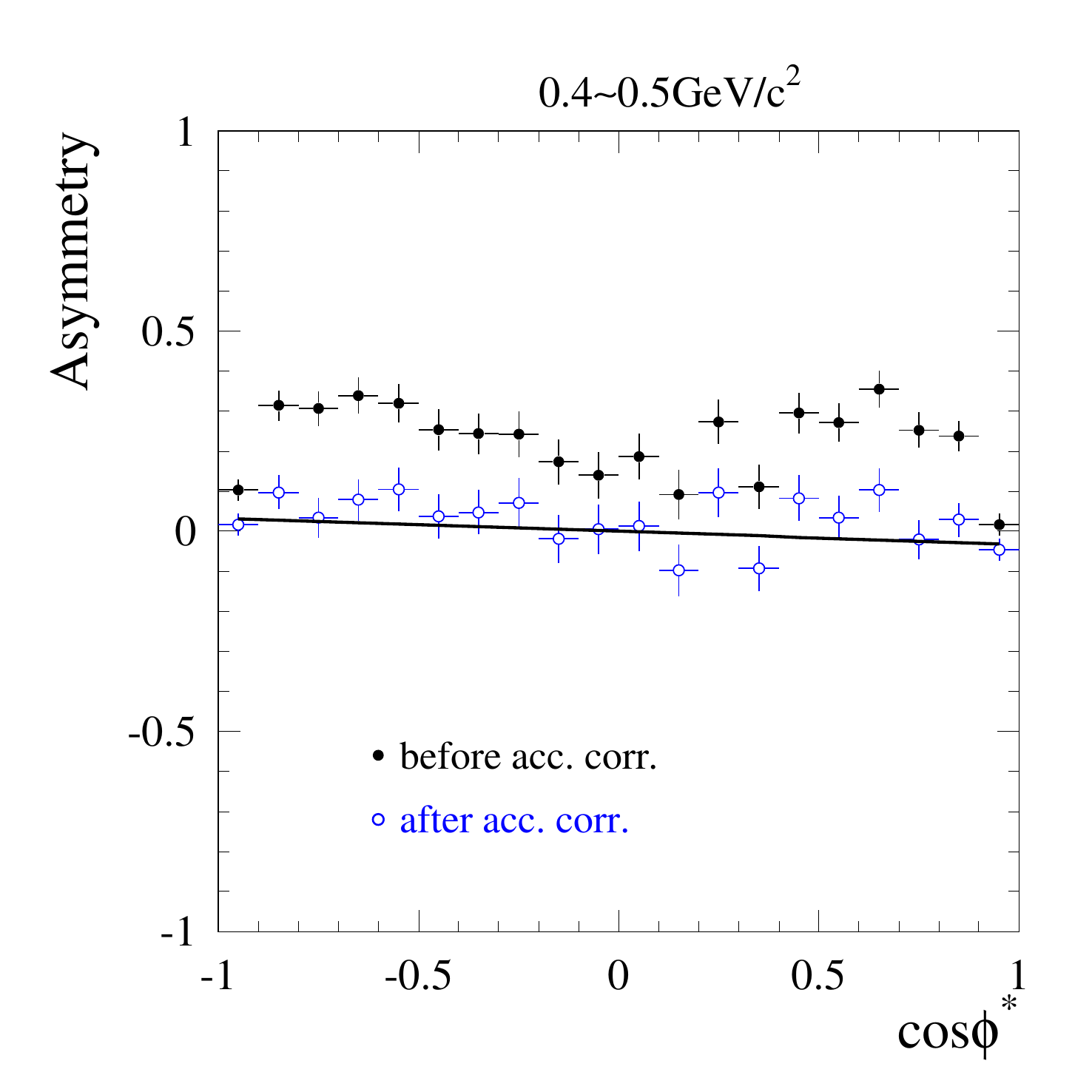}
  \includegraphics[height=0.25\textwidth]{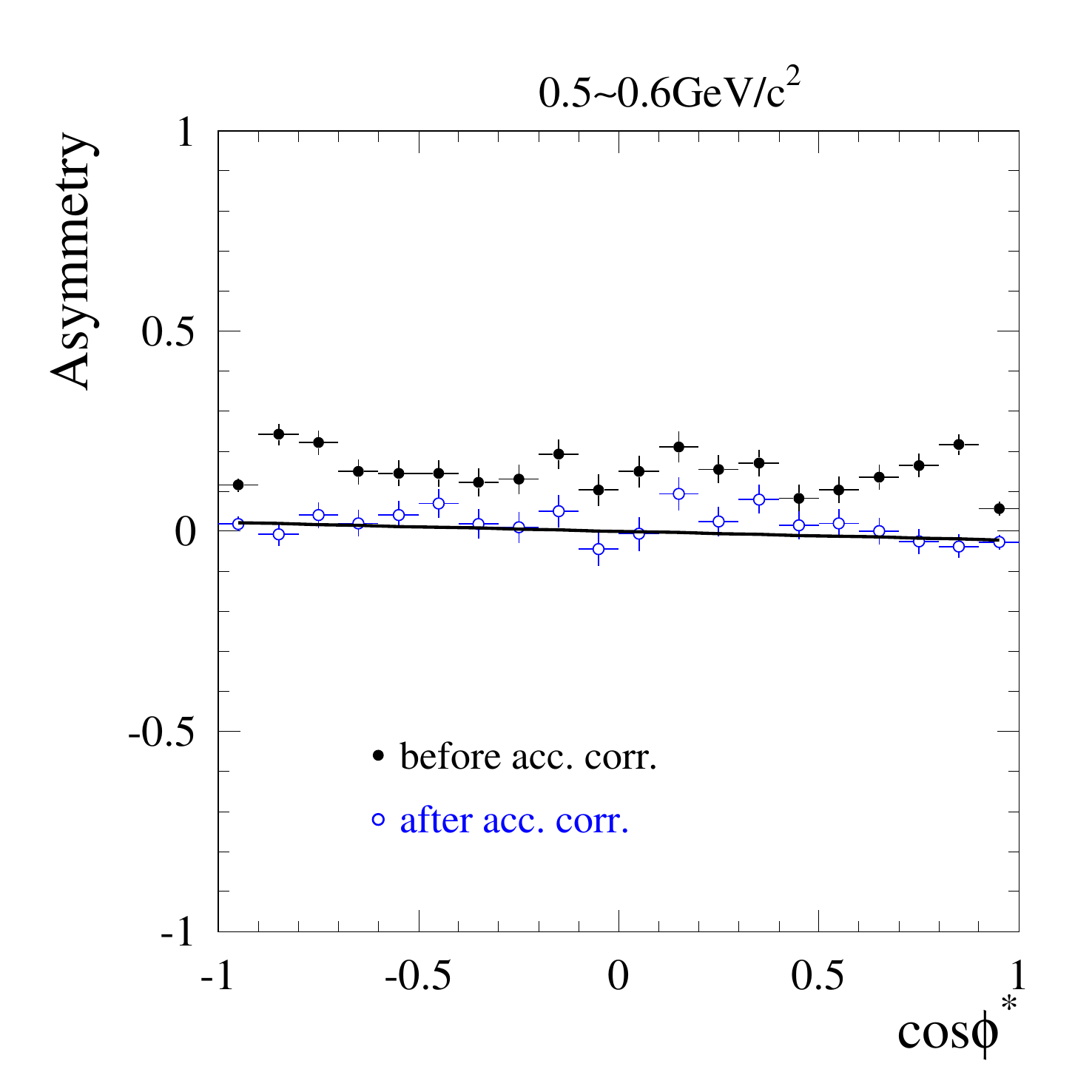}
  \includegraphics[height=0.25\textwidth]{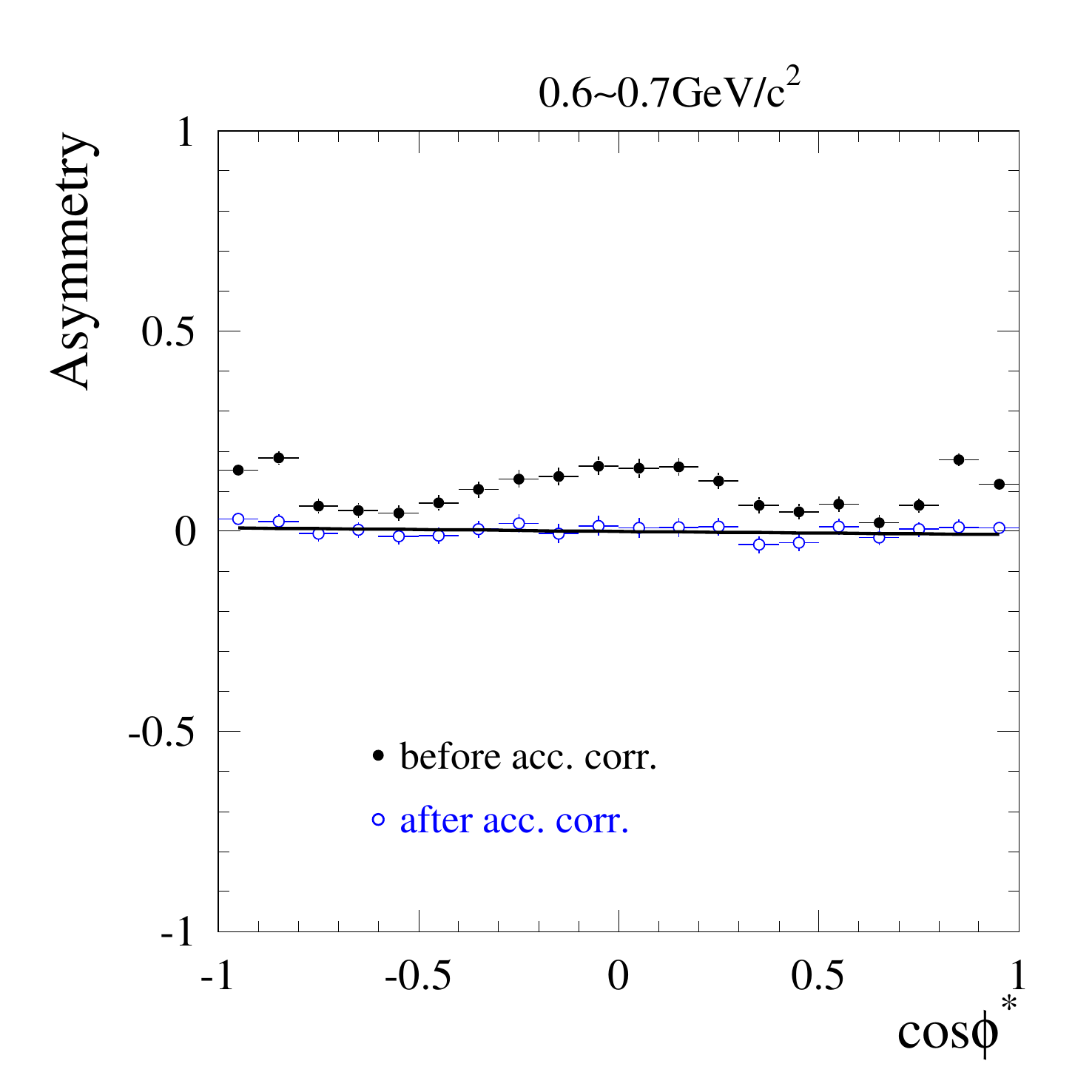}
  \includegraphics[height=0.25\textwidth]{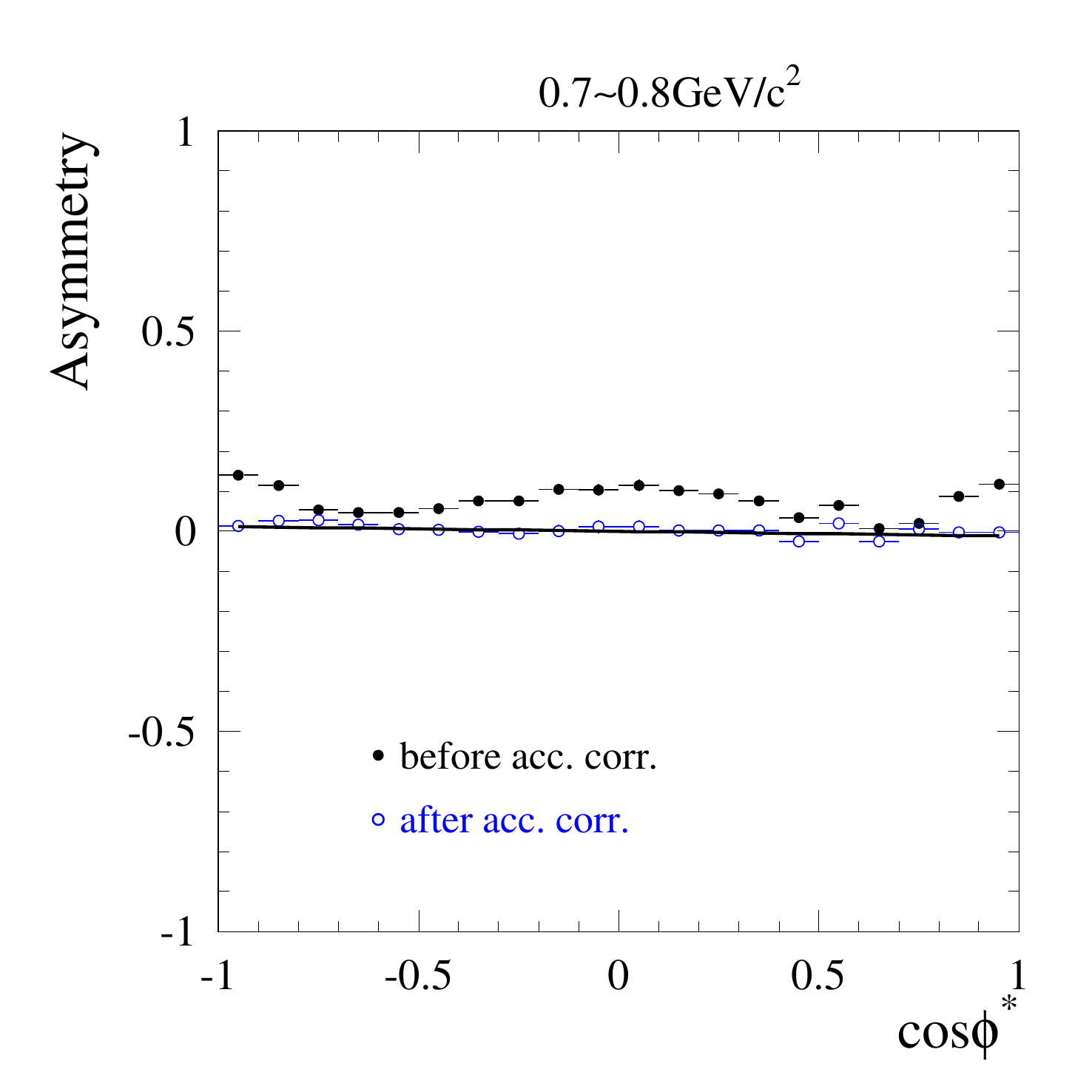}
  \includegraphics[height=0.25\textwidth]{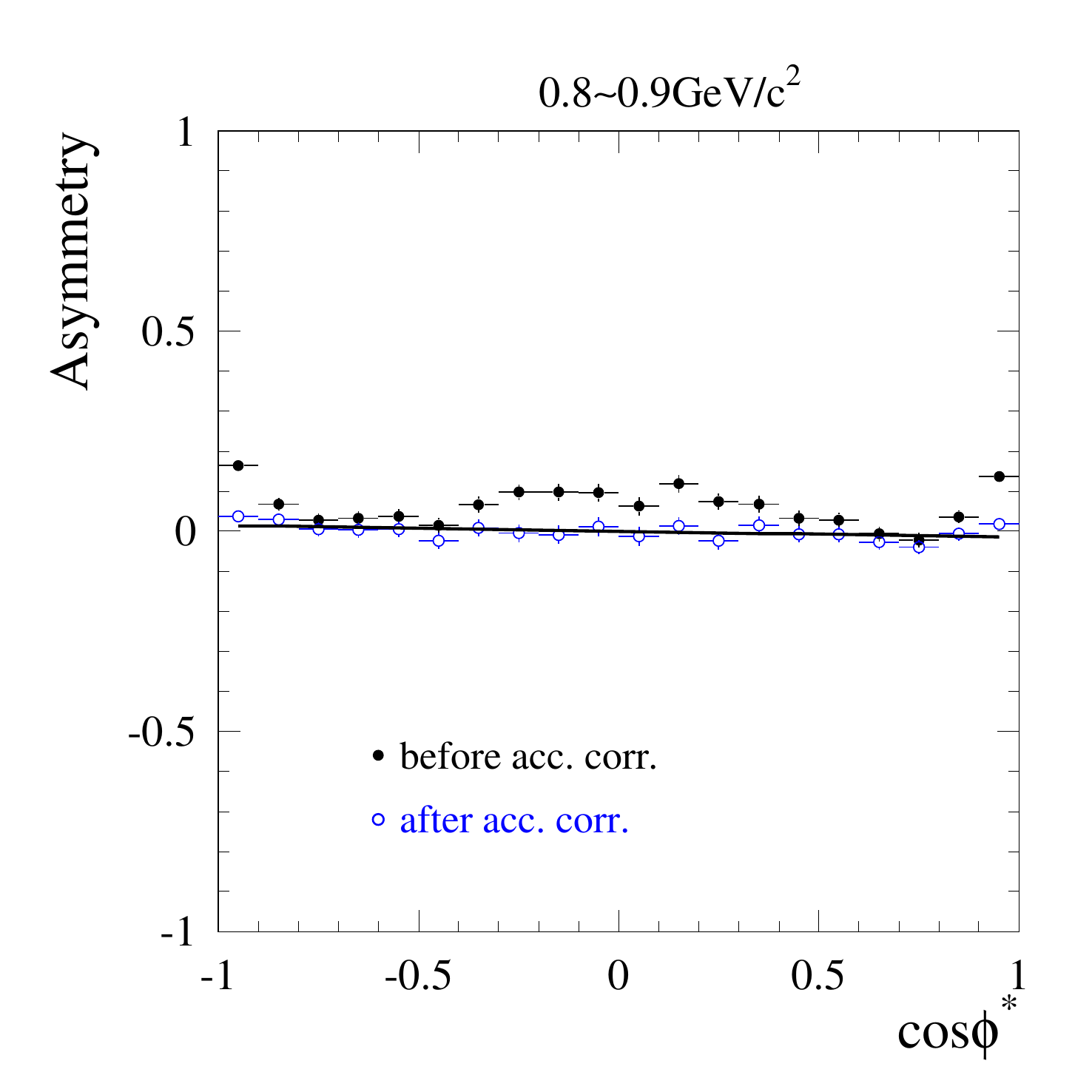}
  \includegraphics[height=0.25\textwidth]{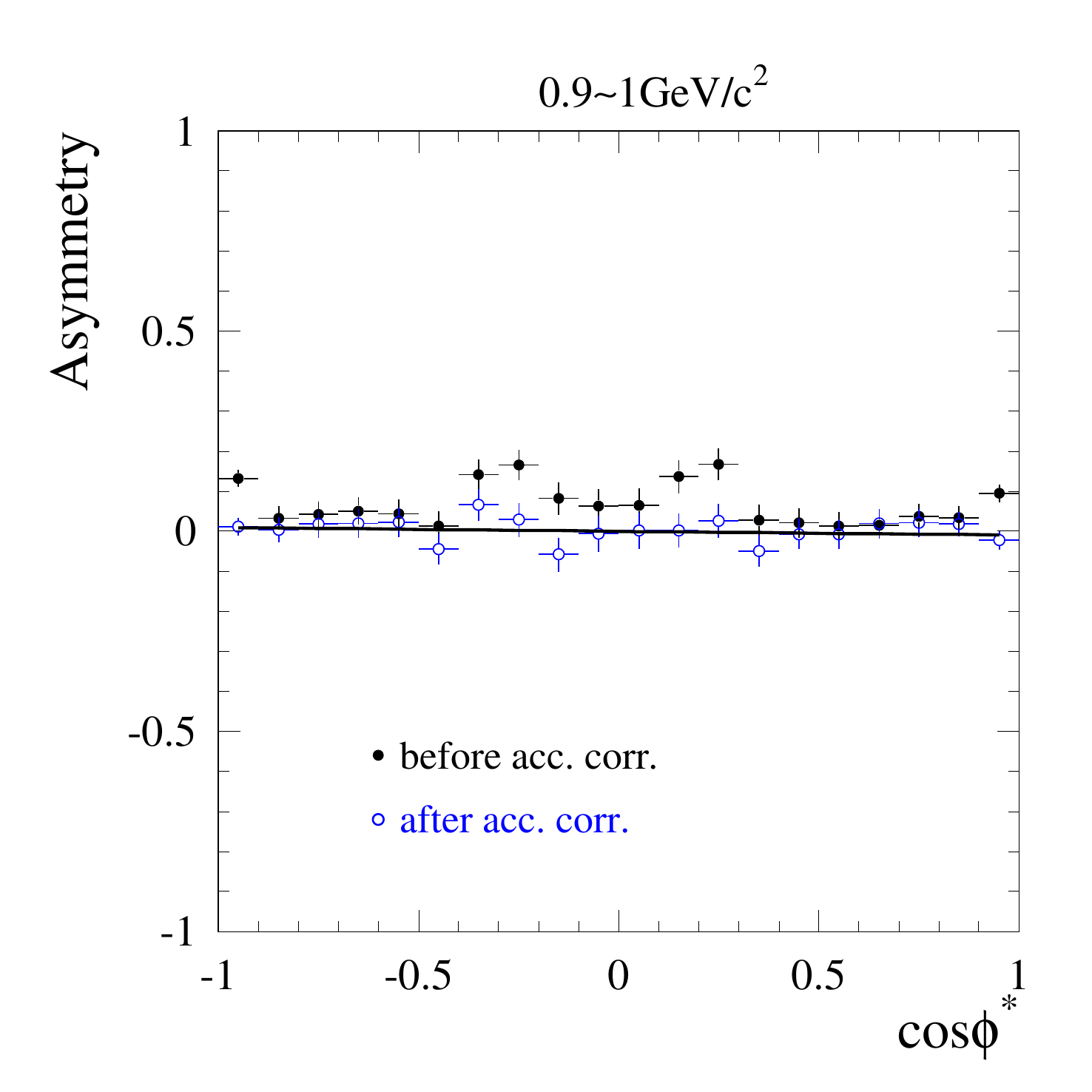}
  \includegraphics[height=0.25\textwidth]{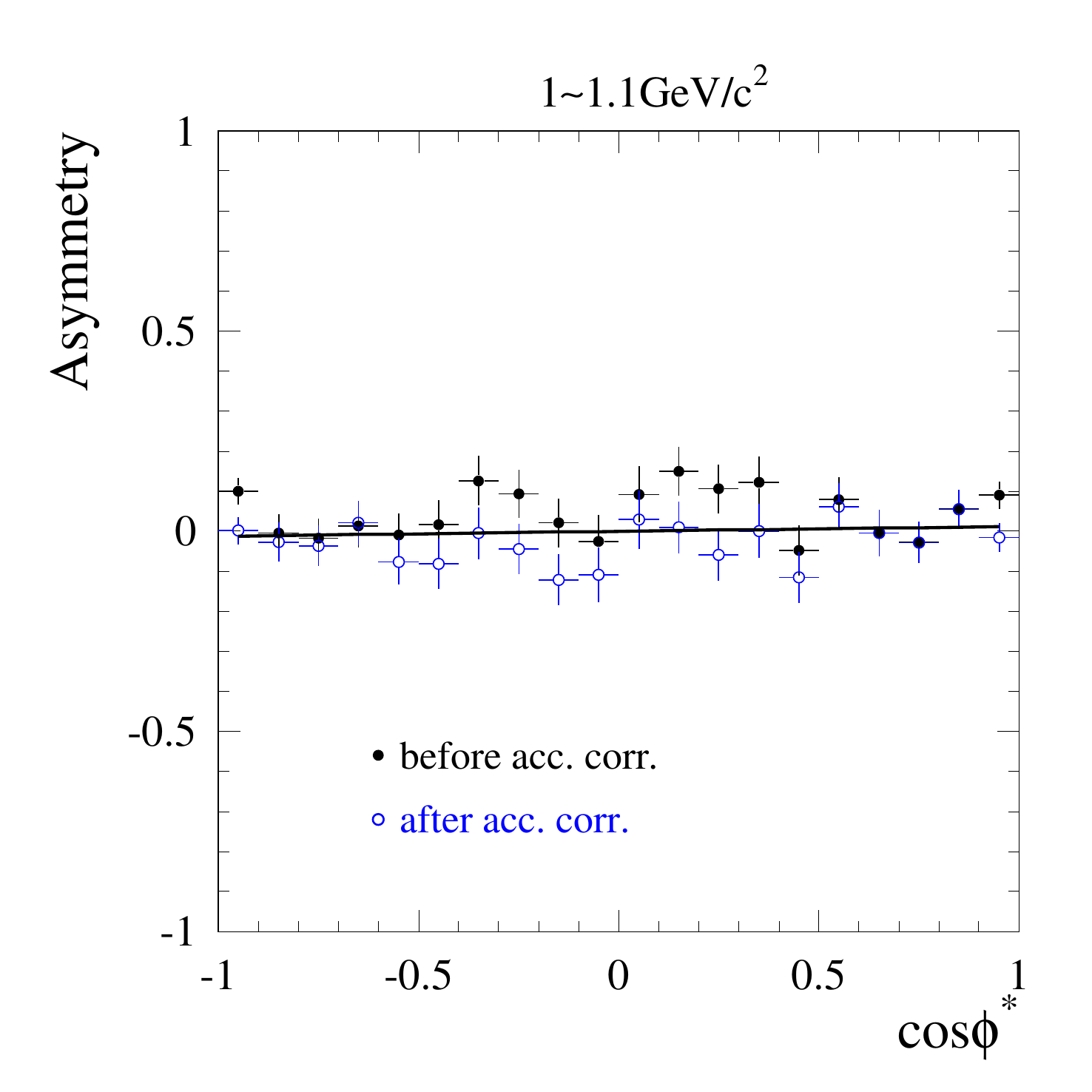}
  \includegraphics[height=0.25\textwidth]{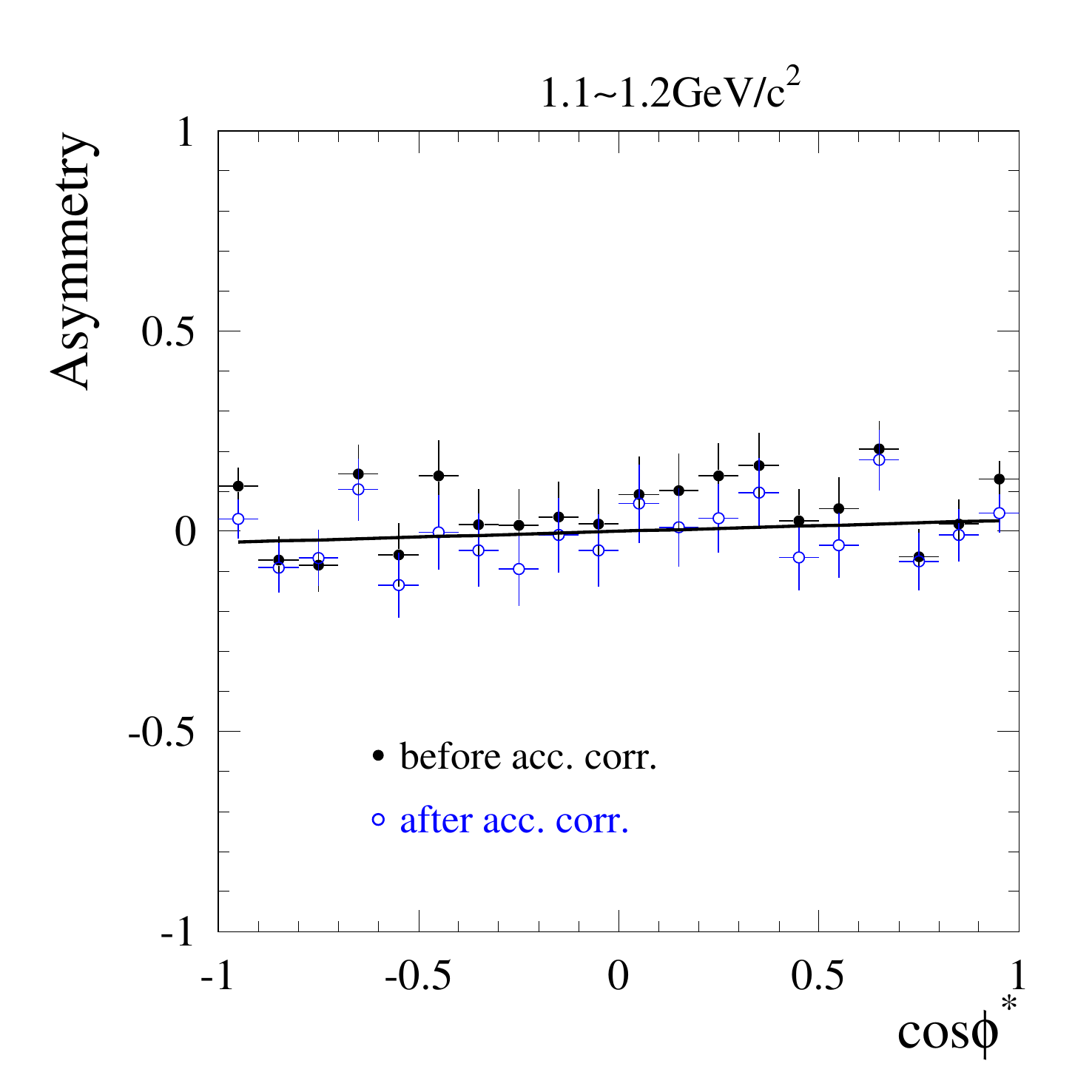}
  \includegraphics[height=0.25\textwidth]{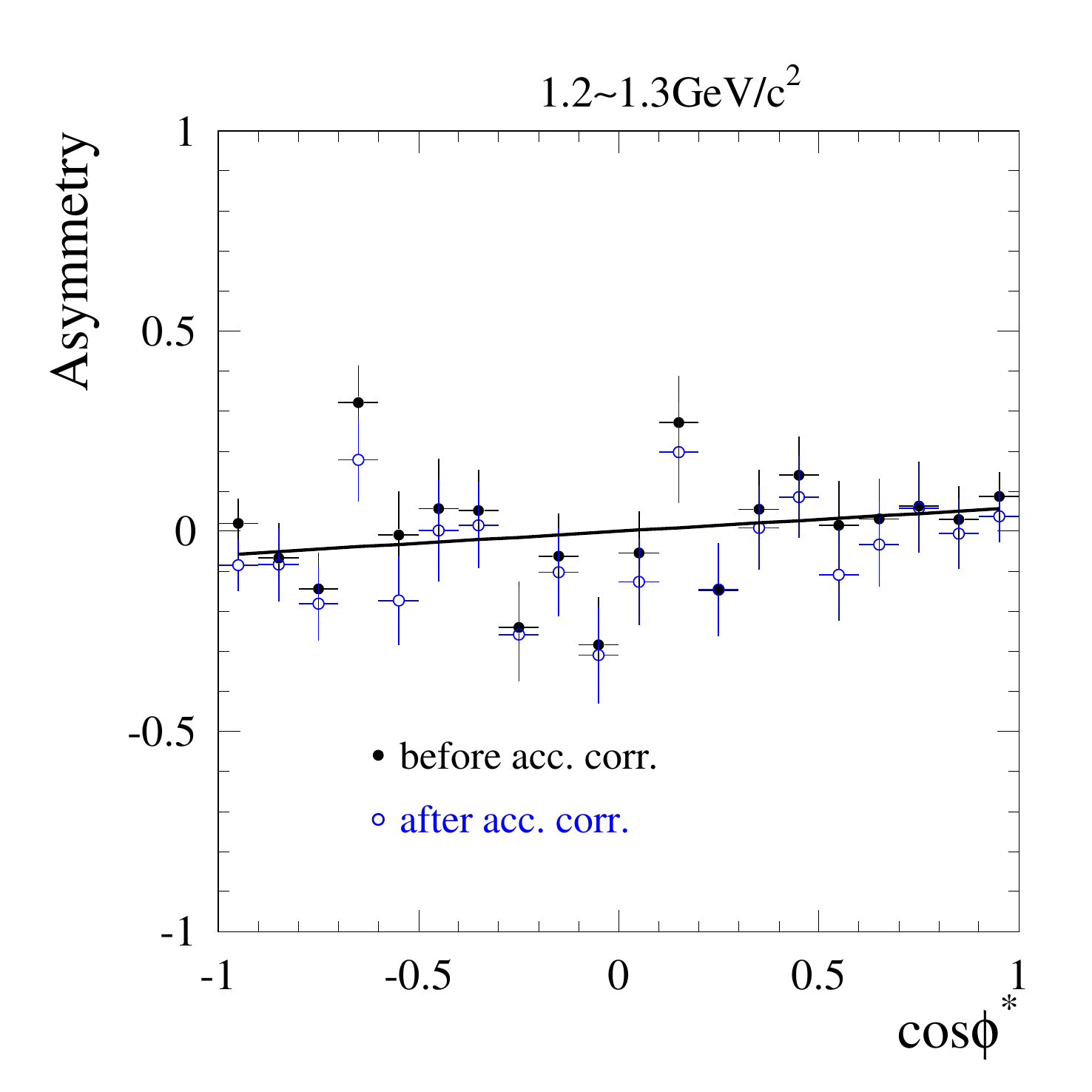}
  \includegraphics[height=0.25\textwidth]{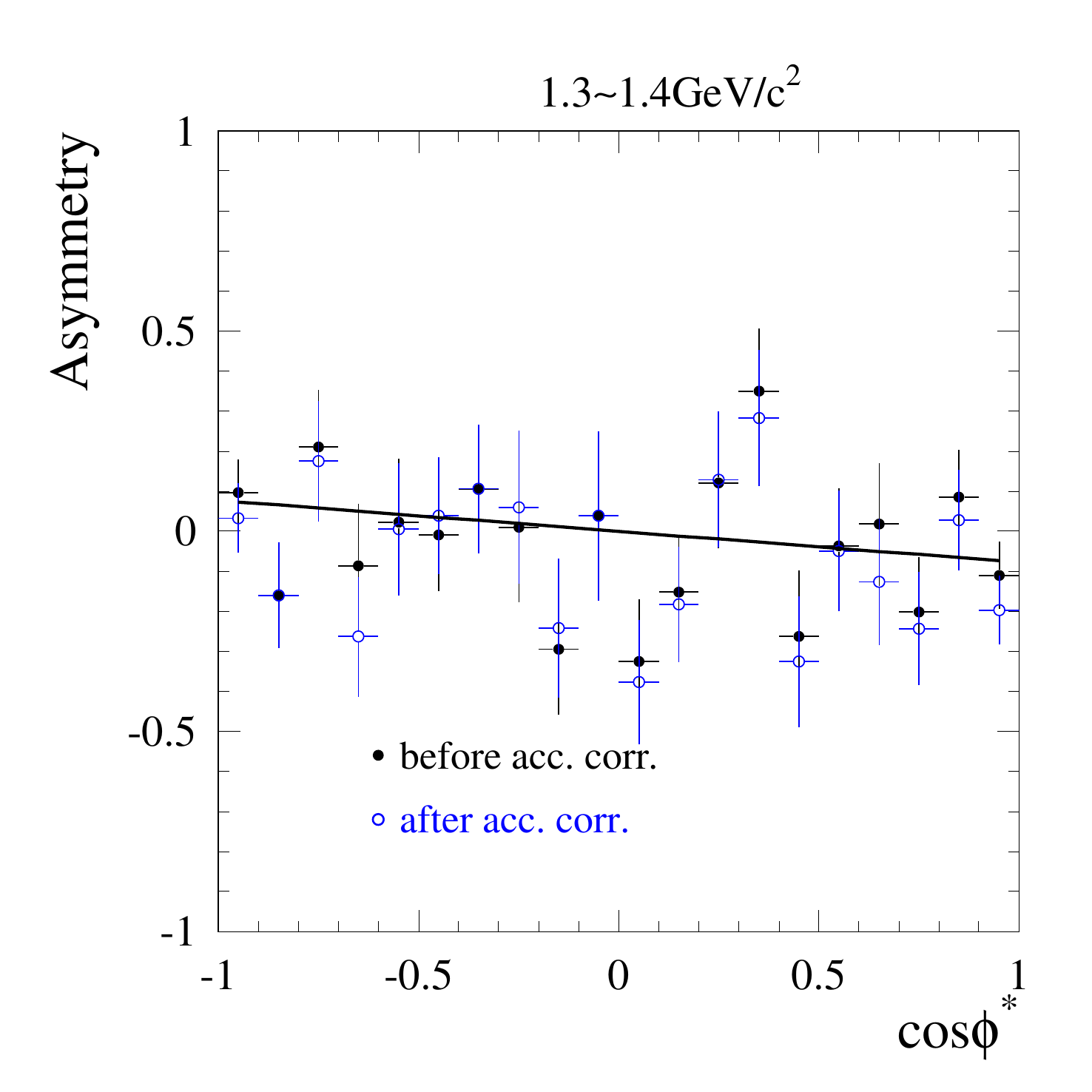}
  \includegraphics[height=0.25\textwidth]{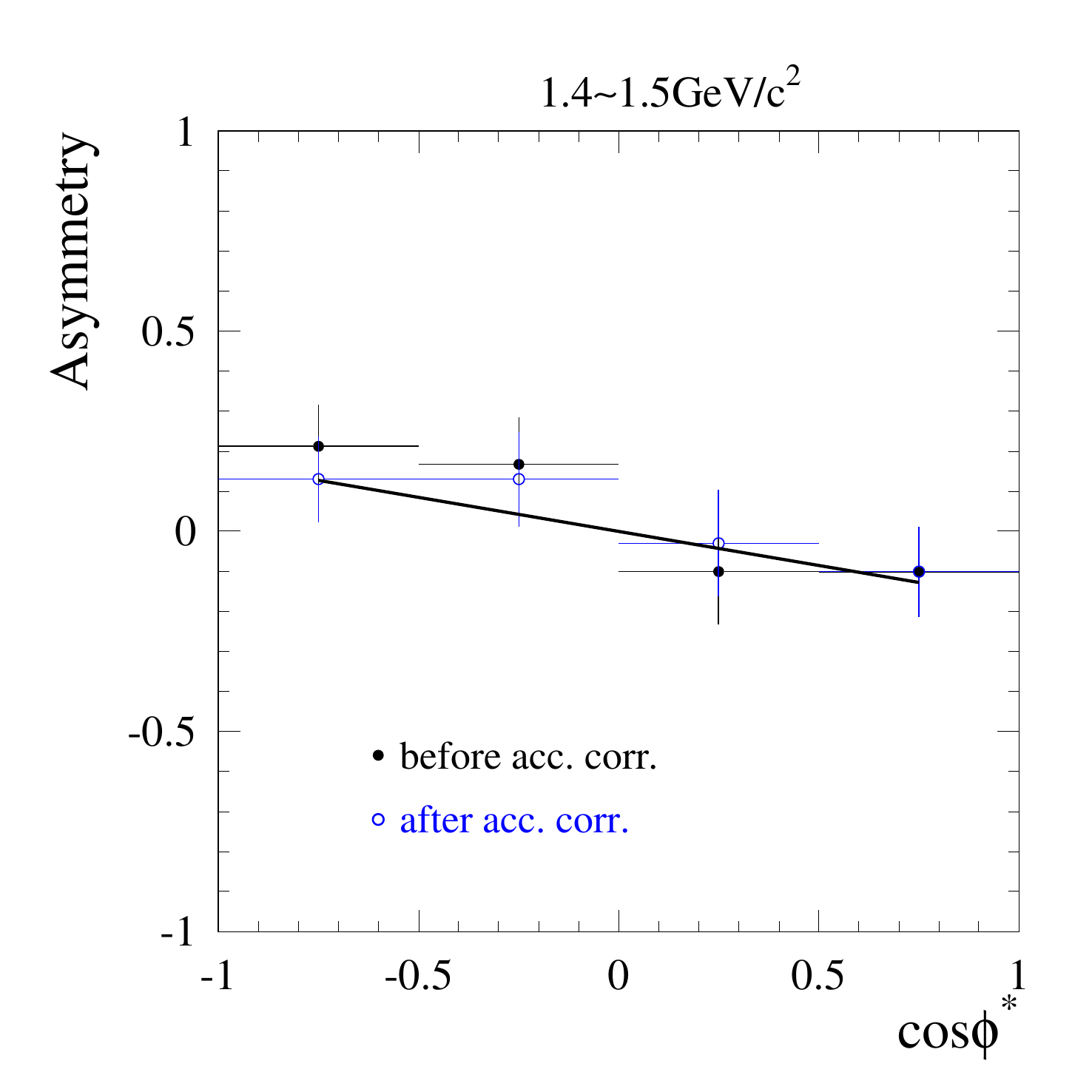}
  \includegraphics[height=0.25\textwidth]{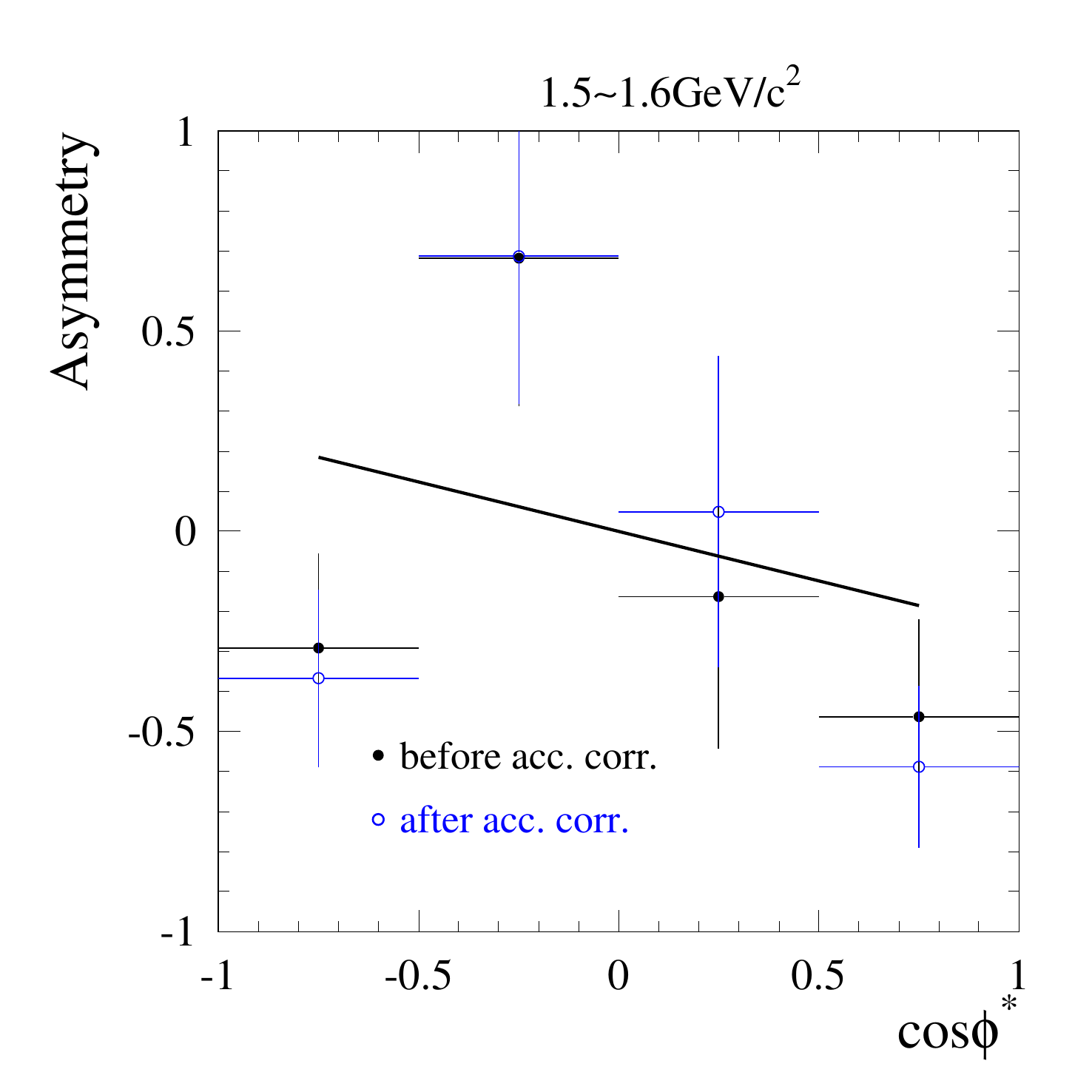}
  \includegraphics[height=0.25\textwidth]{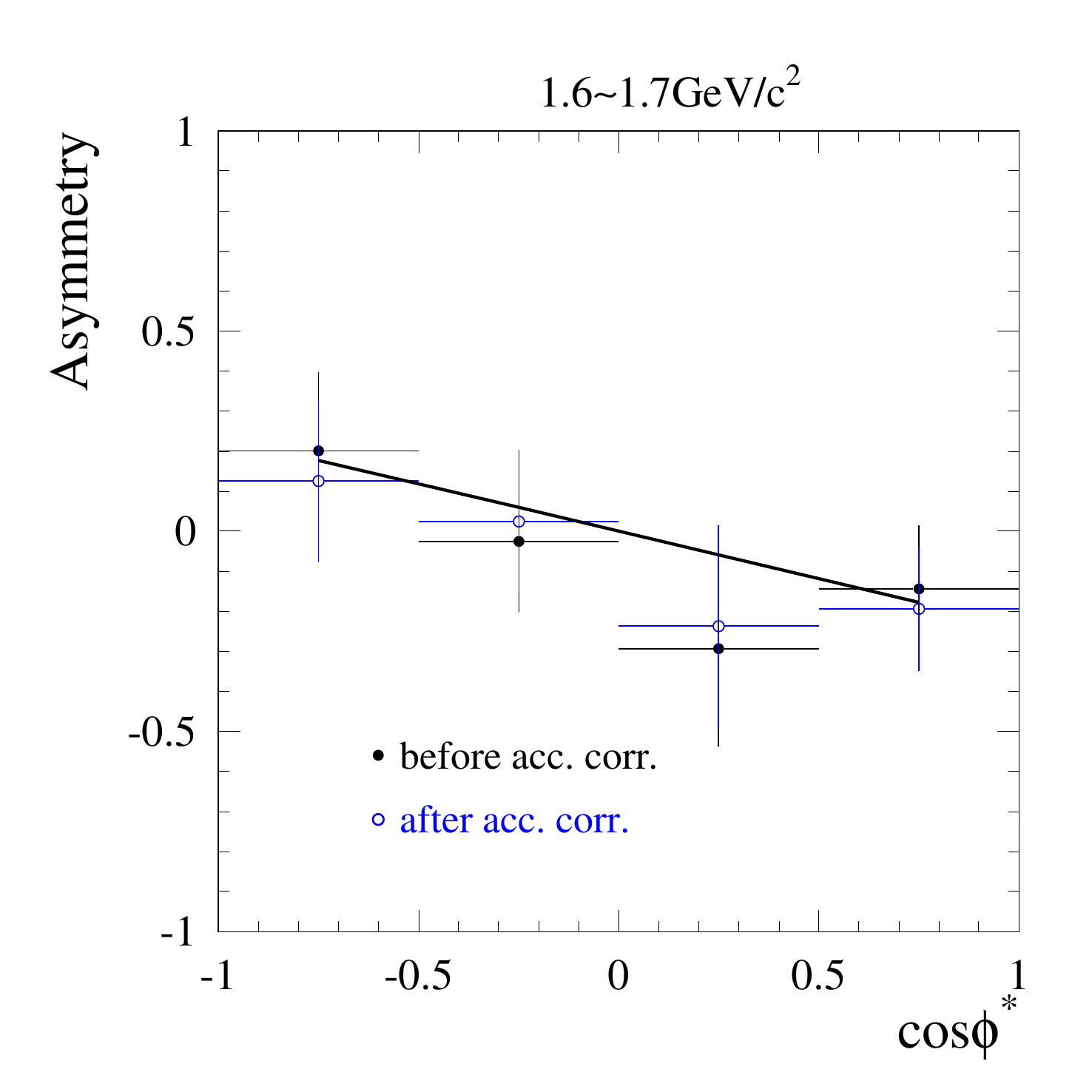}
  \includegraphics[height=0.25\textwidth]{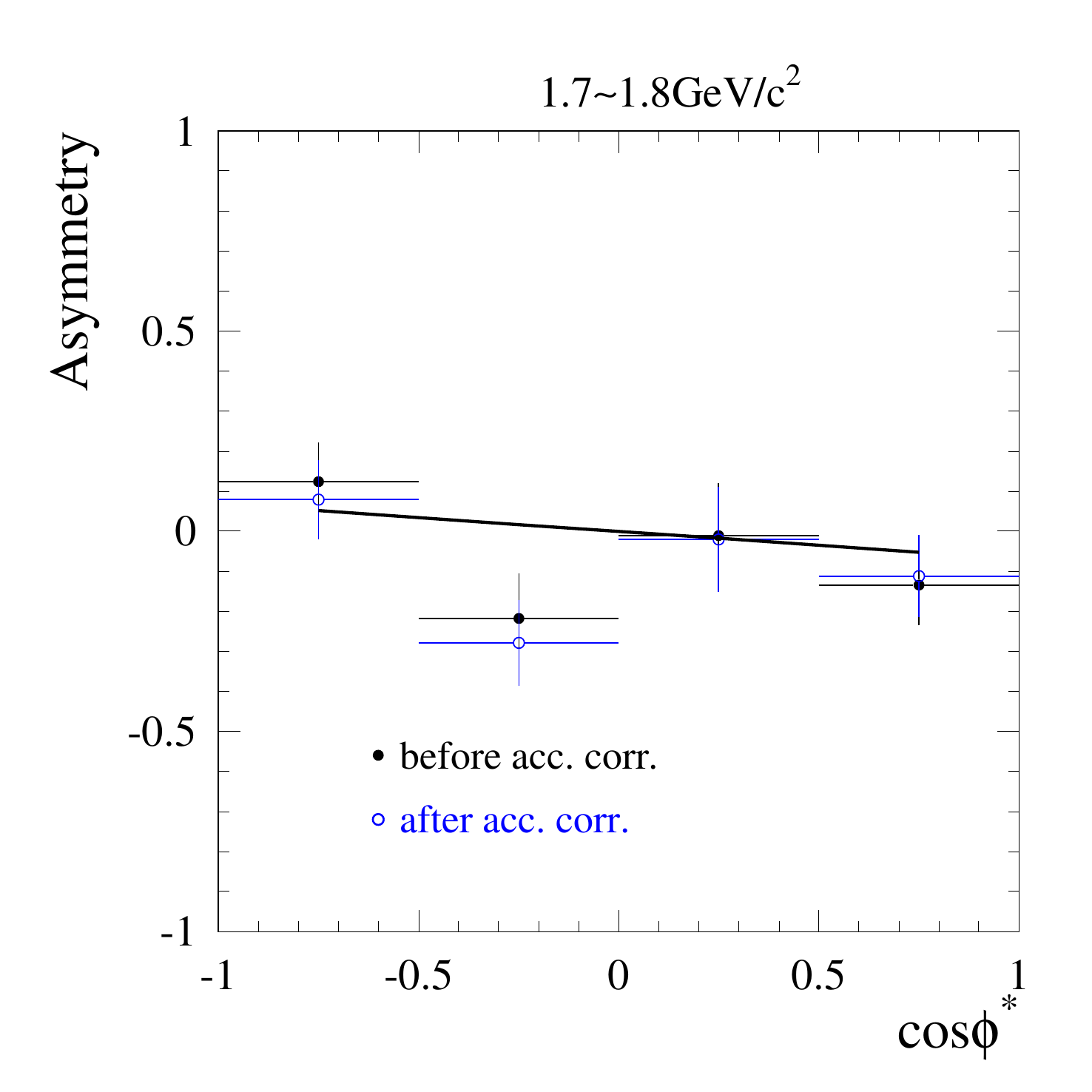}
  \caption{Charge asymmetry for $\epem\to\pipig$ data before ($\bullet$) and after ($\circ$)
  efficiency corrections, in 0.1\gevcc mass intervals from 0.3\gevcc to 1.8\gevcc.
  The line shows the result of the fit to $A_0\cos\phi^*$.
}
\label{fig:A-dataCorr-gpipi}
\end{figure*}

The charge asymmetry for $\epem\to\pipig$ data before and after background subtraction
and efficiency corrections is shown in Fig.~\ref{fig:A-dataCorr-gpipi}.  As for
$\epem\to\mmg$,  the background for the $\pipi\gamma$  process is estimated
with MC, as explained in Sec.~\ref{sec:kineFit_BG}.  The overall efficiency
is obtained with full simulation of $\epem\to\pipig$ events, as a function of
$\cos\phi^*$ at respective $\pi\pi$ masses, and corrected for data/MC differences in
detector response.

\begin{figure}
  \centering 
  \includegraphics[width=0.45\textwidth]{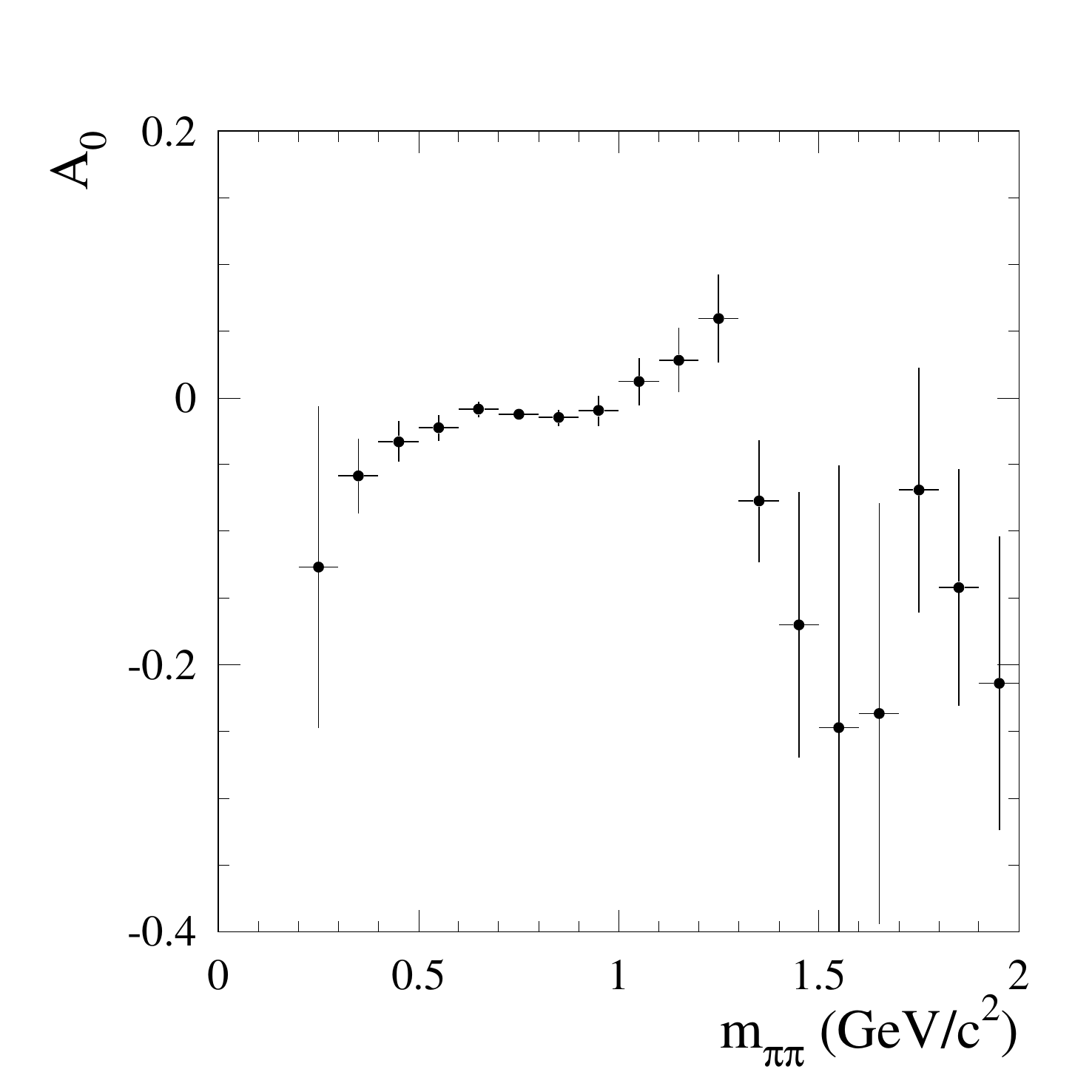}
   \caption{The fitted slope $A_0$ of the charge asymmetry for
  $\epem\to\pipig$ at $\sqrt{s}=10.58\gev$, corrected for efficiency within
  the $20^\circ<\theta^*_\gamma<160^\circ$ acceptance in the $\epem$ c.m.}  
  \label{fig:fit_A0_pi}
\end{figure}

The slopes $A_0$ of charge asymmetry in various $m_{\pi\pi}$  intervals are
obtained by fitting the corrected charge asymmetry distributions to
$A_0\cos\phi^*$. The results for the data are shown in Fig.~\ref{fig:fit_A0_pi} as a
function of $m_{\pi\pi}$. In the $\rho$ resonance region, the measured asymmetry is 
negative, and its magnitude does not exceed $\sim 10^{-2}$. A clear interference pattern is
observed at higher mass.

\subsection{Comparison and fit to models}
\label{sec:piFit}

The magnitude of the charge asymmetry, and its variation with mass,
measured in the $\pipig$ data (Fig.~\ref{fig:fit_A0_pi}) is quite
different from the prediction of the FSR model 1, which treats the
pion as a point-like particle (Sec.~\ref{sec:FSRmd1}). This model is
not considered any further.  Instead the {\it a priori} more realistic quark
FSR model 2 (Sec.~\ref{sec:FSRmd2}), with the modified form of the GDA
formula (Eq.~(\ref{eq:PhiqToFit})), is used to  fit the data. The S-wave
and D-wave magnitudes ($c_{0,2}$) are left free in the fit, while the
mass $m_{f_2}$ and width $\Gamma_{f_2}$ for the $f_2(1270)$ resonance
are fixed to the world averages~\cite{PDG2012}. Because the measured
charge asymmetry loses precision near the $\pi\pi$ production
threshold and above 1.4\gevcc, the fit is performed between 0.3\gevcc
and 1.4\gevcc. The upper limit removes the delicate region around
$1.5\gevcc$ where the pion form factor has a very pronounced dip
leading to a poor knowledge of the ISR  amplitude.

A distinctive interference pattern is observed in Fig.~\ref{fig:fit_A0_pi} at the location of the 
$f_2(1270)$ resonance. In Eq.~(\ref{eq:PhiqToFit}), assuming the dominance of helicity 0 for the $f_2$ 
production, the angular dependence of the interference term in the $\pi\pi$
c.m. is given by the Legendre polynomial $P_2(\cos\theta^*)$, which changes sign
at $|\cos\theta^*|=1/\sqrt{3}$.  As a consequence, the charge
asymmetry is expected to follow the same pattern in the vicinity of the
$f_2(1270)$ resonance.  

\begin{figure}
  \centering 
  \includegraphics[width=0.45\textwidth]{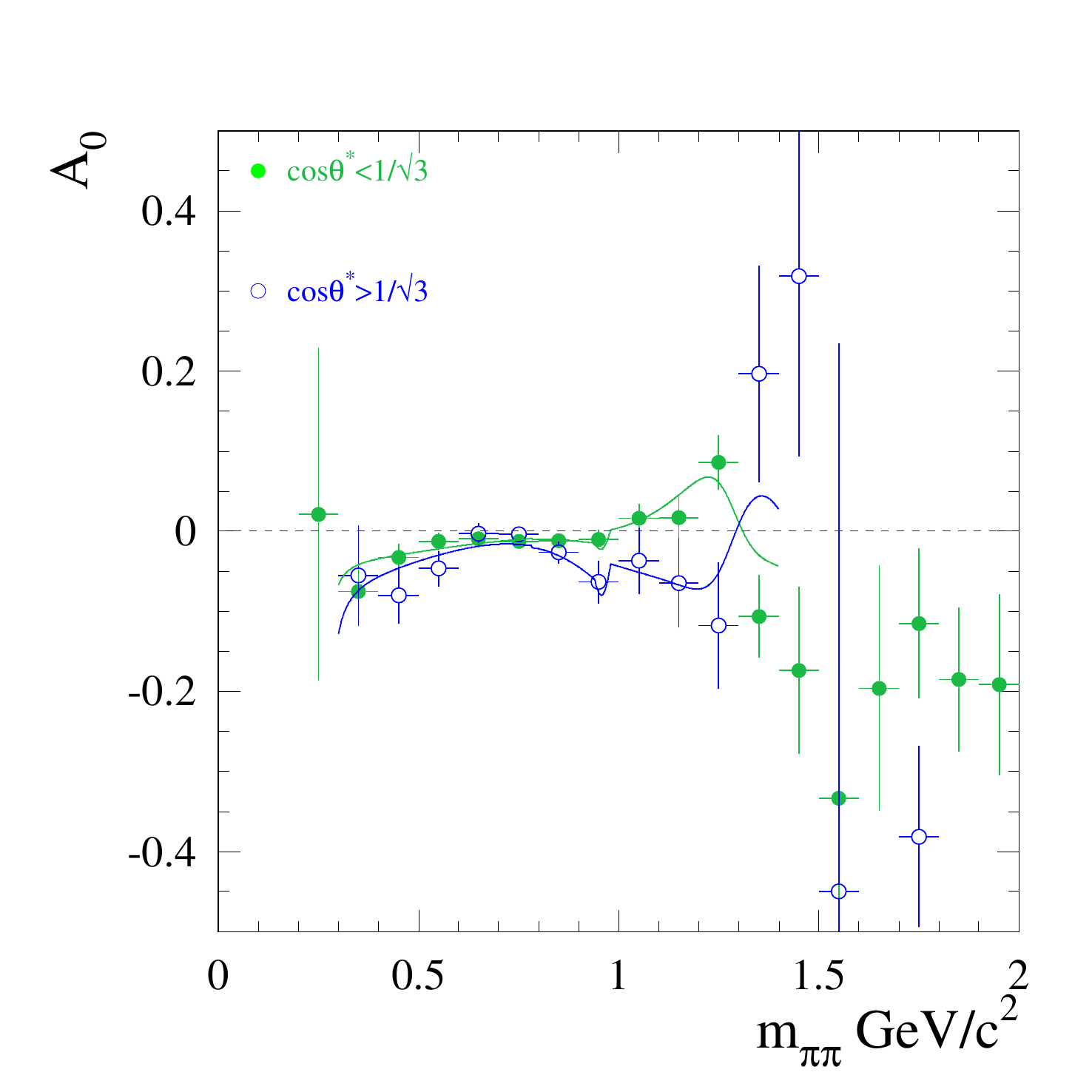}
  \caption{The charge asymmetries measured with $|\cos\theta^*|$ below and
  above $1/\sqrt{3}$. The curves represent the 
  fit results (see text). 
  }
  \label{fig:A0_costhsLtAndGt0.58}
\end{figure}

To check this feature, the charge asymmetries are
measured separately in the phase space below and above
$|\cos\theta^*|=1/\sqrt{3}$.
The data sample is split according to the additional requirement
$|\cos\theta^*|<1/\sqrt{3}$, or $|\cos\theta^*|>1/\sqrt{3}$. To enhance the
efficiency in the high $|\cos\theta^*|$ region, the event selection is loosened,
by removing the $p>1\gevc$ and pion identification requirements on the track with lower
momentum. To keep backgrounds at manageable levels, the higher
momentum pion is required to satisfy the tighter identification criteria of a
`hard $\pi$'~\cite{prd-pipi}. To further reduce the electron contamination, an
enhanced $E_{\rm cal}/p<0.6$ selection is applied to the high momentum track,
and the ionization energy loss in the DCH of the low momentum track
is required to be below the average electron loss ($\dedx_{\rm DCH}<650$). The corresponding efficiencies
are obtained separately from the full simulation of $\epem\to\pipig$ events in
the low and high $|\cos\theta^*|$ regions. While the effective $|\cos\theta^*|$
range is limited to $|\cos\theta^*|<0.8$ with the standard event selection, the
specific selection applied in the high $|\cos\theta^*|$ region allows to extend
the asymmetry measurement up to $|\cos\theta^*|\sim0.95$. Backgrounds are
estimated accordingly using the full simulation of relevant processes. After
background subtraction and overall acceptance correction, the charge asymmetries
obtained for $|\cos\theta^*|$ below and above $1/\sqrt{3}$ are shown in
Fig.~\ref{fig:A0_costhsLtAndGt0.58}.

Since a large fraction of the events in the standard analysis are in the low
$|\cos\theta^*|$ region, the charge asymmetry measured with
$|\cos\theta^*|<1/\sqrt{3}$ is quite close to the one obtained using the full
sample (Fig.~\ref{fig:fit_A0_pi}). Although limited by statistics, the charge
asymmetry measured in the high $|\cos\theta^*|$ region presents the opposite
sign oscillation around the $f_2(1270)$ mass, which is the expected pattern.
The change  of sign between $|\cos\theta^*|<1/\sqrt{3}$ and
$|\cos\theta^*|>1/\sqrt{3}$, and the opposite variation across the resonance
provide a solid validation that the observed charge asymmetry around the
$f_2(1270)$ resonance is indeed due to the interference between the two
amplitudes for $\epem\to\gamma_{\rm ISR}\pipi$ and $\epem\to\gamma_{\rm
FSR}f_2(1270)(\pipi)$, with $f_2$ in the helicity 0  state.  In the mass
range below $1\gevcc$, the asymmetry keeps the same (negative) sign in the two
$|\cos\theta^*|$ regions, as expected for the interference with a scalar amplitude flat in
$|\cos\theta^*|$.

The two independent data samples, with $|\cos\theta^*|$ below and
above $1/\sqrt{3}$, are fitted separately to the model, and the fitted
$c_0$ and $c_2$ amplitudes are obtained in both cases. Since the pure
ISR AfkQed MC used to compute the efficiencies is not expected to
properly correct for unmeasured regions of $|\cos\theta^*|$, the fit
of the data above $1/\sqrt{3}$ is performed in the effective range of
non-null efficiency where asymmetries are measurable.

The model describes the data well, and the two sets of fitted values of $c_0$
and $c_2$ are consistent in sign and magnitude and can be averaged, yielding
$c_0=-1.27\pm0.20$ and $c_2=5.4\pm1.6$.

\subsection{Monte Carlo reweighting and final results}
\label{sec:results-reweightedMC}

\begin{figure*}[htp]
  \centering 
  \includegraphics[width=0.45\textwidth]{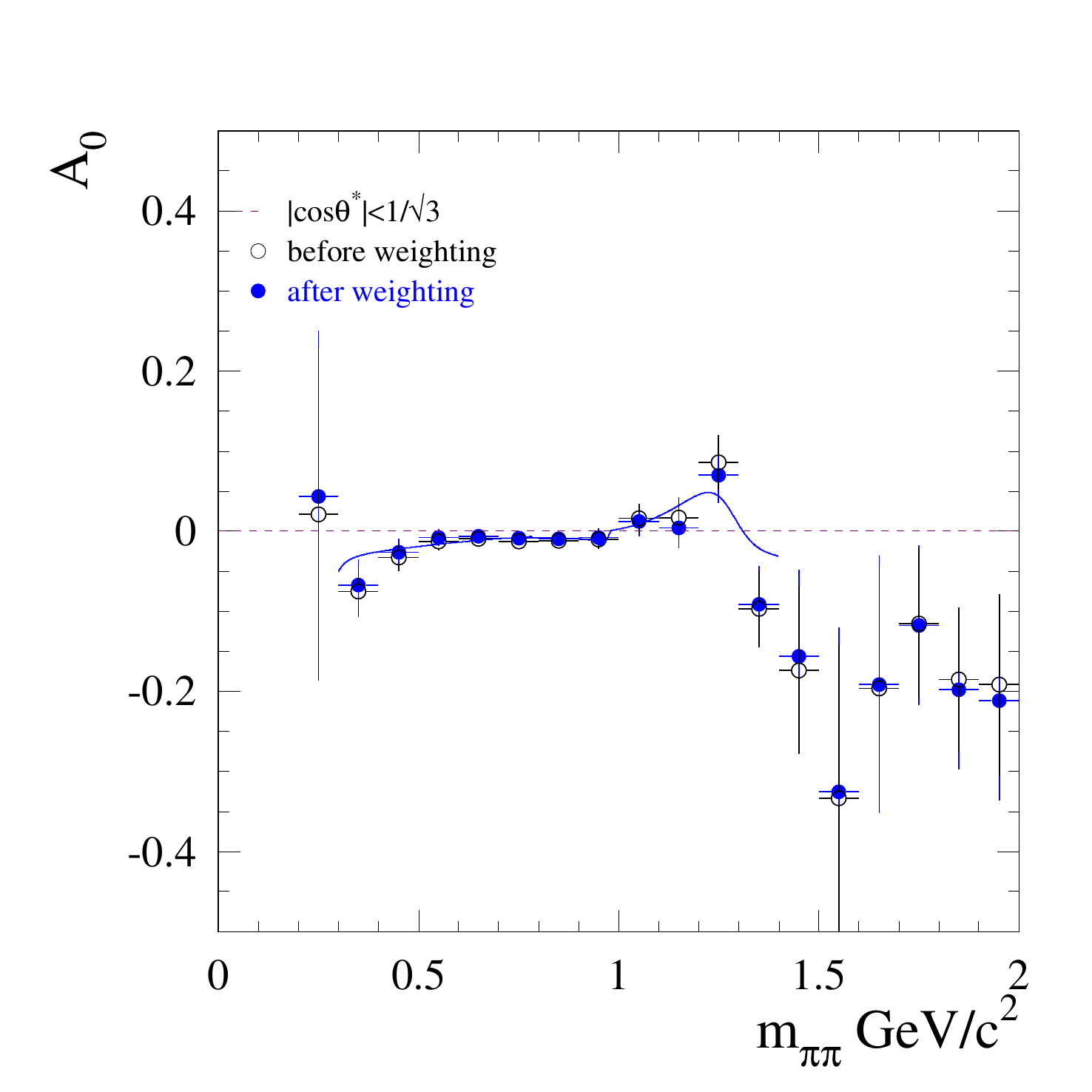}
  \includegraphics[width=0.45\textwidth]{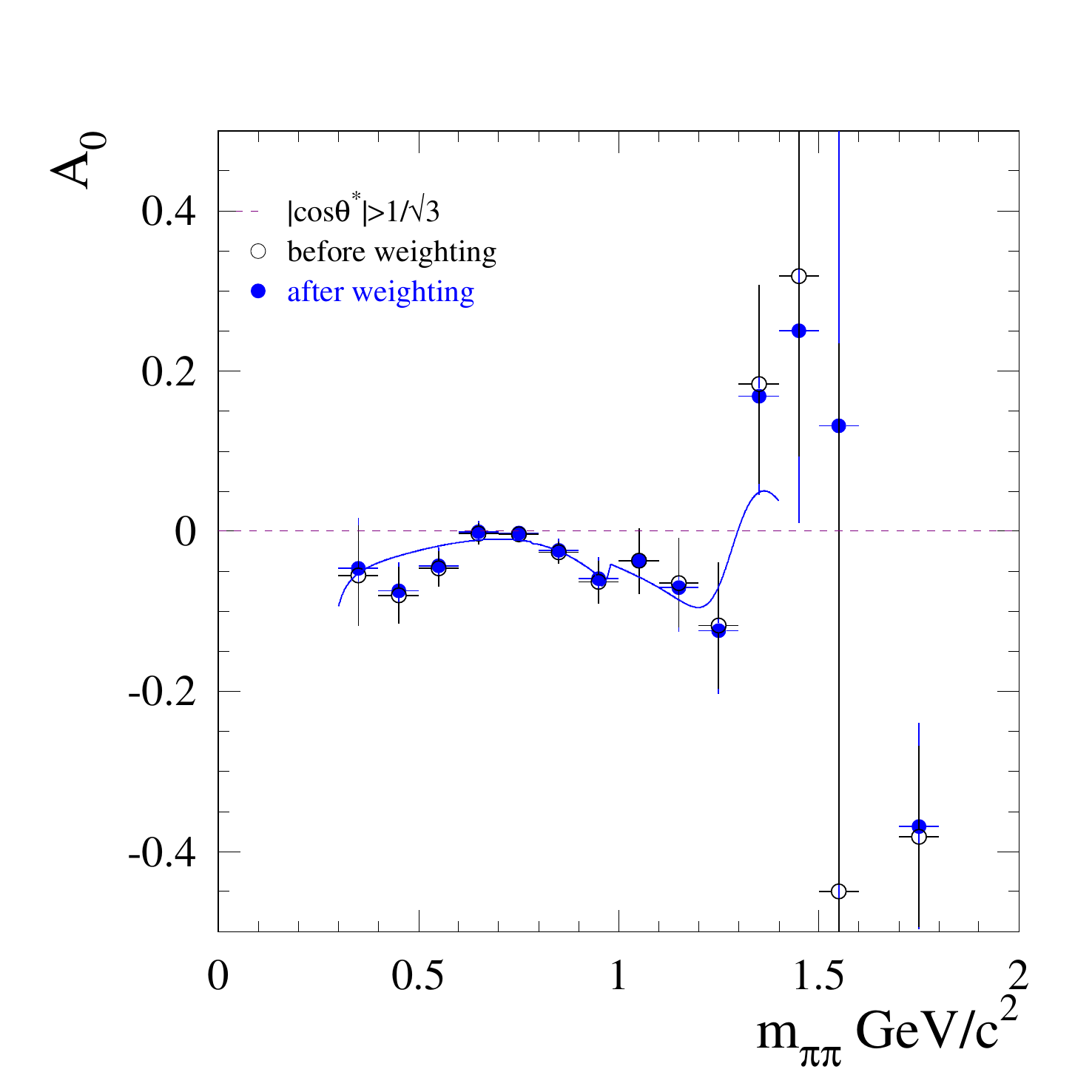}\\
  \includegraphics[width=0.45\textwidth]{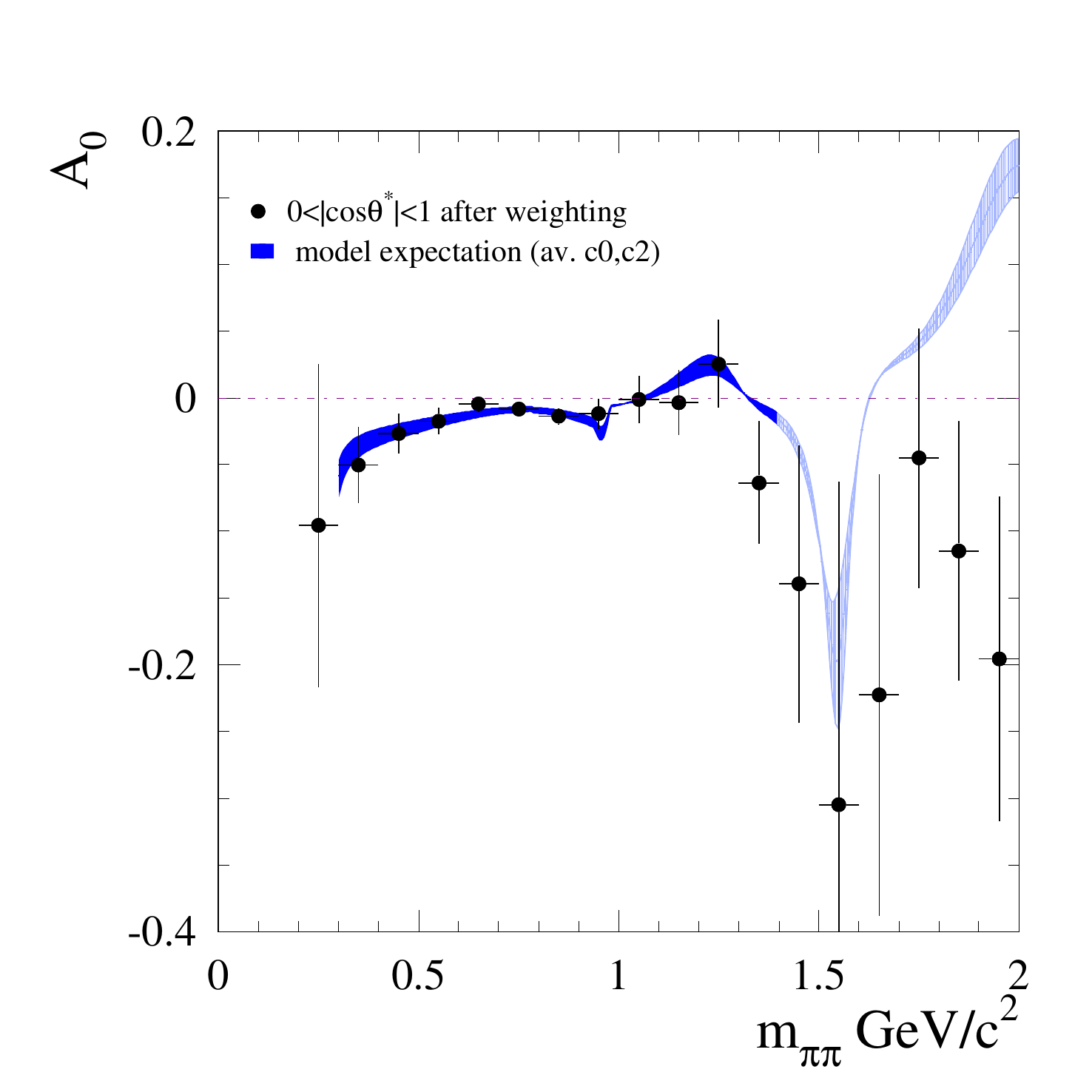}
  \caption{The charge asymmetry slopes for $\epem\to\pipig$ using MC samples 
 with and without reweighting, and fit to the model after reweighting (see text); (top left) for
 $|\cos\theta^*| <1/\sqrt{3}$; (top right) for $|\cos\theta^*| >1/\sqrt{3}$; 
 (bottom) for the full $|\cos\theta^*|$ range; the blue band represents the model-2
 prediction using the average $c_{0,2}$ values after reweighting. 
 The light-blue part corresponds to the extrapolation
 of the model beyond the fitted range. }
\label{fig:A0-pipig-reweighting}
\end{figure*}

Since a significant asymmetry is observed in the data in contrast with AfkQed, which 
does not include LO FSR in the pion channel, the overall efficiencies $\epsilon_\pm$ computed with MC
and used to measure the asymmetry in data are biased through 
the cross effect between the acceptance and the physical asymmetry
 (Sec.\ref{sec:kinAccep}).
This situation calls for an iterative procedure to introduce the
observed interference effect into the MC.

To implement this procedure, new MC samples of reweighted events are produced,
in which the weights are computed event by event as the full cross section value 
including LO FSR divided by the ISR-only cross section, for the values of
$m_{\pi\pi}$, $\cos\theta^*_\gamma$, $\cos\theta^*$ and $\cos\phi^*$ for the 
event. The differential cross sections are given by the model used to fit the data
(Eqs.~(\ref{eq:sigma_pi_ISR}-\ref{eq:sigma_pi_I})). The FSR model is made quantitative by using 
the fitted values for $c_{0,2}$. The studies have been performed separately for the 
regions below and above $1/\sqrt{3}$. 

The fitted values of $c_0$ and $c_2$ are stable after two 
iterations. A third iteration is performed in order to check the stability of
the results. The difference between the last two iterations
is taken as a systematic uncertainty. 

The final $A_0$ values are given in Fig.~\ref{fig:A0-pipig-reweighting}, together with
the FSR model prediction using the fitted $c_{0,2}$ values determined by the iterative
process. The extrapolation of the model beyond the fit
region 0.3-1.4 \gevcc is shown (light-blue band). Although the statistical uncertainty 
of the data is large, there is evidence that the model becomes inadequate above 1.8\gevcc. 
This is not surprising since a constant S-wave amplitude
and the $f_2$ resonance are likely to be insufficient to describe this region,
where many high mass resonances contribute to the $\pipi$ final state.
However, based on the change of asymmetry at the $10^{-3}$ level induced at lower mass, 
the effect of inadequate reweighting in the last few mass bins is expected to be
much smaller than the statistical uncertainty.

Since the two independent sets of parameters agree within their uncertainties,
they can be combined and the weighted average of the fitted values represents
the best information which can be obtained from this interference analysis
(Table~\ref{tab:fit_A0_pi}). An alternative is to fit the overall sample
obtained  with the standard selection, using efficiencies calculated with the
reweighted MC.  In this case, as the data spans over the sign change at
$|\cos\theta^*|=1/\sqrt{3}$, the measured asymmetry is much reduced. Therefore
the combined result from the  two complementary ranges is more sensitive, and
moreover provides a clear confirmation of the helicity 0 $f_2(1270)$ contribution.

\begin{table*}
  \centering  
  \caption{\small The
  parameters obtained from the fit of the charge asymmetry for $\epem\to\pipig$
  at $\sqrt{s}=10.58\gev$ after three iterations, with $20^\circ<\theta^*_\gamma<160^\circ$ in the $\epem$ c.m.,
  in the  $|\cos\theta^*|$ below and above $1/\sqrt{3}$ regions, and the
  weighted average, where the errors are statistical. The results of the direct
  fit over the full range are given in the last column.}   
  \vspace{0.1cm}
  \label{tab:fit_A0_pi}
  \begin{tabular}{|c|c|c|c||c|} \hline\hline\noalign{\vskip2pt} 
  parameter & $|\cos\theta^*|<1/\sqrt{3}$ & $|\cos\theta^*|>1/\sqrt{3}$ & average & all $|\cos\theta^*|$\\
  \hline  
  $c_0$           &  $-0.84\pm0.24$  &  $-1.13\pm0.35$  &  $-0.93\pm0.20$  & $-0.87\pm0.20$\\   
  $c_2$           &  $ 3.82\pm1.81$  &  $ 6.33\pm3.03$  &  $ 4.48\pm1.56$  & $ 3.41\pm4.25$\\
  \hline\hline   
  \end{tabular}   
\end{table*}

\subsection{Systematic uncertainties}
\label{sec:syst_pi}


The difference $\Delta A_0$ between the results with and without background
subtraction is found to be less than $10^{-3}$, except near threshold (1$\times 10^{-2}$) and
above $1.1\gevcc$ (1$\times 10^{-2}$ at $1.25\gevcc$). Except in the dip region at
$1.55\gevcc$ where the ISR cross section has a sharp minimum and statistical
uncertainties are very large, the background level has been
checked~\cite{prd-pipi} with a precision of 20\% in the worst cases. The
systematic uncertainty due to background subtraction is consequently estimated
to be less than 2$\times 10^{-4}$ from 0.4 to $1.1\gevcc$, increasing above
(2$\times 10^{-3}$ at $1.25\gevcc$).


As observed for $\epem\to\mmg$, the selection requirements for
trigger, tracking, and $\pi$-ID have charge-asymmetric efficiencies
for the $\epem\to\pipig$ process. The corrections for the difference
between data and MC on the efficiencies are included in the overall
acceptance.  The difference between the charge asymmetry results with
and without the data-MC corrections is smaller than $10^{-3}$ except
in the dip region (0.5$\times 10^{-2}$).  Since the corrections are determined with
data with a precision of 10\%, the resulting systematic uncertainty is
negligible. 

\begin{figure*}
  \centering 
  \includegraphics[width=0.45\textwidth]{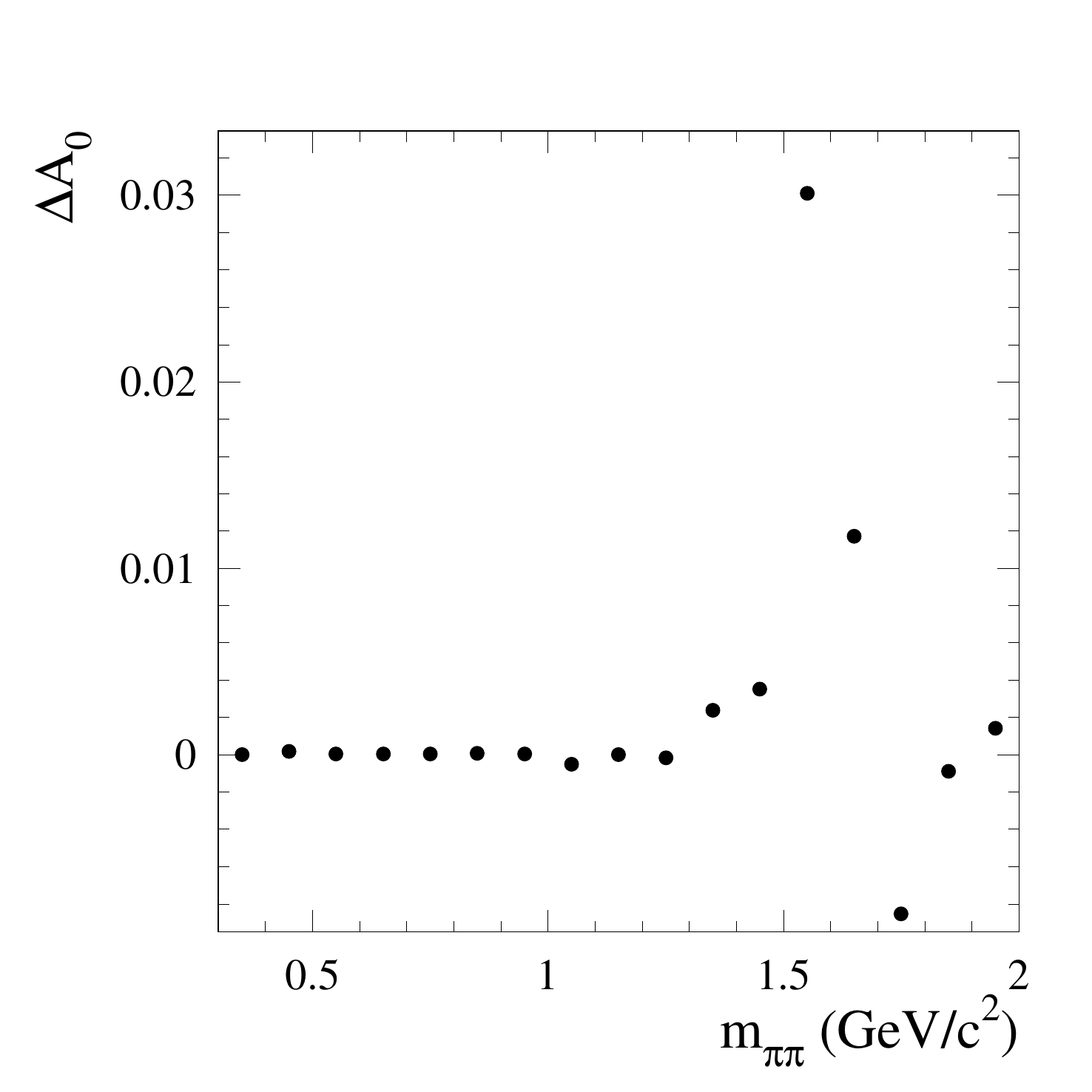}
  \includegraphics[width=0.45\textwidth]{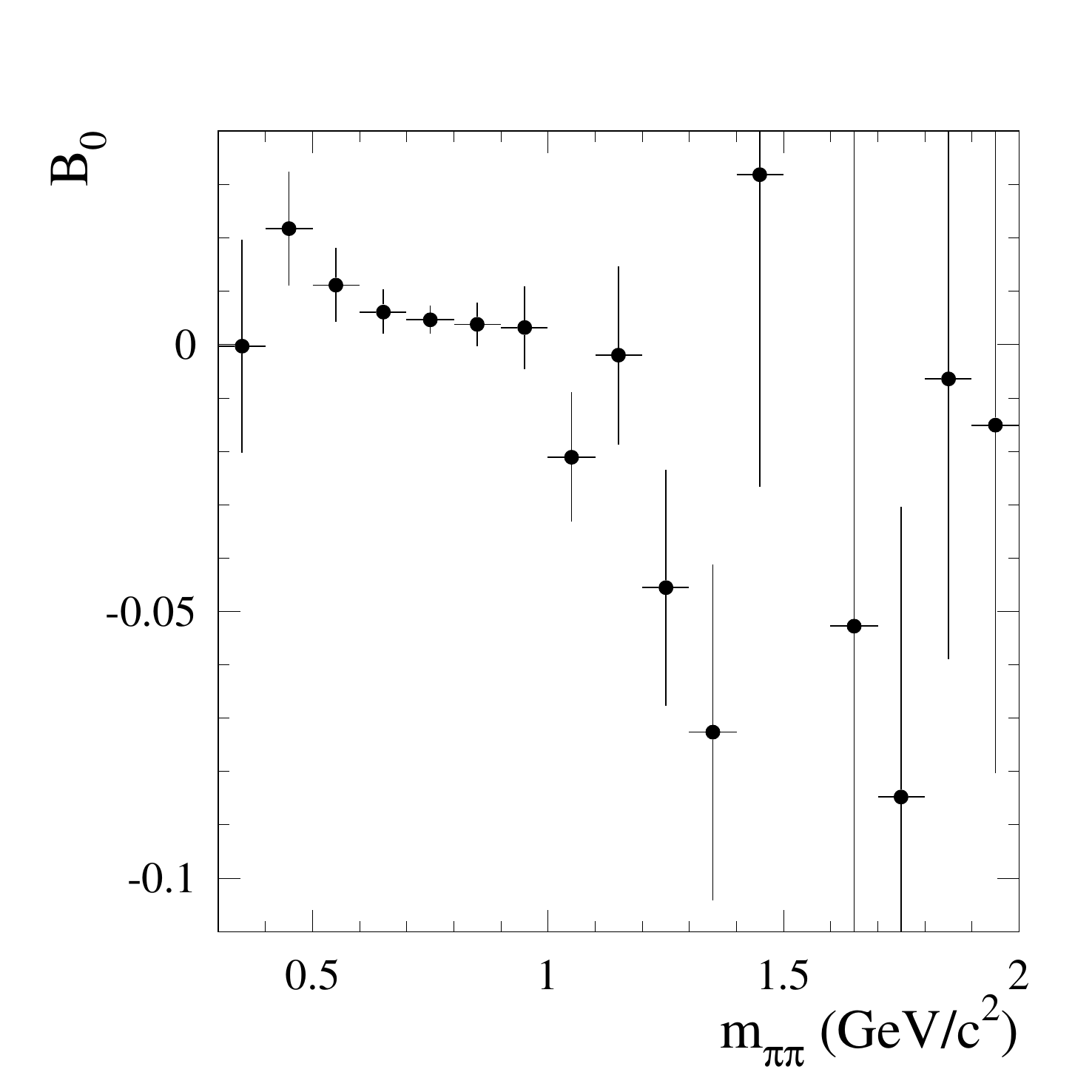} 
  \caption{The change of slope $\Delta A_0$ (left) and constant term $B_0$
  (right) in the fit of charge asymmetry to  $A_0\cos\phi^*+B_0$, as a
  function of $\pi\pi$ mass for $\epem\to\pipig$ .}
  \label{fig:A0B0_pi}
\end{figure*}

As done for $\mmg$, the asymmetry $A(\cos\phi^*)$ in $\pipig$ is
alternatively fitted to  $A_0\cos\phi^*+B_0$ since a bias on $A_0$ and $B_0$ values
inconsistent with zero might disclose an incorrect efficiency
determination, or an incorrect background subtraction. 
As shown in Fig.~\ref{fig:A0B0_pi},
the fitted slopes $A_0$ deviate from the final values by less than $10^{-3}$, 
except in the background dominated dip region of the cross section 
($m_{\pi\pi} \simeq 1.5\gevcc$), where the deviation 
$\Delta A_0$ reaches 3$\times 10^{-2}$. In the $m_{\pi\pi}$ region where the fit of the
theoretical model is performed (0.3--1.4\gevcc), 
the slope $A_0$ does not change by more than 5$\times 10^{-4}$, except in 
the last (1.3--1.4\gevcc) bin, where the deviation is 2$\times 10^{-3}$. 
This is consistent with the estimated background contribution to the systematic error.
The fitted $B_0$ values are within $2.5\sigma$ from zero over the full mass range. 
The average $B_0$ for $m_{\pi\pi}<1\gevcc$ is $(0.41 \pm 0.16)\times 10^{-2}$.

Different interaction rates in the detector material for positive and negative pions 
induce a charge asymmetry. Although such an effect is included in the simulation 
of the detector response
based on {\small GEANT4}, its description and the corresponding track loss are known to be 
somewhat imperfect. Independent studies have shown that data/MC 
discrepancies occur at the 10\% level for both $\pi^+$ and $\pi^-$, in opposite 
directions. 
A residual charge asymmetry is thus expected after applying the MC corrections.
The effect of imperfect simulation of nuclear interactions is investigated 
using the large sample of $\pi^+\pi^-\gamma$ events produced by AfkQed 
at the generator level. A weight is assigned to each track according to its momentum 
and its path length through detector material as a function of the polar angle, using a $\pm10\%$
relative change in the respective $\pi^+$ and $\pi^-$ interaction rates. The charge 
asymmetry obtained after the interaction reweighting is subjected to the two-parameter 
linear fit in the integrated mass range from 0.4 to 1.2 \gevcc. 
The slope $A_0$ changes by only $\Delta A_0=(-0.006\pm0.024)\times 10^{-2}$, which confirms the 
robustness of the $A_0$ observable.
The charge asymmetry itself is however modified as the fitted $B_0$ value is found to be 
displaced significantly, $\Delta B_0=(0.240 \pm 0.016)\times 10^{-2}$, 
in good agreement with the observed $B_0$ value in data in the same mass range.
Imperfect simulation of nuclear interactions thus provides a plausible explanation of the small 
$B_0$ values found in the analysis, while leaving the $A_0$ measurement unaffected.

Summing up all sources, including the estimated cross-effect between 
acceptance and physical asymmetry, the absolute systematic uncertainty on $A_0$ is estimated to 
be less than 0.17\% in the 
$f_2(1270)$ region and less than 0.1\% elsewhere. 

\subsection{\boldmath Searching for an $f_2(1270)$ signal in the $\pi^+\pi^-\gamma$ cross section}

\begin{figure}
  \centering 
  \includegraphics[width=0.45\textwidth]{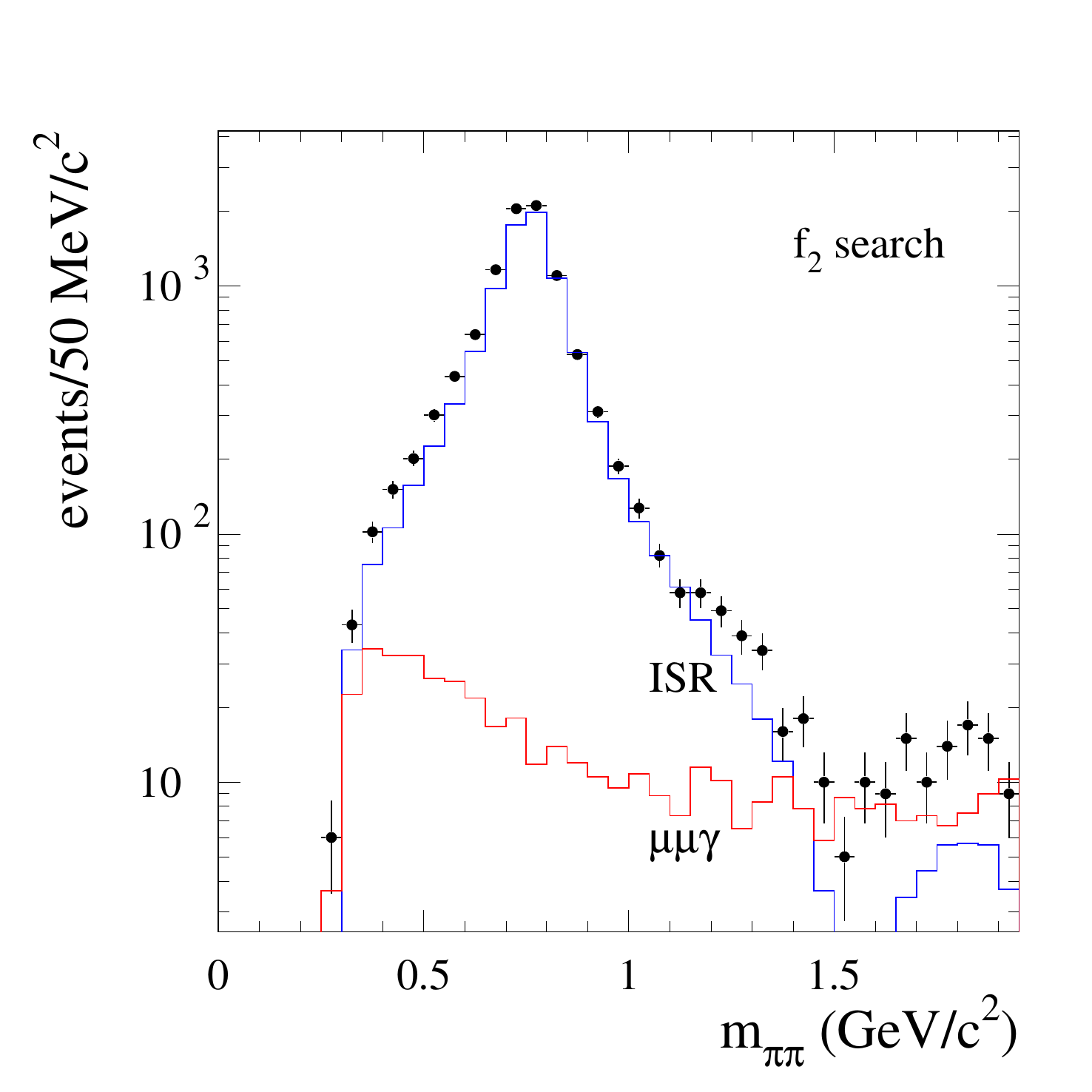} 
  \caption{The
  $\pi\pi$ mass spectrum for $|\cos\theta^*|>0.85$ for the analysis extended to
  low momentum (points), the expected background from misidentified
  $\mu\mu\gamma$ events (red histogram), and the predicted ISR spectrum from the
  standard cross section analysis (blue histogram).}  
\label{fig:f2-search-raw}
\end{figure}
\begin{figure}
  \centering 
  \includegraphics[width=0.45\textwidth]{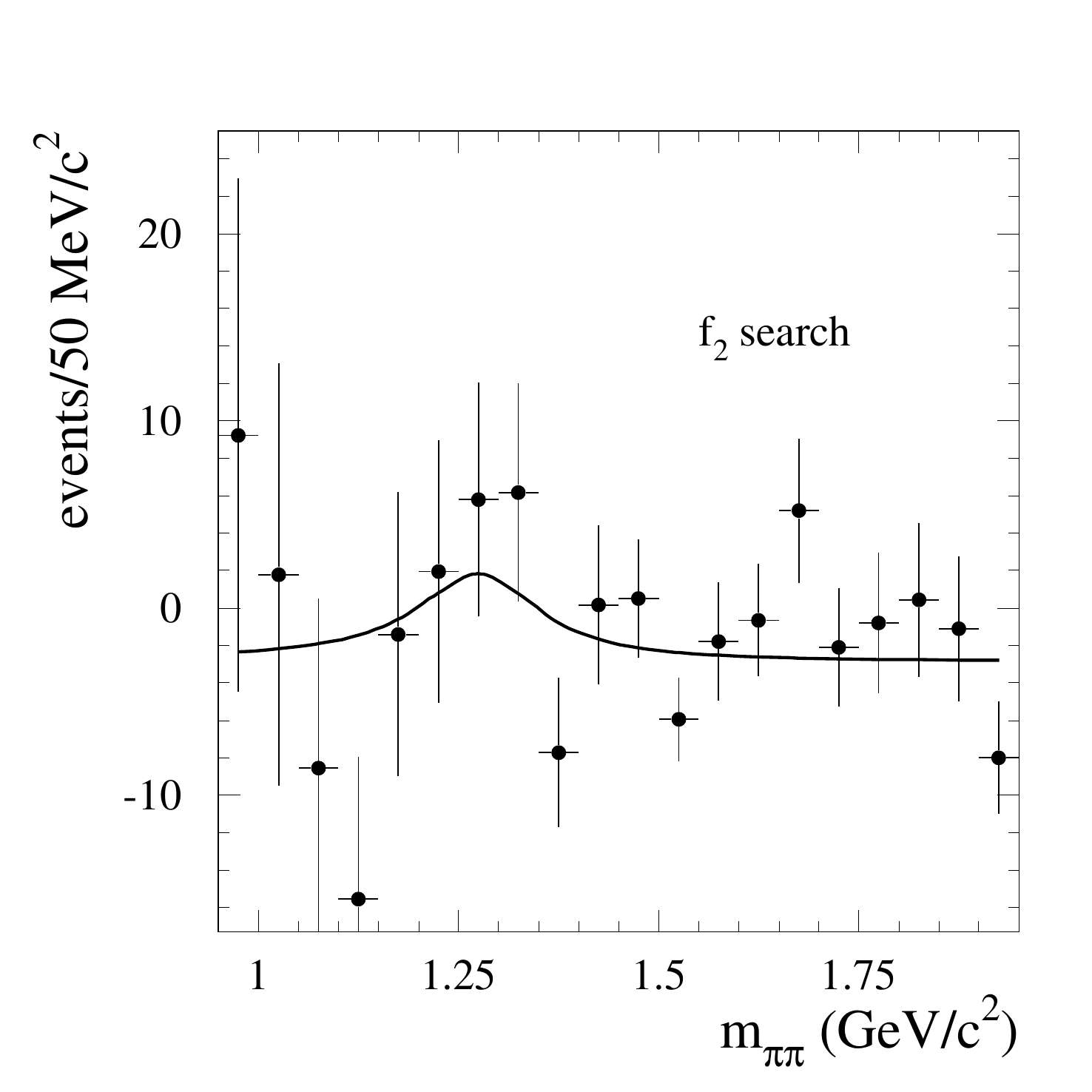} 
  \caption{The $\pi\pi$
  mass spectrum for $|\cos\theta^*|>0.85$ for the analysis extended to low
  momentum (points) after subtraction of the $\mu\mu\gamma$ background and the
  ISR contribution. The curve is the result of a fit to a constant term and a simple
  Breit-Wigner shape for the $f_2$ resonance.}  
\label{fig:f2-search-sub}
\end{figure}

Given the sizeable amplitude $c_2$ of the D-wave contribution to the ISR-FSR
interference obtained from the charge asymmetry measurement, direct evidence of
$f_2(1270)$ production is searched for in the cross section measurement. While the
latter is overwhelmingly dominated by the ISR production of the $\rho$
resonance, the rapid fall off of the pion form factor in the vicinity of the
$f_2(1270)$, and the distinct angular distribution $P_2(\cos\theta^*)$
of the D-wave in the $\pi\pi$ system, are assets used in the direct search.

Since the $P_2(\cos\theta^*)$ distribution exhibits a peak at
$|\cos\theta^*|$ near unity, in contrast with the $\sin^2\theta^*$ dependence of
the ISR cross section, the search is performed in the very high range
$|\cos\theta^*|>0.85$. Because the standard event selection depopulates that region completely, 
due to the momenta of both tracks being required to be larger than $1\gevc$, the direct search uses
the specific selection designed for the charge asymmetry measurement in the
high $|\cos\theta^*|$ region (Sec.~\ref{sec:piFit}), with an even tighter $\dedx_{\rm DCH}<550$ requirement.

The $\pi\pi$ mass spectrum of the reconstructed events in this specific analysis
is displayed in Fig.~\ref{fig:f2-search-raw} with the largest expected
background from $\mu\mu\gamma$ events to be subtracted. The resulting
spectrum is dominated by the ISR production, which is also subtracted.  The
remaining spectrum shown in Fig.~\ref{fig:f2-search-sub} does not present any
significant  excess at the $f_2(1270)$ mass or elsewhere, except for a slow rise above
$2\gevcc$ (not shown) that originates from a residual $ee\gamma$ background. The mass spectrum
is fitted between 0.95 and $1.95\gevcc$  to a constant and a Breit-Wigner lineshape
with the world average $f_2(1270)$ mass and width~\cite{PDG2012} and a free-floating
amplitude. The fitted number of  $f_2$ events in the mass interval at the peak
is found to be $4.7\pm4.2$, to be compared to $26.7\pm1.1$ ISR events in the
same interval.  After correction for the loss of efficiency near
$|\cos\theta^*|=1$ obtained from MC for ISR and $f_2$ candidates, the $f_2(1270)$
fraction $|f_2|^2/(|{\rm ISR}|^2+|f_2|^2)$ in the $f_2$-enhanced range
$0.8<|\cos\theta^*|<1$ is measured  to be $0.22\pm0.15$. This corresponds to a
$|c_2|$ value equal to $4.6\pm2.2$.

\begin{figure}
  \centering 
  \includegraphics[width=0.45\textwidth]{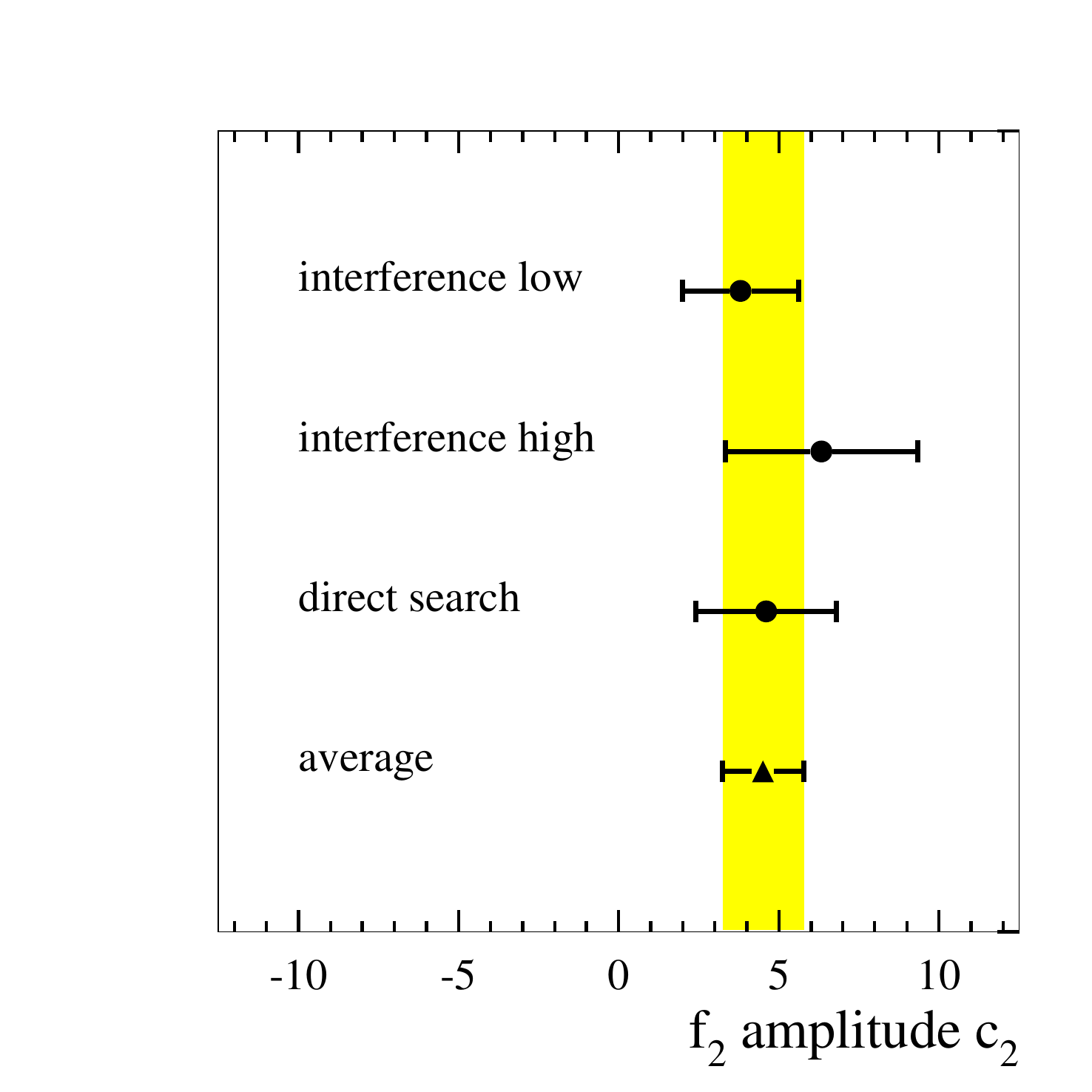} 
  \caption{The results obtained for the $f_2(1270)$ amplitude $c_2$ in the process
$\epem \to \pipig$ from the interference analysis and the
direct $f_2(1270)$ search in the cross section. The labels `low' and `high' refer
to the determination in the two angular ranges $|\cos\theta^*|<1/\sqrt{3}$ and
$1/\sqrt{3}<|\cos\theta^*|$. For the direct search the positive solution is
chosen. The combined value for the three independent analyses is given by the
vertical band.}  
\label{fig:c2-res}
\end{figure}

The three independent determinations of $|c_2|$ (the interference fits in two
$\cos\theta^*$ regions and the direct $f_2(1270)$ search in the cross section) yield
consistent results. Since a positive sign is clearly indicated by the 
interference analysis, the value from the direct search is also taken to be
positive. The three independent values can be combined with the result
$c_2 = 4.5 \pm1.3$, establishing LO FSR production of the $f_2(1270)$ resonance at
the $3.6\sigma$ level.  The corresponding production cross section is
$(37^{+ 24}_{ -18}) \fb$. The results are displayed in Fig.~\ref{fig:c2-res}.
The size of $c_2$ is about a factor of two larger
than the value predicted by Chernyak~\cite{chernyak} with a QCD model giving 
$|c_2^{\rm th}|=2.2$. However the difference only amounts to $1.8\sigma$, not 
including the unknown theoretical uncertainty. The sign is not provided in 
Chernyak's prediction.  

\subsection{ \boldmath Consequences for the cross section measurement 
by \babar\ for $\epem\to\pipi$ and contribution to the anomalous magnetic moment 
of the muon}

In the measurement of the $e^+e^-\to \pi^+\pi^-$ cross section by the
\babar\ collaboration~\cite{prd-pipi} using the ISR method, the lowest-order 
FSR contribution was argued to be negligible, based on theoretical estimates. 
The primary result of the present
interference analysis is to determine the actual size of the  $|\mathcal{M}_{\rm
FSR}|^2$ cross section, misinterpreted as ISR, and its contribution to the total
cross section  ($|\mathcal{M}_{\rm ISR}|^2+|\mathcal{M}_{\rm FSR}|^2$).

\begin{figure}
  \centering 
  \includegraphics[width=0.45\textwidth]{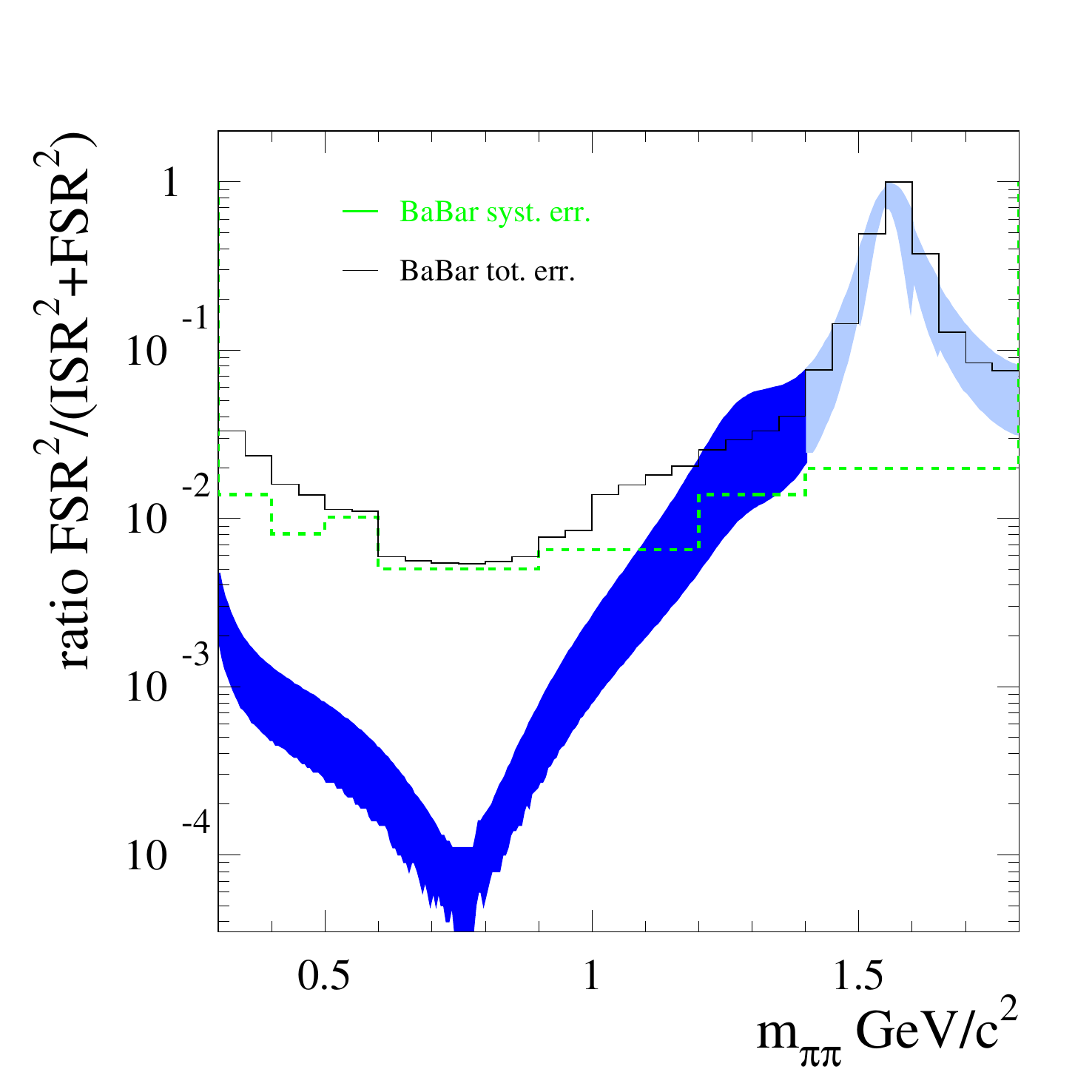} 
  \caption{The FSR fraction in the \babar\ measurement~\protect\cite{prd-pipi} 
  of the $\pi\pi\gamma$ 
  cross section, defined as the ratio 
$\frac {|\mathcal{M}_{\rm FSR}|^2}{|\mathcal{M}_{\rm ISR}|^2+|\mathcal{M}_{\rm FSR}|^2}$ 
obtained in this analysis using FSR model 2 with $c_{0,2}$
parameters fitted to data (blue band). The light-blue part corresponds to the extrapolation
of the model beyond the fitted range. The FSR fraction is compared to the
systematic error of the \babar\ cross section measurement (green dashed histogram) 
and its total error (black histogram). }  
\label{fig:FSR-fraction}
\end{figure}

Using the FSR model 2, which describes well the measured charge asymmetry in the
[0.3--1.4]\gevcc range, the FSR cross section calculated through
Eq.~(\ref{eq:sigma_pi_FSR}) with the fitted $c_{0,2}$ parameters, is extrapolated
to higher masses. The resulting
FSR fraction in the \babar\ cross section is given in
Fig.~\ref{fig:FSR-fraction} as a function of $m_{\pi\pi}$. As expected the FSR fraction
is negligible in the $\rho$ region, but increases significantly above  $1\gevcc$
due to the $f_2(1270)$ contribution and the rapid fall-off of the pion  form
factor. In fact the FSR `background' exceeds the estimated systematic
uncertainty quoted in Ref.~\cite{prd-pipi} (green histogram in 
Fig.~\ref{fig:FSR-fraction}) for mass above
$1.2\gevcc$, while remaining close to the total uncertainty (black histogram).
The FSR contribution is found to be dominant around $1.5\gevcc$, in the region
where the ISR cross section displays a deep dip and is consistent with zero
within the large errors.

The contribution to the muon magnetic anomaly, $a_\mu=(g_\mu-2)/2$, from hadronic
vacuum polarization involves a dispersion integral over the cross section
$e^+e^-\to {\rm hadrons}$ weighted by a known kernel (Ref.~\cite{dehz}
and references therein). The integral is dominated by the $\pi^+\pi^-$ channel
and its most precise determination to date is from \babar\ using the ISR
method~\cite{prd-pipi} with the value 
$a_\mu^{\pi\pi(\gamma),\rm LO}=(514.09\pm 2.22_{\rm stat}\pm 3.11_{\rm syst})\times 10^{-10}$ 
when integrating from threshold to 1.8\gev.
This value is derived under the assumption that the cross section
for $\epem\to\pipig (\gamma)$ has a negligible contribution from LO FSR.

The present measurement of the charge asymmetry allows one to validate this
assumption in a quantitative way. Using the FSR fraction shown in
Fig.~\ref{fig:FSR-fraction}, the contribution to $a_\mu$ from the LO
FSR falsely attributed to the ISR cross section is found to be, in the same
energy range up to 1.8\gev
\beqn
\Delta a_\mu^{\pi\pi}({\rm FSR}) = (0.26\pm 0.12)\times 10^{-10}.
\label{amu-FSR}
\eeqn
This reduces the value of $a_\mu^{\pi\pi}$ by $(5.1\pm 2.3)\times 10^{-4}$ 
relative to the \babar\ determination. The correction is small 
compared to the total \babar\ relative
uncertainty of $7.4\times 10^{-3}$, which justifies its earlier neglect.

\section{Conclusions}

The radiative process $\epem\to X \gamma$, where $X=\mumu$ and $\pipi$ are considered in this
analysis, involves contributions from both LO ISR
and FSR. Because charge parities  of the final state pair are
opposite for ISR and FSR, the interference between ISR and FSR changes sign with
the charge interchange of the two muons (pions). As a consequence, 
investigation of the charge asymmetry of the process gives a way to study the
interference between ISR and FSR, which is sensitive to the relative
contribution of LO FSR.

From QED for $\mmg$, and from FSR models for $\pipig$, we find that the charge asymmetry
$A$ has a strong dependence on the angle $\phi^*$ between the $\mu^-\mu^-$
($\pipi$) plane and the $\epem\gamma$ plane in the $\epem$ c.m. system,
which can be simply represented by a linear function $A=A_0\cos\phi^*$. The
slope $A_0$ quantifies the magnitude of the interference between ISR and FSR.

The acceptance effects on the measured charge asymmetry are studied
with the full simulation of $\epem\to\mmg$ and $\epem\to\pipig$
events. We find that the detector and event selection, including
trigger, tracking, PID, and kinematic fitting, induce nonlinear patterns
on the $\cos\phi^*$ dependence of the charge asymmetry, but have a
small impact on the determined slope $A_0$. Kinematic acceptance --- namely the
angular acceptance, and energy or momentum requirements on the final state
particles --- changes the slope of the observed charge asymmetry
significantly, although the kinematic requirements are
charge-symmetric.   This is due to a cross effect between acceptance
and true interference that produces a bias in the measured asymmetry
if the physical asymmetry differs between data and MC. This bias is
corrected through an iterative procedure in the $\pipig$ analysis, as 
in that case the charge asymmetry is null in the generator.

After background subtraction and correction for the overall
acceptance, which are obtained from the full simulation with
corrections for data/MC differences, we measured the slope $A_0$ of
the charge asymmetry as a function of $m_{\mu\mu}$ ($m_{\pi\pi}$). The
QED test, namely the comparison between the charge asymmetry measured
in the $\mmg$ data and predicted by the simulation, in which the LO
ISR-FSR interference is implemented,
shows an overall
good consistency.   However, some absolute deviation amounting to
$\Delta A_0  = A_0^{\rm data}-A_0^{\rm MC} \simeq 0.03$ in the
$3\gevcc$ region is observed and cannot be fully explained by known
systematic effects, either in the data or in the MC generators, 
which are estimated to be less than 0.014.  

The measured slope $A_0$ of charge asymmetry in the $\epem\to\pipig$ data is
about $-1\%$ and flat around the $\rho$ mass.  Outside of the $\rho$ peak, the data
exhibits the pattern expected from the interference between
$\epem\to\gamma_{\rm ISR}\pipi$ and $\epem\to\gamma_{\rm FSR}f_2(1270)(\pipi)$. The data
shows a good consistency with the predictions of a model of FSR from quarks with
contributions of a scalar widespread mass distribution and the
$f_2(1270)$ tensor resonance. In the $\rho$ region the results are not
consistent with a model based on FSR from point-like pions
(scalar QED), in contrast with the observations at low energies [4].

These results are first measurements of the charge asymmetry in the 
$\epem\to\mmg$ process, and for $\epem\to\pipig$ at high energy ($\sqrt{s}\sim10.58\gev$).
The FSR contribution to $\epem\to\pipig$ derived from this analysis is small and 
this confirms that it is
negligible in the measurement of the cross section obtained by \babar\ assuming pure
ISR~\cite{prd-pipi}. Accordingly this FSR bias translates into a correction to
the muon magnetic anomaly of only $(0.51\pm 0.23)$ per mille of the  $\pi\pi$
hadronic vacuum polarization determined from \babar\ data, small compared to the total quoted
uncertainty of 7.4 per mille.

\vsp
%
We gratefully acknowledge useful discussions on theoretical 
issues with H.\ Czy\.z, Zhun Lu and B.\ Pire, and clarifications by  A.B.\ Arbuzov, S.J.\ Brodsky, R.\ Gastmans, and R.\ Kleiss.
We are grateful for the 
extraordinary contributions of our \pep2\ colleagues in
achieving the excellent luminosity and machine conditions
that have made this work possible.
The success of this project also relies critically on the 
expertise and dedication of the computing organizations that 
support \babar.
The collaborating institutions wish to thank 
SLAC for its support and the kind hospitality extended to them. 
This work is supported by the
US Department of Energy
and National Science Foundation, the
Natural Sciences and Engineering Research Council (Canada),
the Commissariat \`a l'Energie Atomique and
Institut National de Physique Nucl\'eaire et de Physique des Particules
(France), the
Bundesministerium f\"ur Bildung und Forschung and
Deutsche Forschungsgemeinschaft
(Germany), the
Istituto Nazionale di Fisica Nucleare (Italy),
the Foundation for Fundamental Research on Matter (The Netherlands),
the Research Council of Norway, the
Ministry of Education and Science of the Russian Federation, 
Ministerio de Econom\'{\i}a y Competitividad (Spain), the
Science and Technology Facilities Council (United Kingdom),
and the Binational Science Foundation (U.S.-Israel).
Individuals have received support from 
the Marie-Curie IEF program (European Union) and the A. P. Sloan Foundation (USA). 


\end{document}